\documentclass[11pt]{article}
\usepackage[utf8]{inputenc}
\usepackage{xcolor}
\usepackage{hyperref}
\usepackage{amsmath}
\usepackage{amsfonts}
\usepackage{amssymb}
\usepackage{stmaryrd} 
\usepackage{physics}
\usepackage{graphicx}
\usepackage[margin=1in]{geometry}
\usepackage{adjustbox}
\usepackage{appendix}
\usepackage{cite}
\usepackage{braket}
\usepackage{bm}
\allowdisplaybreaks 

\graphicspath{ {./diagrams/} }
\newcommand\qbin[2]{\left[\begin{matrix} #1 \\ #2 \end{matrix} \right]_{q}}
\newcommand\qbini[2]{\left[\begin{matrix} #1 \\ #2 \end{matrix} \right]_{q^{d_i}}}

\newcommand{\rootone}{\bm{\alpha}_1}
\newcommand{\roottwo}{\bm{\alpha}_2}
\newcommand{\weightone}{\bm{w}_1}
\newcommand{\weighttwo}{\bm{w}_2}
\begin{document}

\thispagestyle{empty}

\begin{center}

\Large{Integrability of rank-two web models}

\vskip 1cm

\large{Augustin Lafay$^{1,2}$, Azat M.\ Gainutdinov$^{3}$, and Jesper Lykke Jacobsen$^{2,4,5}$}

\vspace{1.0cm}

{\sl\small $^1$ Department of Mathematics and Systems Analysis, Aalto University, Finland}

{\sl\small $^2$ Laboratoire de Physique de l'\'Ecole Normale Sup\'erieure, ENS, Universit\'e PSL, \\
CNRS, Sorbonne Universit\'e, Universit\'e de Paris, F-75005 Paris, France\\}

{\sl\small $^3$
Institut Denis Poisson, CNRS, Universit\'e de Tours, \\Parc de Grandmont, F-37200 Tours, France\\}

{\sl\small $^4$ Sorbonne Universit\'e, \'Ecole Normale Sup\'erieure, CNRS, \\
Laboratoire de Physique (LPENS), F-75005 Paris, France\\}

{\sl\small $^5$ Universit\'e Paris Saclay, CNRS, CEA, Institut de Physique Th\'eorique, \\ F-91191 Gif-sur-Yvette, France\\}
 
\end{center}

\begin{abstract}

We continue our work on lattice models of webs, which generalise the well-known loop models to allow for various kinds of bifurcations\cite{Lafay:2021scv,Lafay:2021wyf}.
Here we define new web models corresponding to each of the rank-two spiders considered by Kuperberg \cite{Kuperberg_1996}.
These models are based on the $A_2$, $G_2$ and $B_2$ Lie algebras,
and their local vertex configurations are intertwiners of the corresponding $q$-deformed quantum algebras.
In all three cases we define a corresponding model on the hexagonal lattice, and in the case of $B_2$ also on the square lattice.
For specific root-of-unity choices of $q$, we show the equivalence to a number of three- and four-state spin models on the dual lattice.

The main result of this paper is to exhibit integrable manifolds in the parameter spaces of each web model.
For $q$ on the unit circle, these models are critical and we characterise the corresponding conformal field theories via numerical diagonalisation
of the transfer matrix.

In the $A_2$ case we find two integrable regimes. The first one contains a dense and a dilute phase, for which we have
analytic control via a Coulomb gas construction, while the second one is more elusive and likely conceals non-compact physics.
Three particular points correspond to a three-state spin model with plaquette interactions, of which the one in the second regime appears to present a new universality class.
In the $G_2$ case we identify four regimes numerically.
The $B_2$ case is too unwieldy to be studied numerically in the general case, but it found analytically to contain a simpler sub-model based on generators
of the dilute Birman-Murakami-Wenzl algebra.

\end{abstract}
\newpage
\tableofcontents

\section{Introduction}

One of the most striking applications of conformal invariance in two dimension is the study of random geometrical objects.
The specific case of random curves has been the object of much interest in the theoretical physics literature since the 1980s, where
the so-called loop models have been investigated both in their lattice discretisation, notably using quantum integrability and lattice algebras
of the Temperley-Lieb type, and in the continuum limit, via Conformal Field Theories (CFT); see \cite{nienhuis_critical_1984,Nienhuis_review_1987,CGreview} for reviews.
Interestingly, these CFT are in general
not only non-unitary, but also logarithmic \cite{LCFTreview}, and in certain cases of physical interest they may even possess a non-compact target space \cite{Vernier14,vernier_new_2015}.
In parallel with these developments, random curves have been extensively studied in probability theory since the 2000s, where
they have spearheaded the development of frameworks such as Schramm-Loewner Evolution (SLE) \cite{lawler2007schrammloewner,SLEreview} and the Conformal Loop Ensemble (CLE) \cite{CLEreview}.
More recent work on loop models aims at the rigorous construction of the field theory\cite{RV14,KRV20,ang2021integrability} and the computation of correlation functions \cite{Delfino_2010,Picco_2013,JS19,HJS20,GJNRS23}.

Notwithstanding all these interesting developments, it is known that many applications of random geometry need to go beyond the concept
of curves, and focus instead on more general random graphs that allow for branchings and bifurcations. In two recent papers \cite{Lafay:2021scv,Lafay:2021wyf}
we have initiated a study of so-called web models, which are generalisations of lattice models of loops, in which algebraic spiders play
a prominent role. These spiders have their root in the study of invariants for quantum deformations of the classical Lie algebras,
and our first objective has been to show that
they lead naturally to lattice models for random geometries with the required properties. The spiders provide a convenient geometrical or graph-like description of quantum group invariants in tensor product representations, and they are known for a number of root systems, following
the seminal work by Kuperberg \cite{Kuperberg_1996} on the rank-two cases $A_2$, $G_2$ and $B_2$.

Our first paper introduced the $A_n$ web models which gives rise to a lattice model on the hexagonal lattice much like
the well-known Nienhuis loop model, but now allowing for bifurcations at nodes that are incident on three links.
The web configurations are identical to domain walls in an $n$-state spin model, but they carry a non-local weight parameterised
by $q$. At a specific parameter value, $q = e^{i \pi / (n+2)}$, the non-local weights become trivial so that the partition functions of
the web and spin models agree. In general the web model provides a continuous, non-local deformation of the spin model,
just like the well-known $O(N)$ loop model is a non-local deformation of the Ising model (the case $n=1$).

Our second paper focussed on the physics of the simplest member of this family, the $A_2$ model. In particular, we reformulated the geometrical model as a local vertex model with complex weights. This was used to interpret geometrically some of its critical exponents, as well as numerically determining its phase diagram. In this case, the resulting geometrical structures take the form of a collection of mutually avoiding bipartite cubic graphs embedded in the lattice. Each graph has a statistical weight that is a product of local Boltzmann weights for links and nodes, and a non-local weight for each graph component that generalises the usual Boltzmann weight of a loop. The non-local weight depends
on the deformation parameter $q$ in the quantum algebra $U_q(A_2)$, and it can be unambiguously evaluated by the ``reduction'' relations that define the $A_2$ spider. Meanwhile, the usual loop model is recovered in the rank-one $A_1$ case.

The purpose of the present paper is to show that an innocuous modification of the $A_2$ web models makes it {\it quantum integrable}.
Similarly we obtain, in fact, quantum integrability of web models based on all rank-two spiders.
To this end, we first provide corresponding definitions of statistical models for the other rank-two spiders, $G_2$ and $B_2$, and we reformulate them as local vertex models.
We also modify our previous definition of the $A_2$ web model, by assigning an extra weight conjugate to the angle by which a piece of web
bends in-between two successive bifurcations. The resulting models are defined on the hexagonal lattice for all three spiders, while for the $B_2$
case we also define a variant model on the square lattice.

In our first paper \cite{Lafay:2021scv} we showed how the $A_2$ web model possesses a special point, when $q = e^{i \pi / 4}$, for which the partition function becomes
proportional to that of a three-state chiral spin model on the dual (triangular) lattice. With the bending parameter included, this construction
carries through, but the spin model now has an extra three-spin plaquette interaction. We show here that the two other web models also have special points, with $q$ a root of unity,
for which they are dual to three- or four-state spin models with various symmetries.

To study the continuum limit of web models, either analytically or via numerical diagonalisation of the transfer matrix, it turns out that we need an
equivalent formulation as a vertex model with purely local weights. In our second paper \cite{Lafay:2021wyf} we provided this construction for the
(unmodified) $A_2$ model, based on a representation-theoretical analysis of its underlying quantum group $U_q\, \mathfrak{sl}(3)$. We extend here this local reformulation
to the modified-by-bending-weight $A_2$ model, as well as to the $G_2$ and $B_2$ models.
In all cases the local Hilbert space of the vertex models consist of trivial and fundamental representations.

The main result of this piece of work is then that the corresponding vertex models are
quantum integrable, for any value of $q$, provided that the various local weights are carefully adjusted. A general idea behind this discovery was to find a `big' affine quantum group that contains the quantum symmetry of our models: it corresponds in each case to a certain (possibly twisted) affine Dynkin diagram that reduces, after erasing one of its nodes, to the (finite) Dynkin diagram of the web model. This affine quantum group then generates solutions of the spectral-parameter dependent Yang-Baxter equation through analysis of intertwining conditions on the tensor product of its evaluation representations. We found these solutions in all three cases explicitly and decomposed them in terms of spider diagrams. After a fine tuning, we found a complete agreement with the local transfer matrices from the vertex-model formulation (see Section \ref{sec:integrability} for more details). Achieving this connection with integrable models was in fact the principal motivation for the proposed modification of the $A_2$ model by the bending weight. 

In the $A_2$ and $G_2$ cases, the solutions to Yang-Baxter equations already appeared in another context \cite{takacs_quantum_1997,Tak_cs_1997}, but the solution in the $B_2$ case is new to the best of our knowledge. However, in all three cases, the geometrical interpretation was not discussed before. As usual we expect the continuum limit of these integrable models to be critical, and in fact conformally invariant,
when $q$ belongs to the unit circle. The question thus arises, what are the characteristics of the corresponding CFT?

In \cite{Lafay:2021wyf} we already provided some first numerical results on the central charge of the $A_2$ model, based on the scan of a two-dimensional
parameter space of local weights for each choice of $q$. It was found that the model possesses both a dilute and a dense critical phase, just like the
Nienhuis loop model (the $A_1$ model). Based on experience with the loop model, we can hope that each of these universality classes can be retrieved
within the integrable sub-manifold of the (modified) model, hence making the numerical work much easier. This turns out to be the case indeed.
Varying $q$ along the unit circle for the integrable model, the diagonalisation of the transfer matrices for cylinders of circumference up to size $L=9$ leads, in fact,
to the identification of two distinct critical regimes (see Figure~\ref{fig:ccA2}), with distinct analytical behaviour of the central charge.

The first of these regimes encompasses the dilute and dense critical points, confirming our expectations that the $A_2$ models with and without modification belong to the same universality class. The numerically determined
central charge in this regime is in excellent agreement with~\cite{Lafay:2021wyf} and with an analytical Coulomb gas computation, that will be exposed in a separate publication \cite{LGJ}. The ground state of the second regime exists only when $L$ is a multiple of $3$, and its central charge takes larger values ($c \gtrapprox 3$) than expected for a
rank-two model, with a conspicuously slow convergence. All of these observations hint at a larger symmetry, and possibly a non-compact continuum limit, like the
one found for the Nienhuis loop model in regime III \cite{Vernier14,vernier_new_2015}. We did not find this regime in \cite{Lafay:2021wyf}, presumably because we only focussed on real, positive fugacities there. 

Along the integrable line, three particular points correspond to a three-state spin model.
Within the first regime, one finds two such points, with $c=\tfrac45$ and $c=0$, which are the expected results for a three-state Potts model in the
dilute (critical for the spin model) and dense (infinite-temperature) phases, respectively. The third point is in the second regime, and has a value $c \approx 1.5$ that has not
previously been reported for a three-state Potts model, to our best knowledge.

We present a similar numerical  investigation of the $G_2$ model along its integrable sub-manifold (see Figure~\ref{fig:ccG2}), this time up to size $L=8$. We identify here
four different regimes for which the numerical results are well-behaved, leaving one or two further regimes for subsequent analytical investigations.
In addition we find two handfuls of special points for which mappings to other well-understood models can be made. These include dense and dilute $O(N=-1)$ loop models,
three-state spin models, spanning trees and loop-erased random walks.

Finally, for the $B_2$ model, the dimension of the transfer matrix is too large in order for us to obtain exploitable results. However, by adjusting certain local Boltzmann weights, we
obtain a reduction of the size of the Hilbert space, leading to a more amenable model. In fact, we 
demonstrate that in this case the local transfer matrices can be written in terms of the dilute Birman-Murakami-Wenzl (BMW) algebra. Mathematically, the appearance of this algebra is not surprising because the BMW algebra is known to be Schur-Weyl dual to $U_q (B_2)$ in the tensor product of its natural representations, and here we have the diluted version of this well-known $B_2$ type Schur-Weyl duality. The corresponding integrable solution is in agreement with the baxterisation of the dilute BMW algebra \cite{Grimm_1995}.

The paper is structured as follows. In section~\ref{sec:latt} we specify the lattices and boundary conditions needed to define the models of this paper.
The next three sections (\ref{sec:A2}, \ref{sec:G2} and \ref{sec:B2}) deal with respectively the $A_2$, $G_2$ and $B_2$ models.
In each of them we first discuss the corresponding spider, define the lattice model, and establish
the equivalence with spin models on the dual lattice. Section \ref{sec:tm} provides the analysis of the quantum groups necessary to
define the local vertex models and the corresponding transfer matrices. The integrable models are then constructed in section~\ref{sec:integrability}, as sketched above.
Finally we gather some discussion and concluding remarks in section~\ref{sec:disc}. A few technical ingredients are dispatched in four appendices.

\section{Lattices}
\label{sec:latt}

\begin{figure}
\begin{center}
    \includegraphics[scale=0.4]{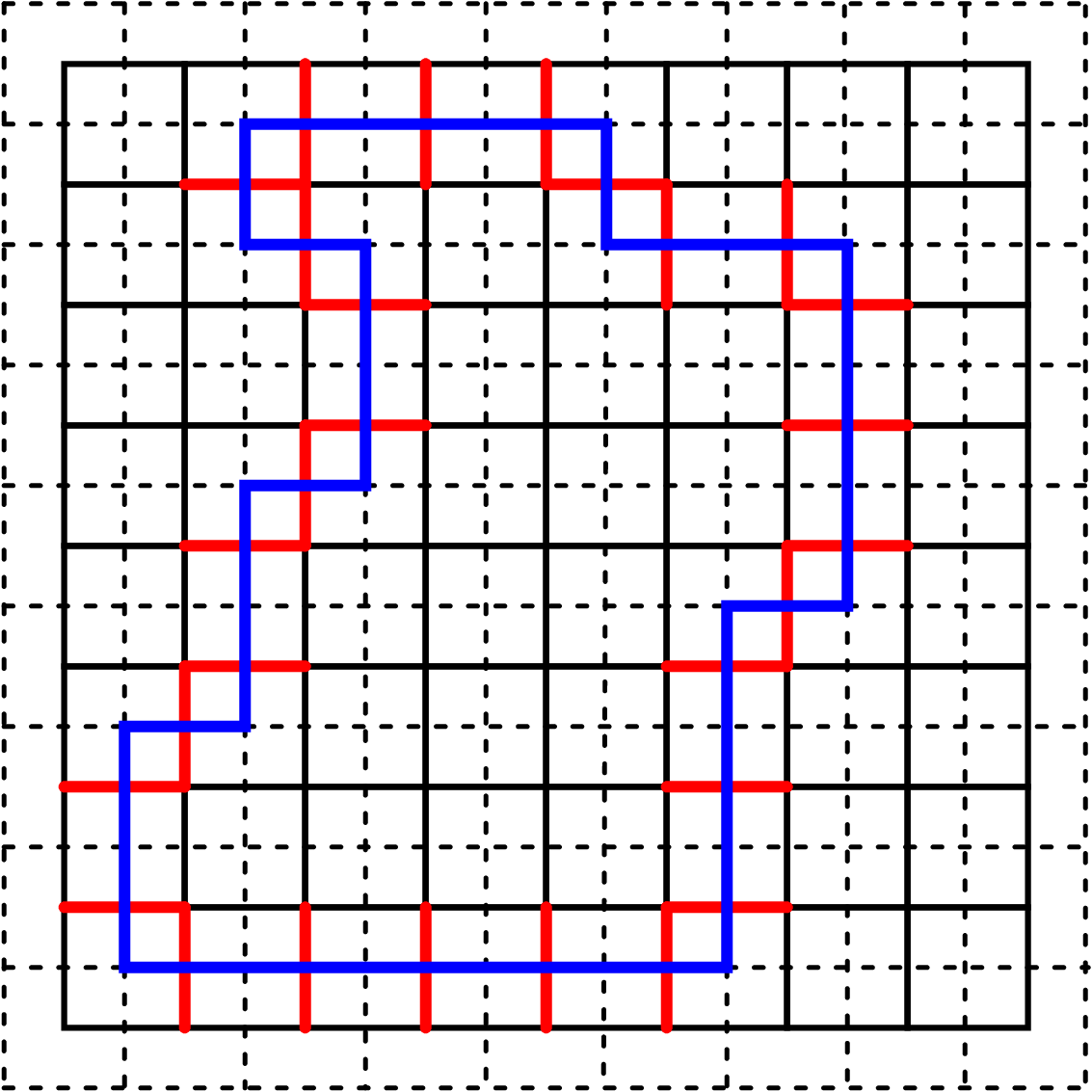} 
\end{center}
    \caption{The primal square lattice (undashed) and its dual (dashed). The path $C$ is drawn in blue and the set $\mathcal{E}_b$ of edges crossed by $C$ is drawn in red.}
    \label{fig:Lattice}
\end{figure}

We will consider lattice models defined on the hexagonal lattice $\mathbb H$, the square lattice $\mathbb S$ and their duals, $\mathbb H ^*=\mathbb T$ and $\mathbb S^*$ respectively. To fix notations, we will say that these lattices are comprised of {\em nodes} and {\em links} connecting them. To be more precise, the primal lattices $\mathbb H$ and $\mathbb S$ will be finite subgraphs of the infinite hexagonal and square lattices respectively, embedded either in the plane or the infinite cylinder as defined in the following way.

In the plane, consider a self avoiding closed path $C$ on the dual infinite lattice, either the triangular lattice, or a shifted square lattice. The path $C$ seperates the plane in two open sets, one of which is bounded and denoted by $\mathring C$. The primal lattice is then comprised of all nodes and links that are inside $\mathring C$, in particular its intersection with $\mathcal{E}_b$---the set of links crossed by $C$---is empty. By dual lattices $\mathbb H ^*$ and $\mathbb S^*$, we mean the subgraphs of the dual infinite lattices that are inside $\mathring C \cup C$. The nodes of the dual lattices that are inside $C$ are called {\em boundary} nodes. Informally, the dual lattices contain an external layer comprised of boundary nodes, surrounding the primal lattices. An example for the plane case is given in Figure \ref{fig:Lattice}.

On the cylinder, we pick two self avoiding and mutually avoiding non contractible paths $C$ and $C'$ on the dual infinite lattices, $C'$ being on top of $C$ when we orient the cylinder vertically. There is a bounded open set $\mathring{CC'}$ between $C$ and $C'$. The primal lattices are defined to be the subgraphs contained in $\mathring{CC'}$. Denote by $\mathcal{E}_b$ the set of links crossed by $C$ or $C'$. The dual lattices are defined to be the subgraphs contained in $\mathring{CC'}\cup C \cup C'$. We will call boundary nodes, the nodes inside $C\cup C'$. Note that, in this case, the boundary nodes split in two connected components, $C$ and $C'$. 

We will consider two types of lattice models. The first of them is the well-known setting of spin models where we assign on each node of the dual lattices a variable, the ``spin'',
which takes a finite number of values, identified with elements of $\mathbb Z_n$ for some integer $n$. We will always consider monochromatic boundary conditions, i.e.\ the spins at the boundary nodes of a given connected component of the boundary are forced to take the same value. One may equivalently imagine the boundary nodes to be contracted
into a single node, one for each connected component of the boundary.

The second type of lattice model is the web model outlined in the Introduction, for which we now recall a few definitions (see also \cite{Lafay:2021scv}).
Its configurations are called {\em webs}. A configuration can be represented abstractly as a graph
with {\em vertices} and {\em edges}, or---for the purpose of defining the lattice model---as embedded in a lattice of {\em nodes} and {\em links}, in which case we shall
call a link covered by the web a {\em bond}, while a link not covered is said to be {\em empty}. Given an embedded web, its abstract representation is obtained by
deleting all the empty links, and contracting any path of consecutive bonds in between two vertices into a single edge. Obviously there will in general be many embedded
webs that correspond to a given abstract one. The properties of the abstract web are described by the corresponding {\em spider}, an algebraic object
which will be defined precisely in the next section. The first step
in the definition of a web model is therefore to ``lift'' the spider to the lattice embedding. This must be done in a suitable way, so that the algebraic properties are preserved,
while keeping the lattice model as simple and elegant as possible.

We here embed the webs in the primal lattice. The Boltzmann weights of a configuration $G$ in the web models will be expressed as the product of two factors.
The first of these is the product of local weights, such as the fugacity of a bond or of a specific arrangement around a given vertex.
The other factor, called the Kuperberg weight $w_{\rm K}(G)$, is a priori non-local and depends only on $G$ seen as an abstract web, regardless of its embedding.

\section{The $A_2$ web models}
\label{sec:A2}

We previously defined and studied the $A_2$ web models on the hexagonal lattice in \cite{Lafay:2021scv}. Here, we give a more general definition that will prove exhibiting integrable points. We begin by recalling the definition of the $A_2$ spider as well as the mapping of webs to spin interfaces of the 3 states Potts models.
\subsection{The $A_2$ spider}
$A_2$ webs are planar oriented graphs embedded in a simply connected domain whose connected components are either closed loops or graphs with trivalent vertices inside the domain or univalent vertices connected to the boundary of the domain. Webs that do not have univalent vertices connected to the boundary of the domain will be called closed webs, otherwise, we will call them open webs. 
There are two types of trivalent vertices, called sinks and sources:
\begin{center}
    \includegraphics[scale=0.2]{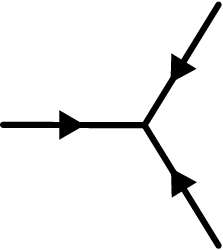}\qquad \qquad \includegraphics[scale=0.2]{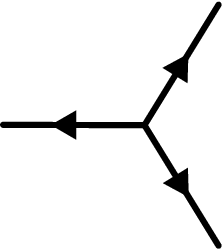}
\end{center}

A number can be assigned unambiguously to any closed $A_2$ web from the following relations\cite{Kuperberg_1996}:
\begin{subequations}\label{eq:Kup-rels}
\begin{align}
    \vcenter{\hbox{\includegraphics[scale=0.2]{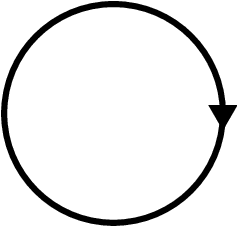}}}&\quad=\quad[3]_q\label{eq:Kup-rule1}\\[5pt]
    \vcenter{\hbox{\includegraphics[scale=0.2]{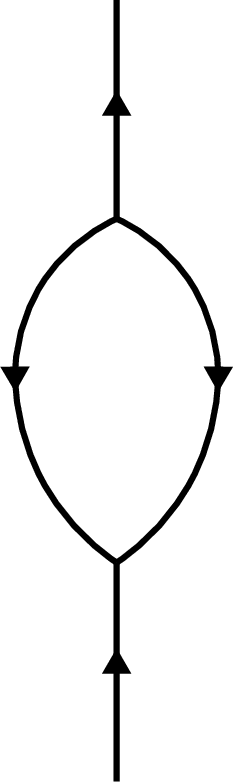}}}&\quad=\quad[2]_q\;\vcenter{\hbox{\includegraphics[scale=0.2]{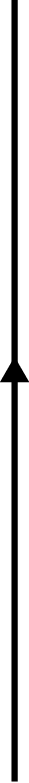}}}\\[5pt]
    \vcenter{\hbox{\includegraphics[scale=0.2]{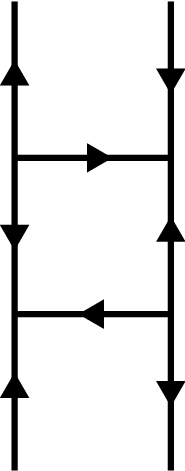}}}&\quad=\quad\vcenter{\hbox{\includegraphics[scale=0.2]{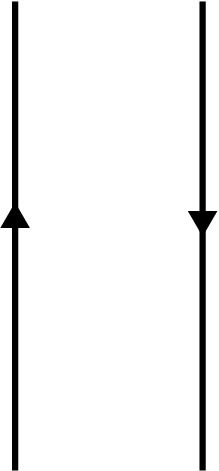}}}\quad +\quad\vcenter{\hbox{\includegraphics[scale=0.2]{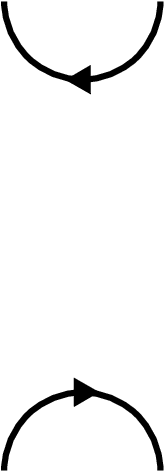}}}\label{kupsquare}
\end{align}
\end{subequations}
Indeed, it is a result of \cite{Kuperberg_1996} that any closed web is proportional to the empty one. For an $A_2$ web $G$, we call the proportionality factor the {\em Kuperberg weight} of $G$ and denote it by $w_{\rm K} (G)$. 

The $q$-numbers $[k]_q$ appearing in \eqref{eq:Kup-rels} and {\em passim} are defined in Appendix~\ref{sec:quantumgroupconventions}.
They depend on a deformation parameter $q \in \mathbb{C}$. We can consider $q$ as parameterising a family of web models
having the symmetry of a given spider.

\subsection{Definition of the models}
\label{sec:defmodels}

Similarly as in \cite{Lafay:2021scv}, we define the $A_2$ web models on the hexagonal lattice $\mathbb{H}$. Configurations are given by closed $A_2$ webs embedded in $\mathbb{H}$. We will denote the configuration space by $\mathcal{K}$. Let $G$ be one such configuration. We assign fugacities $x$ to bonds and fugacities $y$ and $z$ to vertices that are sinks and sources respectively. In addition to the previous local fugacities that were present in the original definition of \cite{Lafay:2021scv}, we give a fugacity to each node of $\mathbb{H}$  that is adjacent to precisely two bonds. The two bonds inherit orientations from the web $G$, so that one of them is directed into the common node and the other goes out of the node. An observer that follows the bonds along that orientation, turns through an angle $\pi/3$ (anticlockwise) or $-\pi/3$ (clockwise) at the node: we call this the {\em bending} at the node. To this we assign a fugacity $e^{i\phi}$ for an anticlockwise bending, or $e^{-i\phi}$ for a clockwise bending. Here, $\phi$ is a new parameter at our disposal: the work \cite{Lafay:2021scv} corresponds to $\phi=0$.

The product of the local fugacities and the non local weight given by the Kuperberg weight defines the Boltzmann weight of a configuration. The partition function then reads:
\begin{align} \label{Z_A2}
    Z_{A_2} =\sum_{c\in K} x^{N}(yz)^{N_V}e^{i\theta}w_{\rm K}(c) \,,
\end{align}
where $N$ is the number of bonds, and $N_V$ is the number of sink/source pairs of vertices.
We have defined $\theta = \sum_i \phi_i$, where the sum is over all nodes and $\phi_i=\pm \phi$ is the bending at node $i$, so that
$e^{i\theta}$ is the total weight given by the bending of edges.

Remark that the Boltzmann weights are invariant under discrete rotations of the lattice but not reflections when $e^{i\phi}\notin \mathbb R$. It is also clear that $Z_{A_2}$ is invariant under the transformation $\phi\rightarrow -\phi$. Assuming, $x$, $y$, $z$ and $\phi$ to be real, the Boltzmann weight of a configuration is sent to its complex conjugate under a spatial reflection, or the reversal of all orientations within a given web. Since the partition function $Z_{A_2}$ comprises a sum over orientations, it follows that it is real. It is also clear that $Z_{A_2}$ is invariant under
\begin{align}
    q\rightarrow q^{-1} \,,
\end{align}
since this transformation keeps the $q$-numbers unchanged.
Furthermore, we show in Appendix \ref{sec:A2sym} that the Boltzmann weights are of the form    $x^{N}(yz)^{N_V}e^{iM\phi}w_{\rm K}(c)$ with
\begin{align*}
 &N+N_V+M= 0 \text{ mod }2\\
 &M= 0 \text{ mod }3.
\end{align*} 
This implies the invariance of $Z_{A_2}$ under the following transformations 
\begin{align}
    e^{i\phi}\rightarrow \tau e^{i\phi}
\end{align}
where $\tau$ is a third root of unity, or
\begin{subequations}
\begin{align}
    &x\rightarrow -x\\
    &y\rightarrow iy\\
    &z\rightarrow iz\\
    &e^{i\phi}\rightarrow -e^{-i\phi}
\end{align}
\end{subequations}
We finally note that the partition function is invariant under the following transformation
\begin{subequations}
\begin{align}
    &q\rightarrow -q\\
    &y\rightarrow iy\\
    &z\rightarrow iz
\end{align}
\end{subequations}
because the factors of $i$ can be absorbed in the Kuperberg rules \eqref{eq:Kup-rels}. This will be explained in more detail in the next section.

\begin{figure}
\begin{center}
    \includegraphics[scale=0.3]{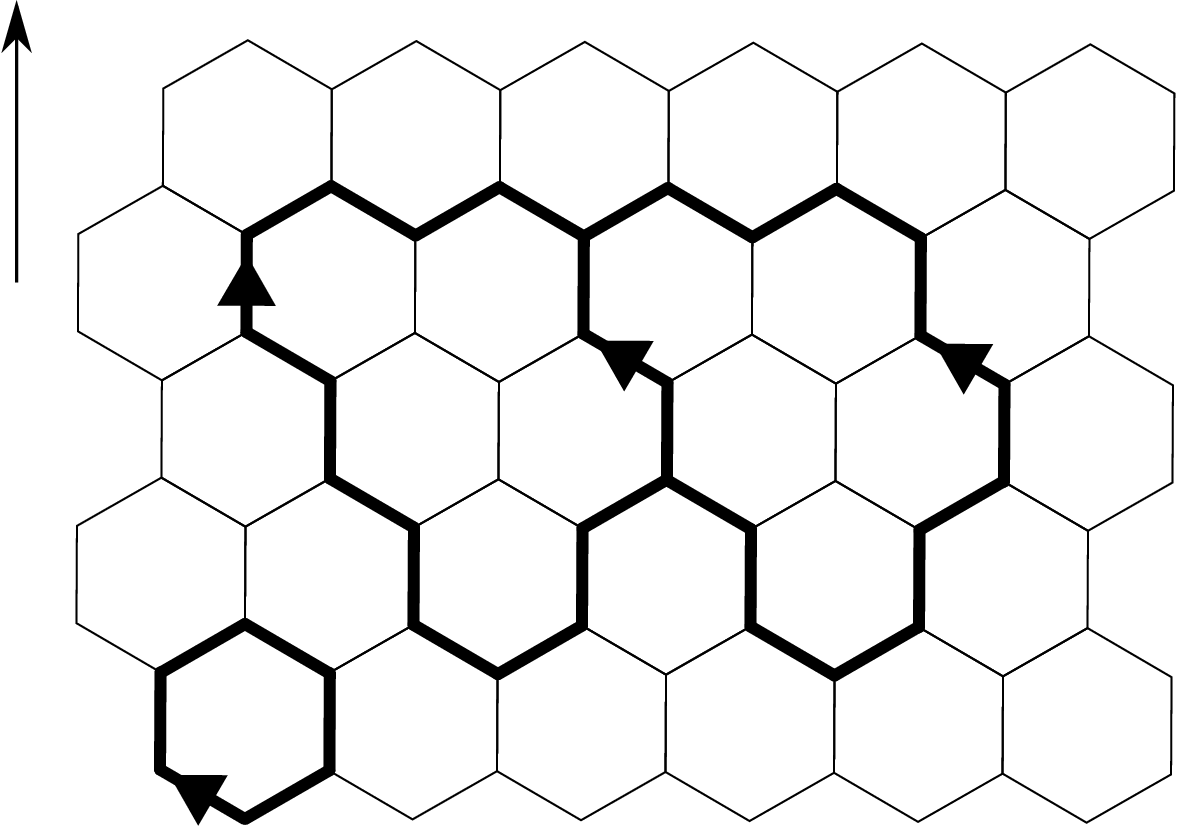} \qquad \qquad \includegraphics[scale=0.3]{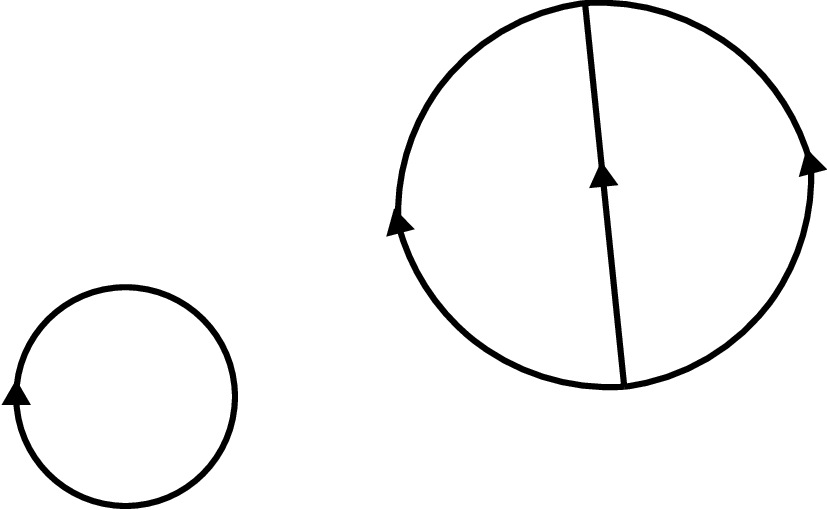}
\end{center}
    \caption{Left panel: A configuration on $\mathbb{H}$ of weight $x^{35}yze^{6i\phi}[2]_q[3]_q^2$. The arrow is parallel to the axis of the cylinder. The left and right sides of the drawing are identified by periodic boundary conditions. Right panel: The same configuration drawn as a web.}
    \label{fig:configA2}
\end{figure}

\subsection{Relation with $\mathbb{Z}_3$ spin interfaces}
\label{sec:A2mapping}
In \cite{Lafay:2021scv}, we defined a mapping from the $A_2$ web model to the 3 states Potts model on the triangular lattice dual to $\mathbb{H}$ with nearest neighbour interactions. Here, we slightly generalise this mapping to account for three site plaquette interactions. Given two spins $\sigma_i$ and $\sigma_j$ in $\mathbb{Z}_3\equiv\{0,1,2\}$ at sites $i$ and $j$ respectively, the nearest neighbour interaction is given by a local Boltzmann weight $x_{|\sigma_i-\sigma_j|}$. It follows that $x_1=x_2 =: x$, and we set $x_0=1$. Given three sites $i$, $j$, $k$ situated in a clockwise manner around a plaquette, the three site plaquette interaction is given by
\begin{align}
    p_{\sigma_i,\sigma_j,\sigma_k}=&\delta_{\sigma_i,\sigma_j}\delta_{\sigma_j,\sigma_k}+2^{\frac{1}{4}}y\delta_{\sigma_i,\sigma_j+1}\delta_{\sigma_j,\sigma_k+1}+2^{\frac{1}{4}}z\delta_{\sigma_i,\sigma_j-1}\delta_{\sigma_j,\sigma_k-1}\nonumber\\
    &+e^{-i\phi}\left(\delta_{\sigma_i,\sigma_j+1}\delta_{\sigma_j,\sigma_k}+\delta_{\sigma_j,\sigma_k+1}\delta_{\sigma_k,\sigma_i}+\delta_{\sigma_k,\sigma_i+1}\delta_{\sigma_i,\sigma_j}\right) \label{pZ3} \\
    &+e^{i\phi}\left(\delta_{\sigma_i,\sigma_j-1}\delta_{\sigma_j,\sigma_k}+\delta_{\sigma_j,\sigma_k-1}\delta_{\sigma_k,\sigma_i}+\delta_{\sigma_k,\sigma_i-1}\delta_{\sigma_i,\sigma_j}\right) \,. \nonumber
\end{align}
Notice that the plaquette interactions reduce the $S_3$ symmetry to $\mathbb{Z}_3$;
in our previous work \cite{Lafay:2021scv,Lafay:2021wyf} we did not include plaquette interactions, so in that case the symmetry was actually $S_3$.
In the present model, $\phi$, $y$ and $z$ are adjustable parameters. Setting $\phi=0$ and $y=z=2^{-1/4}$ the plaquette interaction becomes the identity operator and we recover the previous model \cite{Lafay:2021scv,Lafay:2021wyf}.

The partition function then reads
\begin{align}
    Z_{\mathbb{Z}_3}=\sum_{\sigma}\left(\prod_{\langle ij \rangle}x_{|\sigma_i-\sigma_j|}\right)\left(\prod_{\langle ijk \rangle}p_{\sigma_i,\sigma_j,\sigma_k}\right)
\end{align}
where $\langle ij \rangle$ denotes nearest neighbours pairs of sites and $\langle ijk \rangle$ denotes plaquettes of three sites. We have used the subscript $\mathbb{Z}_3$ to emphasise that the interaction \eqref{pZ3} is invariant
under cyclic permutations of the colours.
The integrable solutions for the $A_2$ web models that will be described in Section \ref{sec:integrabilityA2} contain points that can be mapped to $\mathbb{Z}_3$ spin models only for a non-trivial plaquette interaction.

We now reformulate the partition function in terms of its domain walls. For two neighbouring spins $\sigma_i$ and $\sigma_j$, if $|\sigma_i-\sigma_j|=1$, we draw a bond on the link of $\mathbb{H}$ separating the two spins and we orient the bond such that the when going from the node $i$ to the node $j$, the spin value increases (respectively decreases by) $1$ when traversing a right-pointing (respectively left-pointing) bond. If $|\sigma_i-\sigma_j|=0$ we let the link empty. We obtain this way a closed simple $A_2$ web $G$ embedded in $\mathbb{H}$. The mapping is many to one and onto. The number of spin configurations having $G$ as their domain wall is exactly $3$ corresponding to the choice for the spin of an arbitrary face. We thus have that
\begin{align}
    Z_{\mathbb Z_3}=3\sum_{G\in \mathcal{K}} x^{N}(\sqrt{2}yz)^{N_V}e^{i\theta}
\end{align}

Set $q=e^{i\frac{\pi}{4}}$. In order to relate $Z_{\mathbb{Z}_3}$ to $Z_{A_2}$, we follow the idea of \cite{Lafay:2021scv}. Remark that the product of the vertex fugacities and the Kuperberg weight do not depend on the embedding of the web into $\mathbb{H}$. Given a closed simple web $G$, rewriting the product of vertex fugacities as
\begin{align*}
    (yz)^{N_V}=(\sqrt{2}yz)^{N_V} \left(\frac{1}{\sqrt{2}}\right)^{N_V}
\end{align*}
we call the product $\left(\frac{1}{\sqrt{2}}\right)^{N_V} w_{\rm K}(G)$, the \textit{topological} weight of the web, $w_{\rm top}(G)$. Similarly as in \cite{Lafay:2021scv} for the $A_2$ case, it can be computed thanks to modifications of the relations \eqref{eq:Kup-rels} in order to incorporate the vertex fugacity in the reduction process. This can be seen as a rescaling of the vertices of webs:
\begin{align}
    \vcenter{\hbox{\includegraphics[scale=0.2]{diagrams/A2vertex1.eps}}}\mapsto 2^{-\frac{1}{4}}\ \vcenter{\hbox{\includegraphics[scale=0.2]{diagrams/A2vertex1.eps}}} \qquad  \qquad\vcenter{\hbox{\includegraphics[scale=0.2]{diagrams/A2vertex2.eps}}}\mapsto 2^{-\frac{1}{4}}\ \vcenter{\hbox{\includegraphics[scale=0.2]{diagrams/A2vertex2.eps}}}
\end{align}
The rules computing the topological weight are then 
\begin{subequations}
\begin{align}
    \vcenter{\hbox{\includegraphics[scale=0.2]{diagrams/rel1kup.eps}}}&\quad=\quad1\label{eq:Kup-rule1}\\[5pt]
    \vcenter{\hbox{\includegraphics[scale=0.2]{diagrams/rel2kup1.eps}}}&\quad=\;\vcenter{\hbox{\includegraphics[scale=0.2]{diagrams/rel2kup2.eps}}}\\[5pt]
    \vcenter{\hbox{\includegraphics[scale=0.2]{diagrams/rel3kup1.eps}}}&\quad=\quad\frac{1}{2}\;\vcenter{\hbox{\includegraphics[scale=0.2]{diagrams/rel3kup2.eps}}}\quad +\quad\frac{1}{2}\;\vcenter{\hbox{\includegraphics[scale=0.2]{diagrams/rel3kup3.eps}}}\label{kupsquare}
\end{align}
\end{subequations}
The stochastic nature of these rules, i.e., the fact that the sum of prefactors appearing on each side of any of a given rule is the same implies that \cite{Lafay:2021scv}
\begin{align*}
    w_{\rm top}(G)=1
\end{align*}
We conclude that
\begin{align}
    Z_{\mathbb{Z}_3}=3Z_{A_2}
\end{align}
where the spin interfaces are mapped to $A_2$ webs.

\section{The $G_2$ web models}
\label{sec:G2}

We now turn to web models based on the $G_2$ spider. In contrast to the $A_2$ case there are no orientations involved, and hence no bending. This will also lead
to some other physical consequences, as we shall soon see.

\subsection{The $G_2$ spider}
$G_2$ webs are planar graphs embedded in a simply connected domain whose connected components are either closed loops or graphs with trivalent vertices inside the domain or univalent vertices connected to the boundary of the domain. Webs that do not have univalent vertices connected to the boundary of the domain will be called closed webs, otherwise, we will call them open webs. Edges come in two types and are called simple and double edges, depicted respectively as 
\begin{center}
    \includegraphics[scale=0.2]{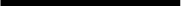}\qquad \qquad \includegraphics[scale=0.2]{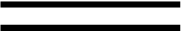}
\end{center}
There are two types of trivalent vertices:
\begin{center}
    \includegraphics[scale=0.2]{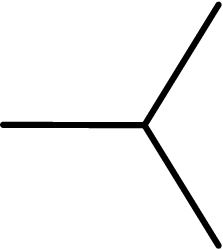}\qquad \qquad \includegraphics[scale=0.2]{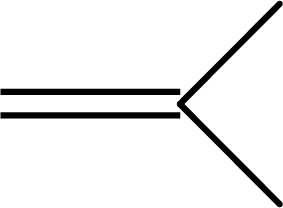}
\end{center}
We will call the first vertex, vertex of type $1$ and the second, vertex of type $2$.

The free vector space spanned by closed $G_2$ webs will be denoted by $\mathcal{FS}p(G_2)$. We then denote by $\mathcal{S}p(G_2)$, the quotient of $\mathcal{FS}p(G_2)$ by the following local relations\footnote{Note that our conventions differ from \cite{Kuperberg_1996} by $q\leftrightarrow q^{\frac{1}{2}}$.} \cite{Kuperberg_1996}:
\begin{subequations}
\label{G2rel}
\begin{align}
    \vcenter{\hbox{\includegraphics[scale=0.12]{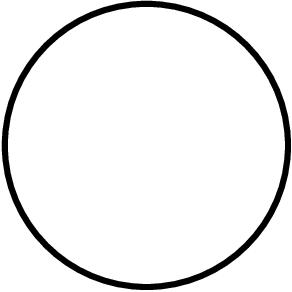}}}\quad=\quad &q^{10}+q^8+q^2+1+q^{-2}+q^{-8}+q^{-10}\label{G2rel1}\\[5pt]
    \vcenter{\hbox{\includegraphics[scale=0.12]{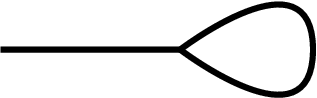}}}\quad=\quad &0\\[5pt]
    \vcenter{\hbox{\includegraphics[scale=0.12]{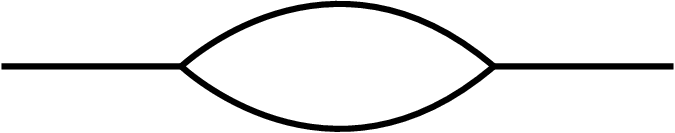}}}\quad=\quad &-(q^6+q^4+q^2+q^{-2}+q^{-4}+q^{-6})\quad \vcenter{\hbox{\includegraphics[scale=0.12]{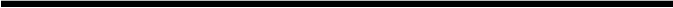}}}\\[5pt]
    \vcenter{\hbox{\includegraphics[scale=0.12]{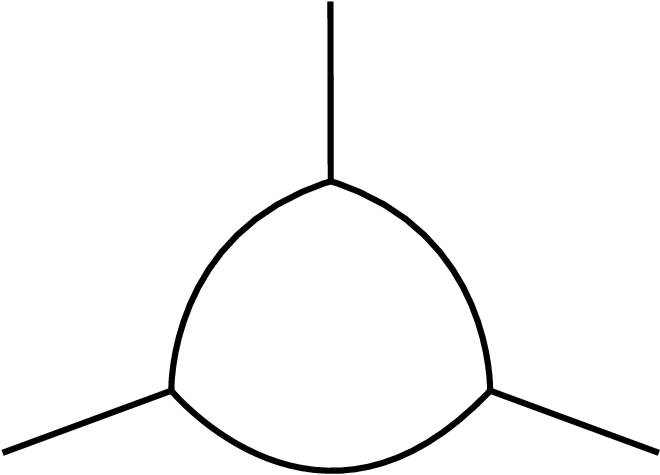}}}\quad=\quad &(q^4+1+q^{-4}) \vcenter{\hbox{\includegraphics[scale=0.12]{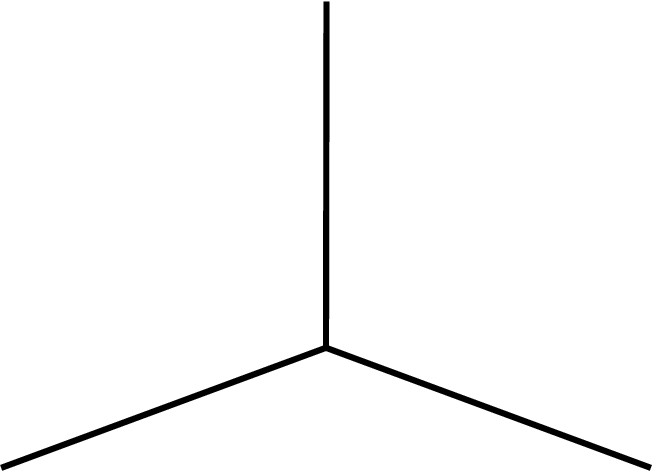}}}\\[5pt]
    \vcenter{\hbox{\includegraphics[scale=0.12]{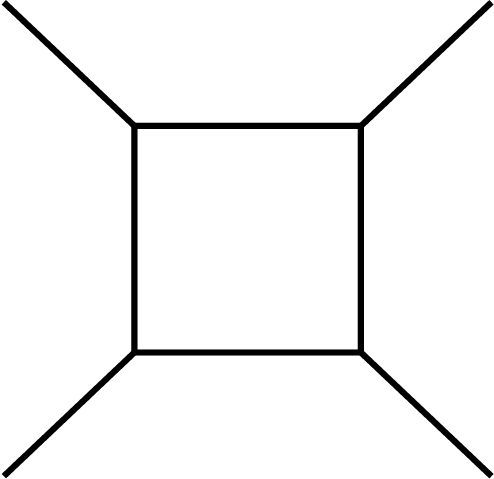}}}\quad=\quad&-(q^2+q^{-2}) \left(\  \vcenter{\hbox{\includegraphics[scale=0.12]{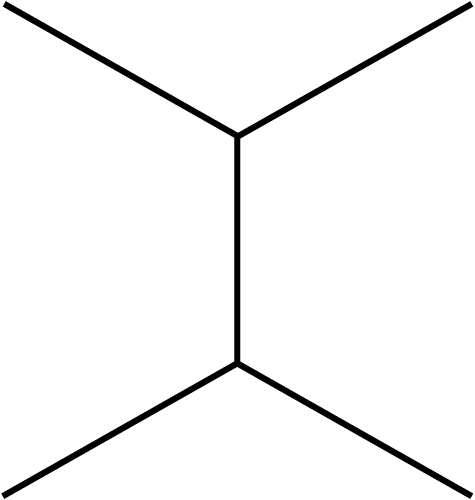}}}+\vcenter{\hbox{\includegraphics[scale=0.12]{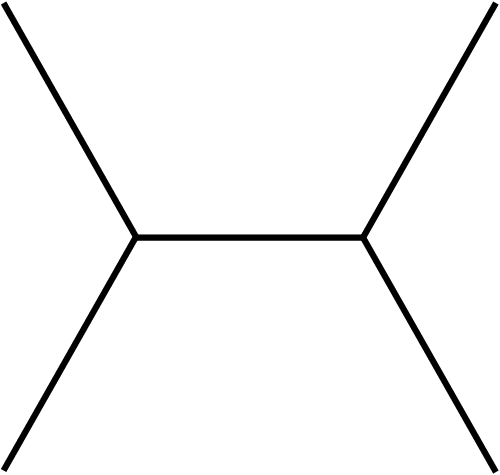}}}\ \right)+(q^2+1+q^{-2})\left(\  \vcenter{\hbox{\includegraphics[scale=0.12]{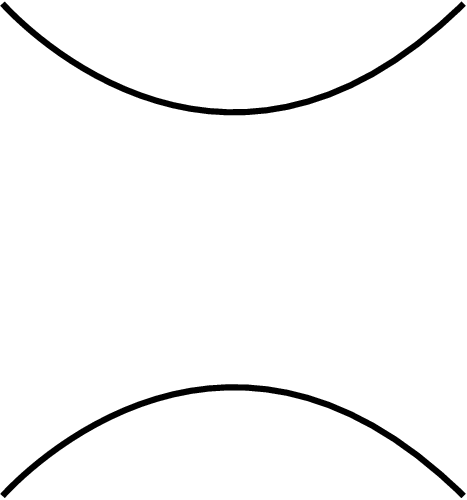}}}+\vcenter{\hbox{\includegraphics[scale=0.12]{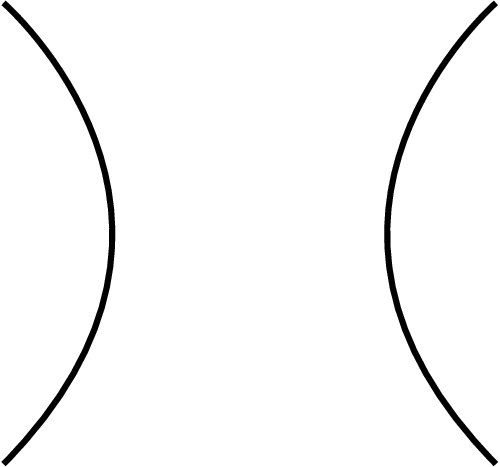}}}\ \right)\\[5pt]
    \vcenter{\hbox{\includegraphics[scale=0.12]{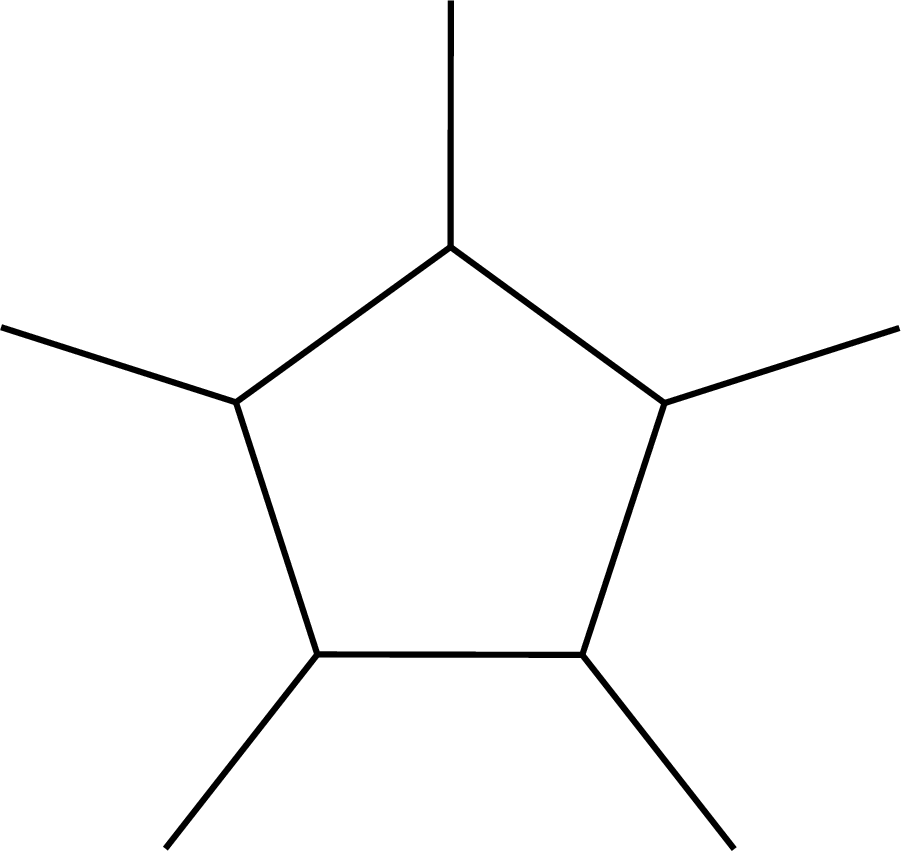}}}\quad=\quad&+ \left(\  \vcenter{\hbox{\includegraphics[scale=0.10]{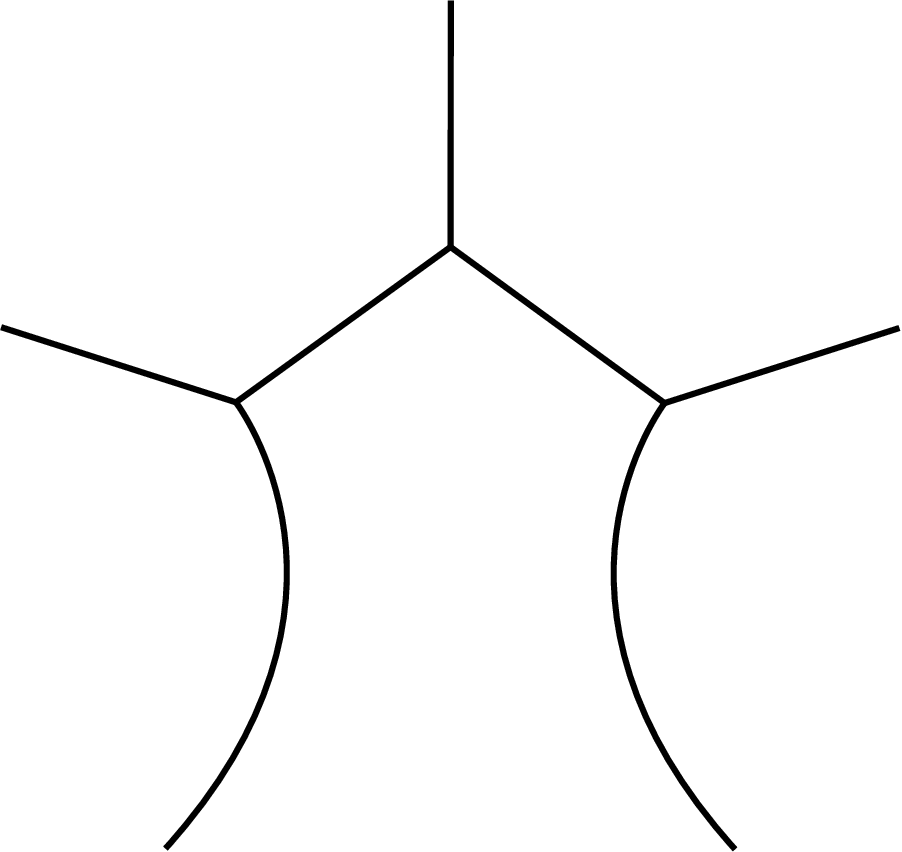}}}+\vcenter{\hbox{\includegraphics[scale=0.10]{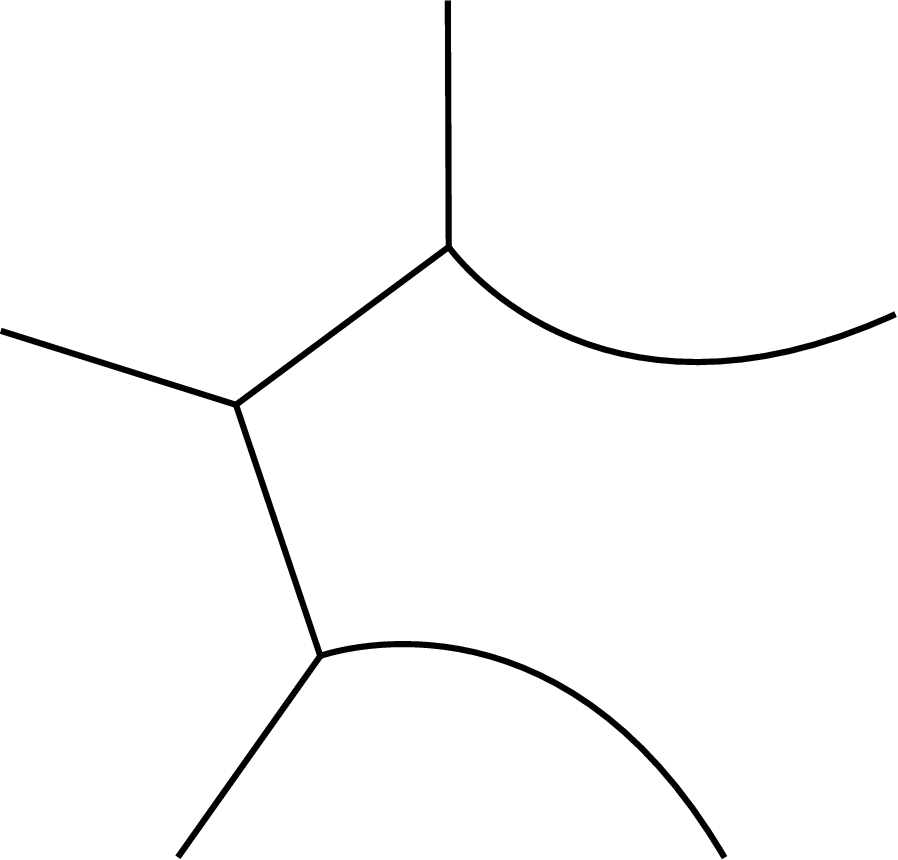}}}+\vcenter{\hbox{\includegraphics[scale=0.10]{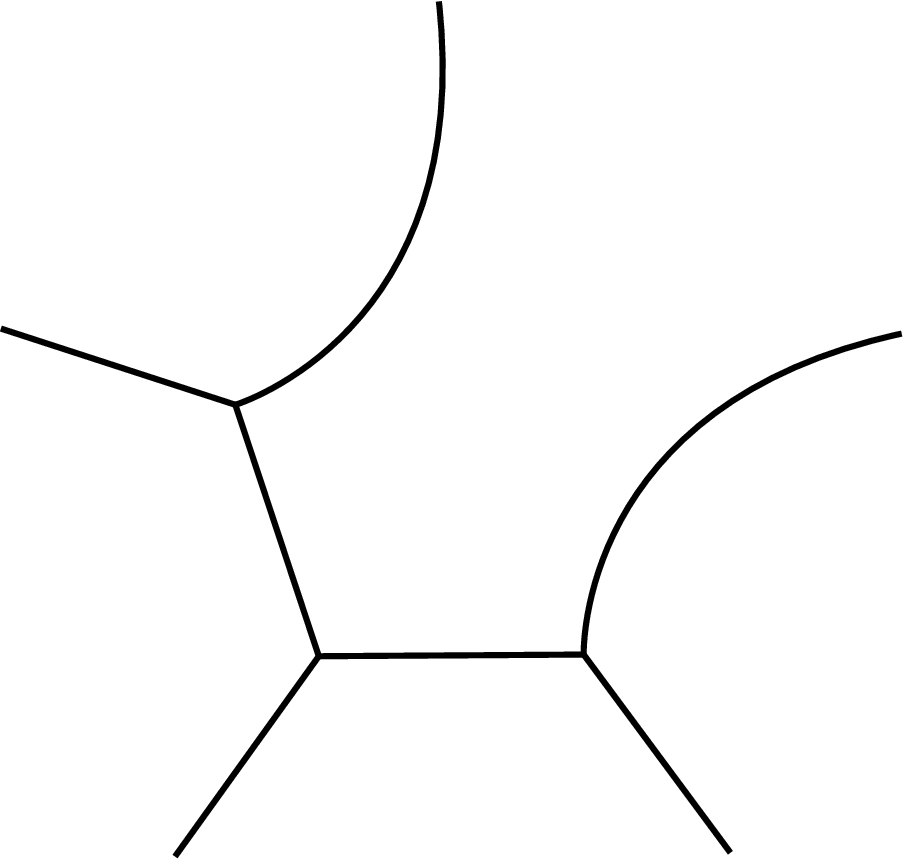}}}+\vcenter{\hbox{\includegraphics[scale=0.10]{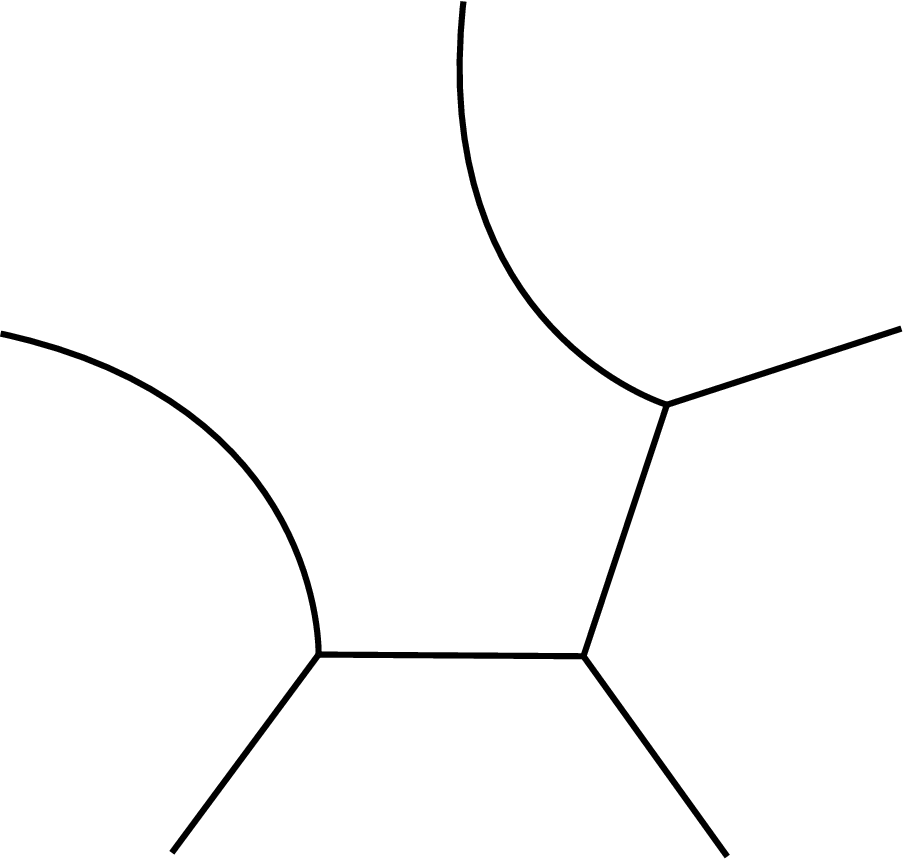}}}+\vcenter{\hbox{\includegraphics[scale=0.10]{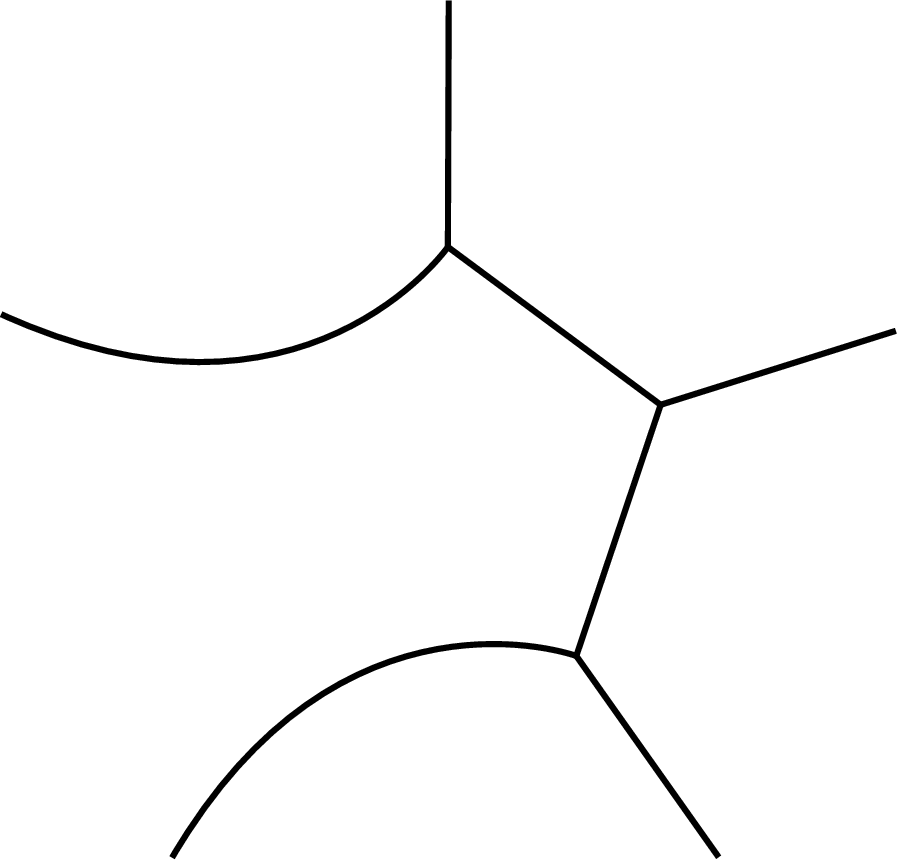}}}\ \right)\nonumber \\[5pt]
    &-\left(\  \vcenter{\hbox{\includegraphics[scale=0.10]{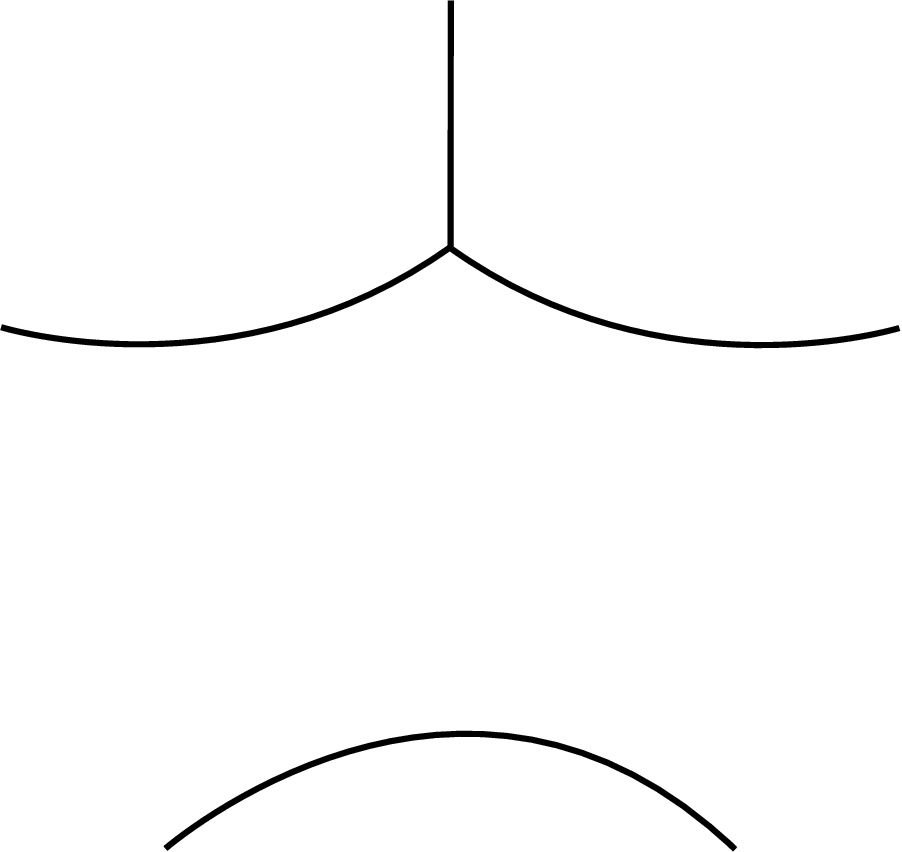}}}+\vcenter{\hbox{\includegraphics[scale=0.10]{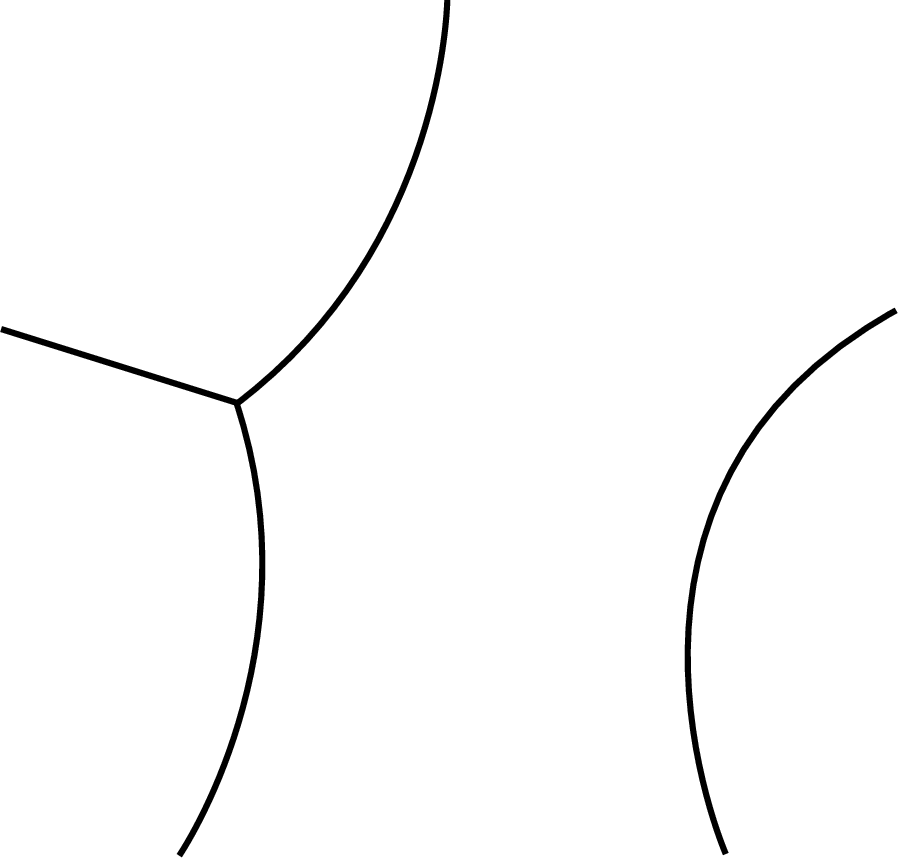}}}+\vcenter{\hbox{\includegraphics[scale=0.10]{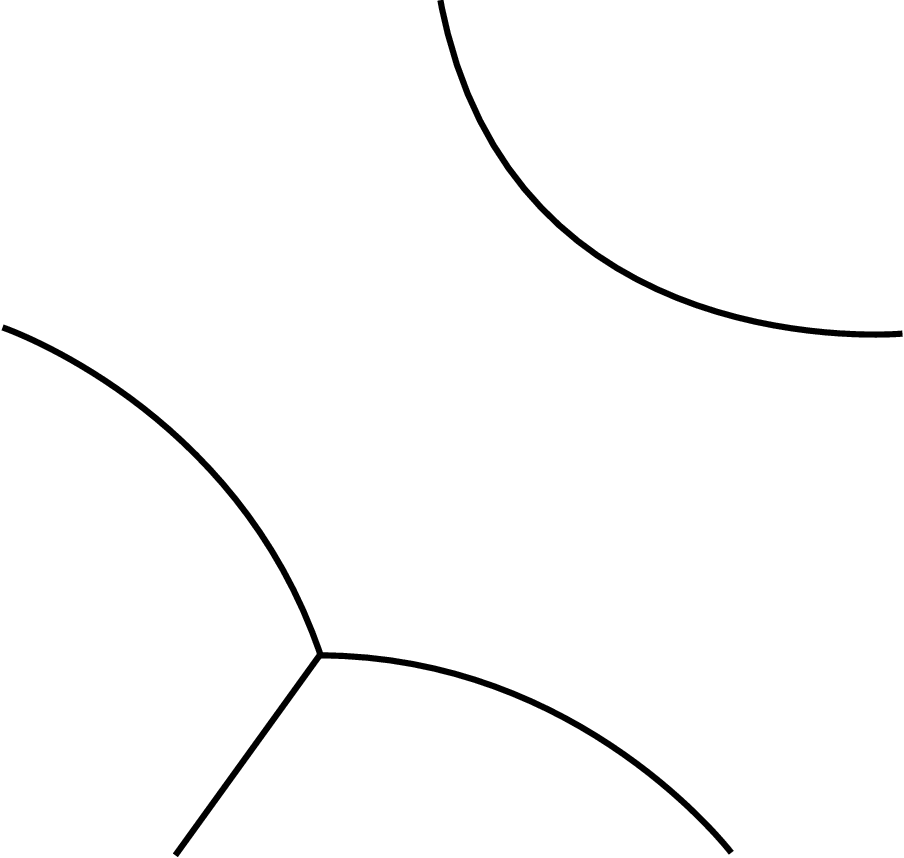}}}+\vcenter{\hbox{\includegraphics[scale=0.10]{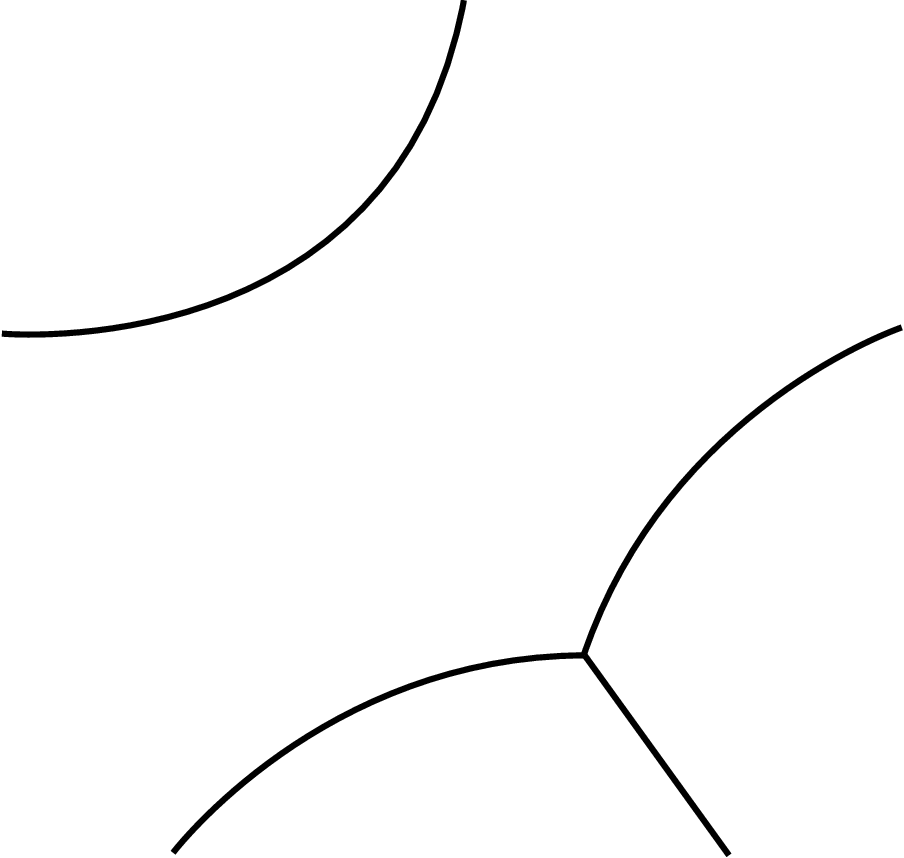}}}+\vcenter{\hbox{\includegraphics[scale=0.10]{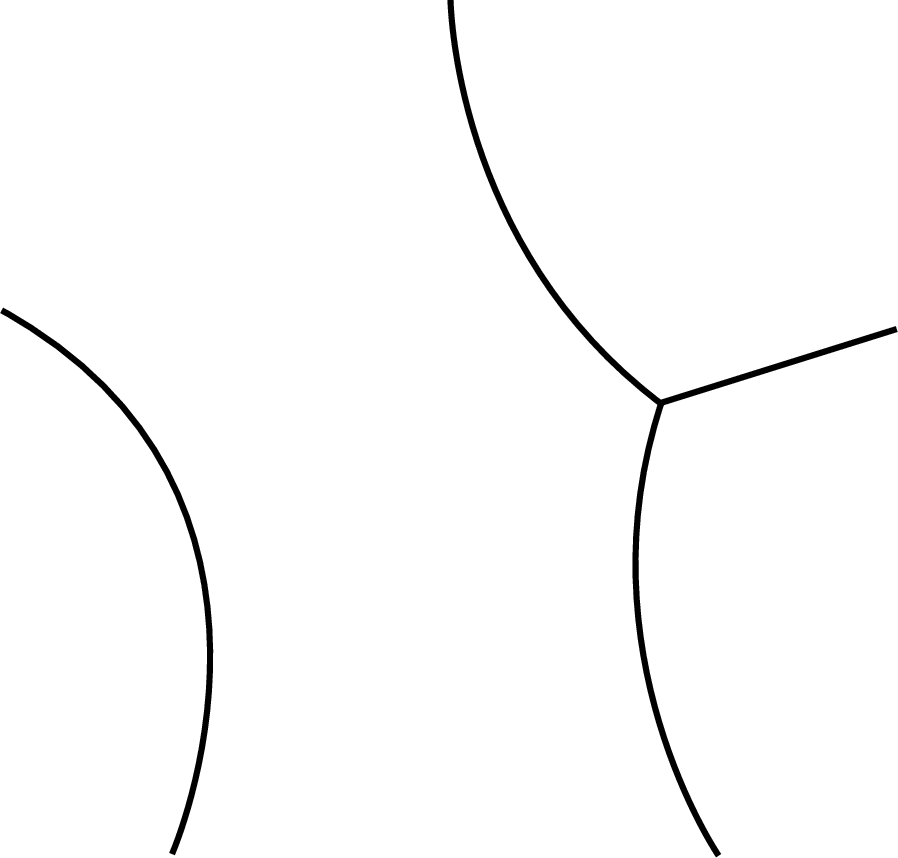}}}\ \right) \label{G2rel6}\\[5pt]
    \vcenter{\hbox{\includegraphics[scale=0.12]{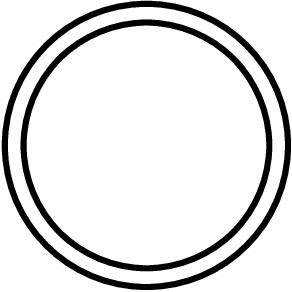}}}\quad=\quad &q^{18}+q^{12}+q^{10}+q^{8}+q^{6}+q^{2}+2+q^{-2}+q^{-6}+q^{-8}+q^{-10}+q^{-12}+q^{-18}\\[5pt]
    \vcenter{\hbox{\includegraphics[scale=0.12]{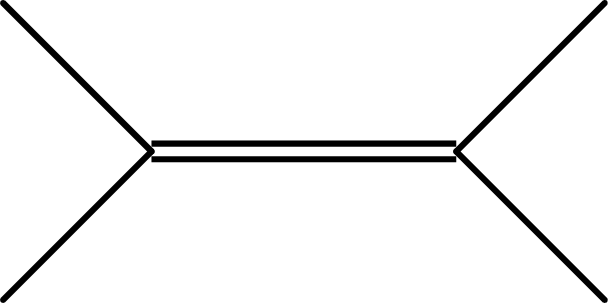}}}\quad = \quad &\vcenter{\hbox{\includegraphics[scale=0.12]{diagrams/G2rel11.eps}}}- \vcenter{\hbox{\includegraphics[scale=0.12]{diagrams/G2rel10.eps}}}- \frac{1}{q^4-1+q^{-4}}\vcenter{\hbox{\includegraphics[scale=0.12]{diagrams/G2rel12.eps}}}+\frac{1}{q^2+1+q^{-2}}\vcenter{\hbox{\includegraphics[scale=0.12]{diagrams/G2rel9.eps}}}
\end{align}
\end{subequations}
A number can be assigned unambiguously to any closed $G_2$ web thanks to these relations. Indeed, it is a result of \cite{Kuperberg_1996} that any closed web is proportional to the empty one. For a $G_2$ web $G$, we call the proportionality factor the {\em Kuperberg weight} of $G$ and denote it by $w_{\rm K} (G)$. Moreover, it is a result of \cite{kuperberg1991quantum} that relations \eqref{G2rel1}-\eqref{G2rel6} are sufficient to reduce unambiguously a closed $G_2$ web made of simple edges only to the empty one. We call such a web a \textit{simple} $G_2$ web. An easy argument to see why any simple web can be reduced is the following. If a web contains a loop or a face surrounded by $n\leq 5$ edges, then it can be reduced in terms of smaller webs. If this always happens, the result follows by induction. Suppose it is not the case for a given web. Denote by $F$, $E$ and $V$ the number of faces, edges and vertices of a this web. By the hand-shake lemma and Euler relation, we have that $2E=3V$ and $F-E+V=2$. Thus $F-\frac{1}{2}V=2$.
The assumption that all faces are of degree at least six means that $6F\leq 2E$, using the hand-shake lemma on the dual graph, so inserting we get $12+3V \leq 2E = 3V$, a contradiction. 

We denote by $\mathcal{FS}p'(G_2)$, the free vector space generated by closed simple webs. The quotient of this space by relations \eqref{G2rel1}-\eqref{G2rel6} will be called $\mathcal{S}p'(G_2)$.

\subsection{Definition of the models}

We now define the $G_2$ web models on the hexagonal lattice $\mathbb{H}$. Configurations are given by closed simple $G_2$ webs embedded in $\mathbb{H}$. We will denote the configuration space by $\mathcal{K}$. We assign fugacities $x$ to bonds and fugacities $y$ to vertices. The product of the local fugacities and the non local weight given by the Kuperberg weight defines the Boltzmann weight of a configuration. The partition function then reads:
\begin{align}
\label{G2PF}
    Z_{G_2}=\sum_{G\in \mathcal{K}} x^{N}y^{M}w_{\rm K}(G)
\end{align}
where  $N$ is the number of bonds and $M$ is the number of of vertices appearing in a given configuration. Remark that a trivalent graph has an even number of vertices, thus the partition function is independent of the sign of $y$.

One could also define a $G_2$ web model using both simple and double edge webs and two types of vertices. In this paper, we chose to focus on the simple case only.

\begin{figure}
\begin{center}
    \includegraphics[scale=0.3]{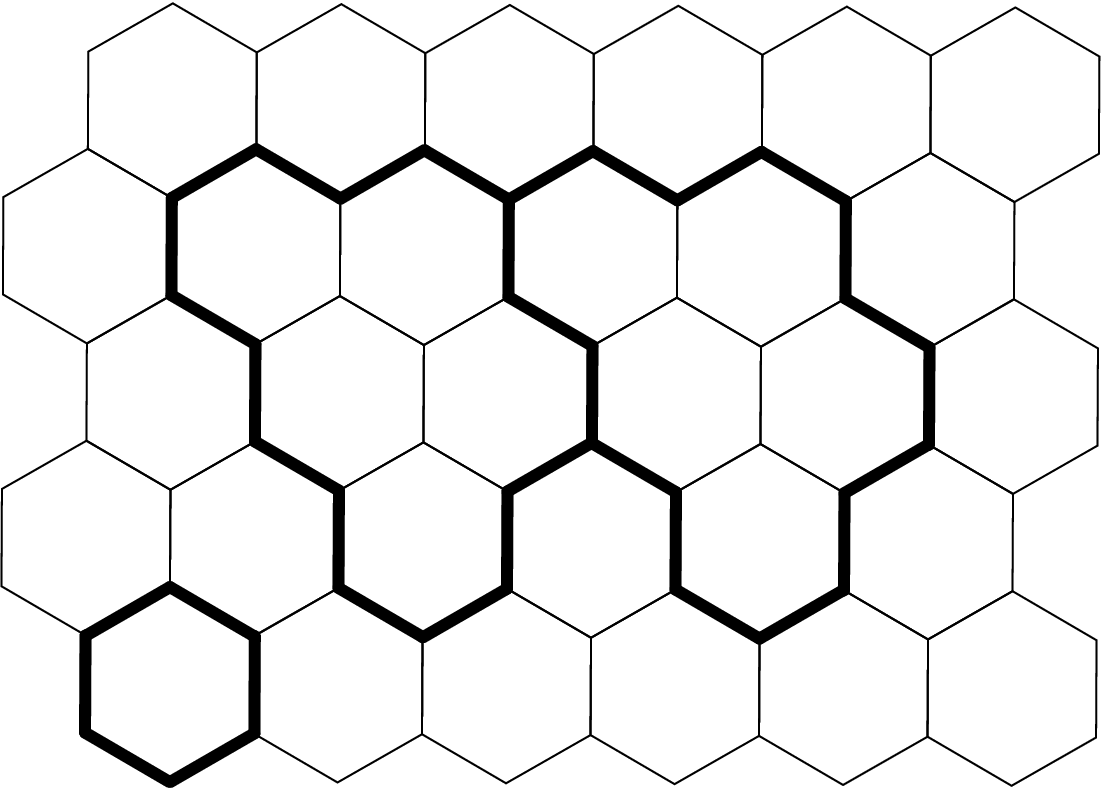} \qquad \qquad \includegraphics[scale=0.3]{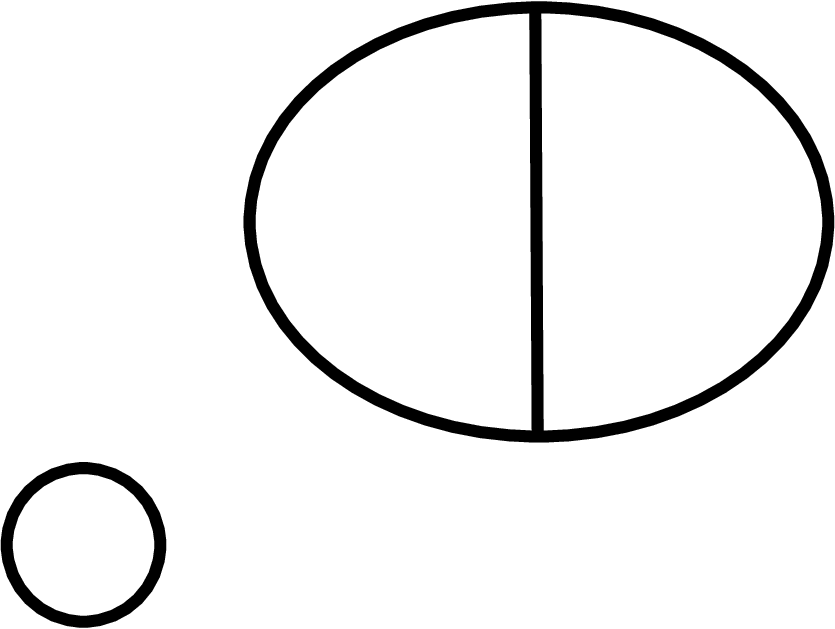}
\end{center}
    \caption{Left panel: A configuration on $\mathbb{H}$ of weight $-x^{35}y^2(q^6+q^4+q^2+q^{-2}+q^{-4}+q^{-6})(q^{10}+q^8+q^2+1+q^{-2}+q^{-8}+q^{-10})^2$. Right panel: The same configuration drawn as a web.}
    \label{fig:configG2}
\end{figure}

\subsection{Relation with an $S_3$ spin model}
\label{sec:G2mapping}

We can formulate an $S_3$ spin model defined on $\mathbb{T}$ in terms of its domain walls\footnote{Remark that this model is in general different than the one defined in Section \ref{sec:A2mapping}.}. Consider spins $\{\sigma_i,\ i\in \mathbb{T}\}$ taking values in $\mathbb{Z}_3= \{0,1 ,2\}$. We define nearest neighbours interactions $x_{|\sigma_i-\sigma_j|}$ for a pair $\langle ij\rangle$ of neighbourings nodes. Here $|\sigma_i-\sigma_j|$ is to be understood modulo $3$ and we normalise interactions such that $x_0=1$. Hence, this interaction depends only on one parameter $x_1$ that we rename $x$ in the following. We also define a $3$-site interaction for each plaquette $\langle ijk \rangle$ as 
\begin{align}
    p_{\sigma_i,\sigma_j,\sigma_k}=&\delta_{\sigma_i=\sigma_j=\sigma_k}+\sqrt{2}y\delta_{\sigma_i \neq \sigma_j \neq \sigma_k \neq \sigma_i} 
    +\delta_{\sigma_i \neq \sigma_j = \sigma_k}+\delta_{\sigma_j \neq \sigma_k = \sigma_i}+\delta_{\sigma_k \neq \sigma_i = \sigma_j} \,.
\end{align}
Notice that this interaction is now invariant under any permutation of the three colours, so the corresponding model has an $S_3$ colour symmetry.
The partition function of the model reads
\begin{align}
\label{partfunctionS3model}
    Z_{S_3}=\sum_{\sigma}\left(\prod_{\langle ij\rangle}x_{|\sigma_i-\sigma_j|}\right)\left(\prod_{\langle ijk\rangle}p_{\sigma_i,\sigma_j,\sigma_k}\right) \,.
\end{align}

We now reformulate the partition functions in terms of its domain walls. For two neighbouring spins $\sigma_i$ and $\sigma_j$, if $|\sigma_i-\sigma_j|=1$, we draw a simple bond on the link of $\mathbb{H}$ separating the two spins whereas if $|\sigma_i-\sigma_j|=0$ we let the link empty. We obtain in this way a closed simple $G_2$ web $G$ embedded in $\mathbb{H}$. The mapping is many to one and onto. The number of spin configurations having $G$ as their domain wall is given by the number of proper $3$-colourings of the dual graph $\hat{G}$. Denoting the chromatic polynomial with $Q$ colours of $\hat{G}$ by $\chi_{\hat{G}}(Q)$, we have that
\begin{align}
    Z_{S_3}=\sum_{G\in \mathcal{K}} x^{N}(\sqrt{2}y)^M \chi_{\hat{G}}(3) \,,
\end{align}
where $N$ denotes the number of bonds, while $M$ is the number of vertices of $G$.
As an example, figure~\ref{fig:configG2} corresponds to $\chi_{\hat{G}}(Q) = Q(Q-1)^2 (Q-2)$.

\medskip

We will now show that $Z_{S_3}$ is equal to the partition function of the $G_2$ web model, up to an overall multiplicative constant, when 
\begin{align}
\label{specialpoint}
    q=&e^{i\frac{\pi}{6}} \quad \text{ or } \quad q = e^{i\frac{5\pi}{6}} \,.
\end{align}
Remark that the product of the vertex fugacities and the Kuperberg weight do not depend on the embedding of the web into $\mathbb{H}$. Given a closed simple web $G$, rewriting the product of vertex fugacities as
\begin{align*}
    y^M=(\sqrt{2}y)^M \left(\frac{1}{\sqrt{2}}\right)^M \,,
\end{align*}
we call the product $\left(\frac{1}{\sqrt{2}}\right)^Mw_{\rm K}(G)$, the \textit{topological} weight of the web, $w_{\rm top}(G)$. Similarly as in the $A_2$ case, it can be computed thanks to modifications of the relations \eqref{G2rel} in order to incorporate the vertex fugacity in the reduction process. This can be seen as a rescaling of the vertices of webs:
\begin{align}
    \vcenter{\hbox{\includegraphics[scale=0.2]{diagrams/Vertex2.eps}}}\mapsto \frac{1}{\sqrt{2}}\ \vcenter{\hbox{\includegraphics[scale=0.2]{diagrams/Vertex2.eps}}}
\end{align}
The relations to compute the topological weight of a simple web at either of the points \eqref{specialpoint} are thus
\begin{subequations}
\label{G2relspec}
\begin{align}
    \vcenter{\hbox{\includegraphics[scale=0.12]{diagrams/B2rel1.eps}}}\quad=\quad &2\label{G2relspec1}\\[5pt]
    \vcenter{\hbox{\includegraphics[scale=0.12]{diagrams/G2rel3.eps}}}\quad=\quad &0\\[5pt]
    \vcenter{\hbox{\includegraphics[scale=0.12]{diagrams/G2rel4.eps}}}\quad=\quad &\quad \vcenter{\hbox{\includegraphics[scale=0.12]{diagrams/G2rel5.eps}}}\\[5pt]
    \vcenter{\hbox{\includegraphics[scale=0.12]{diagrams/G2rel6.eps}}}\quad=\quad &0\\[5pt]
    2\vcenter{\hbox{\includegraphics[scale=0.12]{diagrams/G2rel8.eps}}}\quad=\quad& -\left(\  \vcenter{\hbox{\includegraphics[scale=0.12]{diagrams/G2rel10.eps}}}+\vcenter{\hbox{\includegraphics[scale=0.12]{diagrams/G2rel9.eps}}}\ \right)+\left(\  \vcenter{\hbox{\includegraphics[scale=0.12]{diagrams/G2rel11.eps}}}+\vcenter{\hbox{\includegraphics[scale=0.12]{diagrams/G2rel12.eps}}}\ \right) \label{G2relspec5} \\[5pt]
    4\ \vcenter{\hbox{\includegraphics[scale=0.12]{diagrams/G2rel13.eps}}}\quad=\quad&+ 2\left(\  \vcenter{\hbox{\includegraphics[scale=0.10]{diagrams/G2rel14.eps}}}+\vcenter{\hbox{\includegraphics[scale=0.10]{diagrams/G2rel15.eps}}}+\vcenter{\hbox{\includegraphics[scale=0.10]{diagrams/G2rel16.eps}}}+\vcenter{\hbox{\includegraphics[scale=0.10]{diagrams/G2rel17.eps}}}+\vcenter{\hbox{\includegraphics[scale=0.10]{diagrams/G2rel18.eps}}}\ \right)\nonumber \\[5pt]
    &-\left(\  \vcenter{\hbox{\includegraphics[scale=0.10]{diagrams/G2rel19.eps}}}+\vcenter{\hbox{\includegraphics[scale=0.10]{diagrams/G2rel20.eps}}}+\vcenter{\hbox{\includegraphics[scale=0.10]{diagrams/G2rel21.eps}}}+\vcenter{\hbox{\includegraphics[scale=0.10]{diagrams/G2rel22.eps}}}+\vcenter{\hbox{\includegraphics[scale=0.10]{diagrams/G2rel23.eps}}}\ \right) \label{G2relspec6}
\end{align}
\end{subequations}

\begin{figure}
\begin{center}
    \includegraphics[scale=0.3]{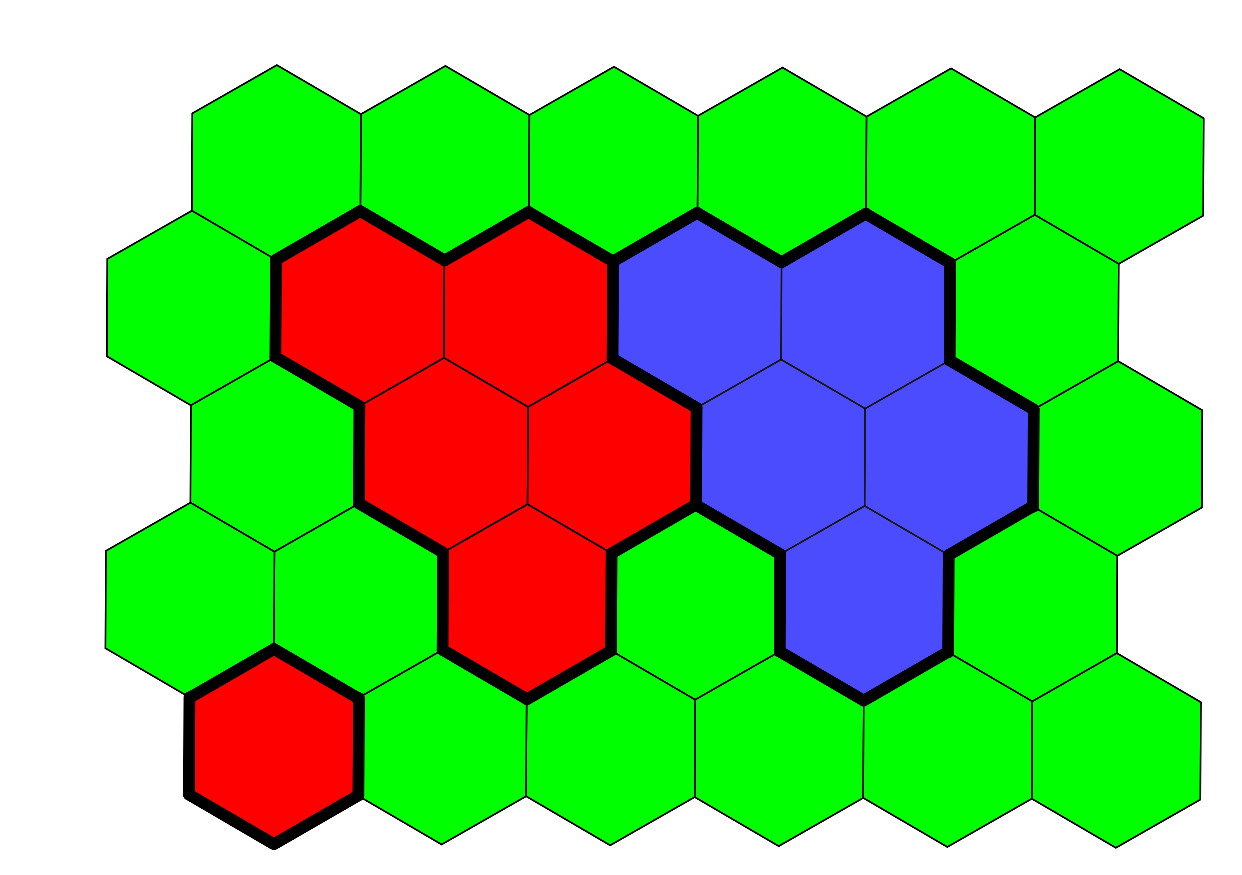} 
    \caption{A spin configuration, with the colours $\{{\rm red},{\rm blue},{\rm green}\}$ representing the spin values $\mathbb{Z}_3 := \{0,1,2\}$.}\label{fig:configspinZ3}
\end{center}
\end{figure}

We will now show that, for any closed simple web $G$, $\chi_{\hat{G}}(3)=3w_{\rm top}(G)$. We begin by sketching a simple intuitive argument and below present a more formal variant using the chromatic algebra. 

We want to show that relations (\ref{G2relspec}) hold
true for domain walls in a 3-colouring problem. To this end we assign colours to the external faces of each relation and check the agreement between weights on the left-
and right-hand sides of each relation. Consider, as an example, relation (\ref{G2relspec5}). Encircling the whole diagram we encounter four domain walls, so the colours on
one or both pairs of opposite external faces have to coincide. In the first case, there is no available colour for the central face on the left-hand side, so this side vanishes.
On the right-hand side, one of the diagrams in the first parenthesis equals $1$ and is compensated by one of the diagrams in the second parenthesis, while the other two diagrams vanish.
In the second case, the left-hand side equals $2$, while one the right-hand side both diagrams in the first parenthesis vanish, while each of the diagrams in the second parenthesis
equals $1$. The other relations are derived similarly.

Now, we move to the more formal argument using the chromatic algebra \cite{fendley2009link}. It will be sufficient to consider the chromatic algebra of degree $0$, denoted $\mathcal{C}_0$, which is defined as follows. Consider the free vector space $\mathcal{F}_0$, spanned by planar graphs possibly containing closed loops, embedded in some simply connected domain. No edges are adjacent to the boundary of the domain, which is the meaning of ``degree~$0$''. $\mathcal{C}_0$ is defined to be the quotient of $\mathcal{F}_0$ by the following local relations:
\begin{enumerate}
 \item[(1)] If $e$ is an edge of a graph $G$ which is not a loop, then $G=G/e- G\setminus e$, where $G/e$ denotes the graph obtained from $G$ by the contraction of $e$.
 \item[(2)] If $G$ contains a loop-edge $e$ (i.e., an edge that connects a vertex to itself%
 \footnote{The standard name for ``loop-edge'' in graph theory is simply {\em loop}, but we already use ``loop'' for what graph theorists would call a {\em cycle}.}), then $G= (Q-1)G\setminus e$.
 \item[(3)] If $G$ contains a 1-valent vertex then $G=0$.\\
\end{enumerate}
Contrarily to webs, we do not consider graphs containing loops without vertices in $\mathcal{C}_0$. The chromatic algebra $\mathcal{C}_0$ depends on a parameter $Q$ which has to be thought as a number of colours. Indeed, it was shown in \cite{fendley2009link} (Proposition 3.4) that a graph $G$ in $\mathcal{C}_0$ is proportional to the empty graph:
\begin{align}
\label{chromaticreduction}
    G=Q^{-1}\chi_{\hat{G}}(Q) \emptyset
\end{align}

For completeness, we recall here the elements of the proof of Proposition 3.4 of \cite{fendley2009link}. Let $e$ be an edge of $G$. Consider $\hat{e}$, the edge crossing $e$ in the dual graph $\hat{G}$. Then, one has $\widehat{G/e}=\hat{G}\setminus \hat{e}$ and $\widehat{G\setminus e}=\hat{G}/\hat{e}$. Relation (1) is then translated to $\hat{G}=\hat{G}\setminus \hat{e} - \hat{G}/\hat{e}$ for the dual graph. This the deletion-contraction relation for the chromatic polynomial---a special case of a similar relation for the $Q$-state Potts model. Relation (3) follows as a $1$-valent vertex in $G$ corresponds to a loop in $\hat{G}$ and there are no proper $Q$-colourings of a graph containing loops. Finally, relation (2) follows because a loop whose interior trivially intersects $G$ corresponds to a 1-valent vertex $v$ in $\hat{G}$ and the number of $Q$-colourings of $\hat{G}$ is $Q-1$ times the number of $Q$-colourings of $\hat{G}\setminus \{v\}$. For a loop whose interior does not intersect $G$ trivially, one can use relation (1) and (3) to reduce the interior of the loop to obtain a collection of nested loops. The interior of innermost of these loops then intersect $G$ trivially. The overall factor of $Q^{-1}$ in \eqref{chromaticreduction} corresponds to the fact that the dual graph of a mere vertex, which has $Q$ ways to be coloured, is the empty graph.

Let us define a map $g$ that sends a web in $\mathcal{FS}p'(G_2)$ to the chromatic algebra of degree $0$, $\mathcal{C}_0$. This map simply identifies any web with its graph in $\mathcal{C}_0$, possibly adding a vertex to a loop if it is one of the components of a web. It is then extended by linearity. We will now show that this map factors through the quotient defined by relations \eqref{G2relspec1}-\eqref{G2relspec6}, i.e. all these relations are satisfied in $\mathcal{C}_0$. The first $3$ relations follow straightforwardly from the relations of $\mathcal{C}_0$. The fourth one is satisfied as there is clearly no $3$ colouring of the dual graph of a graph containing the subgraph of the left hand side. The fifth one follows from repeated application of the relations of $\mathcal{C}_0$. To show that the last one is satisfied, we first remark that the left hand side is zero in $\mathcal{C}_0$ as there is no $3$ colouring of the dual graph of a graph containing the left hand side as a subgraph. On the other hand, by repeated application of the relations of $\mathcal{C}_0$, one has that
\begin{align}
    \vcenter{\hbox{\includegraphics[scale=0.12]{diagrams/G2rel13.eps}}}\quad=\quad&-2\quad \vcenter{\hbox{\includegraphics[scale=0.12]{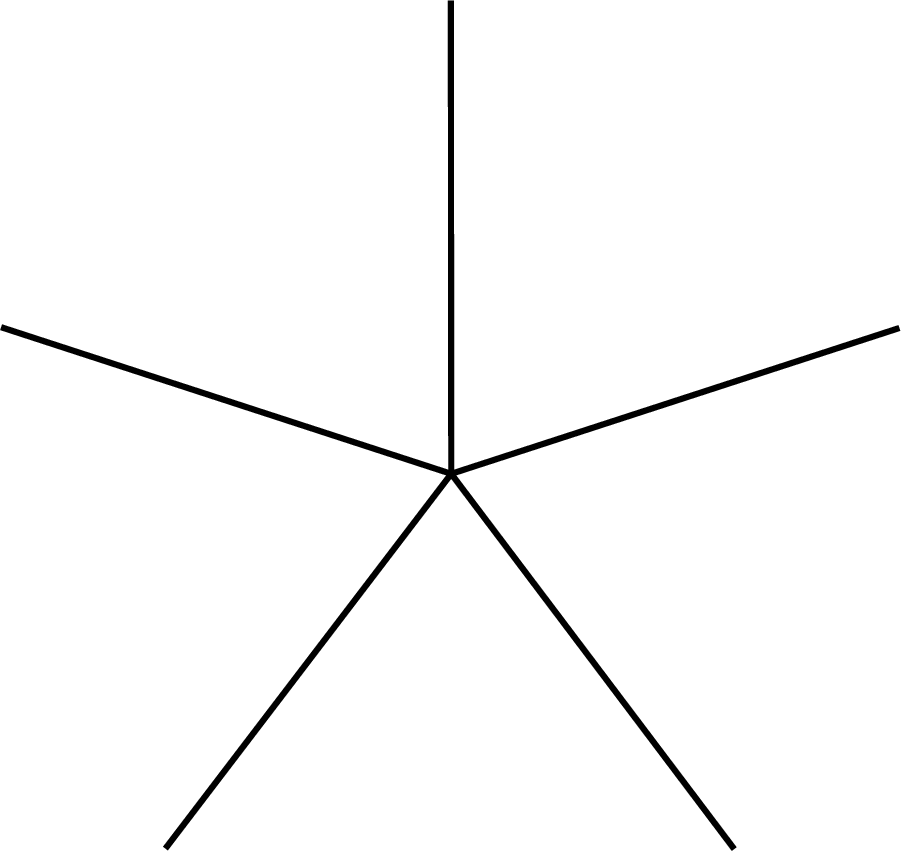}}}  +\left(\  \vcenter{\hbox{\includegraphics[scale=0.12]{diagrams/G2rel19.eps}}}+\text{cycl.\ perm.}\ \right) 
\end{align}
where by cycl(ic) perm(utations) we mean the $4$ graphs obtained by discrete rotations, as in \eqref{G2relspec6}. This linear combination is thus zero in $\mathcal{C}_0$. Applying the relations to the right hand side of \eqref{G2relspec6}, we obtain 
\begin{align}
    10\quad \vcenter{\hbox{\includegraphics[scale=0.12]{diagrams/G2rel24.eps}}} -5\quad \left(\  \vcenter{\hbox{\includegraphics[scale=0.12]{diagrams/G2rel19.eps}}}+\text{cycl.\ perm.}\ \right) 
\end{align}
which is then vanishing as well in $\mathcal{C}_0$. Hence \eqref{G2relspec6} holds in $\mathcal{C}_0$.

We have thus shown that $g$ defines a well defined linear map $\Tilde{g}$ from $\mathcal{S}p'(G_2)$ to $\mathcal{C}_0$. We have then
\begin{align}
    \Tilde{g}(G)&=w_{\rm top}(G)\ \emptyset\\
    &=3^{-1}\chi_{\hat{G}}(3)\ \emptyset
\end{align} leading to $\chi_{\hat{G}}(3)=3w_{\rm top}(G)$ as claimed.%

We thus conclude that:
\begin{align}
    Z_{S_3}=3Z_{G_2}
\end{align}
where domain walls of spin configurations are mapped to webs.

\subsection{Relation with spanning trees}
\label{sec:G2mapping2}
In the previous subsection, we define a $S_3$ symmetric spin model on the triangular lattice with nearest-neighbour and plaquette interactions. Let us focus for a moment on the $S_Q$ symmetric spin model, or $Q$-state Potts models, with nearest-neighbour interactions only. The partition function reads
\begin{align}
    Z_{\text{Potts}}=\sum_{\sigma}\prod_{\langle ij\rangle}(x+(1-x)\delta_{\sigma_i \sigma_j})
\end{align}
where $\sigma : \mathbb T \rightarrow \{1,\ldots, Q\}$. This model is known to be critical when \cite{BTA78}
\begin{align}
\label{Pottscriticalman}
    &Q=v^3+3v^2\\
    &v=x^{-1}-1 \nonumber
\end{align}
We can again express the partition function in terms of the domain walls between spin clusters. These subgraphs of $\mathbb H$ are again understood as simple $G_2$ webs. We thus obtain
\begin{align*}
    Z_{\text{Potts}}=\sum_{G\in \mathcal{K}} x^{N} \chi_{\hat{G}}(Q)
\end{align*}
where $N$ denotes the number of bonds. The naive limit $Q=0$ is not interesting, as $\chi_{\hat{G}}(0)=0$ for any non-empty graph. Instead, we can focus on the polynomial 
\begin{align}
    \kappa_{\hat{G}}(Q)=Q^{-1}\chi_{\hat{G}}(Q)
\end{align}
Then the limit 
\begin{align}
    Z_{\text{tree}}=\lim_{Q\rightarrow 0}Q^{-1}Z_{\text{Potts}}=\sum_{G\in \mathcal{K}} x^{N} \kappa_{\hat{G}}(0)
\end{align} defines a model of spanning trees on $\mathbb T$ \cite{JSS05}. The uniform probability measure on the set of spanning trees is obtained at $x=1$ and is critical by \eqref{Pottscriticalman}. We will now show that $ Z_{\text{tree}}$ is a special case of the $G_2$ web models for
\begin{align*}
    q=i \,, \qquad
    y=i \,.
\end{align*}
The proof is entirely analogous to the one for the $S_3$ spin model. First, we incorporate the vertex fugacity $y$ into the Kuperberg relations to obtain a topological weight given by the modified relations
\begin{subequations}
\label{G2relspecc}
\begin{align}
    \vcenter{\hbox{\includegraphics[scale=0.12]{diagrams/B2rel1.eps}}}\quad=\quad &-1\label{G2relspecc1}\\[5pt]
    \vcenter{\hbox{\includegraphics[scale=0.12]{diagrams/G2rel3.eps}}}\quad=\quad &0\\[5pt]
    \vcenter{\hbox{\includegraphics[scale=0.12]{diagrams/G2rel4.eps}}}\quad=\quad &-2\quad \vcenter{\hbox{\includegraphics[scale=0.12]{diagrams/G2rel5.eps}}}\\[5pt]
    \vcenter{\hbox{\includegraphics[scale=0.12]{diagrams/G2rel6.eps}}}\quad=\quad &-3\vcenter{\hbox{\includegraphics[scale=0.12]{diagrams/G2rel7.eps}}}\\[5pt]
    \vcenter{\hbox{\includegraphics[scale=0.12]{diagrams/G2rel8.eps}}}\quad=\quad& -2\left(\  \vcenter{\hbox{\includegraphics[scale=0.12]{diagrams/G2rel10.eps}}}+\vcenter{\hbox{\includegraphics[scale=0.12]{diagrams/G2rel9.eps}}}\ \right)-\left(\  \vcenter{\hbox{\includegraphics[scale=0.12]{diagrams/G2rel11.eps}}}+\vcenter{\hbox{\includegraphics[scale=0.12]{diagrams/G2rel12.eps}}}\ \right)  \\[5pt]
    \ \vcenter{\hbox{\includegraphics[scale=0.12]{diagrams/G2rel13.eps}}}\quad=\quad&-\left(\  \vcenter{\hbox{\includegraphics[scale=0.10]{diagrams/G2rel14.eps}}}+\vcenter{\hbox{\includegraphics[scale=0.10]{diagrams/G2rel15.eps}}}+\vcenter{\hbox{\includegraphics[scale=0.10]{diagrams/G2rel16.eps}}}+\vcenter{\hbox{\includegraphics[scale=0.10]{diagrams/G2rel17.eps}}}+\vcenter{\hbox{\includegraphics[scale=0.10]{diagrams/G2rel18.eps}}}\ \right)\nonumber \\[5pt]
    &-\left(\  \vcenter{\hbox{\includegraphics[scale=0.10]{diagrams/G2rel19.eps}}}+\vcenter{\hbox{\includegraphics[scale=0.10]{diagrams/G2rel20.eps}}}+\vcenter{\hbox{\includegraphics[scale=0.10]{diagrams/G2rel21.eps}}}+\vcenter{\hbox{\includegraphics[scale=0.10]{diagrams/G2rel22.eps}}}+\vcenter{\hbox{\includegraphics[scale=0.10]{diagrams/G2rel23.eps}}}\ \right) 
\end{align}
\end{subequations}

Straightforward computations show that all the above relations hold in $\mathcal C_0$ for $Q=0$. Hence the topological weight of $G$ is nothing but $\kappa_{\hat{G}}(0)$ and we obtain
\begin{align}
     Z_{\text{tree}}=Z_{G_2}
\end{align}

\section{The $B_2$ web models}
\label{sec:B2}

For the last spider, the $B_2$ one, we can define two kinds of web models and establish relations with two kinds of spin models.

\subsection{The $B_2$ spider relations}
$B_2$ webs are planar graphs embedded in a simply connected domain whose connected components are either closed loops or graphs with trivalent vertices inside the domain or univalent vertices connected to the boundary of the domain. We again denote by open webs, the ones that possess univalent vertices connected to the boundary of the domain and closed webs, otherwise. Edges come again in two types and are called simple and double edges, depicted respectively as 
\begin{center}
    \includegraphics[scale=0.2]{diagrams/SimpleEdge.eps}\qquad \qquad \includegraphics[scale=0.2]{diagrams/DoubleEdge.eps}
\end{center}
Any trivalent vertex is required to be of the form
\begin{equation}
 \label{B2vertex}
    \includegraphics[scale=0.2]{diagrams/Vertex.eps}
\end{equation}

The free vector space spanned by closed $B_2$ webs will be called $\mathcal{FS}p(B_2)$. Then, $\mathcal{S}p(B_2)$ is the quotient of $\mathcal{FS}p(B_2)$ by the following local relations\footnote{Note that our conventions differ from \cite{Kuperberg_1996} by $q\leftrightarrow q^{\frac{1}{2}}$ and a rescaling of vertices by $i$.} \cite{Kuperberg_1996}:
\begin{subequations}
\label{B2rel}
\begin{align}
    \vcenter{\hbox{\includegraphics[scale=0.2]{diagrams/B2rel1.eps}}}&\quad=\quad -(q^4+q^2+q^{-2}+q^{-4})\\[5pt]
    \vcenter{\hbox{\includegraphics[scale=0.2]{diagrams/B2rel2.eps}}}&\quad=\quad q^6+q^2+1+q^{-2}+q^{-6}\\[5pt]
    \vcenter{\hbox{\includegraphics[scale=0.2]{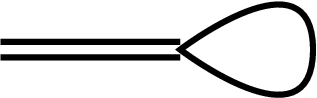}}}&\quad=\quad 0\\[5pt]
    \vcenter{\hbox{\includegraphics[scale=0.2]{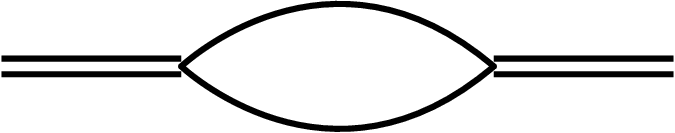}}}&\quad=\quad (q^2+2+q^{-2})\vcenter{\hbox{\includegraphics[scale=0.2]{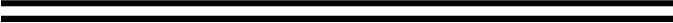}}}\\[5pt]
    \vcenter{\hbox{\includegraphics[scale=0.2]{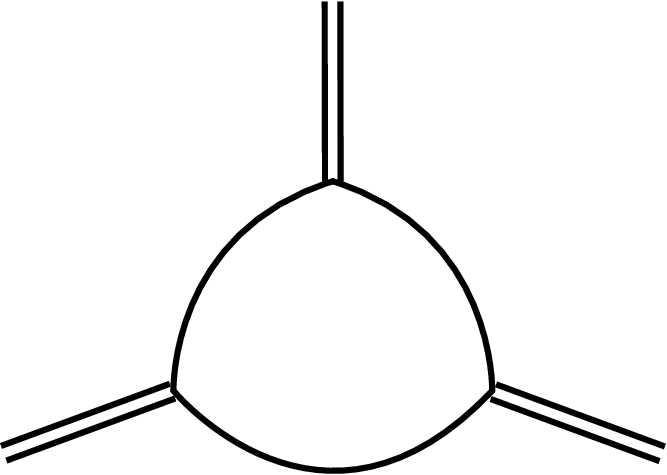}}}&\quad=\quad 0\\[5pt]
    \vcenter{\hbox{\includegraphics[scale=0.2]{diagrams/B2rel7.eps}}}&-\vcenter{\hbox{\includegraphics[scale=0.2]{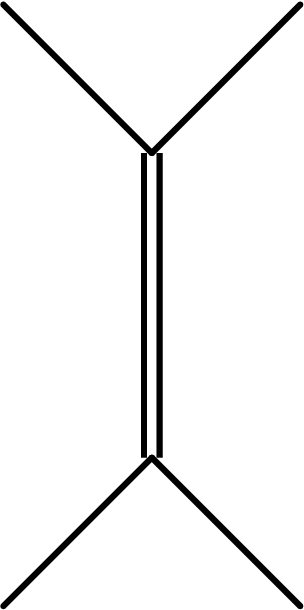}}}\quad=\quad \vcenter{\hbox{\includegraphics[scale=0.2]{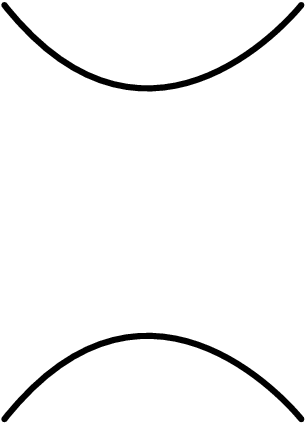}}}-\vcenter{\hbox{\includegraphics[scale=0.2]{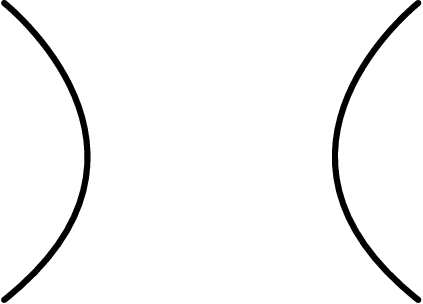}}} \label{B2rel-f}
\end{align}
\end{subequations}
A number can be assigned unambiguously to any closed $B_2$ web thanks to these relations. Indeed, it is a result of \cite{Kuperberg_1996} that any closed web is proportional to the empty one. For a $B_2$ web $G$, we call the proportionality factor, the Kuperberg weight of $G$ and denote it by $w_{\rm K} (G)$.

\subsection{Definition of the models on the hexagonal lattice $\mathbb{H}$}
The configuration space $\mathcal{K}$ is given by $B_2$ webs embedded in $\mathbb{H}$. We say a bond is \textit{simple} (respectively \textit{double}) when it is covered by a simple edge (respectively double). To a configuration $G$, we give a weight (or fugacity) $x_{t;1}$ (respectively $x_{v;1}$) to any tilted (respectively vertical) simple bond and a weight $x_{t;2}$ (respectively $x_{v;2}$) to any tilted (respectively vertical) double bond. We also give a fugacity $y$ to any vertex. This determines the local part of the weight of $G$. The non local part is given by the Kuperberg weight $w_{\rm K} (G)$.

The partition function then reads:
\begin{align}
\label{B2PFHex}
    Z_{B_2}=\sum_{G\in \mathcal{K}} x_{t;1}^{N_{t;1}}x_{v;1}^{N_{v;1}}x_{t;2}^{N_{t;2}}x_{v;2}^{N_{v;2}}y^{N_V}w_{\rm K}(G)
\end{align}
where $N_{t;1}$ (resp.\ $N_{v;1}$) is the number of tilted (resp.\ vertical) simple bonds, $N_{t;2}$ (resp.\ $N_{v;2}$) is the number of tilted (resp.\ vertical) double bonds and $N_V$ is the number of of vertices.

\subsection{Definition of the models on the square lattice $\mathbb{S}$}\label{sec:B2websquare}
Here we define $B_2$ web models on the square lattice. Their definition is motivated by connections with spin models that will be exposed in section \ref{sec:B2mappingsSquare}.

First, let us augment $\mathcal{FS}p(B_2)$ by allowing $4$-valent vertices whose adjacent edges are all simple. Let us denote the vector space spanned by such graphs by $\mathcal{FS}p'(B_2)$. We then quotient this space by the original $B_2$ relations \eqref{B2rel} as well as:
\begin{align}
\label{4valentvertex}
    \vcenter{\hbox{\includegraphics[scale=0.2]{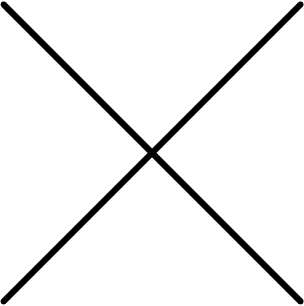}}}\quad=\quad \vcenter{\hbox{\includegraphics[scale=0.2]{diagrams/B2rel7.eps}}}+\vcenter{\hbox{\includegraphics[scale=0.2]{diagrams/B2rel9.eps}}}\quad=\quad \vcenter{\hbox{\includegraphics[scale=0.2]{diagrams/B2rel8.eps}}}+\vcenter{\hbox{\includegraphics[scale=0.2]{diagrams/B2rel10.eps}}}
\end{align}
where the second equality is just a repetition of the defining relation \eqref{B2rel-f}. This four-valent vertex was originally defined by Kuperberg \cite{Kuperberg_1996};
notice that \eqref{B2rel-f} guarantees that it is indeed
invariant under rotations through $\pm \frac{\pi}{2}$.
The quotient space will be denoted by $\mathcal{S}p'(B_2)$.

First define a configuration on $\mathbb{S}\cup \mathcal{E}_b$ to be the replacement of each of its nodes by any of the following local states with the corresponding Boltzmann weights $b_1,\ldots,b_{14}$
\begin{center}
    \includegraphics[scale=0.1]{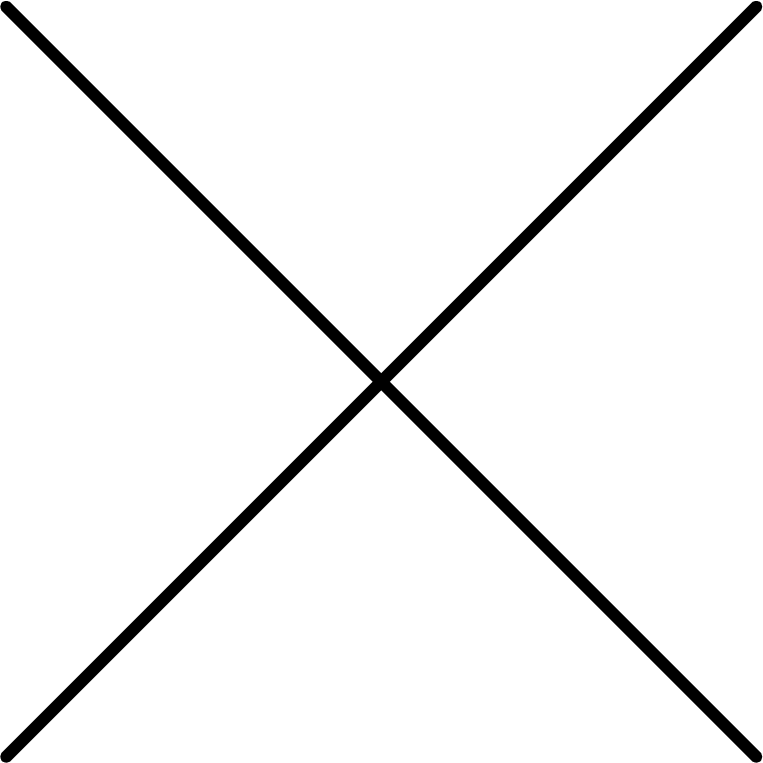} \qquad \qquad \includegraphics[scale=0.1]{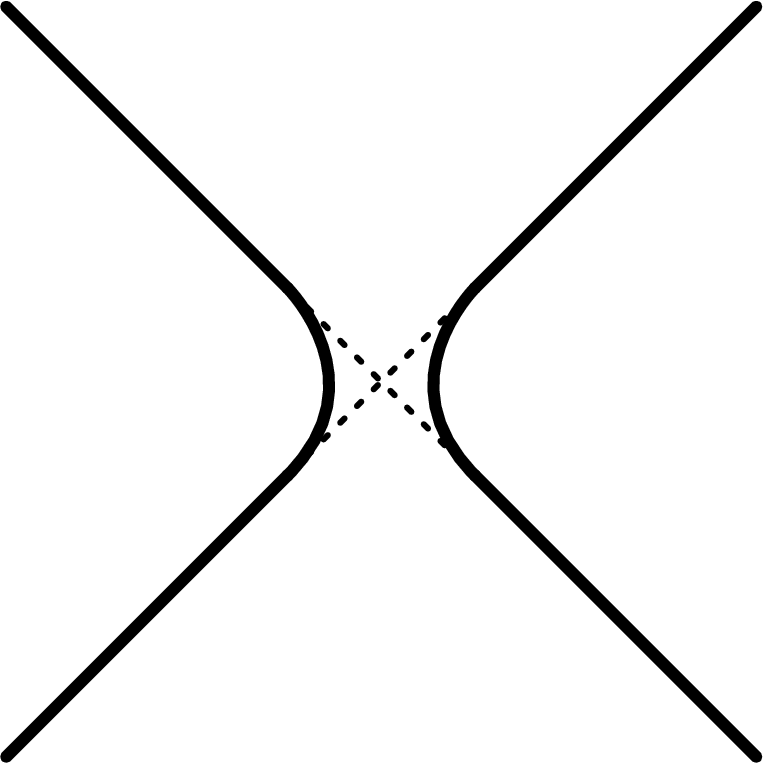}\qquad \qquad \includegraphics[scale=0.1]{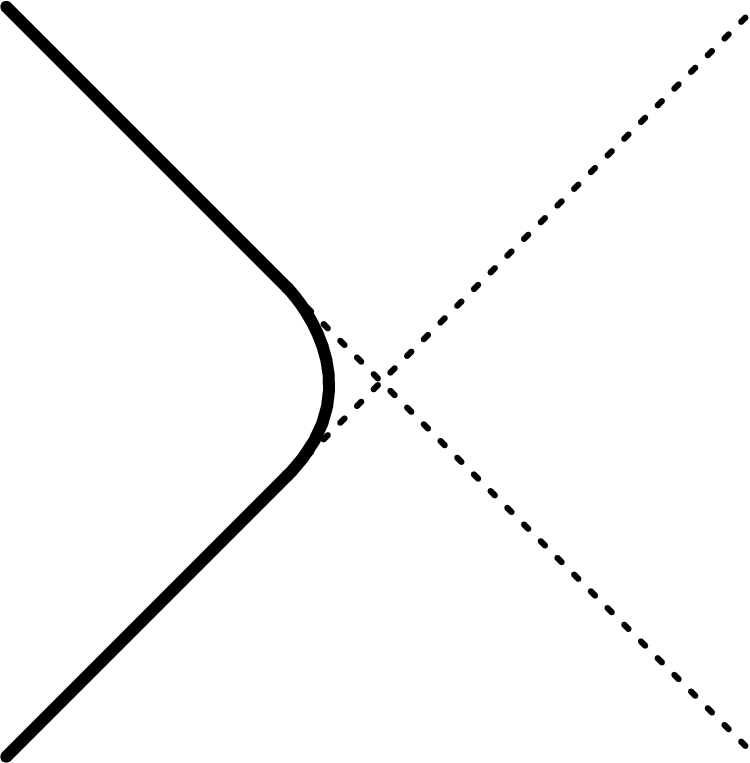}\qquad \qquad \includegraphics[scale=0.1]{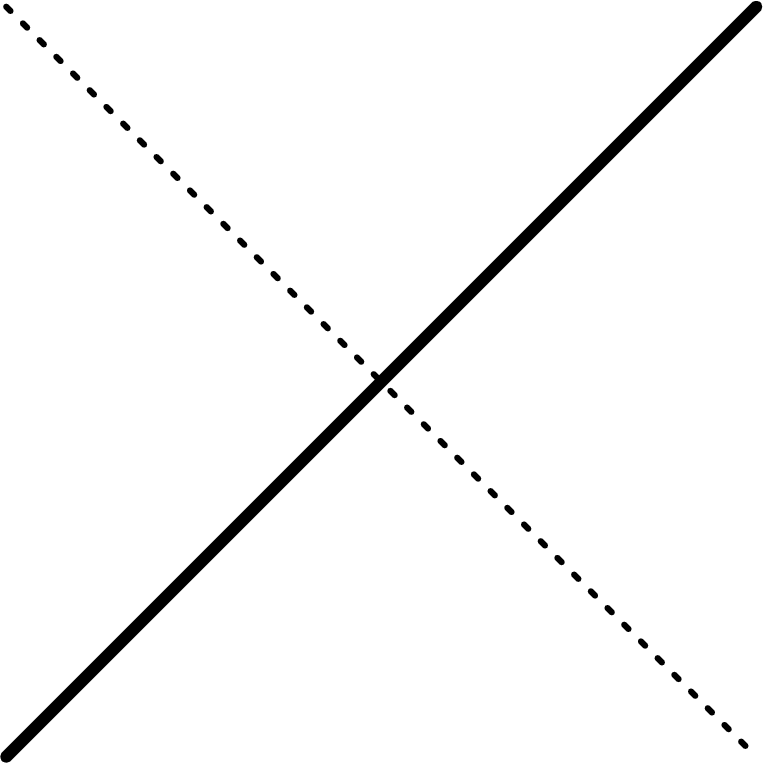}\\[5pt]
    $b_1$\qquad \qquad \qquad \; $b_2$\qquad \qquad \qquad\;  $b_3$\qquad \qquad  \qquad \;$b_4$\\[10pt]
    \includegraphics[scale=0.1]{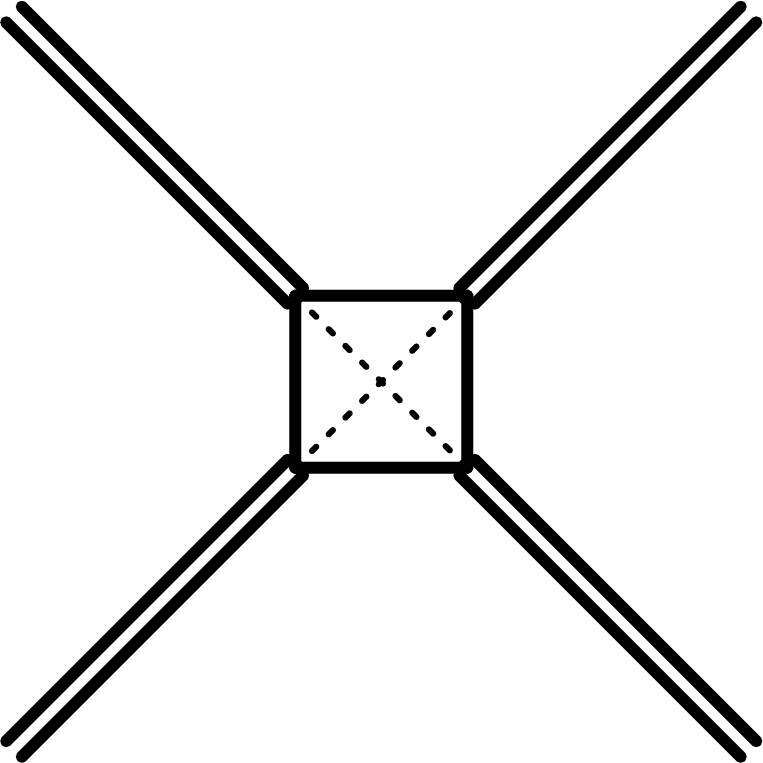} \qquad \qquad \includegraphics[scale=0.1]{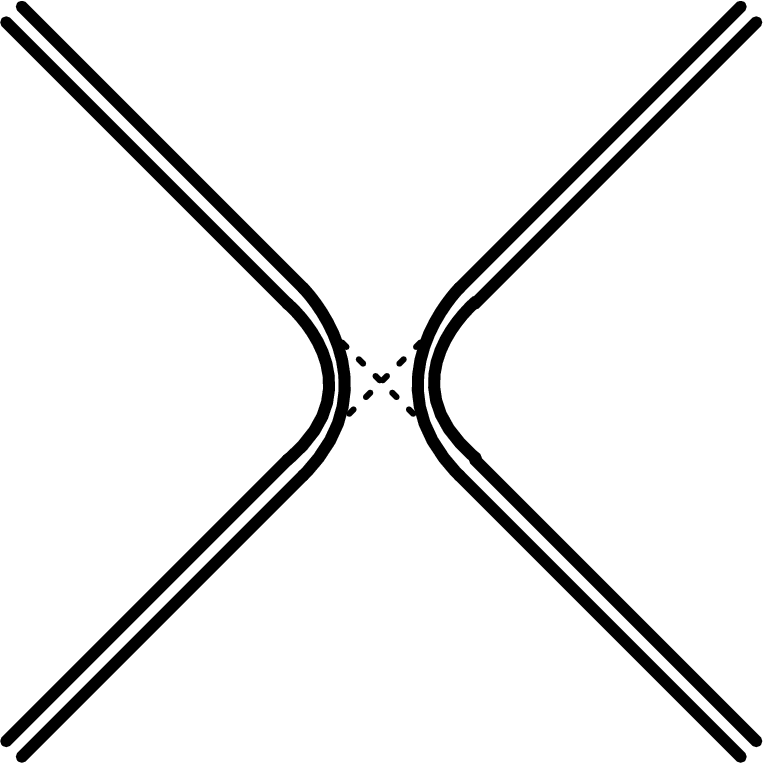}\qquad \qquad \includegraphics[scale=0.1]{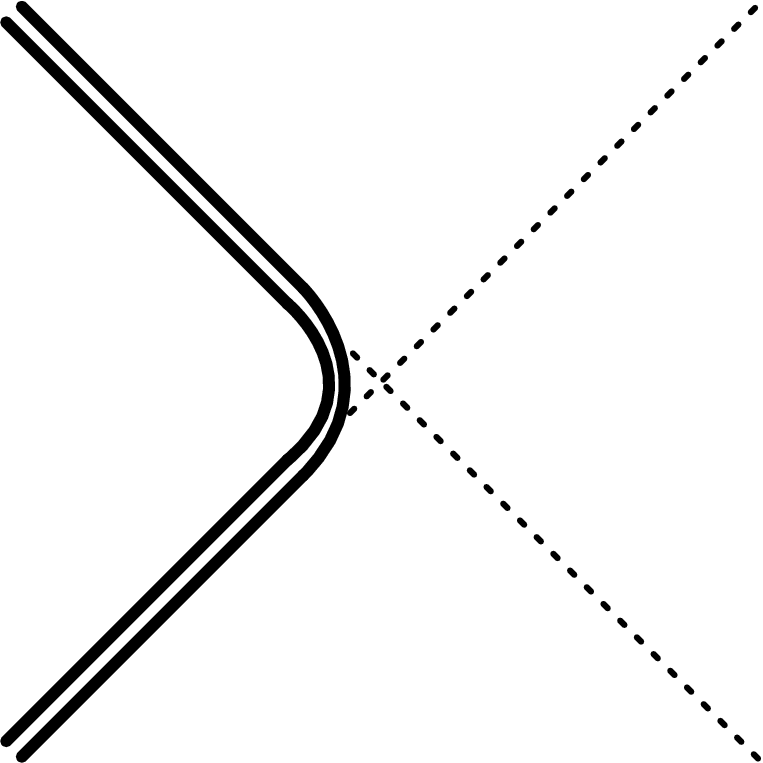}\qquad \qquad \includegraphics[scale=0.1]{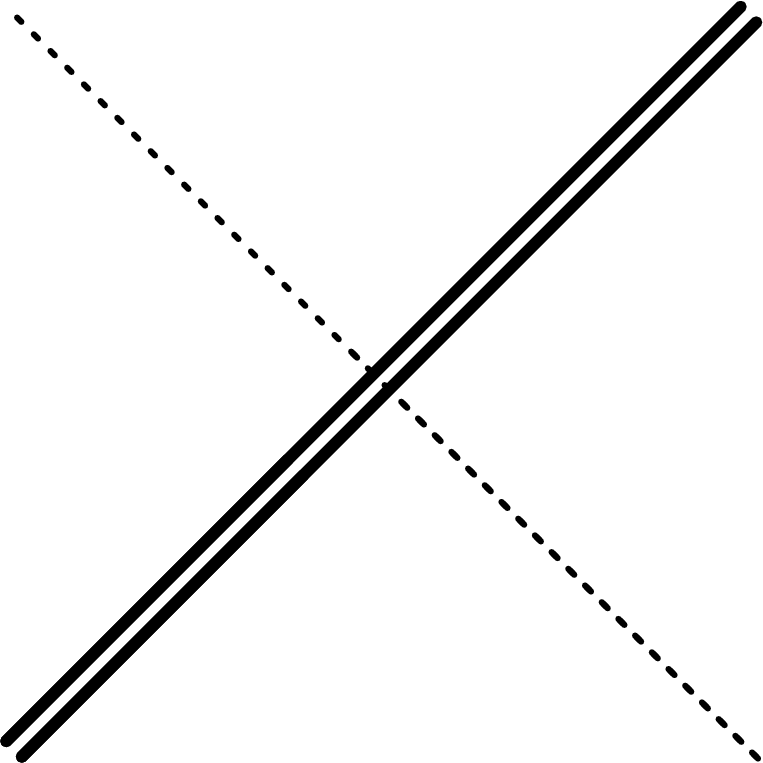}\\[5pt]
    $b_5$\qquad \qquad \qquad \;  $b_6$\qquad \qquad \qquad \; $b_7$\qquad \qquad  \qquad \; $b_8$\\[10pt]
    \includegraphics[scale=0.1]{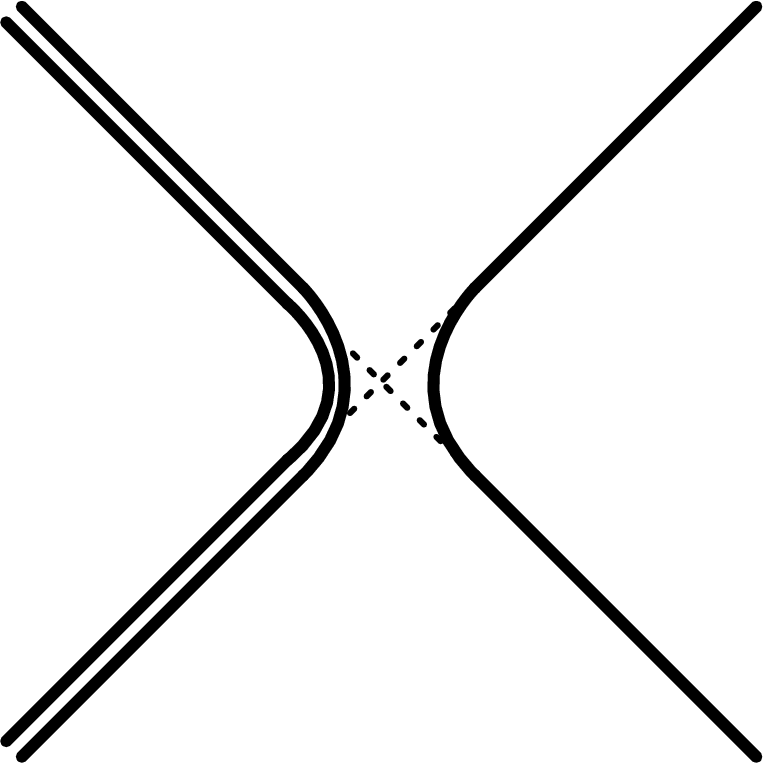} \qquad \qquad \includegraphics[scale=0.1]{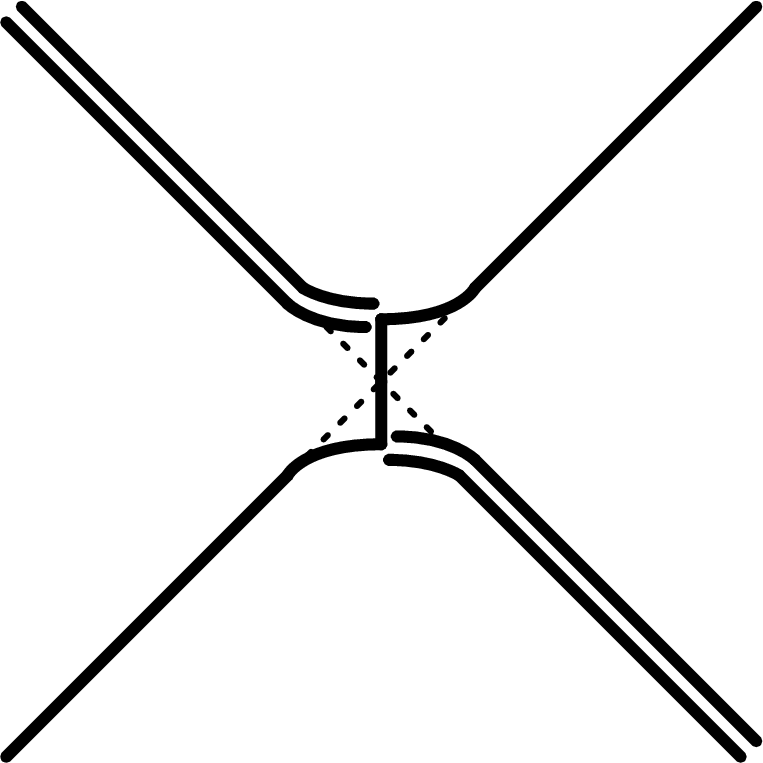}\qquad \qquad \includegraphics[scale=0.1]{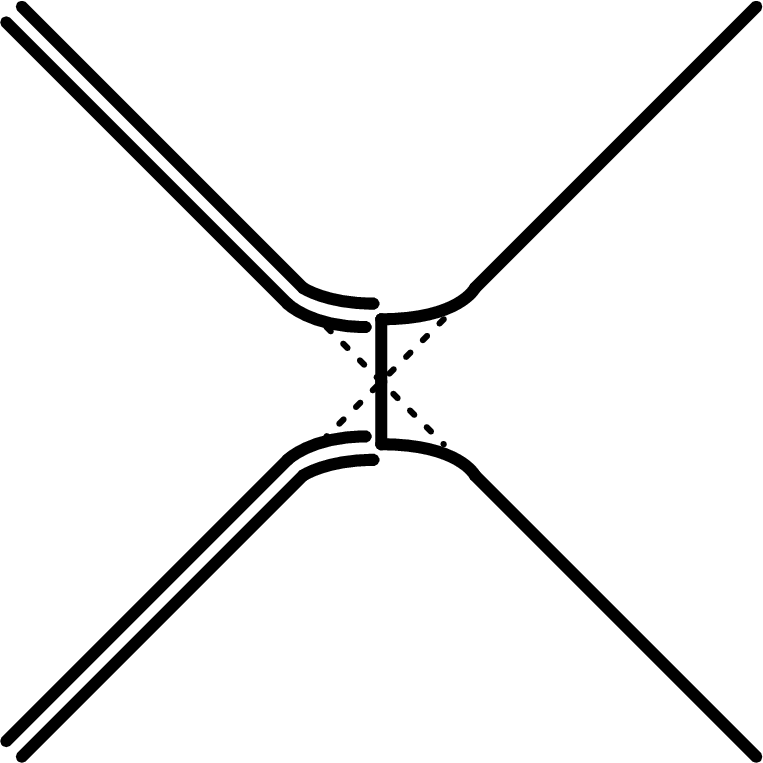}\qquad \qquad \includegraphics[scale=0.1]{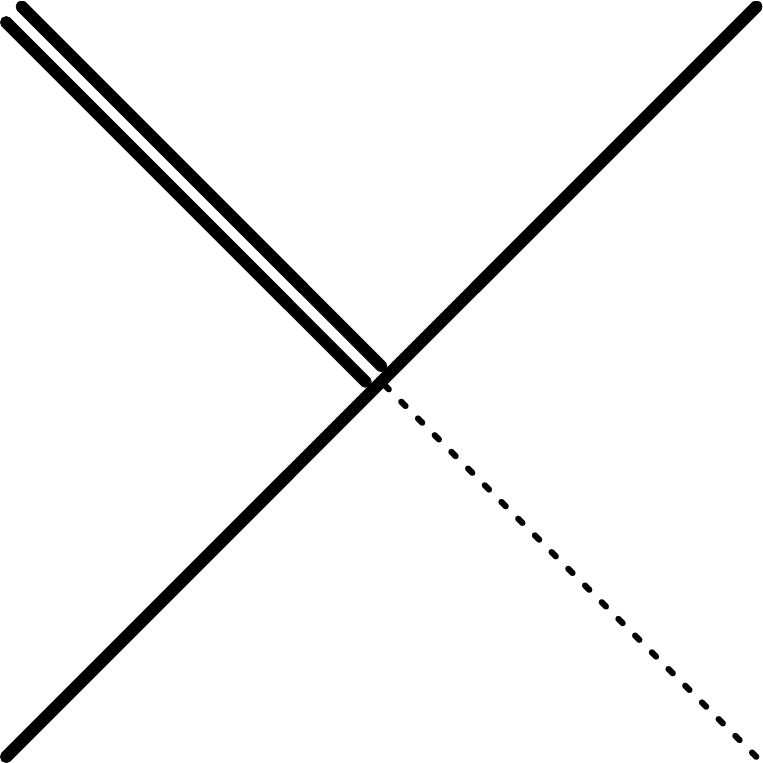}\\[5pt]
    $b_9$ \qquad \qquad \qquad   $b_{10}$\qquad \qquad \qquad  $b_{11}$ \qquad \qquad  \qquad  $b_{12}$\\[10pt]
    \includegraphics[scale=0.1]{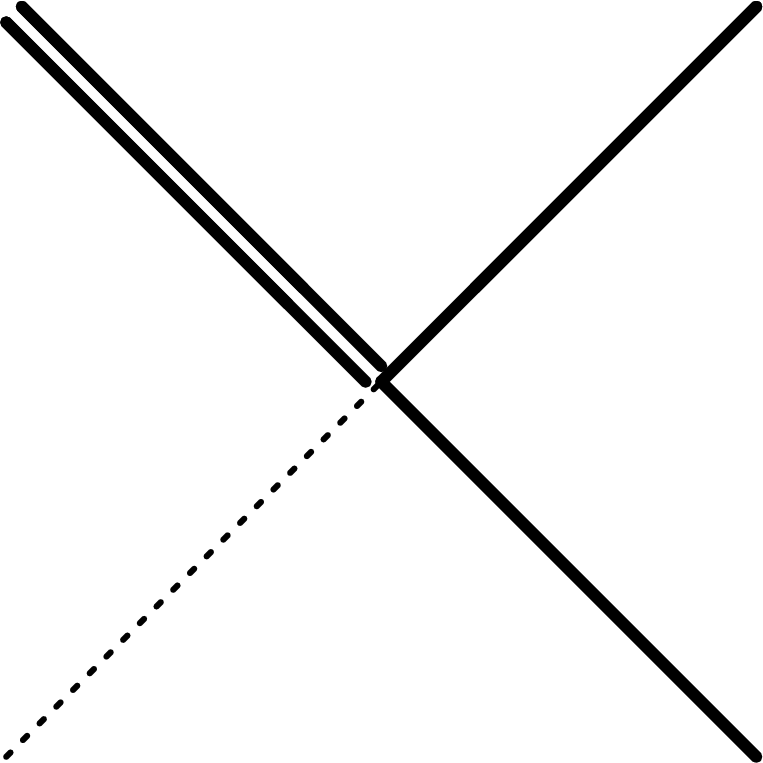} \qquad \qquad \includegraphics[scale=0.1]{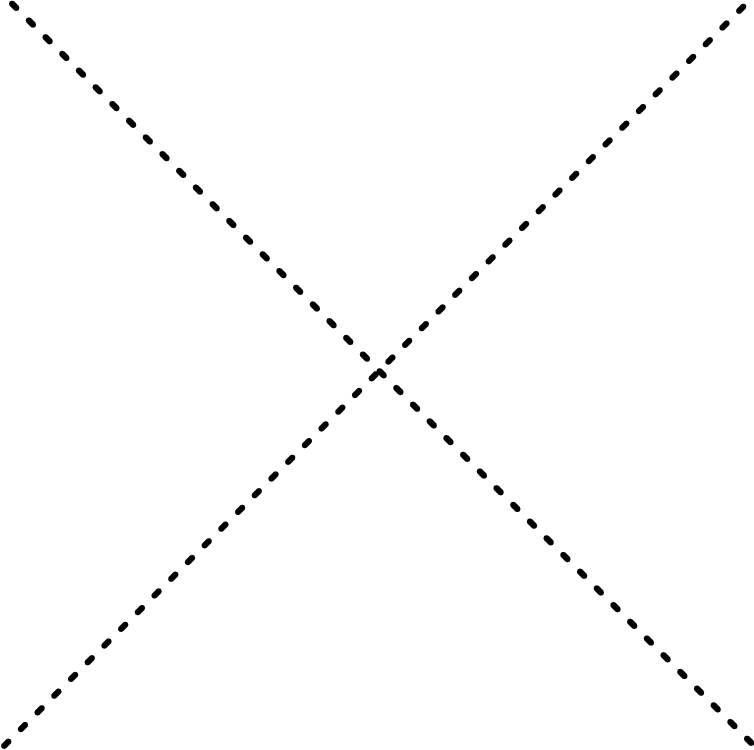}\\[5pt]
    $b_{13}$ \qquad \qquad \qquad   $b_{14}$
\end{center}
where we show only states up to rotations and reflections (the Boltzmann weights are taken to be invariant under these transformations). Moreover, each configuration is
subject to the constraints that any link in $\mathcal{E}_b$ is empty and that any link has the same occupancy (empty, simple or double) with respect to each of the two nodes of $\mathbb{S}$ on which it is incident.
Then delete from a given configuration the set of empty links.
The result is a graph having only trivalent vertices of the type \eqref{B2vertex} and four-valent vertices surrounded
by simple edges. In other words, it is a graph in $\mathcal{FS}p'(B_2)$.

The weight of a configuration $G$ is again the product of a local and a non-local part. The local part is a product of the local Boltzmann weights $b_i$. The non-local part is again $w_{\rm K}(G)$. It is computed by using the relations \eqref{B2rel} once all $4$-valent vertices, i.e. the ones with local weight $b_1$, have been resolved thanks to \eqref{4valentvertex}.

The partition function then reads:
\begin{align}
\label{B2PFSquare}
    Z_{B_2}=\sum_{G\in \mathcal{K}} \left(\prod_{i=1}^{14} b_i^{N_i}\right)w_{\rm K}(G)
\end{align}
where  $N_i$ is the number of local patterns of weight $b_i$.

Remark that we could have introduced models on the square lattice for the $A_2$ and $G_2$ cases as well. Here we consider only the hexagonal lattice for the latter as the integrable solutions described below contain the hexagonal lattice models with {\em isotropic} local weights. In contrast, this is not the case in the $B_2$ case which is why we defined the model on the hexagonal lattice with bond fugacities depending on the bond orientation. We can however define the above square lattice model with isotropic local weights for which there are integrable points. Finally, note that the local weights in the square lattice model are less constrained than in the hexagonal lattice case where they factorise in terms of bond and vertex fugacities. We could have defined the $b_i$s as products of bond and vertex fugacities but, again, the more general non-factorised local weights $b_i$ are needed in order for the model to contain integrable points.

\subsection{Relation with $S_3$ and $D_4$ spin models}
\label{sec:B2mappings}
In this section, we show how the $B_2$ web models for some specific values of $q$ are equivalent, at the level of partition functions, to three- and four-state spin models
with global symmetries $S_3$ and $D_4$, respectively. In both cases the webs are identified with the corresponding spin clusters.

\subsubsection{$B_2$ webs in $\mathbb{H}$ and the $D_4$ spin model}

Consider the lattice dual to $\mathbb{H}$ embedded in the strip (respectively the cylinder), that is, a triangular lattice $\mathbb{T}$ with one (respectively two) point at infinity. 

We begin by formulating the $D_4$ spin model in terms of its domain walls. Consider spins $\{\sigma_i,\ i\in \mathbb{T}\}$ taking values in $\mathbb{Z}_4\cong \{0,1,2,3\}$. We define nearest neighbours interactions $x_{v,|\sigma_i-\sigma_j|}$ (respectively $x_{t,|\sigma_i-\sigma_j|}$) for two neighbouring sites $i$ and $j$ horizontally separated (respectively diagonally separated). Here $|\sigma_i-\sigma_j|$ is to be understood modulo $4$ and we normalise interactions such that $x_0=1$. Hence, the nearest neighbour interaction term contains four parameters $x_{t;1}$, $x_{v;1}$, $x_{t;2}$ and $x_{v;2}$. We also add a plaquette interaction term 
\begin{align*}
    p_{\sigma_i,\sigma_j,\sigma_k}=&1+y\left(\delta_{\sigma_i,\sigma_j\pm1}\delta_{\sigma_j,\sigma_k\pm1}\delta_{\sigma_k,\sigma_i\pm2}+\delta_{\sigma_i,\sigma_j\pm1}\delta_{\sigma_j,\sigma_k\pm2}\delta_{\sigma_k,\sigma_i\pm1}+\delta_{\sigma_i,\sigma_j\pm2}\delta_{\sigma_j,\sigma_k\pm1}\delta_{\sigma_k,\sigma_i\pm1}-1\right)
\end{align*}
which preserves $D_4$ symmetry.
The partition function of the model reads
\begin{align}
    Z_{D_4}=\sum_{\sigma}\left(\prod_{\langle ij\rangle^t}x_{t,|\sigma_i-\sigma_j|}\right)\left(\prod_{\langle ij\rangle^v}x_{v,|\sigma_i-\sigma_j|}\right)\left(\prod_{\langle ijk\rangle}p_{\sigma_i,\sigma_j,\sigma_k}\right)
\end{align}
where $\langle ij\rangle^t$ (respectively $\langle ij\rangle^v$) denotes nearest neighbour pairs of sites diagonally separated (respectively horizontally separated).

Now we will rewrite the partition function in terms of its domain walls. Consider a spin configuration $\{\sigma_i,\ i\in \mathbb{T}\}$. For two neighbouring spins $\sigma_i$ and $\sigma_j$, if $|\sigma_i-\sigma_j|=1$, we draw a simple bond on the link of $\mathbb{H}$ separating the two spins. If $|\sigma_i-\sigma_j|=2$, we draw a double bond, whereas if $|\sigma_i-\sigma_j|=0$ we let the link empty. We obtain this way a Kuperberg $B_2$ web $G$ embedded in $\mathbb{H}$, i.e. $G\in \mathcal{K}$. Clearly, the mapping is onto and many to one. That is, any $G\in \mathcal{K}$ is reached as a domain wall but different spin configurations may have the same set of domain walls $G$. Note that $G$ contains, not only the information about domain walls between different spins but also what type of difference there is between spins, i.e a difference of $\pm 1$ or $\pm 2$. 

All configuration having the same domain wall $G$ have the same weight. Hence, in order to write the partition function, it suffices to count how many spin configurations have the same domain walls. For later convenience we define this number in the following way. For a graph $G$ whose connected components can be closed loops, such that all of its edges (or loops) are simple or double edges, consider the graph dual to $G$ having its edges labelled by $1$ (respectively $2$) if they cross simple edges (respectively double edges). We say that its edges are of type $1$ or type $2$. We call such graphs \textit{decorated}. Remark that $G$ could be a $B_2$ web here, but it can be a more general graph with $4$-valent vertices for instance. We will denote the dual graph of $G$ by $\hat{G}$ and we stress that we consider $\hat{G}$ as a decorated graph. We call a \textit{proper colouring} of a decorated graph $H$, a colouring of its vertices with colours in $\mathbb{Z}_4$ such that two colours connected by an edge of type $1$ (respectively type $2$) differ by $\pm 1$ modulo $4$ (respectively $\pm 2$ modulo $4$). We denote the number of proper colourings of $H$ by $\psi_H$. 

It is clear that, given a domain wall configuration $G\in \mathcal{K}$, the number of spin configurations that have $G$ as its domain wall configuration is equal to $\psi_{\hat{G}}$. Hence the partition function of the spin model can be written as:
\begin{align}
    Z_{D_4}=\sum_{G\in \mathcal{K}} x_{t;1}^{N_{t;1}}x_{v;1}^{N_{v;1}}x_{t;2}^{N_{t;2}}x_{v;2}^{N_{v;2}}y^{N_V}\psi_{\hat{G}}
\end{align}

We will now show that $\psi_{\hat{G}}=4w_{\rm K}(G)$ when $q=e^{i\frac{\pi}{4}}$, establishing the claimed equivalence. First, observe that we can extend by linearity the map $\psi$ to $\mathcal{FS}p(B_2)$ obtaining a linear form. We now claim that this map factors through the relations \eqref{B2rel} to a well defined map on $\mathcal{S}p(B_2)$. For $q=e^{i\frac{\pi}{4}}$, the $B_2$ relations read:
\begin{subequations}
\begin{align}
    \vcenter{\hbox{\includegraphics[scale=0.2]{diagrams/B2rel1.eps}}}\quad-2\quad&=\quad 0\\[5pt]
    \vcenter{\hbox{\includegraphics[scale=0.2]{diagrams/B2rel2.eps}}}\quad-1\quad&=\quad 0\\[5pt]
    \vcenter{\hbox{\includegraphics[scale=0.2]{diagrams/B2rel3.eps}}}\quad&=\quad 0\\[5pt]
    \vcenter{\hbox{\includegraphics[scale=0.2]{diagrams/B2rel4.eps}}}\quad-2\quad \vcenter{\hbox{\includegraphics[scale=0.2]{diagrams/B2rel5.eps}}}\quad&=\quad 0\\[5pt]
    \vcenter{\hbox{\includegraphics[scale=0.2]{diagrams/B2rel6.eps}}}\quad&=\quad 0\\[5pt]
    \vcenter{\hbox{\includegraphics[scale=0.2]{diagrams/B2rel7.eps}}}-\vcenter{\hbox{\includegraphics[scale=0.2]{diagrams/B2rel8.eps}}}-\quad \vcenter{\hbox{\includegraphics[scale=0.2]{diagrams/B2rel10.eps}}}+\vcenter{\hbox{\includegraphics[scale=0.2]{diagrams/B2rel9.eps}}}\quad&=\quad 0
\end{align}
\end{subequations}
We must verify that, for all relations, the left hand side is in the kernel of $\psi$. It is clear that this holds for the first $5$ relations. To show that it holds for the last one, we can extend the definition of $\psi$ to $\mathcal{FS}p'(B_2)$ and show that it factors to a well defined map on $\mathcal{S}p'(B_2)$. We thus have to show that the linear combinations
\begin{subequations}\label{B2Z4proof1}
\begin{align}
        &\vcenter{\hbox{\includegraphics[scale=0.2]{diagrams/B2rel11.eps}}}\quad-\quad \vcenter{\hbox{\includegraphics[scale=0.2]{diagrams/B2rel7.eps}}}-\vcenter{\hbox{\includegraphics[scale=0.2]{diagrams/B2rel9.eps}}}\\[5pt]
        &\vcenter{\hbox{\includegraphics[scale=0.2]{diagrams/B2rel11.eps}}}\quad-\quad \vcenter{\hbox{\includegraphics[scale=0.2]{diagrams/B2rel8.eps}}}-\vcenter{\hbox{\includegraphics[scale=0.2]{diagrams/B2rel10.eps}}}
\end{align}
\end{subequations}
are in the kernel of $\psi$. The proof being similar for the two expressions, let us detail it for the first one. Consider three webs $G_1$, $G_2$, $G_3$ that are the same except inside a disk where they look like:
\begin{subequations}
\begin{align}
        G_1=\quad&\vcenter{\hbox{\includegraphics[scale=0.3]{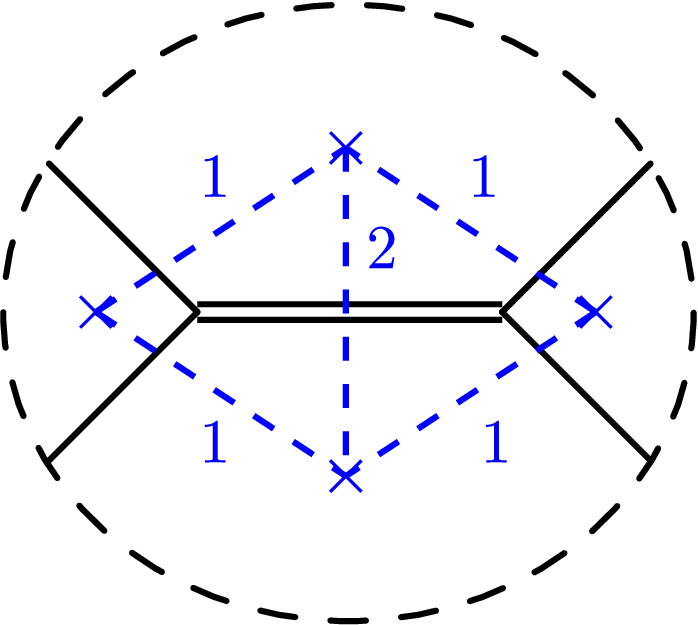}}}\\[5pt]
        G_2=\quad&\vcenter{\hbox{\includegraphics[scale=0.3]{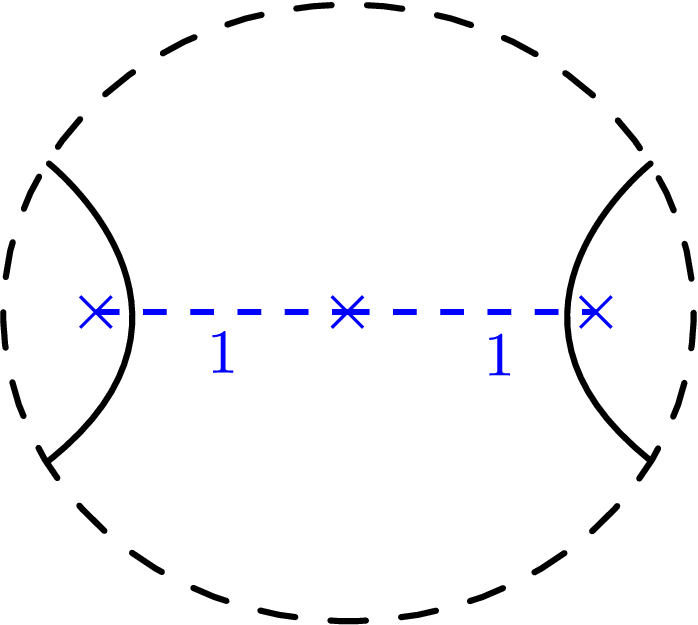}}}\\[5pt]
        G_3=\quad&\vcenter{\hbox{\includegraphics[scale=0.3]{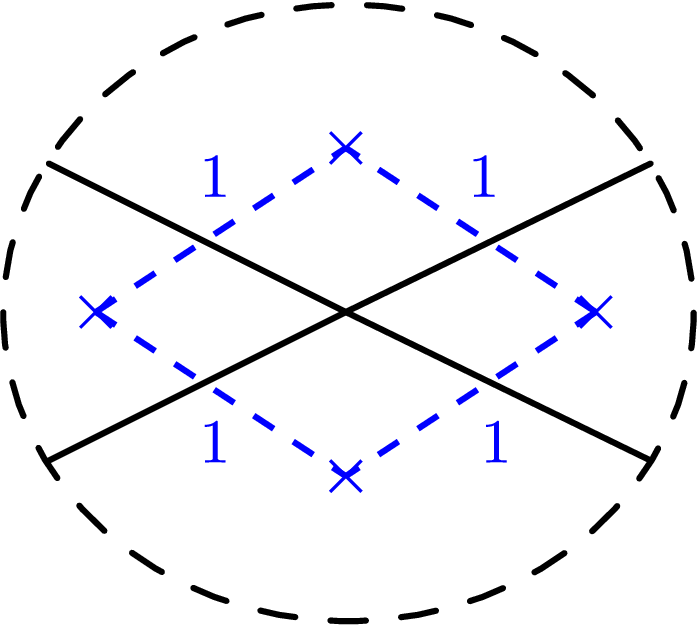}}}
\end{align}
\end{subequations}
where we have drawn in blue the parts of the dual graphs $\hat{G}_1$, $\hat{G}_2$ and $\hat{G}_3$ that are totally contained inside the disk. It is understood that there could be blue edges connecting vertices inside the disk to vertices outside the disk. It is clear that the number of proper colourings of $\hat{G}_3$ is the sum of the number of proper colourings where the top and bottom vertices are the same and the number of proper colourings where the the vertices are different. Here, different implies differing by $\pm 2$, hence we have $\psi_{\hat{G}_3}=\psi_{\hat{G}_1}+\psi_{\hat{G}_2}$. This shows that \eqref{B2Z4proof1} is inside the kernel of $\psi$.

Since $\psi$ is a well defined linear form on $\mathcal{S}p'(B_2)$ and every web $G$ is proportional to the empty web by a factor $w_{\rm K}(G)$, we have that
\begin{align}
    \psi_{\hat{G}}=w_{\rm K}(G)\psi_{\hat{\emptyset}}=4w_{\rm K}(G)
    \label{psihatB2}
\end{align}
We thus have that 
\begin{align}
    Z_{D_4}=4Z_{B_2}
\end{align}
where domain walls and webs are identified in the mapping.

\subsubsection{$B_2$ webs in $\mathbb{S}$, a $D_3$ and a $D_4$ spin model}
\label{sec:B2mappingsSquare}
Let us define a $D_n$ spin model, with $n=3$ or $4$, on the lattice $\mathbb{S}^*$ dual to the square lattice $\mathbb{S}$. The local Boltzmann weights $p_{\sigma_i,\sigma_j,\sigma_k,\sigma_l}$ of the model depend of the values of the spins $\sigma_i$, $\sigma_j$, $\sigma_k$, $\sigma_l\in \mathbb{Z}_n$ (clockwise) cyclically ordered around a plaquette $\langle ijkl \rangle$, in the following way
\begin{align*}
    p_{\sigma_i,\sigma_j,\sigma_k,\sigma_l}=&b_1\delta_{\sigma_i,\sigma_j\pm 1}\delta_{\sigma_j,\sigma_k\pm 1}\delta_{\sigma_k,\sigma_l\pm 1}\delta_{\sigma_l,\sigma_i\pm 1}\\
    &+b_2\delta_{\sigma_i,\sigma_j\pm 1}\delta_{\sigma_j,\sigma_k\pm 1}\delta_{\sigma_k,\sigma_l\pm 1}\delta_{\sigma_l,\sigma_i\pm 1}(\delta_{\sigma_i,\sigma_k}+\delta_{\sigma_j,\sigma_l})\\
    &+b_3(\delta_{\sigma_i,\sigma_j}\delta_{\sigma_j,\sigma_k}\delta_{\sigma_k,\sigma_l\pm 1}\delta_{\sigma_l,\sigma_i\pm 1}+\delta_{\sigma_i,\sigma_j\pm 1}\delta_{\sigma_j,\sigma_k}\delta_{\sigma_k,\sigma_l}\delta_{\sigma_l,\sigma_i\pm 1}\\
    &+\delta_{\sigma_i,\sigma_j\pm 1}\delta_{\sigma_j,\sigma_k\pm 1}\delta_{\sigma_k,\sigma_l}\delta_{\sigma_l,\sigma_i}+\delta_{\sigma_i,\sigma_j}\delta_{\sigma_j,\sigma_k\pm 1}\delta_{\sigma_k,\sigma_l\pm 1}\delta_{\sigma_l,\sigma_i})\\
    &+b_4(\delta_{\sigma_i,\sigma_j\pm 1}\delta_{\sigma_j,\sigma_k}\delta_{\sigma_k,\sigma_l\pm 1}\delta_{\sigma_l,\sigma_i}+\delta_{\sigma_i,\sigma_j}\delta_{\sigma_j,\sigma_k\pm 1}\delta_{\sigma_k,\sigma_l}\delta_{\sigma_l,\sigma_i\pm 1})\\
    &+2b_5\delta_{\sigma_i,\sigma_j\pm 2}\delta_{\sigma_j,\sigma_k\pm 2}\delta_{\sigma_k,\sigma_l\pm 2}\delta_{\sigma_l,\sigma_i\pm 2}\\
    &+b_6\delta_{\sigma_i,\sigma_j\pm 2}\delta_{\sigma_j,\sigma_k\pm 2}\delta_{\sigma_k,\sigma_l\pm 2}\delta_{\sigma_l,\sigma_i\pm 2}(\delta_{\sigma_i,\sigma_k}+\delta_{\sigma_j,\sigma_l})\\
    &+b_7(\delta_{\sigma_i,\sigma_j}\delta_{\sigma_j,\sigma_k}\delta_{\sigma_k,\sigma_l\pm 2}\delta_{\sigma_l,\sigma_i\pm 2}+\delta_{\sigma_i,\sigma_j\pm 2}\delta_{\sigma_j,\sigma_k}\delta_{\sigma_k,\sigma_l}\delta_{\sigma_l,\sigma_i\pm 2}\\
    &+\delta_{\sigma_i,\sigma_j\pm 2}\delta_{\sigma_j,\sigma_k\pm 2}\delta_{\sigma_k,\sigma_l}\delta_{\sigma_l,\sigma_i}+\delta_{\sigma_i,\sigma_j}\delta_{\sigma_j,\sigma_k\pm 2}\delta_{\sigma_k,\sigma_l\pm 2}\delta_{\sigma_l,\sigma_i})\\
    &+b_8(\delta_{\sigma_i,\sigma_j\pm 2}\delta_{\sigma_j,\sigma_k}\delta_{\sigma_k,\sigma_l\pm 2}\delta_{\sigma_l,\sigma_i}+\delta_{\sigma_i,\sigma_j}\delta_{\sigma_j,\sigma_k\pm 2}\delta_{\sigma_k,\sigma_l}\delta_{\sigma_l,\sigma_i\pm 2})\\
    &+b_9(\delta_{\sigma_i,\sigma_j\pm 1}\delta_{\sigma_j,\sigma_k\pm 1}\delta_{\sigma_k,\sigma_l\pm 2}\delta_{\sigma_l,\sigma_i\pm 2}\delta_{\sigma_i,\sigma_k}+\delta_{\sigma_i,\sigma_j\pm 2}\delta_{\sigma_j,\sigma_k\pm 1}\delta_{\sigma_k,\sigma_l\pm 1}\delta_{\sigma_l,\sigma_i\pm 2}\delta_{\sigma_j,\sigma_l}\\
    &+\delta_{\sigma_i,\sigma_j\pm 2}\delta_{\sigma_j,\sigma_k\pm 2}\delta_{\sigma_k,\sigma_l\pm 1}\delta_{\sigma_l,\sigma_i\pm 1}\delta_{\sigma_i,\sigma_k}+\delta_{\sigma_i,\sigma_j\pm 1}\delta_{\sigma_j,\sigma_k\pm 2}\delta_{\sigma_k,\sigma_l\pm 2}\delta_{\sigma_l,\sigma_i\pm 1}\delta_{\sigma_j,\sigma_l})\\
    &+b_{10}(\delta_{\sigma_i,\sigma_j\pm 1}\delta_{\sigma_j,\sigma_k\pm 2}\delta_{\sigma_k,\sigma_l\pm 1}\delta_{\sigma_l,\sigma_i\pm 2}\delta_{\sigma_i,\sigma_k\pm 1}+\delta_{\sigma_i,\sigma_j\pm 1}\delta_{\sigma_j,\sigma_k\pm 2}\delta_{\sigma_k,\sigma_l\pm 1}\delta_{\sigma_l,\sigma_i\pm 2}\delta_{\sigma_j,\sigma_l\pm 1}\\
    &+\delta_{\sigma_i,\sigma_j\pm 2}\delta_{\sigma_j,\sigma_k\pm 1}\delta_{\sigma_k,\sigma_l\pm 2}\delta_{\sigma_l,\sigma_i\pm 1}\delta_{\sigma_i,\sigma_k\pm 1}+\delta_{\sigma_i,\sigma_j\pm 2}\delta_{\sigma_j,\sigma_k\pm 1}\delta_{\sigma_k,\sigma_l\pm 2}\delta_{\sigma_l,\sigma_i\pm 1}\delta_{\sigma_j,\sigma_l\pm 1})\\
    &+b_{11}(\delta_{\sigma_i,\sigma_j\pm 1}\delta_{\sigma_j,\sigma_k\pm 1}\delta_{\sigma_k,\sigma_l\pm 2}\delta_{\sigma_l,\sigma_i\pm 2}\delta_{\sigma_j,\sigma_l\pm 1}+\delta_{\sigma_i,\sigma_j\pm 2}\delta_{\sigma_j,\sigma_k\pm 1}\delta_{\sigma_k,\sigma_l\pm 1}\delta_{\sigma_l,\sigma_i\pm 2}\delta_{\sigma_i,\sigma_k\pm 1}\\
    &+\delta_{\sigma_i,\sigma_j\pm 2}\delta_{\sigma_j,\sigma_k\pm 2}\delta_{\sigma_k,\sigma_l\pm 1}\delta_{\sigma_l,\sigma_i\pm 1}\delta_{\sigma_j,\sigma_l\pm 1}+\delta_{\sigma_i,\sigma_j\pm 1}\delta_{\sigma_j,\sigma_k\pm 2}\delta_{\sigma_k,\sigma_l\pm 2}\delta_{\sigma_l,\sigma_i\pm 1}\delta_{\sigma_i,\sigma_k\pm 1})\\
    &+b_{12}(\delta_{\sigma_i,\sigma_j\pm 1}\delta_{\sigma_j,\sigma_k}\delta_{\sigma_k,\sigma_l\pm 1}\delta_{\sigma_l,\sigma_i\pm 2}+\delta_{\sigma_i,\sigma_j\pm 2}\delta_{\sigma_j,\sigma_k\pm 1}\delta_{\sigma_k,\sigma_l}\delta_{\sigma_l,\sigma_i\pm 1}\\
    &+\delta_{\sigma_i,\sigma_j\pm 1}\delta_{\sigma_j,\sigma_k\pm 2}\delta_{\sigma_k,\sigma_l\pm 1}\delta_{\sigma_l,\sigma_i}+\delta_{\sigma_i,\sigma_j}\delta_{\sigma_j,\sigma_k\pm 1}\delta_{\sigma_k,\sigma_l\pm 2}\delta_{\sigma_l,\sigma_i\pm 1})\\
    &+b_{13}(\delta_{\sigma_i,\sigma_j\pm 1}\delta_{\sigma_j,\sigma_k\pm 1}\delta_{\sigma_k,\sigma_l}\delta_{\sigma_l,\sigma_i\pm 2}+\delta_{\sigma_i,\sigma_j\pm 2}\delta_{\sigma_j,\sigma_k\pm 1}\delta_{\sigma_k,\sigma_l\pm 1}\delta_{\sigma_l,\sigma_i}\\
    &+\delta_{\sigma_i,\sigma_j}\delta_{\sigma_j,\sigma_k\pm 2}\delta_{\sigma_k,\sigma_l\pm 1}\delta_{\sigma_l,\sigma_i\pm 1}+\delta_{\sigma_i,\sigma_j\pm 1}\delta_{\sigma_j,\sigma_k}\delta_{\sigma_k,\sigma_l\pm 2}\delta_{\sigma_l,\sigma_i\pm 1}\\
    &+\delta_{\sigma_i,\sigma_j}\delta_{\sigma_j,\sigma_k\pm 1}\delta_{\sigma_k,\sigma_l\pm 1}\delta_{\sigma_l,\sigma_i\pm 2}+\delta_{\sigma_i,\sigma_j\pm 2}\delta_{\sigma_j,\sigma_k}\delta_{\sigma_k,\sigma_l\pm 1}\delta_{\sigma_l,\sigma_i\pm 1}\\
    &+\delta_{\sigma_i,\sigma_j\pm 1}\delta_{\sigma_j,\sigma_k\pm 2}\delta_{\sigma_k,\sigma_l}\delta_{\sigma_l,\sigma_i\pm 1}+\delta_{\sigma_i,\sigma_j\pm 1}\delta_{\sigma_j,\sigma_k\pm 1}\delta_{\sigma_k,\sigma_l\pm 2}\delta_{\sigma_l,\sigma_i}\\
    &+b_{14}\delta_{\sigma_i,\sigma_j}\delta_{\sigma_j,\sigma_k}\delta_{\sigma_k,\sigma_l}\delta_{\sigma_l,\sigma_i}
\end{align*}
One may check that $p_{\sigma_i,\sigma_j,\sigma_k,\sigma_l}$ respects $D_4$ symmetry.
Note that the factor $2$ in front of $b_5$ is due to the two ways of colouring the inside of the square in the corresponding plaquette.

The partition function then reads
\begin{align}
    Z_{D_4}=\sum_{\{\sigma\}} \prod_{<ijkl>}  p_{\sigma_i,\sigma_j,\sigma_k,\sigma_l}
\end{align}
where the usual nearest-neighbour interactions have now been absorbed into the $b_i$.
When $n=4$, each summand in $p_{\sigma_i,\sigma_j,\sigma_k,\sigma_l}$ can be graphically expressed thanks to the corresponding web in Section \ref{sec:B2websquare} where simple (respectively double) bonds separate spins differing by $\pm 1$ (respectively $\pm 2$). Thus, the only non vanishing summands in the product over all plaquettes $\prod_{<ijkl>}  p_{\sigma_i,\sigma_j,\sigma_k,\sigma_l}$ are given by webs embedded in $\mathbb{S}$ as in Section \ref{sec:B2websquare} . Given such a web $G$, the corresponding summand will be equal to, after summing over spin configurations, 
\begin{align*}
     \left(\prod_{i=1}^{14} b_i^{N_i}\right)\psi_{\hat{G}}
\end{align*}
When $q=e^{i\frac{\pi}{4}}$, we have seen in \eqref{psihatB2} that $\psi_{\hat{G}}=4w_{\rm K}(G)$, hence
\begin{align}
        Z_{D_4}=4Z_{B_2}
\end{align}

When $n=3$, summands in $p_{\sigma_i,\sigma_j,\sigma_k,\sigma_l}$ do not determine uniquely a web as $\sigma \pm 1= \sigma \pm 2$. Yet, if we forbid double-edge configurations by setting
\begin{equation}
 b_i=0 \,, \text{ for } i\in \llbracket 5,13\rrbracket \,,
 \label{nodoubleedge}
\end{equation}
there is again a one-to-one correspondence. The corresponding webs involve only simple edges. Denote by $\mathcal{K}'$ the corresponding set of webs embedded in $\mathcal{S}$. We have that 
\begin{align}
    Z_{D_3}=\sum_{G\in \mathcal{K}'} \left(b_{14}^{N_{14}}\prod_{i=1}^{4} b_i^{N_i}\right)\chi_{\hat{G}}(3)
\end{align}

We will now show that $\chi_{\hat{G}}(3)=3w_{\rm K}(G)$ when $q=e^{i\frac{\pi}{3}}$. 

We define a morphism from $\mathcal{S}p'(B_2)$ at $q=e^{i\frac{\pi}{3}}$ to the chromatic algebra $\mathcal{C}_0$ whose definition was given in Section \ref{sec:G2mapping}. Consider first the map $f$ that sends a web $G$ in $\mathcal{FS}p'(B_2)$ to the graph in $\mathcal{C}_0$ obtained by forgetting the information of edges being simple or double, possibly adding a vertex to a loop if present. Extend $f$ by linearity to $\mathcal{FS}p'(B_2)$. We want to show that $f$ factors through the quotient of $B_2$ relations \eqref{B2rel} and \eqref{4valentvertex} to a map $\Tilde{f}$ from $\mathcal{S}p'(B_2)$ to $\mathcal{C}_0$. We then need to show that the following linear combinations are in the kernel of $f$:
\begin{subequations}
\begin{align}
    \vcenter{\hbox{\includegraphics[scale=0.2]{diagrams/B2rel1.eps}}}&\quad-2\\[5pt]
    \vcenter{\hbox{\includegraphics[scale=0.2]{diagrams/B2rel2.eps}}}&\quad-2\\[5pt]
    \vcenter{\hbox{\includegraphics[scale=0.2]{diagrams/B2rel3.eps}}}&\quad\\[5pt]
    \vcenter{\hbox{\includegraphics[scale=0.2]{diagrams/B2rel4.eps}}}&\quad-\quad \vcenter{\hbox{\includegraphics[scale=0.2]{diagrams/B2rel5.eps}}}\quad\\[5pt]
    \vcenter{\hbox{\includegraphics[scale=0.2]{diagrams/B2rel6.eps}}}&\quad\\[5pt]
    \vcenter{\hbox{\includegraphics[scale=0.2]{diagrams/B2rel7.eps}}}&-\vcenter{\hbox{\includegraphics[scale=0.2]{diagrams/B2rel8.eps}}}-\quad \vcenter{\hbox{\includegraphics[scale=0.2]{diagrams/B2rel10.eps}}}+\vcenter{\hbox{\includegraphics[scale=0.2]{diagrams/B2rel9.eps}}}\quad\\[5pt]
    \vcenter{\hbox{\includegraphics[scale=0.2]{diagrams/B2rel11.eps}}}&\quad-\quad \vcenter{\hbox{\includegraphics[scale=0.2]{diagrams/B2rel7.eps}}}-\vcenter{\hbox{\includegraphics[scale=0.2]{diagrams/B2rel9.eps}}}\\[5pt]
    \vcenter{\hbox{\includegraphics[scale=0.2]{diagrams/B2rel11.eps}}}&\quad-\quad \vcenter{\hbox{\includegraphics[scale=0.2]{diagrams/B2rel8.eps}}}-\vcenter{\hbox{\includegraphics[scale=0.2]{diagrams/B2rel10.eps}}}
\end{align}
\end{subequations}
This follows from a straightforward application of the relations of $\mathcal{C}_0$. Hence, for a given web $G$ we have that 
\begin{align}
    \Tilde{f}(G)&=w_{\rm K}(G)\ \emptyset 
    =3^{-1}\chi_{\hat{G}}(3)\ \emptyset
\end{align}
leading to $\chi_{\hat{G}}(3)=3w_{\rm K}(G)$.

Thus, we have
\begin{align}
        Z_{D_3}=3Z_{B_2}
\end{align}

\section{Transfer matrices}
\label{sec:tm}

From now on, we add to the discussion the $A_1$ case, as it will serve as a simpler and well-known example, the dilute loop model, which will guide the discussion of the rank $2$ web models.
Let $X$ denote one of our algebras of interest, $A_1$, $A_2$, $G_2$ or $B_2$. As in \cite{Lafay:2021wyf}, we define local transfer matrices thanks to the identification given by the spiders between diagrams and intertwiners of quantum group representations\cite{Kuperberg_1996}. To each link of the lattice is associated a local space of state $\mathcal{H}_{X}$ that carries a particular representation of the quantum group, a direct sum of trivial and fundamental representations. The trivial representation $\mathbb{C}$ corresponds to the link not being covered by a web. We denote the vacuum vector $1\in \mathbb{C}$ by $\ket{\ }$. Let us build the row to row transfer matrices by composing smaller, local transfer matrices.
\begin{figure}
\begin{center}
    \includegraphics[scale=0.4]{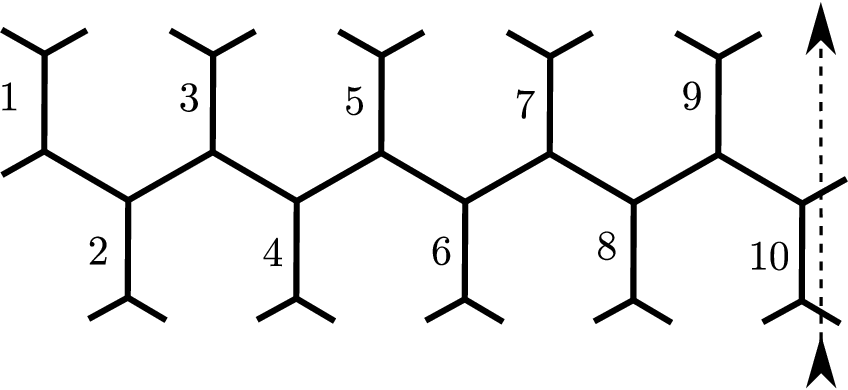}
\end{center}
    \caption{The row to row transfer matrix in the case of periodic boundary conditions with $2L=10$.}
    \label{fig:transfermatrix}
\end{figure}

In the the hexagonal lattice $\mathbb H$ case, we shall call node of type $1$ (respectively type $2$) a node situated at the bottom (respectively top) of a vertical link. Denote by $t^{X}_{(k)}$ the local transfer matrices propagating through a node of type $k \in \{1,2\}$. They are linear maps:
\begin{subequations}
\begin{eqnarray}
  t^{X}_{(1)} &:& \mathcal{H}_{X}\otimes \mathcal{H}_{X} \to \mathcal{H}_{X} \,, \\
  t^{X}_{(2)} &:& \mathcal{H}_{X} \to \mathcal{H}_{X}\otimes \mathcal{H}_{X} \,,
\end{eqnarray}
\end{subequations}
and we use their pictorial notation $\vcenter{\hbox{\includegraphics[scale=0.05]{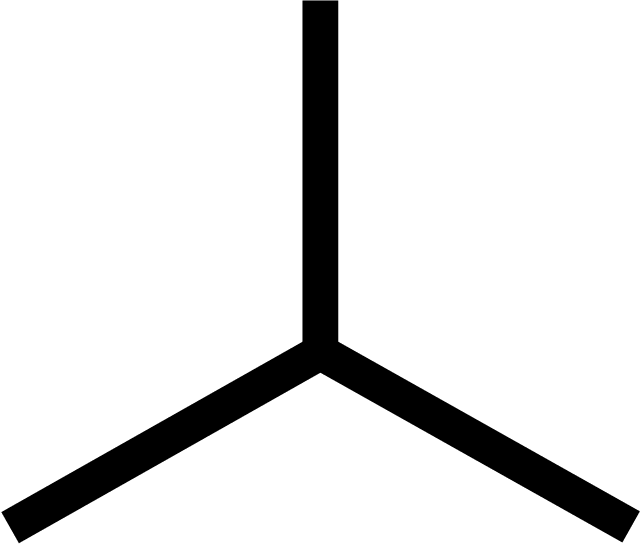}}}$ and $\vcenter{\hbox{\includegraphics[scale=0.05]{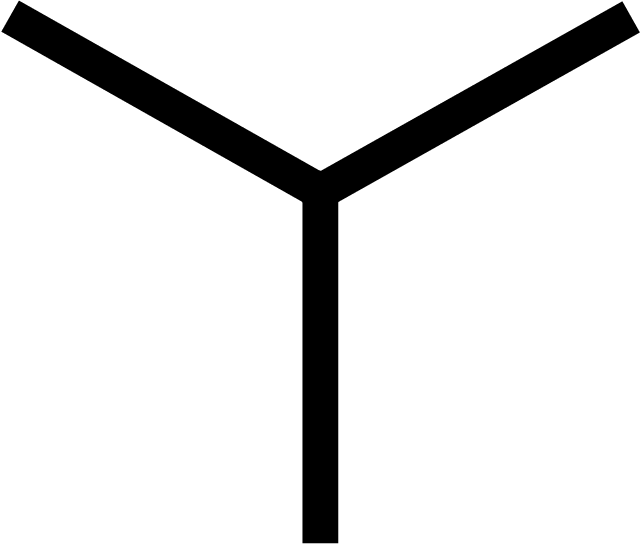}}}$, respectively, in Figure \ref{fig:transfermatrix}. We will show how to obtain these linear maps in the next sections.
Their composition $t^{X}=t^{X}_{(2)}t^{X}_{(1)}$ is a linear map from $\mathcal{H}_{X}\otimes \mathcal{H}_{X}$ to itself (i.e., an endomorphism of $\mathcal{H}_{X}\otimes \mathcal{H}_{X}$).%
\footnote{Remark that $t^{X}$ corresponds to summing over the state of a vertical link, so that a pair of vertices on $\mathbb{H}$ is effectively transformed into a single vertex on a (tilted) square lattice.}
We index the copies $t^{X}_i$ of these operators by their position $i$ in a row as in Figure~\ref{fig:transfermatrix}. The square lattice case is analogous but local transfer matrices are now defined as operators from $\mathcal{H}_{X}\otimes \mathcal{H}_{X}$ to itself from the beginning.

In case of open boundary conditions the row-to-row%
\footnote{Note that with our definition, the row-to-row transfer matrix propagates states through {\em two} rows of the lattice.}
transfer matrix then reads
\begin{align}
\label{looptransfermatrixstrip}
    T_{X}=\left(\prod_{k=0}^{L-1}t^{X}_{2k+1}\right)\left(\prod_{k=1}^{L-1}t^{X}_{2k}\right) \,.
\end{align}
It is an endomorphism of $\mathcal{H}_{X}^{\otimes 2L}$. On appropriate lattices, the partition functions are then recovered as the vacuum expectation values of powers of the
transfer matrix:
\begin{align}
\label{loopPF-TM}
    Z_{X}= \left\langle T_{X}^M \right\rangle \,.
\end{align}
By the vacuum expectation value, we mean the matrix element from $\ket{\ }^{\otimes 2L}$ to itself. To be precise, the right-hand-side of \eqref{loopPF-TM} expresses the partition function $Z_{X}$ on a hexagonal lattice with $2M-2$ rows, because while $T_{X}^M$ builds configurations on a lattice with $2M$ rows, the vertices in the first and last row and their adjacent edges are all constrained to be empty due to our choice of vacuum state.

However, when the web model is embedded in the cylinder we need twisted periodic boundary conditions to give the correct weights to webs that wrap the periodic direction \cite{Lafay:2021wyf}. In the $A_1$ and $A_2$ cases this is obtained by the action of a twist operator
\begin{align}
\label{twistoperator2}
    S_{X}=q^{2H_{\bm{\rho}}}
\end{align}
leading to the following modified transfer matrix \cite{Lafay:2021wyf}
\begin{align}
\label{looptransfermatrixcyl}
    T_{X}=\left(\prod_{k=0}^{L-1}t^{X}_{2k+1}\right)\left(\prod_{k=1}^{L-1}t^{X}_{2k}\right)S_{X}t^{X}_{2L}S_{X}^{-1} \,,
\end{align}
where $S_{X}$ acts non-trivially on site $1$ only.

In the $G_2$ and $B_2$ cases, the transfer matrix is still given by \eqref{looptransfermatrixcyl}, but the twist operator is now
\begin{align}
\label{twistoperator}
    S_{X}=(-1)^{2H_{\bm{\rho}^\vee}} q^{2H_{\bm{\rho}}}
\end{align}
where $\bm{\rho}$ and $\bm{\rho}^\vee$ denote the Weyl vector and the dual Weyl vector respectively (see Appendix \ref{sec:appendix1}).

Remark that the twist operator could also be chosen differently from \eqref{twistoperator2} and \eqref{twistoperator}; see \cite{Lafay:2021scv}. This is useful in order to define modified partition functions that are lattice analogs of two-point functions of electric operators in Coulomb Gas conformal field theories.

\subsection{A reminder on the dilute loop model}
Before going on the discussion of web models, let us remind the well known and similar case of the dilute loop model on the hexagonal lattice. The local transfer matrices of the model can be written graphically as
\begin{align}
\label{LoopLocTM}
    t^{\text{loop}}=&x^2\vcenter{\hbox{\includegraphics[scale=0.15]{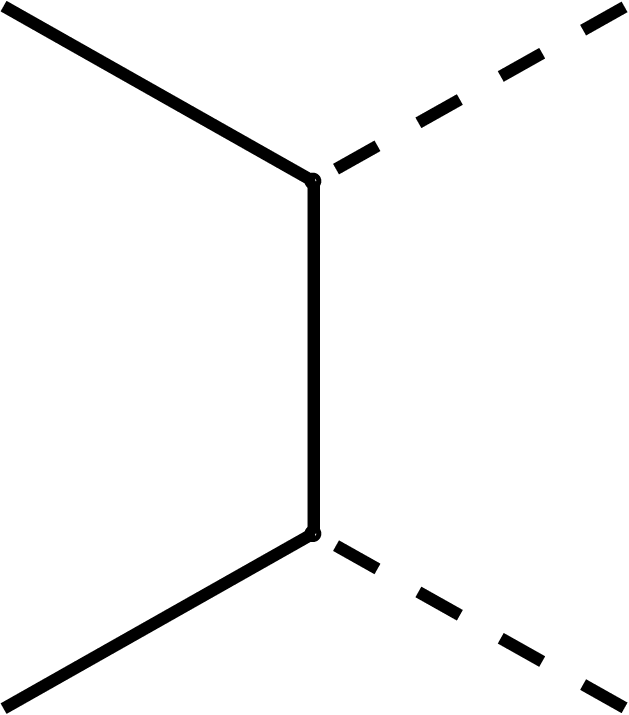}}}+x^2\vcenter{\hbox{\includegraphics[scale=0.15]{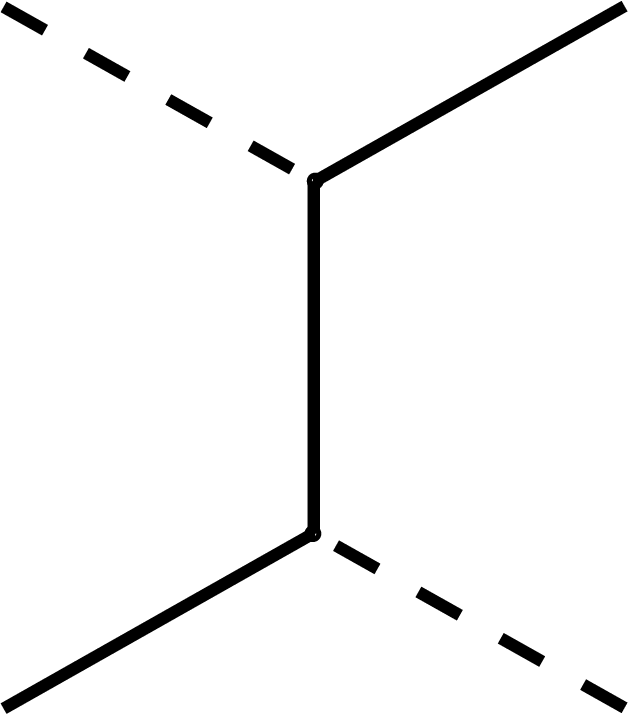}}}+x^2\vcenter{\hbox{\includegraphics[scale=0.15]{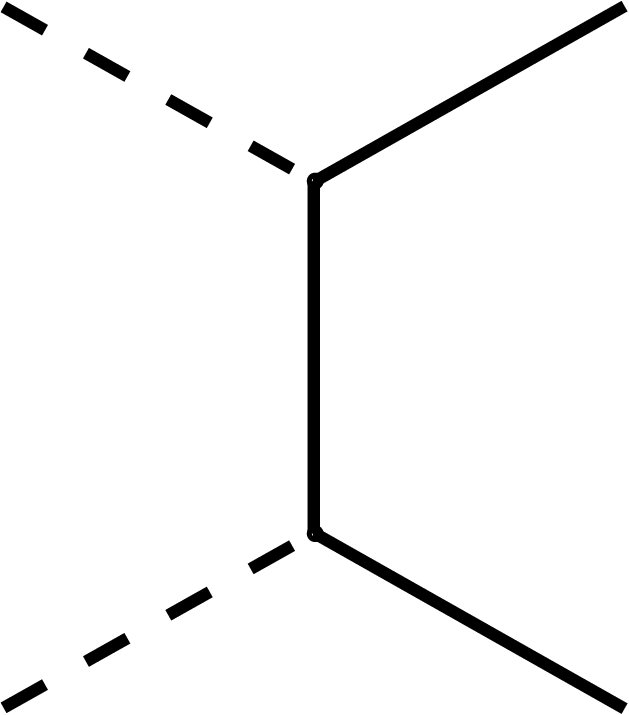}}}+x^2\vcenter{\hbox{\includegraphics[scale=0.15]{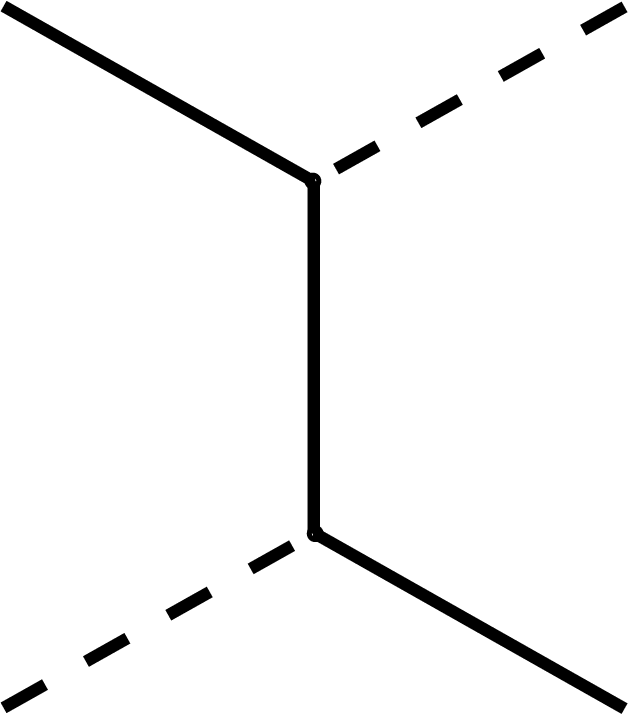}}}\nonumber\\[10pt]
    &+x^2\vcenter{\hbox{\includegraphics[scale=0.15]{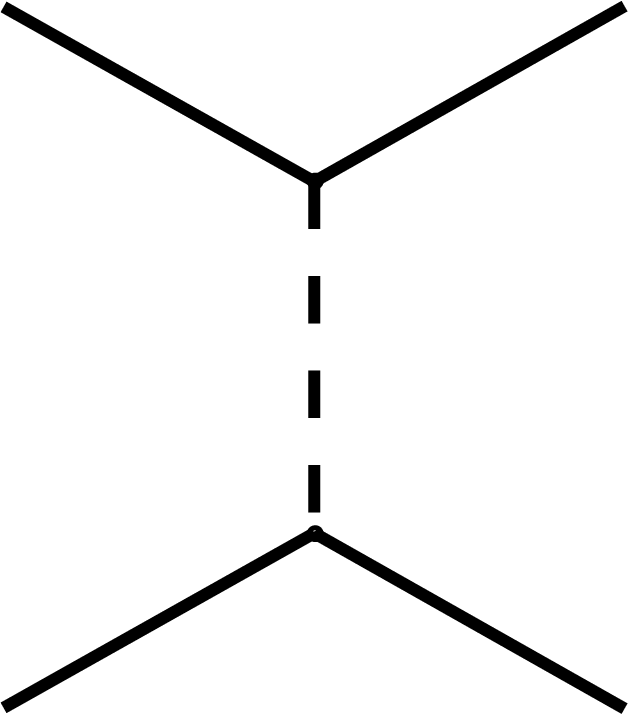}}}+x\vcenter{\hbox{\includegraphics[scale=0.15]{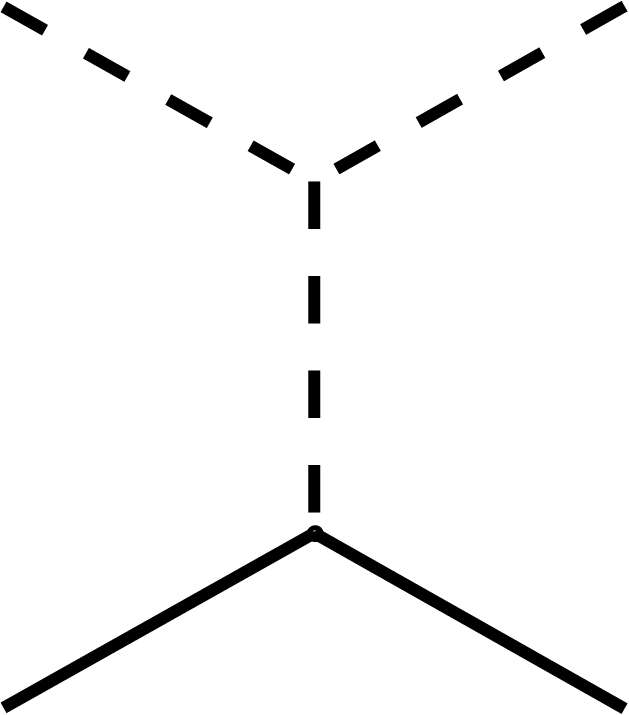}}}+x\vcenter{\hbox{\includegraphics[scale=0.15]{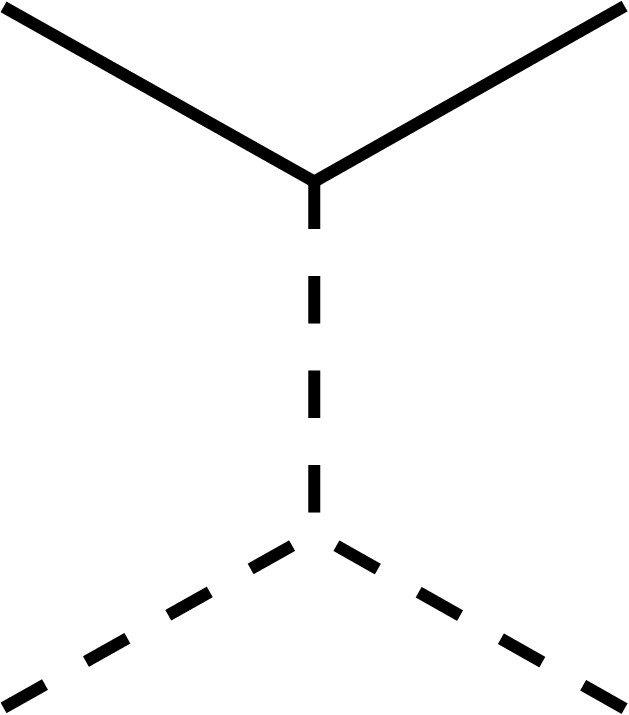}}}+\vcenter{\hbox{\includegraphics[scale=0.15]{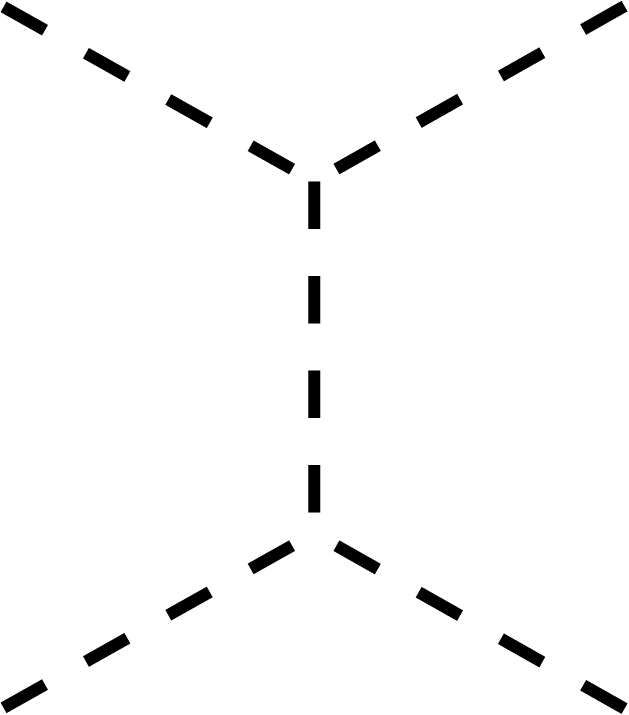}}}
\end{align}
where the dashed line represent an empty link and $x$ is the bond fugacity, i.e, the local weight assigned to a link covered by a loop.

Each diagram can be understood as an intertwiner of $U_{-q}(A_1)$ representations if we set the loop weight to $[2]_q$. These diagrams are called (dilute) Temperley-Lieb (TL) diagrams. The local space of states of the loop model is given by $\mathcal{H}_{A_1}=\mathbb{C}\oplus V_1$ where $V_1$ denotes the fundamental representation. Any diagram can be expressed by concatenating vertically and horizontally juxtaposing the following elementary diagrams (and possibly identity strands joining the top and bottom boundaries)
\begin{align*}
    \text{cup} \;=\;\vcenter{\hbox{\includegraphics[scale=0.2]{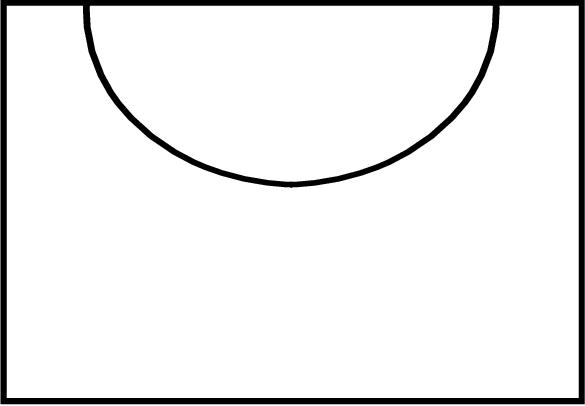}}} \qquad
    \text{cap} \;=\;\vcenter{\hbox{\includegraphics[scale=0.2]{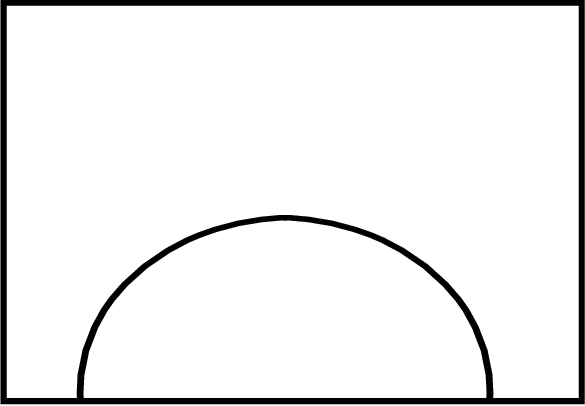}}}
\end{align*}
where cup and cap are embedded in the lattice as, respectively, the second- and third-last diagram of \eqref{LoopLocTM}.

We use a general convention that these `string' diagrams are read from bottom to top, the strings are labeled by the fundamental representation $V_1$ and the empty source/target of a diagram corresponds to the trivial representation $\mathbb{C}$. For example, the above $\text{cap}$ diagram is an intertwiner $V_1\otimes V_1 \rightarrow \mathbb{C}$ that can be the best described in a basis.
Let $\{e_1,e_2\}$ denote the standard basis of $V_1$, and $\{f_1,f_2\}$ be its dual. Then, the corresponding intertwiners are 
\begin{align*}
    \text{cup}:&\quad \mathbb{C}\rightarrow V_1\otimes V_1\nonumber\\
	&\quad 1 \mapsto e_1\otimes e_2 + q^{-1}e_2\otimes e_1\\
	\text{cap}=&\ q f_1\otimes f_2+f_2\otimes f_1 
\end{align*}
where in the last equality we used the obvious identification of $V_1^*\otimes V_1^*$ with the space of linear maps $V_1\otimes V_1 \rightarrow \mathbb{C}$.
The maps $\text{cup}$ and $\text{cap}$ were obtained  by calculation of $U_{-q}(A_1)$ invariant vectors in $V_1\otimes V_1$ and in its dual space. That is, we find a vector annihilated by the action of $E$ and $F$ given by the coproduct \eqref{eq:coprod}. 

We furthermore remark that the invariant vectors, and thus the corresponding maps, are defined up to a multiplicative constant that can be fixed in the following way. The maps cup and cap are assumed to satisfy the zigzag rules 
\begin{align}
    \vcenter{\hbox{\includegraphics[scale=0.2]{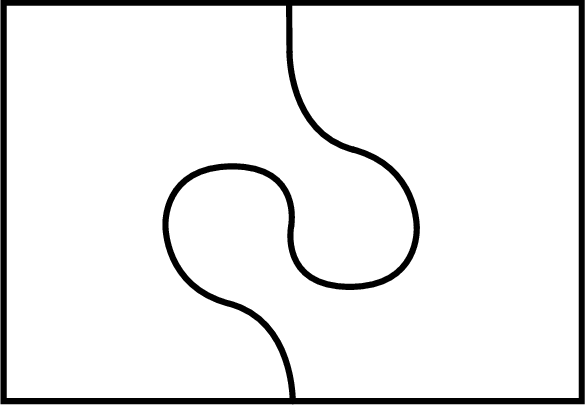}}}=\vcenter{\hbox{\includegraphics[scale=0.2]{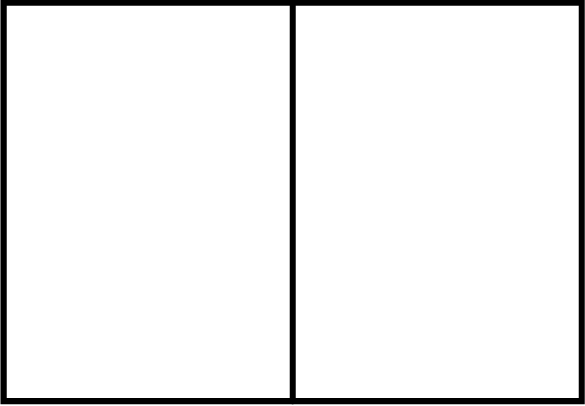}}}=\vcenter{\hbox{\includegraphics[scale=0.2]{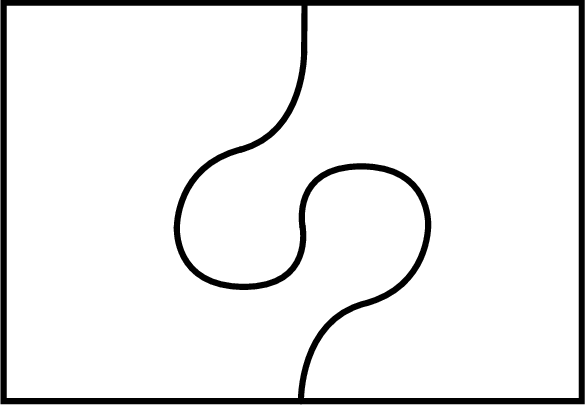}}}
\end{align}
which reflect the fact that our diagrams are considered up to an isotopy, or equivalently, this is an implication of the Temperley-Lieb relations.
These rules reduce the choices of constants to one ``gauge'' degree of freedom: one may multiply the cup intertwiner by some factor $\eta$ and the cap intertwiner by $\eta^{-1}$.
In the expressions above we have chosen a definite value of the gauge factor~$\eta$.

\subsection{The $A_2$ case}
\label{sec:TMA2}

Let $V_1$ be the fundamental representation of $U_{-q}(A_2)$ of highest weight $\bm{w_1}$. In $V_1$, pick a highest weight vector, $u_1$. Then we obtain a basis $\{u_i, i\in \llbracket 1, 3 \rrbracket\}$ by applying lowering operators:
\begin{subequations}
\begin{align}
	u_2=&F_1u_1\\
	u_3=&F_2F_1u_1
\end{align}
\end{subequations}
The action of the quantum group generators in our bases are given in appendix \ref{sec:explicit}.

Let $V_2$ be the fundamental representation of $U_{-q}(A_2)$ of highest weight $\bm{w_2}$. In $V_2$, pick a highest weight vector, $v_1$. Then we obtain a basis $\{v_i, i\in \llbracket 1, 3 \rrbracket\}$ by applying lowering operators:
\begin{subequations}
\begin{align}
	v_2=&F_2v_1\\
	v_3=&F_1F_2v_1
\end{align}
\end{subequations}

Let $\{e_i,i\in \llbracket1,7\rrbracket\}=\{u_1,u_2,u_3,v_1,v_2,v_3,1\}$ be a basis of $ \mathcal{H}_{A_2}=V_1\oplus V_2\oplus \mathbb{C}$.

Any $A_2$ web can be expressed as the vertical concatenation and horizontal juxtaposition of the following elementary webs (and possibly identity strands connecting the bottom and top boundaries)

\begin{alignat}{3}
\label{A2gendiagrams}
    \text{coev}&\;=\;\vcenter{\hbox{\includegraphics[scale=0.2]{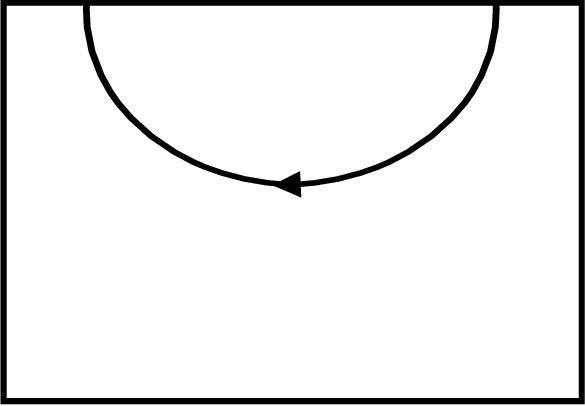}}}\qquad
    \widetilde{\text{coev}}&&\;=\;\vcenter{\hbox{\includegraphics[scale=0.2]{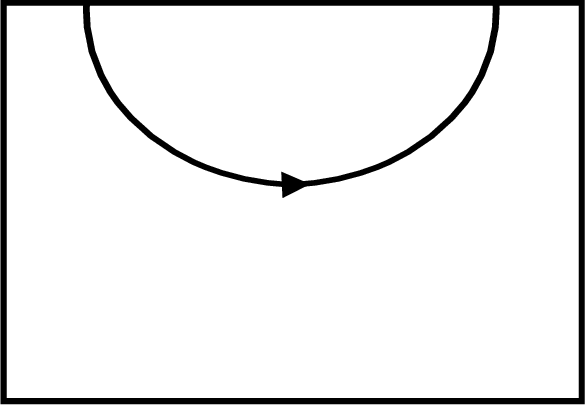}}}\nonumber\\
    \text{ev}&\;=\;\vcenter{\hbox{\includegraphics[scale=0.2]{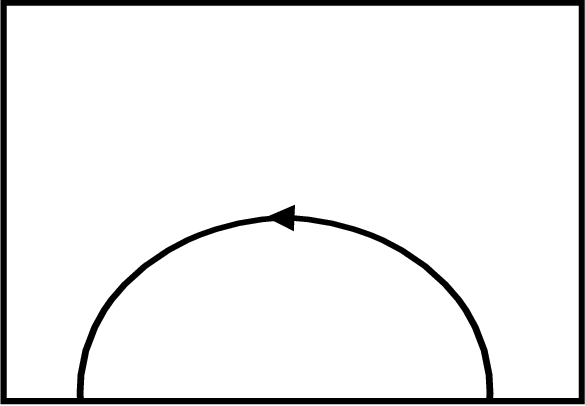}}}\qquad
    \widetilde{\text{ev}}&&\;=\;\vcenter{\hbox{\includegraphics[scale=0.2]{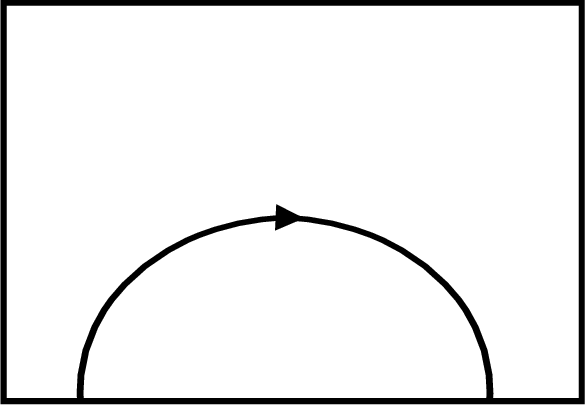}}}\\
    \text{Y}_1&\;=\;\vcenter{\hbox{\includegraphics[scale=0.2]{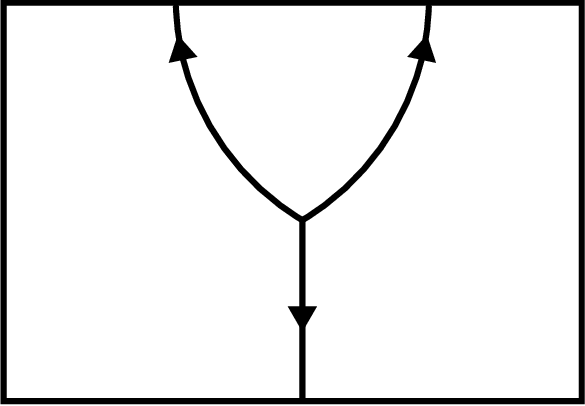}}}\qquad
    \text{Y}_2&&\;=\;\vcenter{\hbox{\includegraphics[scale=0.2]{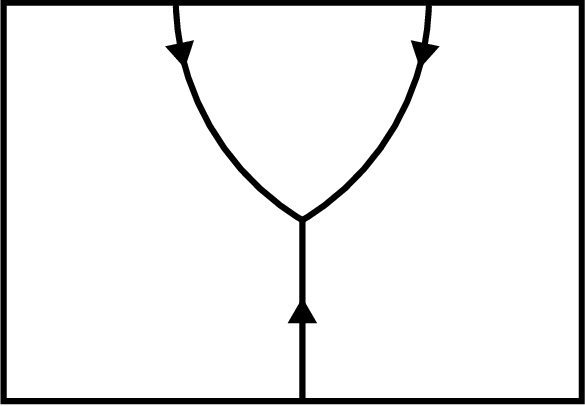}}}\nonumber
\end{alignat}

Here is an illustration on how one can obtain any open web from the above set 
\begin{align}
\label{A2compex}
    \vcenter{\hbox{\includegraphics[scale=0.2]{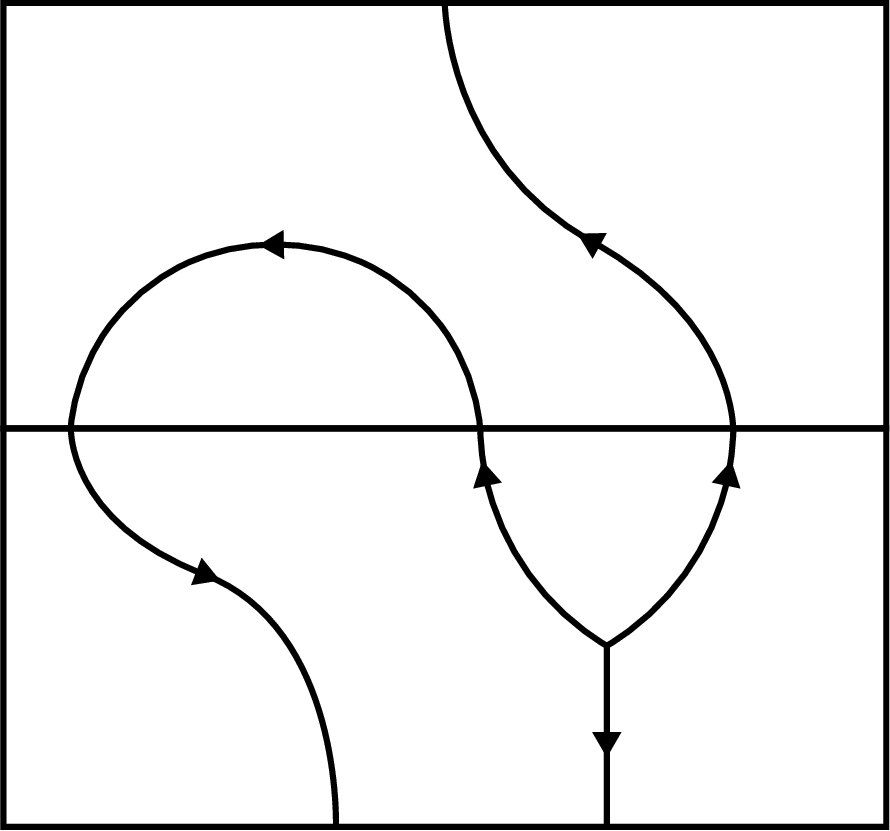}}}=\vcenter{\hbox{\includegraphics[scale=0.2]{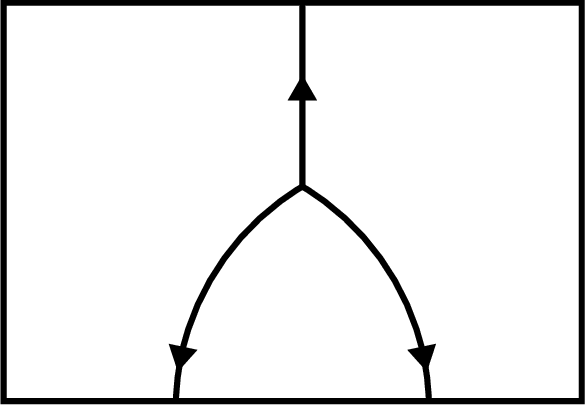}}}
\end{align}
In the left hand side, the top diagram is a juxtaposition of ev and an identity strand whereas the bottom one is a juxtaposition of an identity strand and $\text{Y}_1$. By concatenating them vertically, we get the web on the right hand side.

These webs represent the following intertwiners:
\begin{align*}
	\text{coev}: &\quad \mathbb{C}\rightarrow V_1\otimes V_2\nonumber\\
	&\quad e_7 \mapsto qe_1\otimes e_6 + e_2\otimes e_5 + q^{-1}e_3\otimes e_4\\[10pt]
	\widetilde{\text{coev}}: &\quad\mathbb{C}\rightarrow V_2\otimes V_1\nonumber\\
	&\quad e_7 \mapsto  q^{-1}e_6\otimes e_1 + e_5\otimes e_2 + q e_4\otimes e_3\\[10pt]
	\text{ev}= &\; q^{-1}e_6^*\otimes e_1^* + e_5^*\otimes e_2^* + qe_4^*\otimes e_3^*\\[10pt]
	\widetilde{\text{ev}}=&\; qe_1^*\otimes e_6^* + e_2^*\otimes e_5^* + q^{-1}e_3^*\otimes e_4^* \\[10pt]
	\text{Y}_1:&\quad V_2\rightarrow V_1\otimes V_1\\
	&e_4\mapsto e_1\otimes e_2 + q^{-1} e_2\otimes e_1 \\
	&e_5\mapsto e_1\otimes e_3 + q^{-1} e_3\otimes e_1 \\
	&e_6\mapsto e_2\otimes e_3 + q^{-1} e_3\otimes e_2 \\[10pt]
	\text{Y}_2:&\quad V_1\rightarrow V_2\otimes V_2\\
	&e_1\mapsto q e_4\otimes e_5 +  e_5\otimes e_4 \\
	&e_2\mapsto q e_4\otimes e_6 +  e_6\otimes e_4 \\
	&e_3\mapsto q e_5\otimes e_6 +  e_6\otimes e_5 
\end{align*}
Horizontal juxtaposition of webs corresponds to taking the tensor product of intertwiners and vertical concatenation corresponds to composition. For instance the web in \eqref{A2compex} represents the following intertwiner:
\begin{align*}
    \left(\text{ev}\otimes \text{Id}_{V_1}\right)\circ \left(\text{Id}_{V_2} \otimes \text{Y}_1\right)
\end{align*}

Similarly to the $U_{-q}(A_1)$ case, the maps ev and coev are defined up to  multiplicative scalars from an invariant vector in $V_1\otimes V_2$, i.e. a vector in $V_1\otimes V_2$ that is annihilated by the action of $E_1$, $E_2$, $F_1$ and $F_2$ using the coproduct defined in~\eqref{eq:coprod} from Appendix \ref{sec:quantumgroupconventions}. Similarly, $\widetilde{\text{ev}}$ and $\widetilde{\text{coev}}$ are defined up to multiplicative scalars from an invariant vector in $V_2\otimes V_1$. The freedom on multiplicative scalars is reduced to one degree of freedom once we impose that the maps satisfy the loop rule as well as the zigzag rules
\begin{align}
    \vcenter{\hbox{\includegraphics[scale=0.2]{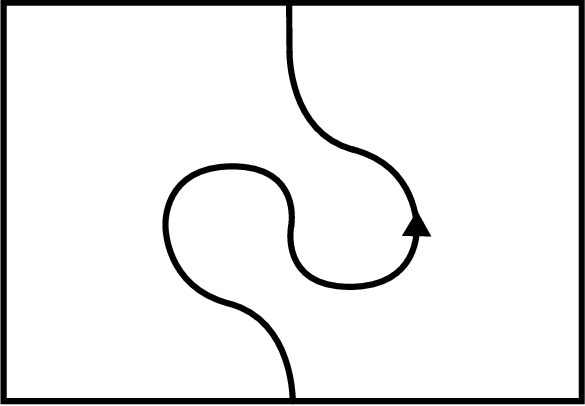}}}=\vcenter{\hbox{\includegraphics[scale=0.2]{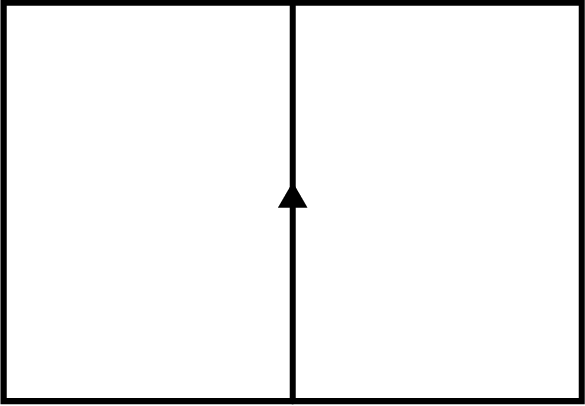}}}=\vcenter{\hbox{\includegraphics[scale=0.2]{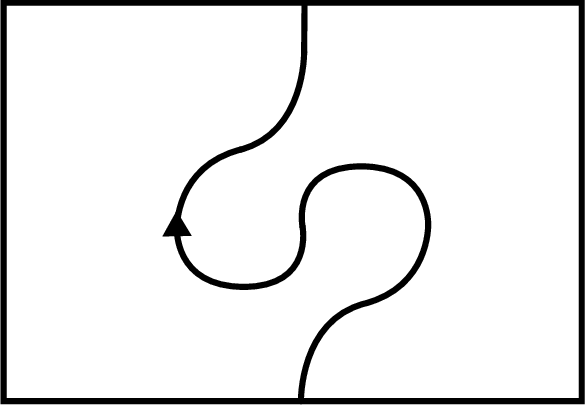}}}
\end{align}
To construct $\text{Y}_1$, one looks for a highest weight vector $v$ of weight $\bm{w}_2$ inside $V_1\otimes V_1$ using the actions of $E_1$, $E_2$ determined by the coproduct~\eqref{eq:coprod}. Note that this specify $v$ only up to a multiplicative scalar. The map $\text{Y}_1$ is then defined as mapping $e_4$ to $v$, $e_5$ to $F_2v$ and $e_6$ to $F_1F_2v$ where the actions of $F_1$, $F_2$ on $V_1\otimes V_1$ are again determined by the coproduct. This defines $\text{Y}_1$ up to a multiplicative scalar. Similarly, one defines $\text{Y}_2$ up to a multiplicative scalar. Asking that the maps satisfy the digon and square rules reduce the freedom to one degree of freedom. So, in total we have two free parameters that we have chosen to fix to some values.

The local transfer matrices of the $A_2$ web model are then given by 
\begin{subequations}
\label{A2transfermatrix}
\begin{align}
    t^{A_2}_{(1)}=&zx^{\frac{3}{2}}\vcenter{\hbox{\includegraphics[scale=0.2]{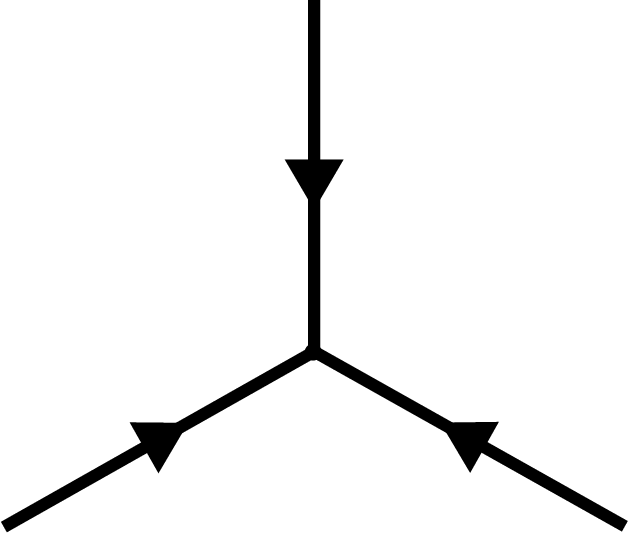}}}+ yx^{\frac{3}{2}} \vcenter{\hbox{\includegraphics[scale=0.2]{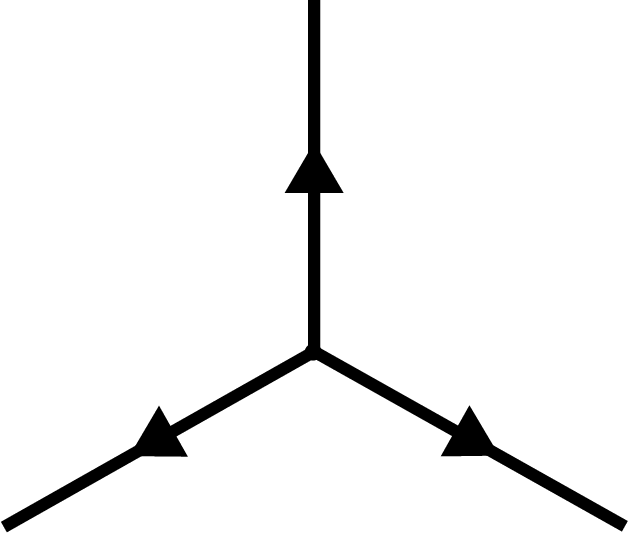}}}+ xe^{-i\phi}\vcenter{\hbox{\includegraphics[scale=0.2]{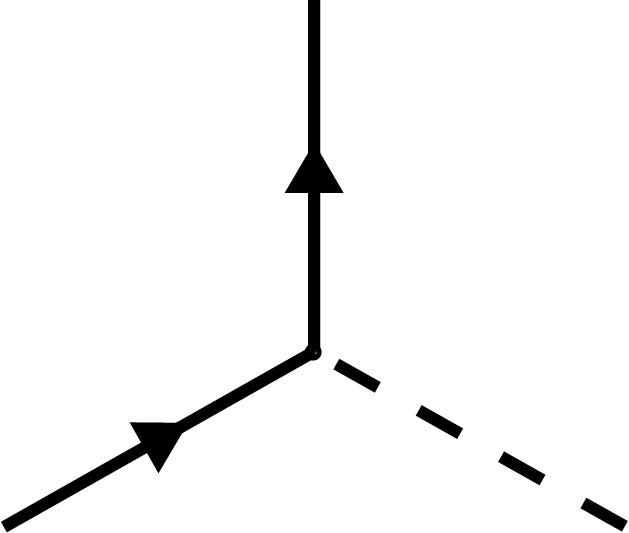}}}+ xe^{i\phi}\vcenter{\hbox{\includegraphics[scale=0.2]{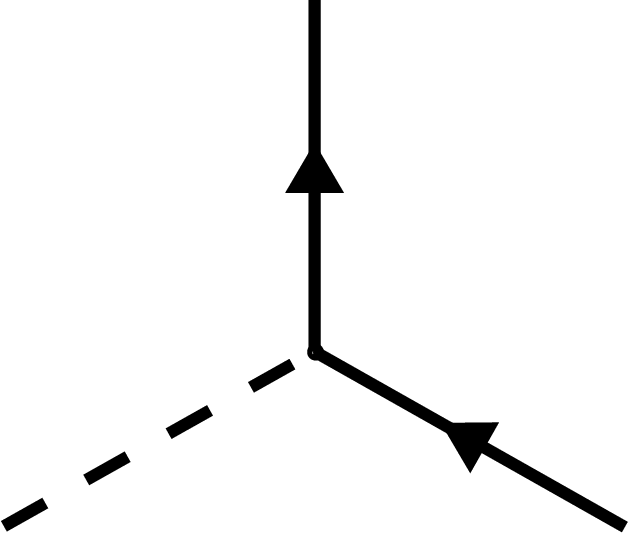}}}+ xe^{i\phi}\vcenter{\hbox{\includegraphics[scale=0.2]{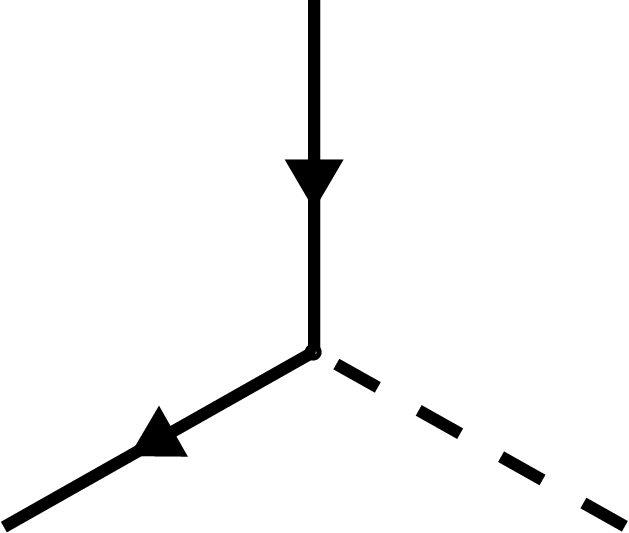}}}\nonumber\\ &+ xe^{-i\phi}\vcenter{\hbox{\includegraphics[scale=0.2]{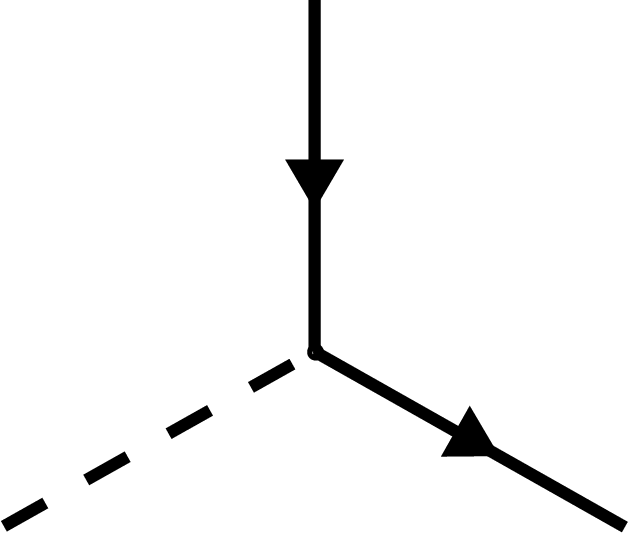}}}+ xe^{i\phi}\vcenter{\hbox{\includegraphics[scale=0.2]{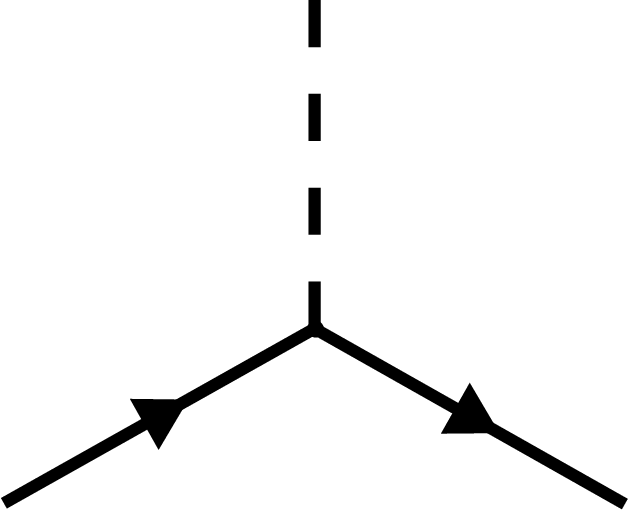}}}+ xe^{-i\phi}\vcenter{\hbox{\includegraphics[scale=0.2]{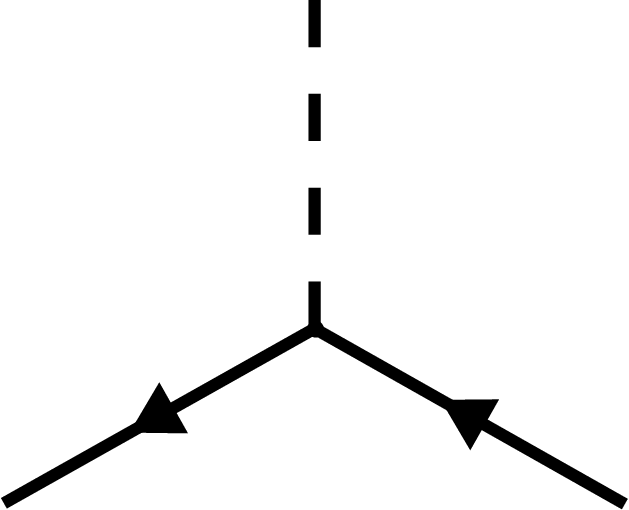}}}+ \vcenter{\hbox{\includegraphics[scale=0.2]{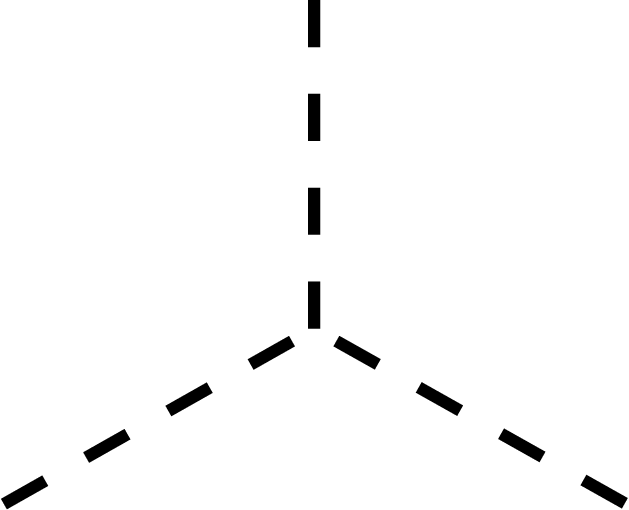}}}\label{A2transfermatrix1}
\end{align}
\begin{align}
    t^{A_2}_{(2)}=&zx^{\frac{3}{2}}\vcenter{\hbox{\includegraphics[scale=0.2]{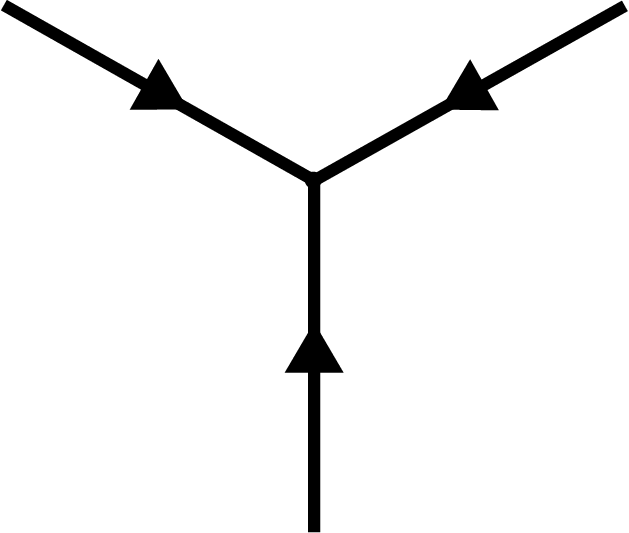}}}+ yx^{\frac{3}{2}}\vcenter{\hbox{\includegraphics[scale=0.2]{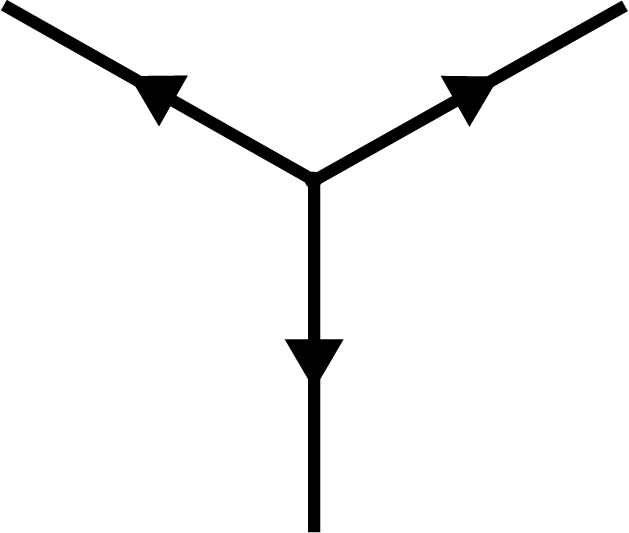}}}+ xe^{-i\phi}\vcenter{\hbox{\includegraphics[scale=0.2]{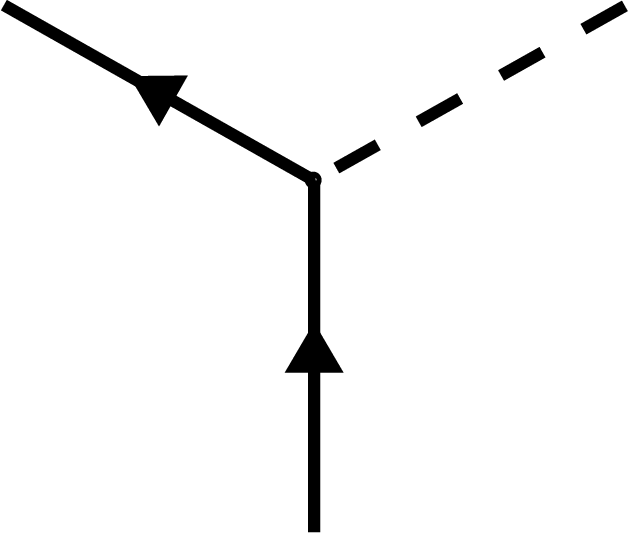}}}+ xe^{i\phi}\vcenter{\hbox{\includegraphics[scale=0.2]{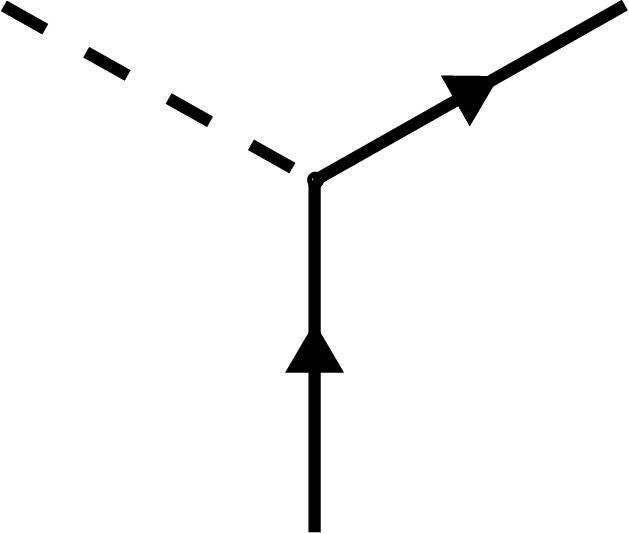}}}+ xe^{i\phi}\vcenter{\hbox{\includegraphics[scale=0.2]{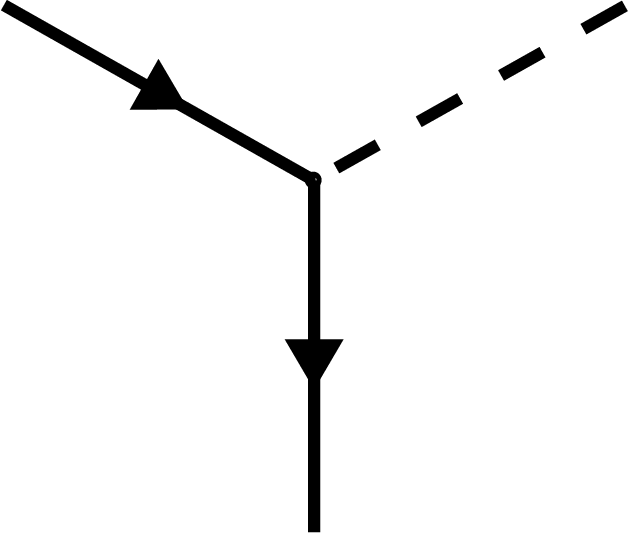}}}\nonumber\\ &+ xe^{-i\phi}\vcenter{\hbox{\includegraphics[scale=0.2]{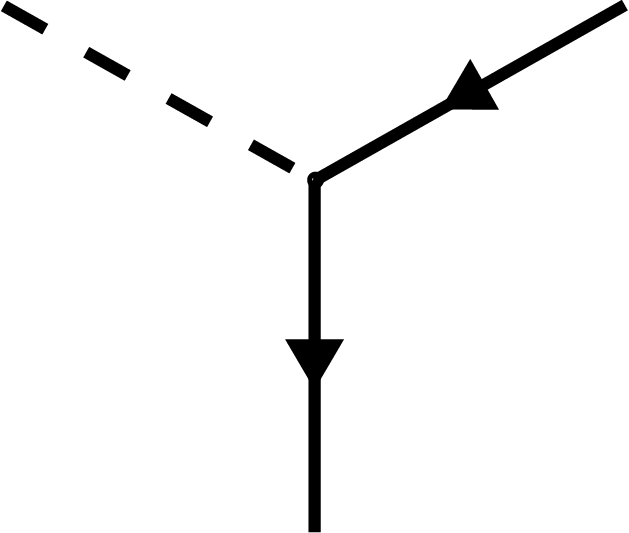}}}+ xe^{i\phi}\vcenter{\hbox{\includegraphics[scale=0.2]{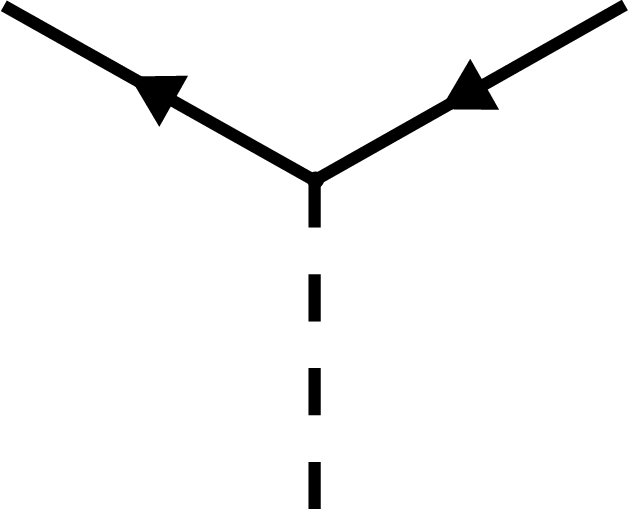}}}+ xe^{-i\phi}\vcenter{\hbox{\includegraphics[scale=0.2]{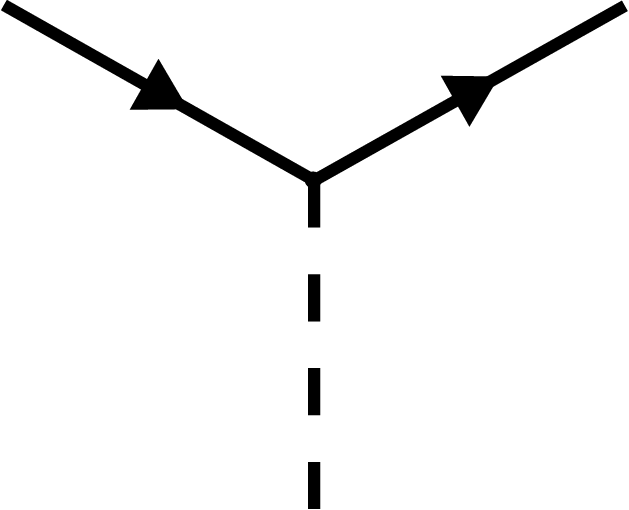}}}+ \vcenter{\hbox{\includegraphics[scale=0.2]{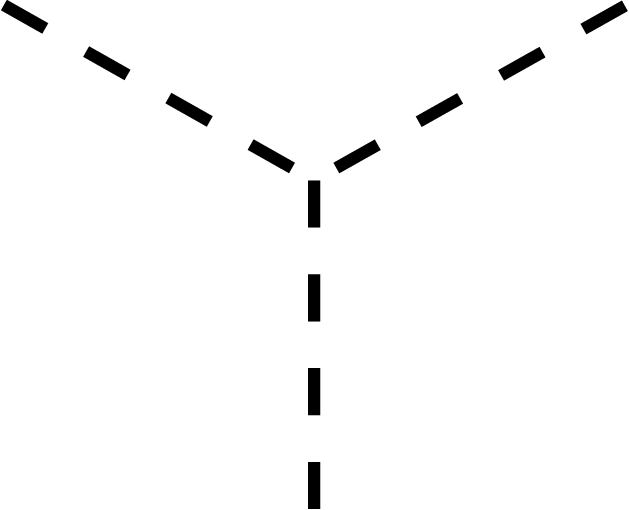}}}\label{A2transfermatrix2}
\end{align}
\end{subequations}
Above, diagrams must be understood as webs when dashed lines are forgotten. 
For instance in the expression of $t^{A_2}_{(1)}$, the first term is defined by \eqref{A2compex}, while the third to the sixth therms are given by identity lines. The seventh and eighth terms are given by ev and $\widetilde{\text{ev}}$ respectively, and the ninth one is the empty web, i.e., the identity on the trivial representation $\mathbb{C}$.
\subsection{The $G_2$ case}
\label{sec:TMG2}
Let $V$ be the fundamental representation of $U_q(G_2)$ of highest weight $\bm{w_1}$. In $V$, pick a highest weight vector, $e_1$. Then we obtain a basis $\{e_i, i\in \llbracket 1, 7 \rrbracket\}$ by applying lowering operators:
\begin{alignat*}{3}
	e_2=&F_1e_1 &&
	e_3=F_2F_1e_1 \\
	e_4=&F_1F_2F_1e_1 &&
	e_5=F_1^2F_2F_1e_1 \\
	e_6=&F_2F_1^2F_2F_1e_1 &\qquad&
	e_7=F_1F_2F_1^2F_2F_1e_1
\end{alignat*}
Denote by $\{f_i,i\in \llbracket 1,7\rrbracket\}$, the dual basis in $V^*$.
Any $G_2$ web can be expressed as the vertical concatenation and horizontal juxtaposition of the following elementary webs (and possibly identity strands)

\begin{alignat}{4}
\label{G2gendiagrams}
    \text{cup}&\;=\;\vcenter{\hbox{\includegraphics[scale=0.2]{diagrams/gen1G2spider.eps}}}\qquad \text{cap}&&\;=\;\vcenter{\hbox{\includegraphics[scale=0.2]{diagrams/gen2G2spider.eps}}}\qquad
    \text{Y}&&\;=\;\vcenter{\hbox{\includegraphics[scale=0.2]{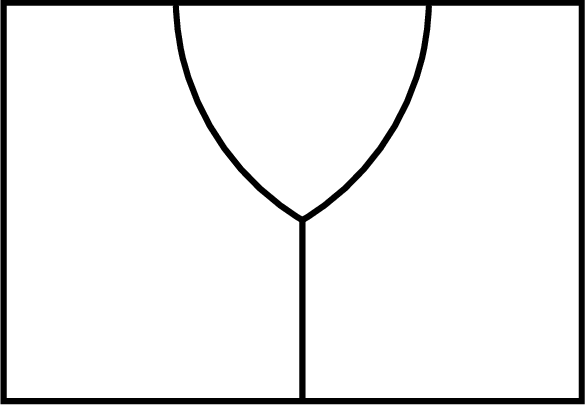}}}
\end{alignat}
These webs represent the following intertwiners:
\begin{align*}
	\text{cup}:\ &\mathbb{C}\rightarrow V\otimes V\nonumber\\
	&1 \mapsto e_7\otimes e_1 + q^{10}e_1\otimes e_7 -q e_6\otimes e_2 -q^9 e_2\otimes e_6 +q^4 e_5\otimes e_3 +q^6 e_3\otimes e_5 - q^6 e_4\otimes e_4\\[10pt]
	\text{cap}=&\ q^{-10}f_7\otimes f_1 + f_1\otimes f_7 -q^{-9} f_6\otimes f_2 -q^{-1} f_2\otimes f_6 +q^{-6} f_5\otimes f_3 +q^{-4} f_3\otimes f_5 - q^{-6} f_4\otimes f_4\\
	\text{Y}:\ &V\rightarrow V\otimes V\nonumber\\[10pt]
	&e_1\mapsto q^6 e_1\otimes e_4 - e_4\otimes e_1 - [2]_q q^4 e_2\otimes e_3 +[2]_q q e_3\otimes e_2\nonumber\\
	&e_2\mapsto -q^4e_2\otimes e_4+q^2e_4\otimes e_2 +q^5 e_1\otimes e_5 - e_5\otimes e_1\nonumber\\
	&e_3\mapsto -q^4e_3\otimes e_4 +q^2e_4\otimes e_3 +q^5 e_1\otimes e_6 - e_6\otimes e_1\nonumber\\
	&e_4\mapsto (q^2 - q^4) e_4\otimes e_4 - q^2 e_3\otimes e_5 + q^2 e_5\otimes e_3 \\ 
	&\quad\quad + q^5 e_2\otimes e_6 - q^{-1}e_6\otimes e_2 +q^4 e_1\otimes e_7 - e_7\otimes e_1\nonumber\\
	&e_5\mapsto q^2 e_5\otimes e_4 - q^4 e_4\otimes e_5 + [2]_q q^5 e_2\otimes e_7 - [2]_q e_7\otimes e_2\nonumber\\
	&e_6\mapsto q^2 e_6\otimes e_4 - q^4 e_4\otimes e_6 +[2]_q q^5 e_3\otimes e_7 -[2]_q e_7\otimes e_3\nonumber\\
	&e_7\mapsto -e_7\otimes e_4 + q^6 e_4\otimes e_7 +q e_6\otimes e_5 - q^4e_5\otimes e_6\nonumber
\end{align*}
Horizontal juxtaposition of webs corresponds to taking the tensor product of intertwiners and vertical concatenation corresponds to composition. The above maps were obtained similarly as in the $A_2$ case.

The local transfer matrices of the $G_2$ web model are then given by 
\begin{subequations}\begin{align}
    t^{G_2}_{(1)}&=x^{3/2}y\vcenter{\hbox{\includegraphics[scale=0.15]{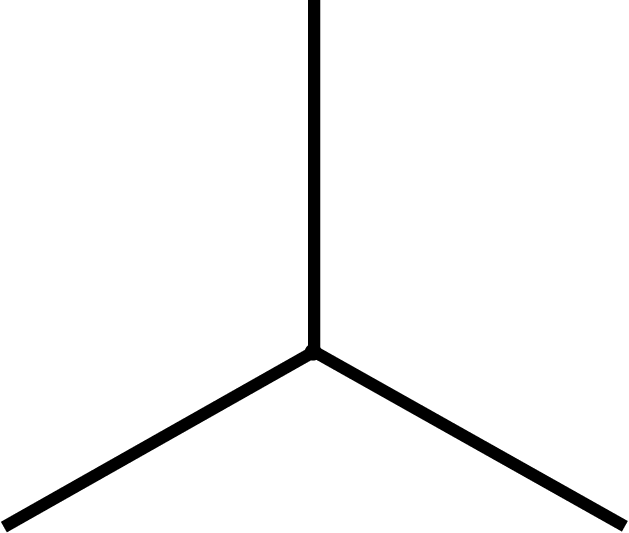}}}+x\vcenter{\hbox{\includegraphics[scale=0.15]{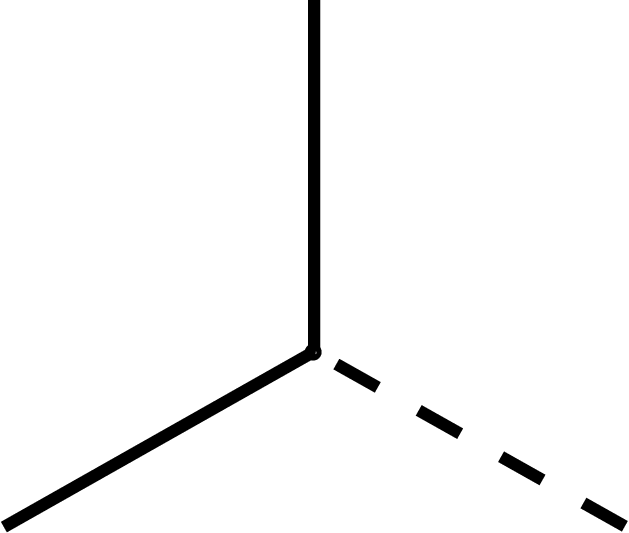}}}+x\vcenter{\hbox{\includegraphics[scale=0.15]{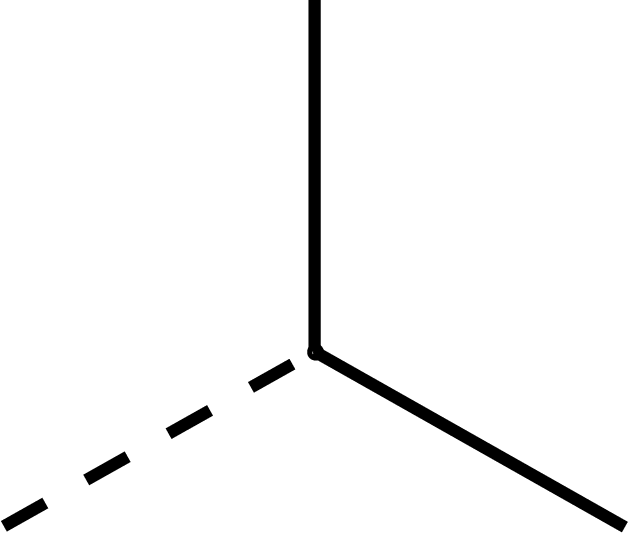}}}+x\vcenter{\hbox{\includegraphics[scale=0.15]{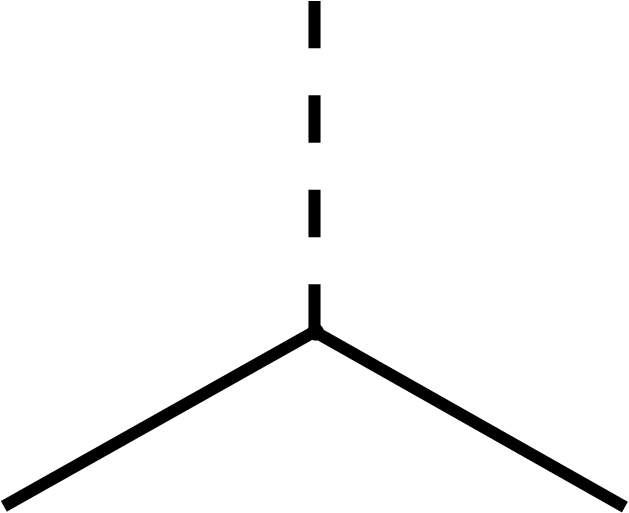}}}+\vcenter{\hbox{\includegraphics[scale=0.15]{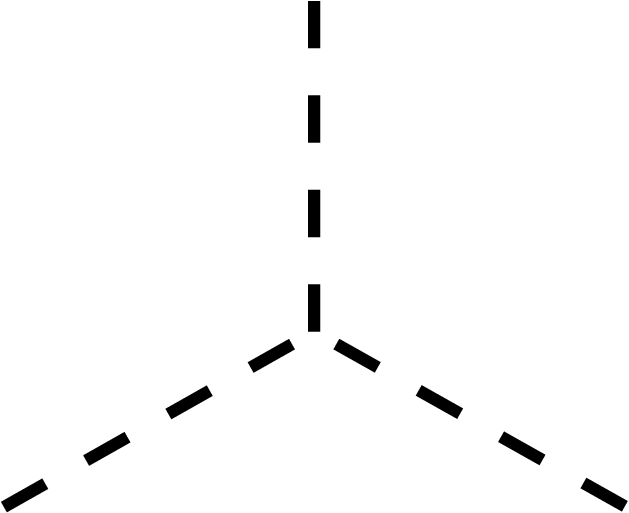}}}\label{G2transfermatrix1}\\[10pt]
    t^{G_2}_{(2)}&=x^{3/2}y\vcenter{\hbox{\includegraphics[scale=0.15]{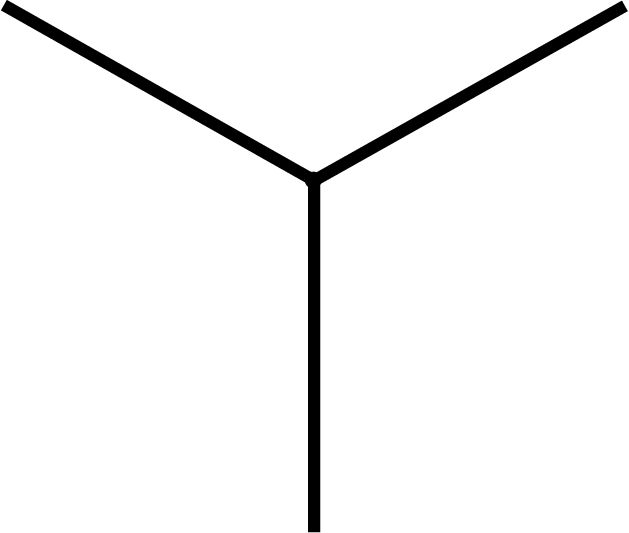}}}+x\vcenter{\hbox{\includegraphics[scale=0.15]{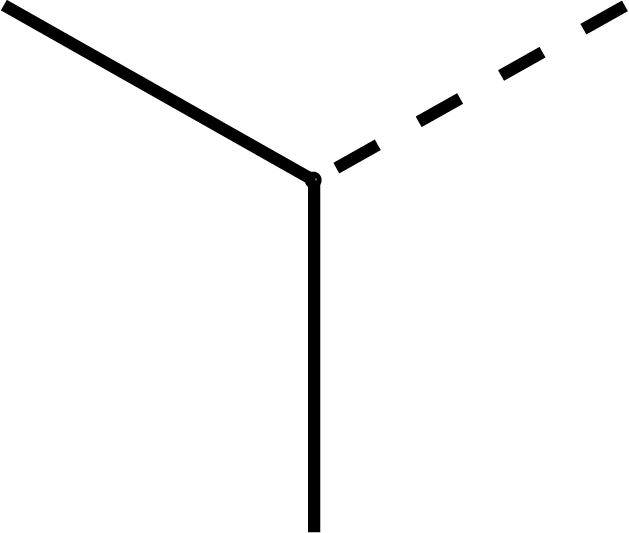}}}+x\vcenter{\hbox{\includegraphics[scale=0.15]{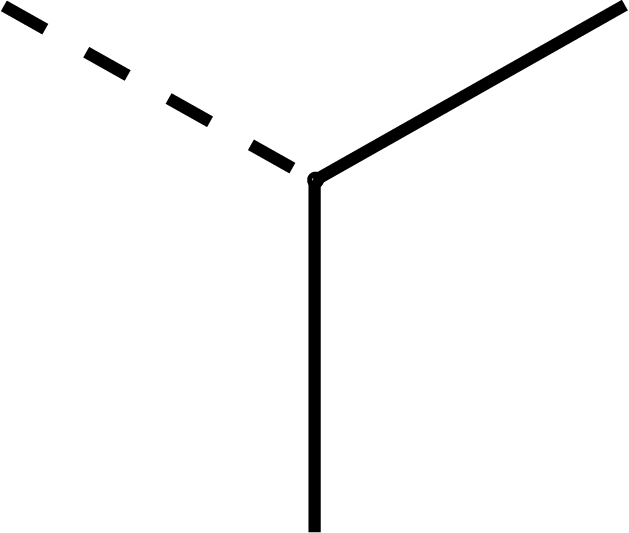}}}+x\vcenter{\hbox{\includegraphics[scale=0.15]{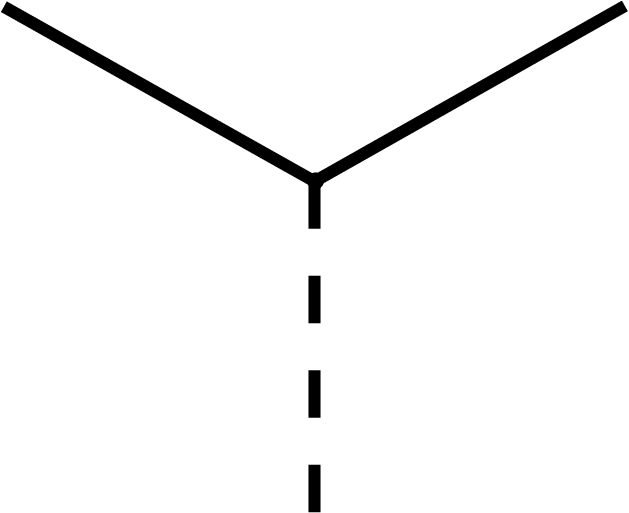}}}+\vcenter{\hbox{\includegraphics[scale=0.15]{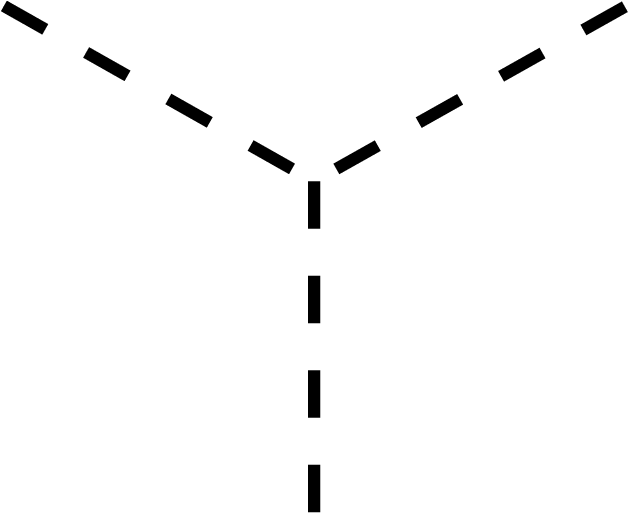}}}\label{G2transfermatrix2}
\end{align}
\label{transfermatrixG2}
\end{subequations}
\subsection{The $B_2$ case}
\label{sec:B2TM}
Let $V_1$ and $V_2$ be the fundamental representations of $U_q(B_2)$ of highest weights $\bm{w_1}$ and $\weighttwo$ respectively. In $V_1$, pick a highest weight vector, $e_1$. Then we obtain a basis $\{e_i, i\in \llbracket 1, 4 \rrbracket\}$ by applying lowering operators:
\begin{subequations}
\begin{align}
	e_2=&F_1e_1\\
	e_3=&F_2F_1e_1\\
	e_4=&F_1F_2F_1e_1
\end{align}
\end{subequations}
Denote by $\{f_i,i\in \llbracket 1,4\rrbracket\}$, the dual basis in $V_1^*$.

In $V_2$, pick a highest weight vector, $v_1$. Then we obtain a basis $\{v_i, i\in \llbracket 1, 5 \rrbracket\}$ by applying lowering operators:
\begin{subequations}
\begin{align}
	v_2=&F_2v_1\\
	v_3=&F_1F_2v_1\\
	v_4=&F_1^2F_2v_1\\
	v_5=&F_2F_1^2F_2v_1
\end{align}
\end{subequations}
Denote by $\{g_i,i\in \llbracket 1,5\rrbracket\}$, the dual basis in $V_2^*$.

Any $B_2$ web can be expressed as the vertical concatenation and horizontal juxtaposition of the following elementary webs (and possibly identity strands) 
\begin{alignat}{3}
\label{B2gendiagrams}
    \text{cup}_1&\;=\;\vcenter{\hbox{\includegraphics[scale=0.2]{diagrams/gen1G2spider.eps}}}\qquad \text{cap}_1&&\;=\;\vcenter{\hbox{\includegraphics[scale=0.2]{diagrams/gen2G2spider.eps}}}\nonumber\\
    \text{cup}_2&\;=\;\vcenter{\hbox{\includegraphics[scale=0.2]{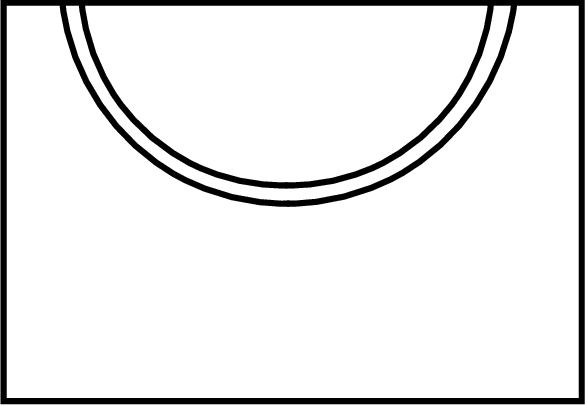}}}\qquad \text{cap}_2&&\;=\;\vcenter{\hbox{\includegraphics[scale=0.2]{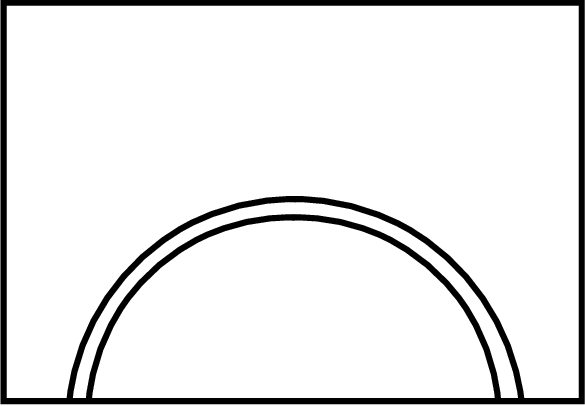}}}\\
    \text{Y}&\;=\;\vcenter{\hbox{\includegraphics[scale=0.2]{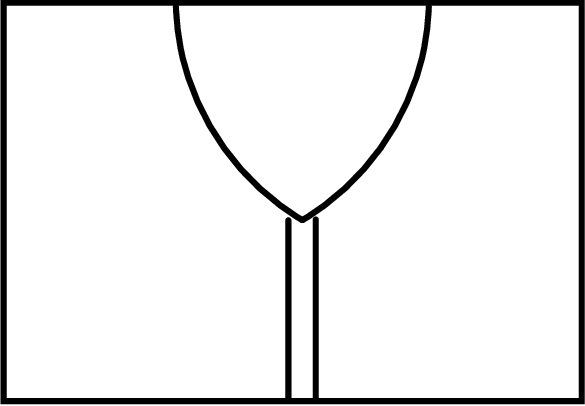}}}&&\nonumber
\end{alignat}
These webs represent the following intertwiners:
\begin{align*}
	\text{cup}_1:&\mathbb{C}\rightarrow V\otimes V\nonumber\\
	&1 \mapsto e_4\otimes e_1 - q^{4}e_1\otimes e_4  -q e_3\otimes e_2+ q^3 e_2\otimes e_3\\[10pt]
	\text{cap}_1=& f_1\otimes f_4-q^{-4}f_4\otimes f_1  -q^{-1} f_2\otimes f_3 +q^{-3} f_3\otimes f_2\\[10pt]
	\text{cup}_2:&\mathbb{C}\rightarrow V\otimes V\nonumber\\
	&1 \mapsto v_5\otimes v_1 + q^{6}v_1\otimes v_5  -q^2 v_4\otimes v_2 -q^4 v_2\otimes v_4 +q^4 v_3\otimes v_3\\[10pt]
	\text{cap}_2=& g_1\otimes g_5 +q^{-6} g_5\otimes g_1  -q^{-2} g_2\otimes g_4 -q^{-4} g_4\otimes g_2 + q^{-4} g_3\otimes g_3\\[10pt]
	\text{Y}:&V_2\rightarrow V_1\otimes V_1\\
	&v_1\mapsto iq e_1\otimes e_2 - ie_2\otimes e_1 \\
	&v_2\mapsto iq e_1\otimes e_3 - ie_3\otimes e_1 \\
	&v_3\mapsto iq e_2\otimes e_3 - iq^{-1} e_3\otimes e_2 + ie_1\otimes e_4 - ie_4\otimes e_1 \\
	&v_4\mapsto i[2]_q q e_2\otimes e_4 - i[2]_q e_4\otimes e_2 \\
	&v_5\mapsto i[2]_q q e_3\otimes e_4 - i[2]_q e_4\otimes e_3 
\end{align*}
Horizontal juxtaposition of webs corresponds to taking the tensor product of intertwiners and vertical concatenation corresponds to composition. The above maps were obtained similarly as in the $A_2$ case.

The local transfer matrices of the $B_2$ web model are then given by 
\begin{subequations}\begin{align}
    t^{B_2}_{(1)}=&x_{t;1}x_{v;2}^{1/2}y\vcenter{\hbox{\includegraphics[scale=0.15]{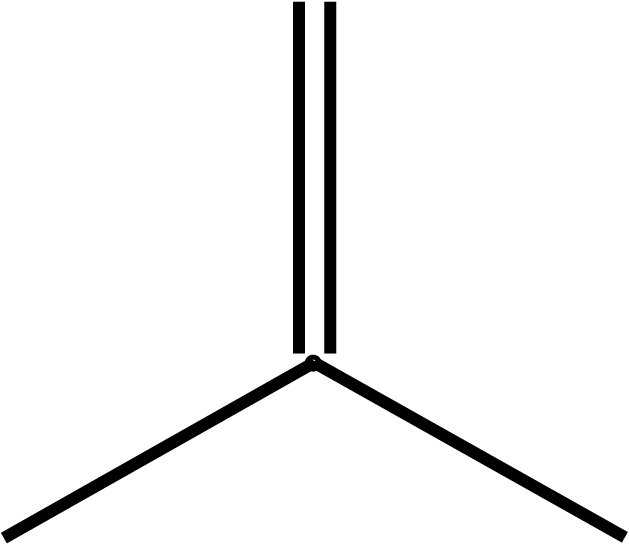}}}+x_{t;1}^{1/2}x_{v;1}^{1/2}x_{t;2}^{1/2}y\vcenter{\hbox{\includegraphics[scale=0.15]{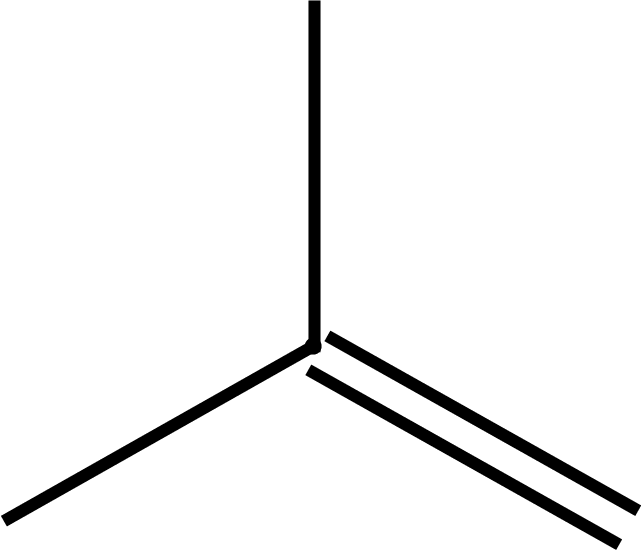}}}+x_{t;1}^{1/2}x_{v;1}^{1/2}x_{t;2}^{1/2}y\vcenter{\hbox{\includegraphics[scale=0.15]{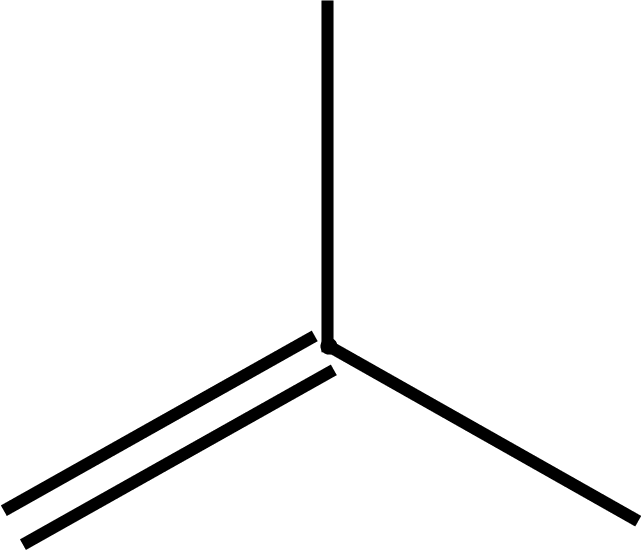}}}\nonumber\\
    &++x_{t;2}^{1/2}x_{v;2}^{1/2}\vcenter{\hbox{\includegraphics[scale=0.15]{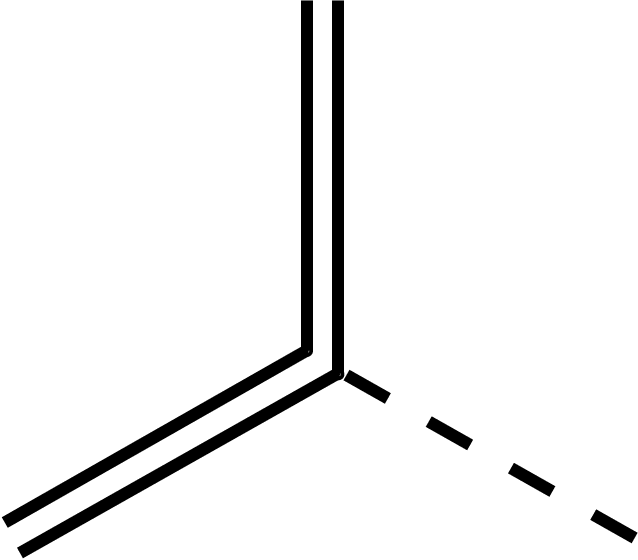}}}+x_{t;2}^{1/2}x_{v;2}^{1/2}\vcenter{\hbox{\includegraphics[scale=0.15]{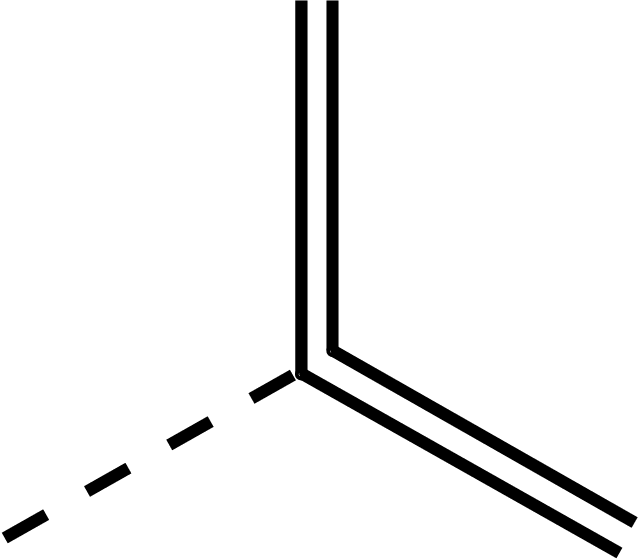}}}+x_{t;2}\vcenter{\hbox{\includegraphics[scale=0.15]{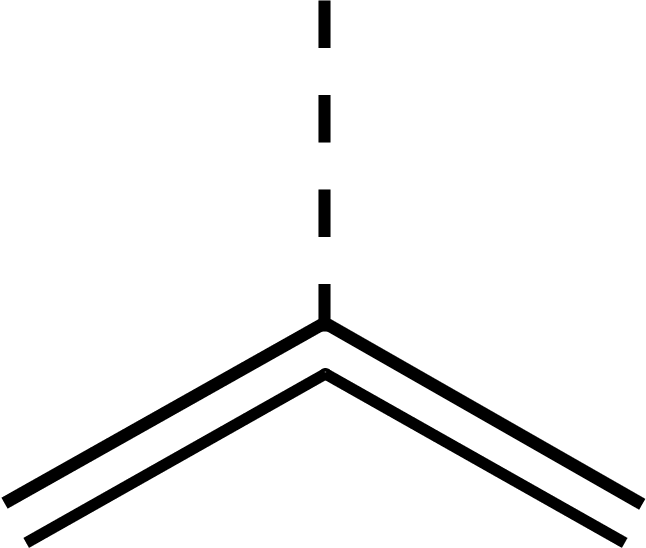}}}\\
    &+x_{t;1}^{1/2}x_{v;1}^{1/2}\vcenter{\hbox{\includegraphics[scale=0.15]{diagrams/G2vertex2.eps}}}+x_{t;1}^{1/2}x_{v;1}^{1/2}\vcenter{\hbox{\includegraphics[scale=0.15]{diagrams/G2vertex3.eps}}}+x_{t;1}\vcenter{\hbox{\includegraphics[scale=0.15]{diagrams/G2vertex4.eps}}}+\vcenter{\hbox{\includegraphics[scale=0.15]{diagrams/G2vertex5.eps}}}\label{B2transfermatrix1}\nonumber\\[10pt]
    t^{B_2}_{(2)}=&x_{t;1}x_{v;2}^{1/2}y\vcenter{\hbox{\includegraphics[scale=0.15]{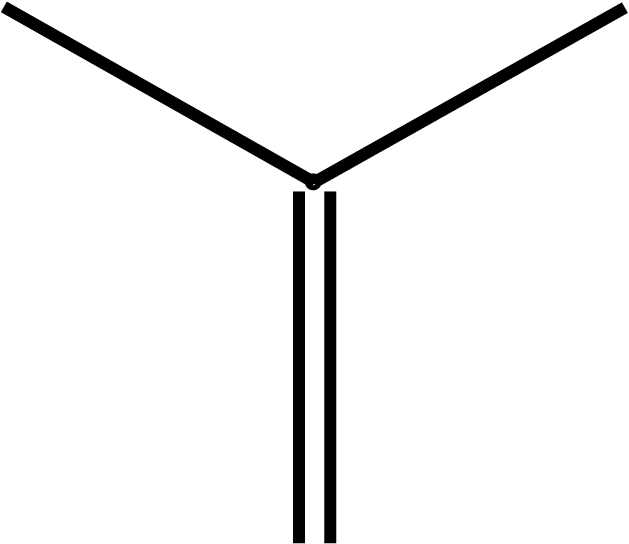}}}+x_{t;1}^{1/2}x_{v;1}^{1/2}x_{t;2}^{1/2}y\vcenter{\hbox{\includegraphics[scale=0.15]{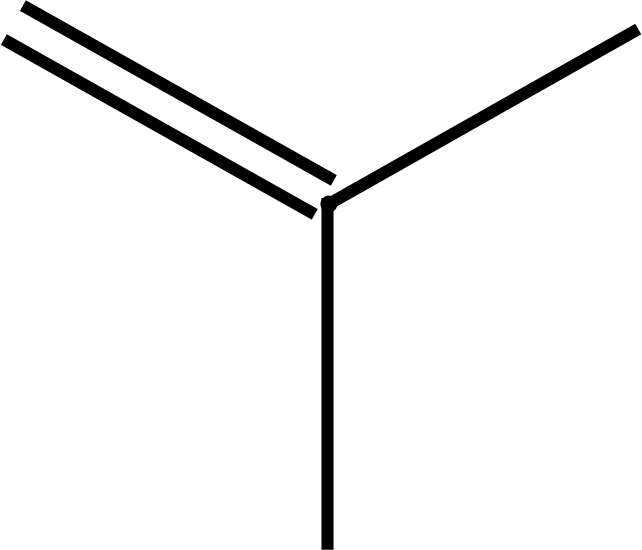}}}+x_{t;1}^{1/2}x_{v;1}^{1/2}x_{t;2}^{1/2}y\vcenter{\hbox{\includegraphics[scale=0.15]{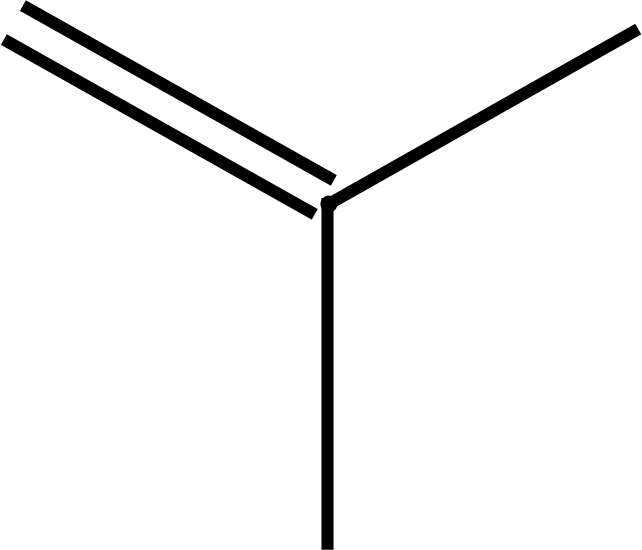}}}\nonumber\\
    &+x_{t;2}^{1/2}x_{v;2}^{1/2}\vcenter{\hbox{\includegraphics[scale=0.15]{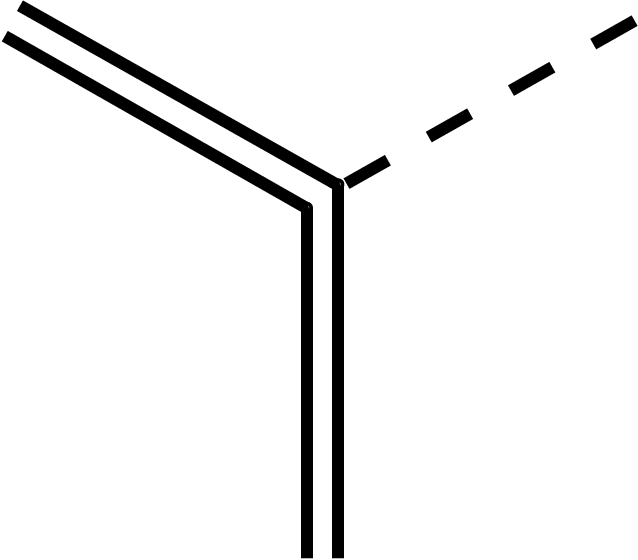}}}+x_{t;2}^{1/2}x_{v;2}^{1/2}\vcenter{\hbox{\includegraphics[scale=0.15]{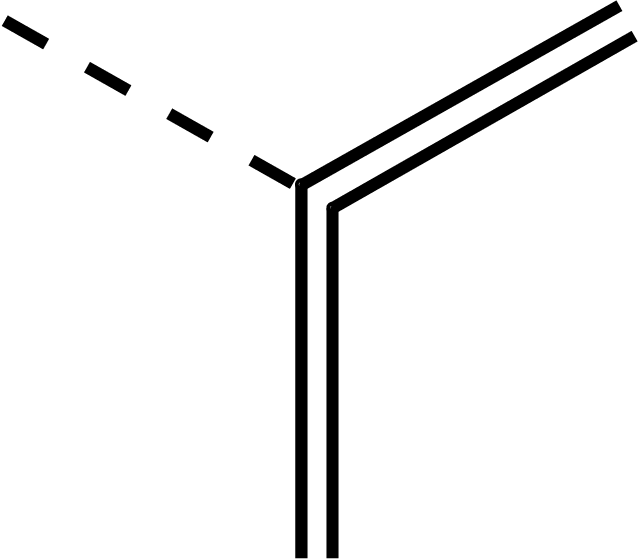}}}+x_{t;2}\vcenter{\hbox{\includegraphics[scale=0.15]{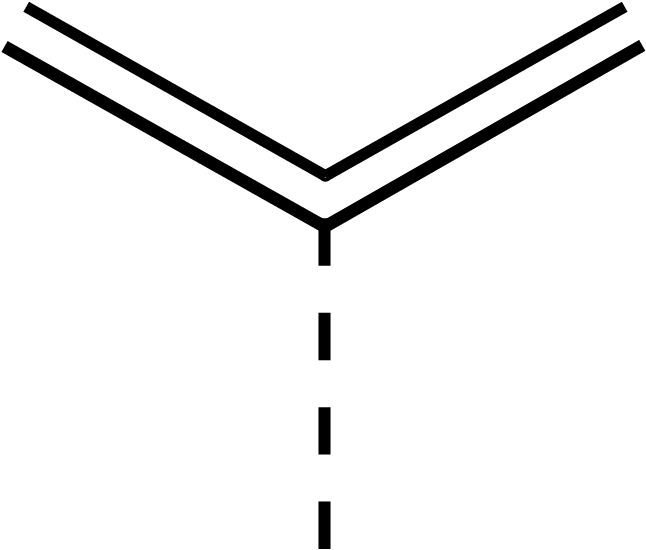}}}\\
    &+x_{t;1}^{1/2}x_{v;1}^{1/2}\vcenter{\hbox{\includegraphics[scale=0.15]{diagrams/G2vertex7.eps}}}+x_{t;1}^{1/2}x_{v;1}^{1/2}\vcenter{\hbox{\includegraphics[scale=0.15]{diagrams/G2vertex8.eps}}}+x_{t;1}\vcenter{\hbox{\includegraphics[scale=0.15]{diagrams/G2vertex9.eps}}}+\vcenter{\hbox{\includegraphics[scale=0.15]{diagrams/G2vertex10.eps}}}\nonumber
\end{align}
\label{transfermatrixB2}
\end{subequations}

The local transfer matrices for the square lattice case can be defined analogously. That is, they are linear operators in $\text{End}(\mathcal{H}_X\otimes \mathcal{H}_X)$ given by the linear combination of diagrams (and their reflections and rotations) from below \eqref{4valentvertex} weighted by the corresponding factors $b_i$'s. Each diagram is again understood as a linear operator.

\section{Integrability}
\label{sec:integrability}

In this section, we exhibit integrable manifolds in the parameter spaces of web models. Let us first give the steps of the general strategy we follow.
\begin{itemize}
    \item First, we look for an affine Dynkin diagram $\Tilde{X}$ that reduces to the finite type Dynkin diagram $X$ when we erase one of its nodes $n_0$. This implies that the Hopf subalgebra $U_t(\Tilde{X})$ generated by the Chevalley generators $E_i$, $F_i$, $H_i$, $i\neq n_0$ is isomorphic to $U_q(X)$ for some $q(t)$. We will call $U_t(\Tilde{X})$ and $U_q(X)$ the ``big" and ``small" quantum groups respectively.\footnote{It should not be confused with Lusztig's small quantum group, defined at roots of unity.} 
    
 Here we list the big and small quantum groups we will consider:\footnote{Given $X$, there might be several choices for $\Tilde{X}$ and $n_0$. For instance, if $X=B_2$, $\Tilde{X}=A_4^{(2)}$, then $n_0=0$ and $n_0=2$ are both valid. The second step actually fixes such choices.}
\begin{table}[h!]
\centering
\setlength\tabcolsep{15pt}
\begin{center}
\def\arraystretch{2.5}
\begin{tabular}{|c|ccc|}
    \hline
    Case & $U_t(\Tilde{X})$ & $U_q(X)$ & $\Tilde{X}$, {\color{blue}$X$}\\
    \hline
    1 & $U_t(A_2^{(2)})$ & $U_{t^4}(A_1)$ & $\vcenter{\hbox{\includegraphics[scale=0.4]{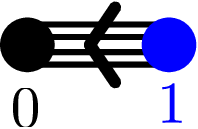}}}$\\
    2 & $U_t(G_2^{(1)})$ & $U_{t^3}(A_2)$ & $\vcenter{\hbox{\includegraphics[scale=0.4]{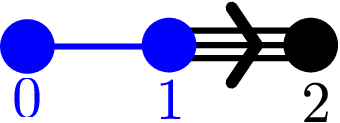}}}$\\
    3 & $U_t(D_4^{(3)})$ & $U_{t}(G_2)$ & $\vcenter{\hbox{\includegraphics[scale=0.4]{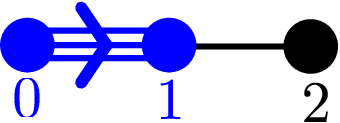}}}$\\
    4 & $U_t(A_4^{(2)})$ & $U_{t^2}(B_2)$ & $\vcenter{\hbox{\includegraphics[scale=0.4]{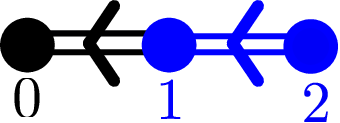}}}$\\
    \hline
\end{tabular}
\end{center}
\caption{Big and small quantum groups.}
\label{table:qg}
\end{table}

The definitions of these quantum groups are recalled in Appendix \ref{sec:quantumgroupconventions}, with the corresponding Cartan matrices given in Appendix \ref{sec:appendix1}.
    \item We then look for an irreducible ``evaluation''\footnote{Strictly speaking, evaluation representations might not exist because the evaluation morphism on the quantum group level exists for $A_n$ types only.} representation $V_u$, $u\in \mathbb{C}^*$ of $U_t(\Tilde{X})$ that decomposes under the subalgebra $U_q(X)$ as the local space of states $\mathcal{H}_X$,  independently of the evaluation, or spectral, parameter $u$. We will denote the representation map $\rho_u : U_t(\Tilde{X})\rightarrow \mathrm{End}(V_u)$.
    \item We then follow Jimbo's strategy\cite{jimbo_quantumr_1986} to find a solution of the spectral parameter dependant Yang Baxter equation. Let us recall it. Suppose that the tensor product $\rho_u\otimes \rho_v$, $u,v\in \mathbb{C}^*$ is irreducible. We are looking for an operator $\check{R}(u,v)$ intertwining $\rho_u\otimes \rho_v$ and $\rho_v\otimes \rho_u$, i.e.
\begin{align}
\label{JimboEq}
    \check{R}(u,v)(\rho_u\otimes \rho_v)(a)=(\rho_v\otimes \rho_u)(a)\check{R}(u,v) ,\quad a\in U_t(\Tilde{X})
\end{align}

Because $\rho_u\otimes \rho_v$ is irreducible, if \eqref{JimboEq} admits a non-zero solution, it is unique up to a multiplicative constant. Moreover, since $U_t(\Tilde{X})$ has a universal R matrix \cite{tolstoy_universal_1992,damiani2011rmatrix}, it follows that $\check{R}(u,v)$ is non-zero and satisfies the spectral parameter dependent Yang-Baxter equation:
\begin{align}
\label{YB}
    \check{R}_{23}(u,v)\check{R}_{12}(u,w)\check{R}_{23}(v,w)=\check{R}_{12}(v,w)\check{R}_{23}(u,w)\check{R}_{12}(u,v) \quad \text{on }V_u\otimes V_v \otimes V_w
\end{align}
where the subscript in $\check{R}_{ij}$ indicates that it acts as $\check{R}$ on the $i$th and $j$th tensorands and as identity elsewhere.

In order to find $\check{R}(u,v)$, we first notice that, under the small quantum group
\begin{align*}
    V_u\otimes V_v\cong V_v\otimes V_u\cong \mathcal{H}_X\otimes \mathcal{H}_X
\end{align*}
Thus, we can expand $\check{R}(u,v)$ as a sum of intertwiners of $U_{q}(X)$ from $\mathcal{H}_X\otimes \mathcal{H}_X$ to itself
\begin{align}
\label{RmatExp}
    \check{R}(u,v)=\sum_{i}a_i(u,v) W_i
\end{align}
where the sum is taken over an index set of the web basis $\{W_i\}$ of $\text{End}_{U_{q}(X)}(\mathcal{H}_X\otimes \mathcal{H}_X)$ and $a_i(u,v)$ are unknown functions. The system \eqref{JimboEq} is then reduced to a system of linear equations on the unknowns $a_i(u,v)$
\begin{align}
\label{LinearSys}
    \check{R}(u,v)(\rho_u\otimes \rho_v)(a)=(\rho_v\otimes \rho_u)(a)\check{R}(u,v) ,\, \quad a=E_{n_0},\, F_{n_0}
\end{align}
which is much simpler to solve than the cubic Yang-Baxter equations \eqref{YB}.

In the case the representation $\mathcal{H}_X$ of the small quantum group $U_{q}(X)$ is irreducible and $\mathcal{H}_X\otimes \mathcal{H}_X$ is a direct sum of irreducible representations of multiplicity $1$, techniques have been developed to solve \eqref{JimboEq} \cite{zhang_representations_1991,delius_construction_1994,delius_twisted_1996}. However, this is not our case because $\mathcal{H}_X$ is reducible and multiplicities are sometimes higher than $1$. Instead, we solve the linear system \eqref{LinearSys} directly by using software.

\item Finally, we identify values $(u_0,v_0)$ of the spectral parameters $(u,v)$ such that
\begin{align*}
    a_i(u_0,v_0)=0
\end{align*}
for all $i$ corresponding to webs $W_i$ that do not appear in the local transfer matrix of the given web model. We then reach the local transfer matrix by making a gauge transformation
\begin{align*}
    t^{X}_{(2)}t^{X}_{(1)}= \left(D_2^{-1}\otimes D_1^{-1}\right) \check{R}(u_0,v_0) \left(D_1\otimes D_2 \right)
\end{align*}
for some diagonal matrices $D_1$ and $D_2$ commuting with $U_{q}(X)$ on $\mathcal{H}_X$.
\end{itemize}

In the case of $A_2$ and $G_2$, the evaluation representations we consider were previously studied in a different context \cite{takacs_quantum_1997,Tak_cs_1997}. They are given respectively in appendices \ref{G21rep} and \ref{sec:evalrepG2}. However, in the case of $B_2$, the two evaluation representations of $U_t(A_4^{(2)})$ 
defined in \ref{sec:evalrepB2} and \ref{sec:evalrepB22} are new to the best of our knowledge.

\subsection{A reminder on the dilute loop model}
Consider the quantum affine algebra $U_t(A_2^{(2)})$ (see Appendix \ref{sec:quantumgroupconventions} for definitions) with Cartan matrix
\begin{align*}
    \begin{pmatrix}
2 & -4 \\
-1 & 2 
\end{pmatrix}
\end{align*}
There is a $3$-dimensional evaluation representation $V_u$ given by the following matrices in the basis $\{1,e_1,e_2\}$, where $\{1\}$ (respectively $\{e_1,e_2\}$) denotes the basis of the trivial (respectively fundamental) representation of $U_{-q}(A_1)$
\begin{alignat*}{2}
    &E_0= u\begin{bmatrix}
0 & \sqrt{[2]_t} & 0 \\
0 & 0 & 0 \\
\sqrt{[2]_t} & 0 & 0 
\end{bmatrix} \qquad && E_1=\begin{bmatrix}
0 & 0 & 0 \\
0 & 0 & 1 \\
0 & 0 & 0 
\end{bmatrix}\\
&F_0= \frac{1}{u}\begin{bmatrix}
0 & 0 & \sqrt{[2]_t} \\
\sqrt{[2]_t} & 0 & 0 \\
0 & 0 & 0
\end{bmatrix} \qquad && F_1= \begin{bmatrix}
0 & 0 & 0 \\
0 & 0 & 0 \\
0 & 1 & 0 
\end{bmatrix}\\
 &H_0= \begin{bmatrix}
0 & 0 & 0 \\
0 & -2 & 0 \\
0 & 0 & 2 
\end{bmatrix} \qquad && H_1= \begin{bmatrix}
0 & 0 & 0 \\
0 & 1 & 0 \\
0 & 0 & -1 
\end{bmatrix}
\end{alignat*}

$U_t(A_2^{(2)})$ contains a $U_{t^4}(A_1)$ Hopf subalgebra generated by $E_1$, $F_1$ and $H_1$. Setting $q=-t^4$, we have that $V_u=\mathbb{C}\oplus V_1$ as representations of this $U_{t^4}(A_1)$ subalgebra. We can write a basis of $\text{End}_{U_{t^4}(A_1)}\left((\mathbb{C}\oplus V_1)^2\right)$ in terms of the TL diagrams defined above. The $R$-matrix can then be decomposed as 
\begin{align}
    \check{R}(u,v)=&\ a_1(u,v)\quad\vcenter{\hbox{\includegraphics[scale=0.12]{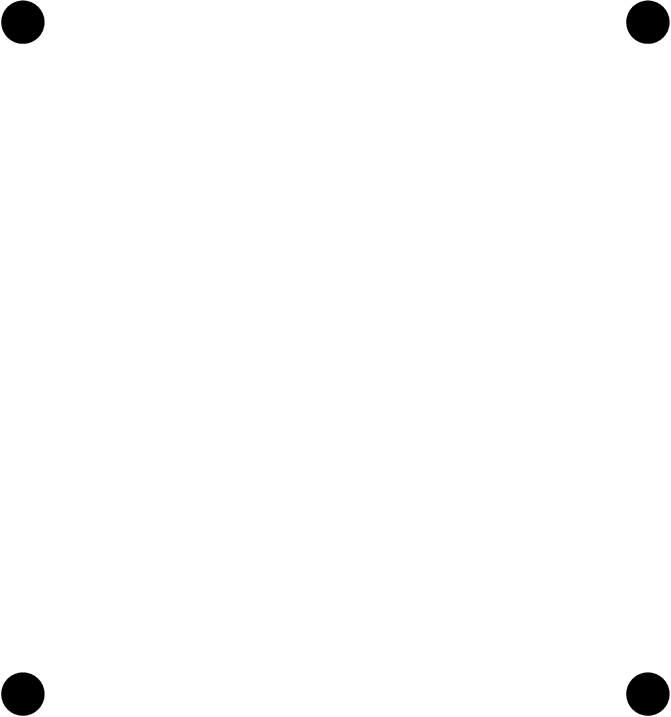}}}\quad+a_2(u,v)\quad\vcenter{\hbox{\includegraphics[scale=0.12]{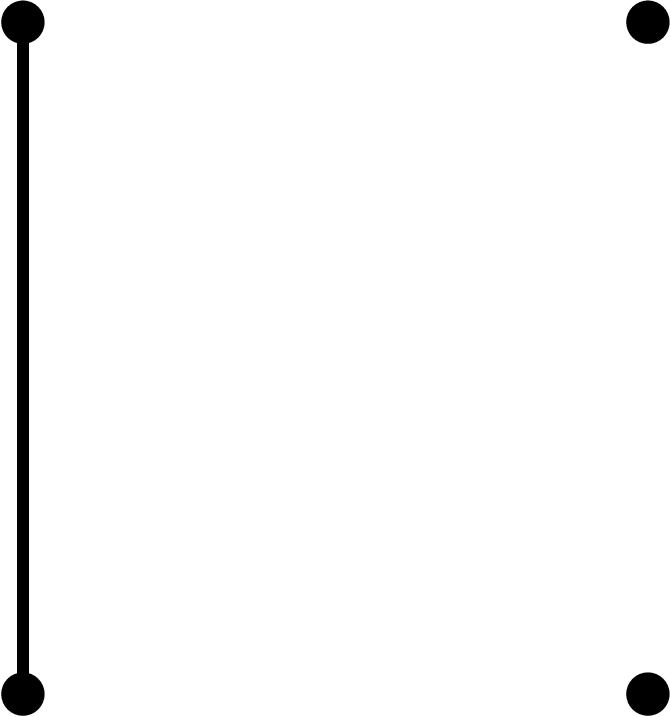}}}\quad+a_3(u,v)\quad\vcenter{\hbox{\includegraphics[scale=0.12]{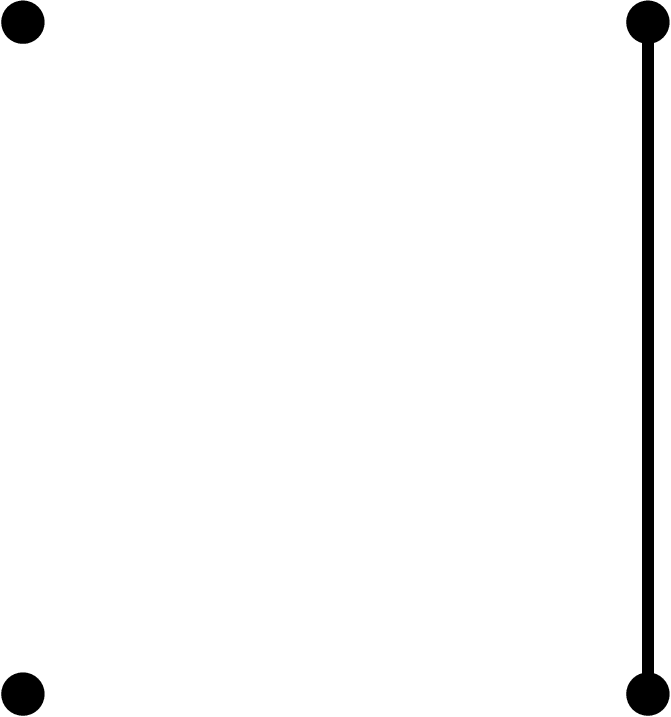}}}\nonumber\\[10pt]
    +&a_4(u,v)\quad\vcenter{\hbox{\includegraphics[scale=0.12]{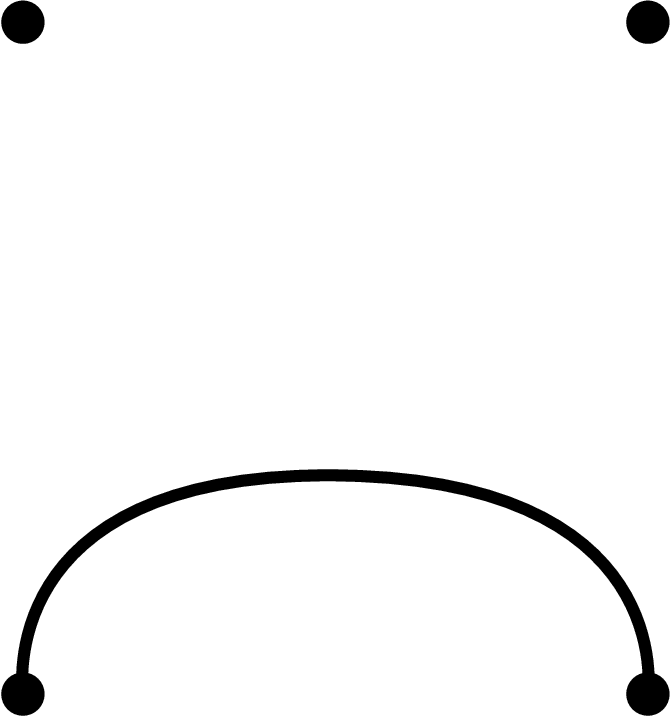}}}\quad+a_5(u,v)\quad\vcenter{\hbox{\includegraphics[scale=0.12]{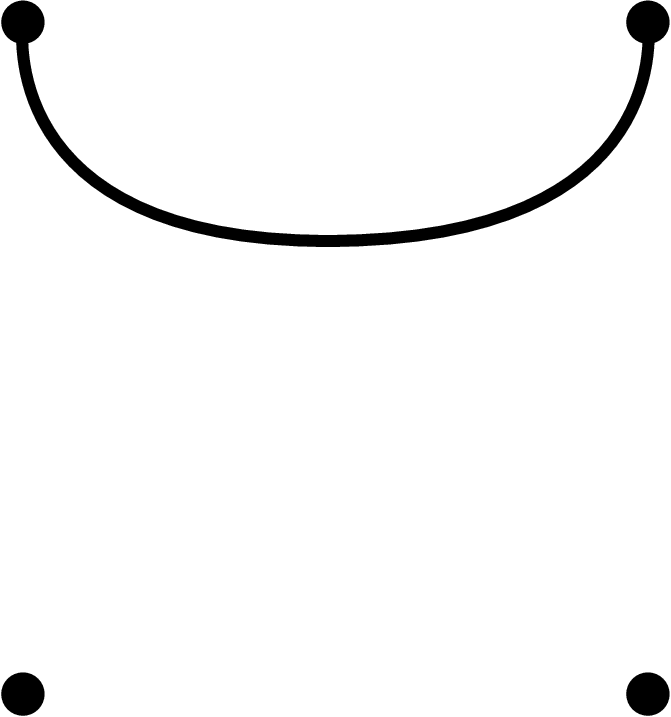}}}\quad+a_6(u,v)\quad\vcenter{\hbox{\includegraphics[scale=0.12]{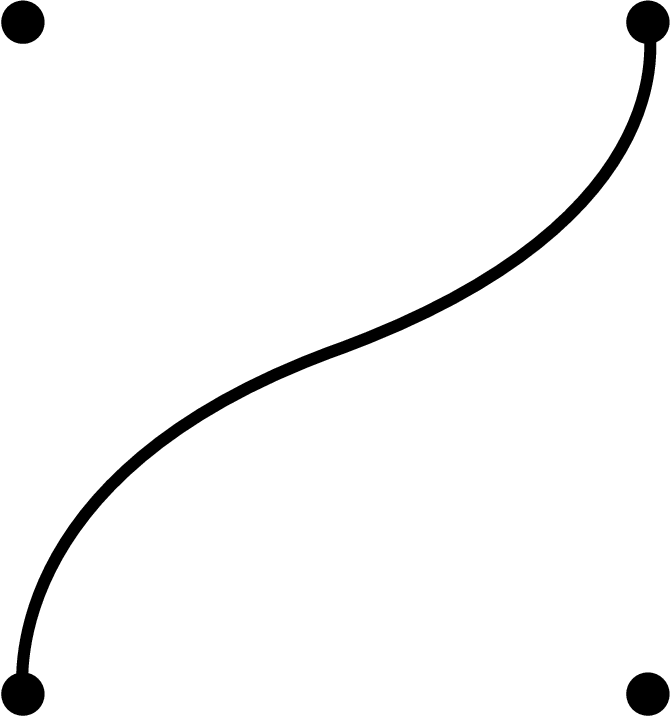}}}\\[10pt]
    +&a_7(u,v)\quad\vcenter{\hbox{\includegraphics[scale=0.12]{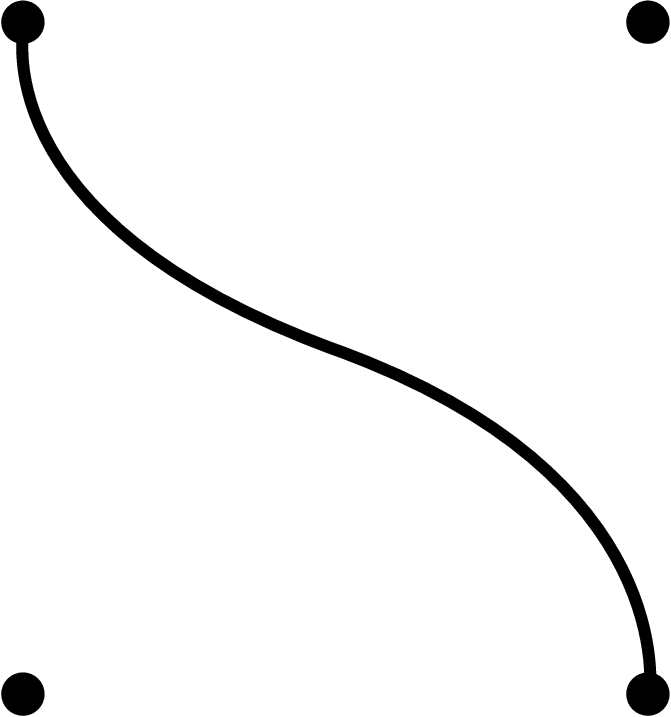}}}\quad+a_8(u,v)\quad\vcenter{\hbox{\includegraphics[scale=0.12]{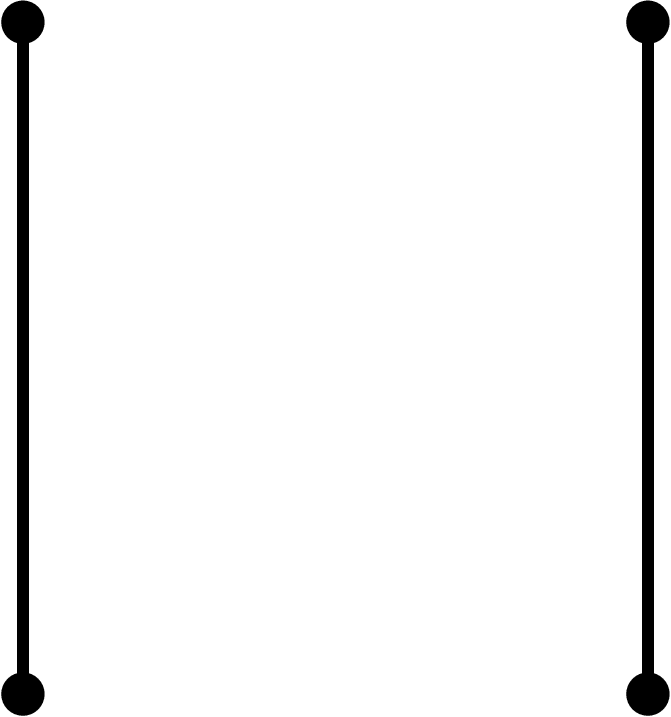}}}\quad+a_9(u,v)\quad\vcenter{\hbox{\includegraphics[scale=0.12]{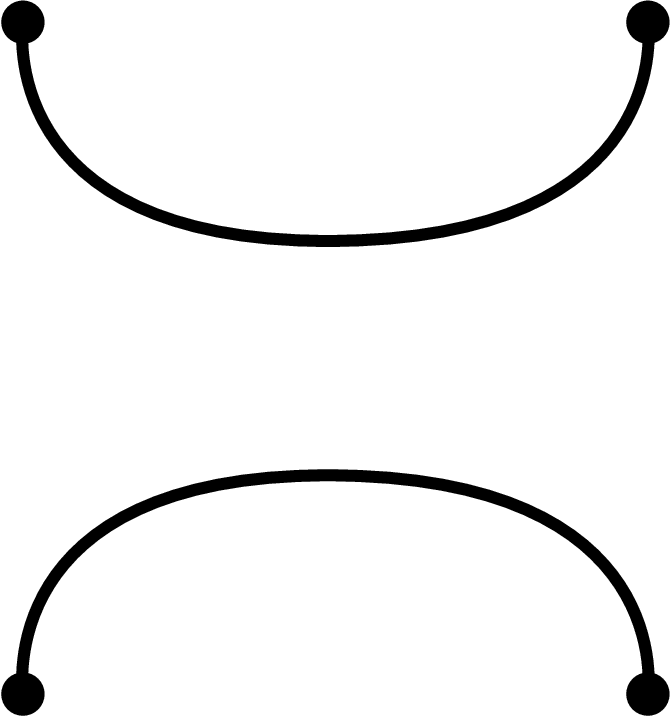}}}\nonumber
\end{align}

Asking that the $R$-matrix commute with the $0$th labeled generators gives a system of linear equations for the coefficients $a_i(u,v)$. From the spectral parameter dependence of the representatives of the $0$th labeled generators, we see that the coefficients depend only on the ratio $s=\frac{u}{v}$ and we write $a_i(u,v)=a_i(s)$. The unique solution, up to a multiplicative constant, is given by
\begin{subequations}
\begin{align}
    a_1(s)=&t^8-t^2 s^4+(t-1) (t+1) \left(t^8-t^4+1\right) s^2\\
    a_2(s)=&\left(t^4-1\right) s \left(t^6+s^2\right)\\
    a_3(s)=&\left(t^4-1\right) s \left(t^6+s^2\right)\\
    a_4(s)=&t^2 \left(t^4-1\right) s \left(s^2-1\right)\\
    a_5(s)=&-t^4 \left(t^4-1\right) s \left(s^2-1\right)\\
    a_6(s)=&-t^2 \left(s^2-1\right) \left(t^6+s^2\right)\\
    a_7(s)=&-t^2 \left(s^2-1\right) \left(t^6+s^2\right)\\
    a_8(s)=&\left(t^4-s^2\right) \left(t^6+s^2\right)\\
    a_9(s)=&-t^4 \left(s^2-1\right) \left(t^2+s^2\right)
\end{align}
\end{subequations}
It was originally found in \cite{nienhuis_critical_1990}.

We see that, if we want to recover \eqref{LoopLocTM}, we need to tune the spectral parameter such that $a_8(s)=0$ and $a_i(s)\neq0$ for $i\neq 8$. This happens for $s=-t^2$. Then, by renormalizing the R matrix and taking the following gauge transformation 
\begin{align}
    D&=\text{Diag}(1,\alpha,\alpha)\nonumber\\
    \check{R}(s)&\mapsto \left(D^{-1}\otimes D^{-1}\right) \check{R}(s) \left(D\otimes D \right)
\end{align}
for well chosen $\alpha$, we recover \eqref{LoopLocTM} with 
\begin{subequations}
\begin{align}
    q=&-e^{4i\psi}\\
    x=&\frac{1}{2\sin(\psi)}
\end{align}
\end{subequations}
Remark that, after a gauge transformation, the $R$-matrix still satisfies the Yang-Baxter equation. The gauge transformation is equivalent to the choice of multiplicative constant in defining the cup and cap maps.

In the next sections, we will employ the same strategy for the rank $2$ web models.

\subsection{The $A_2$ web model}
\label{sec:integrabilityA2}
We now consider the second line of the table \ref{table:qg}. Let $V_u$, $u\in \mathbb{C}^*$, be the representation of $U_t(G_2^{(1)})$ given in Appendix \ref{G21rep} which is actually isomorphic to the one considered in \cite{takacs_quantum_1997}. We are looking for an operator $\check{R}(u,v)$ intertwining $V_u\otimes V_v$ and $V_v\otimes V_u$. Remark that, in $U_t(G_2^{(1)})$, $E_i$, $F_i$ and $H_i$ for $i=0,1$ generate a Hopf subalgebra isomorphic to $U_{t^3}(A_2)$. Under the action of this subalgebra, $V_u$ decomposes as:
\begin{align}
    V_u=V_1\oplus V_2 \oplus \mathbb{C}
   \label{Vu_A2_decomp}
\end{align}
which can be seen from the explicit matrix expressions given in section \ref{sec:UqA2reps} after some relabelling of the nodes in the Dynkin diagram.

Hence $\check{R}(u,v)$ will decompose as a sum of $U_{t^3}(A_2)$ intertwiners:
\begin{align}
    \check{R}(u,v)=&\ a_1(u,v)\quad\vcenter{\hbox{\includegraphics[scale=0.10]{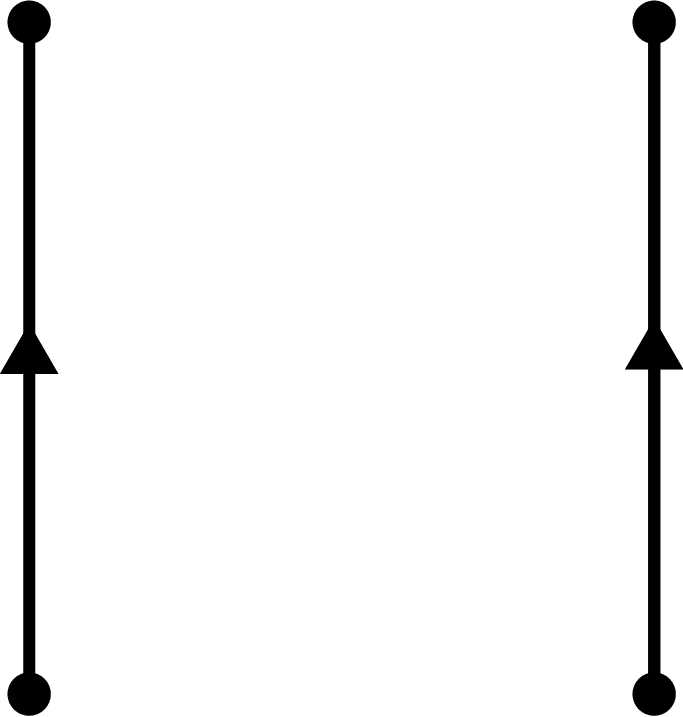}}}\quad+a_2(u,v)\quad\vcenter{\hbox{\includegraphics[scale=0.10]{diagrams/A2Rmat2.eps}}}\quad+a_3(u,v)\quad\vcenter{\hbox{\includegraphics[scale=0.10]{diagrams/A2Rmat3.eps}}}\quad
    +a_4(u,v)\quad\vcenter{\hbox{\includegraphics[scale=0.10]{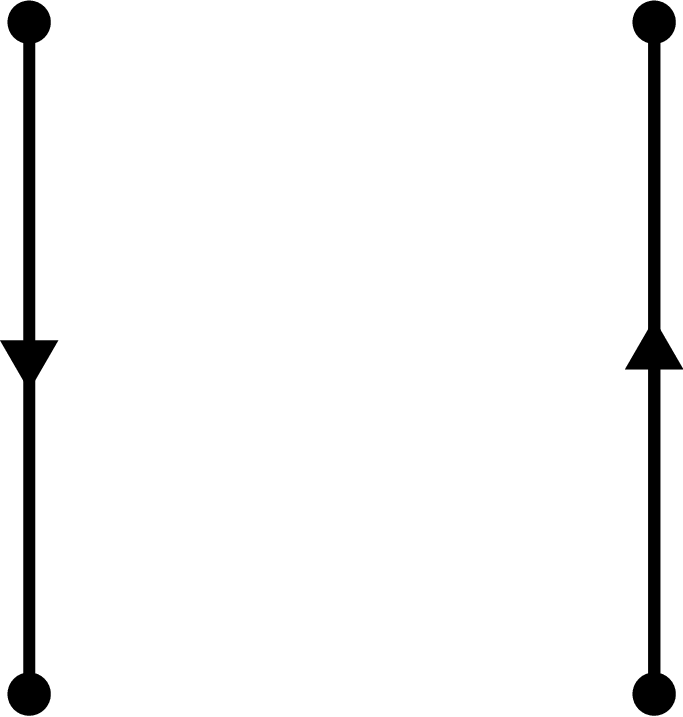}}}\nonumber\\[10pt]
    &+a_5(u,v)\quad\vcenter{\hbox{\includegraphics[scale=0.10]{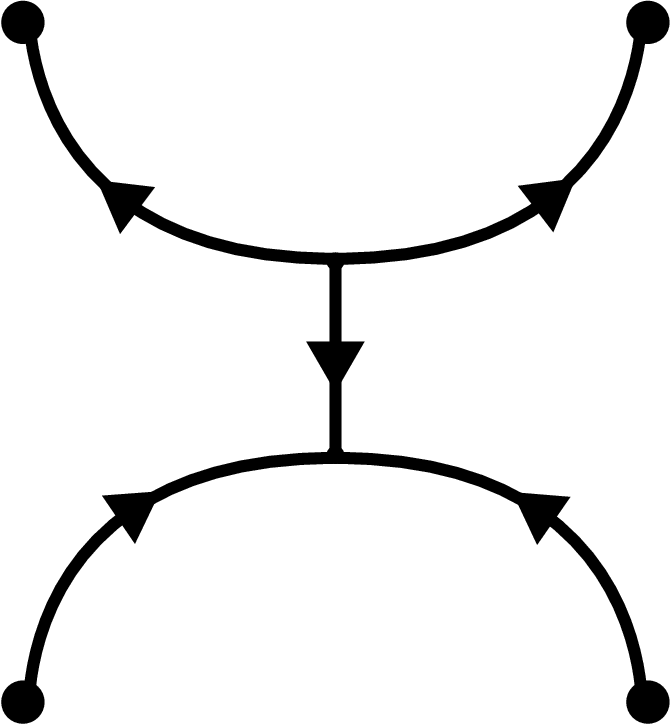}}}\quad+a_6(u,v)\quad\vcenter{\hbox{\includegraphics[scale=0.10]{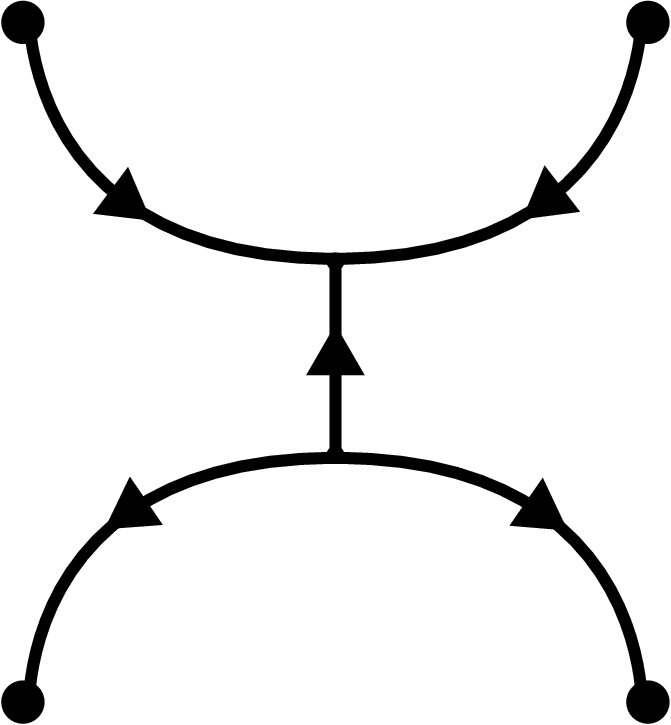}}}\quad +a_7(u,v)\quad\vcenter{\hbox{\includegraphics[scale=0.10]{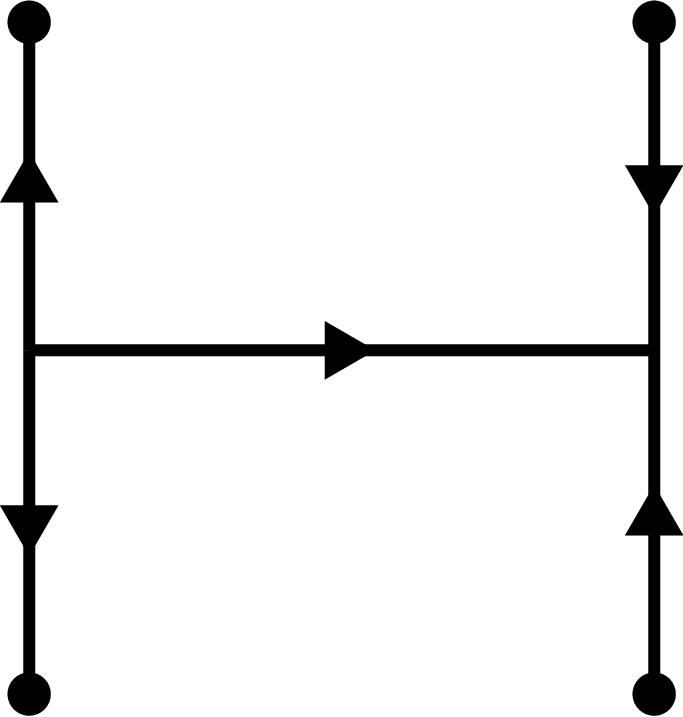}}} \quad +a_8(u,v)\quad\vcenter{\hbox{\includegraphics[scale=0.10]{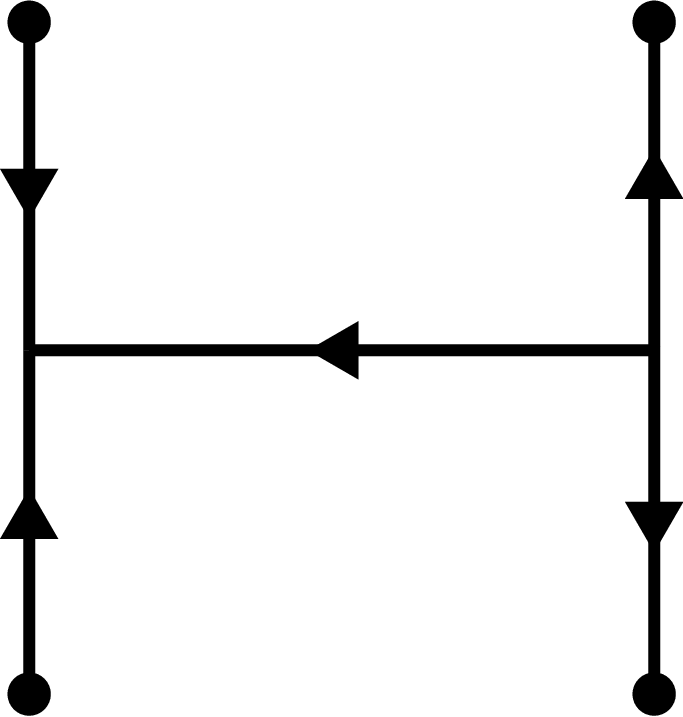}}}\nonumber\\[10pt]
    &+a_9(u,v)\quad\vcenter{\hbox{\includegraphics[scale=0.10]{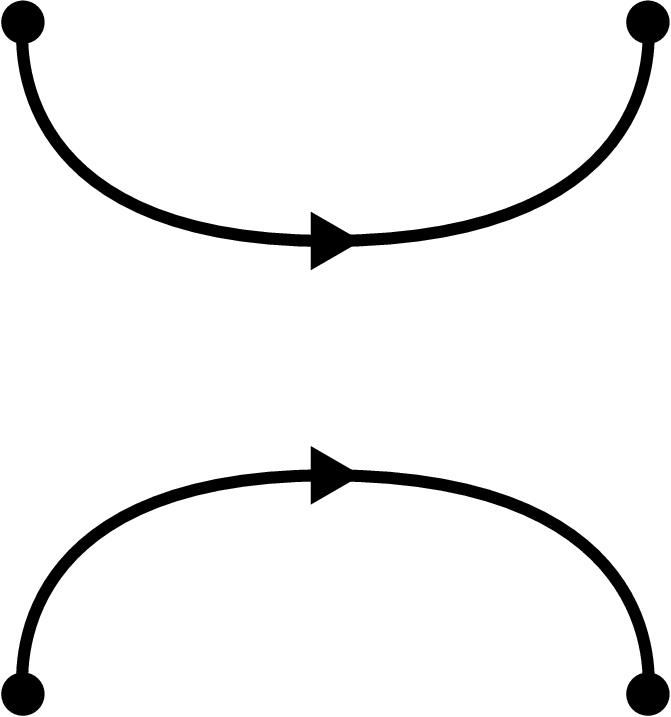}}}\quad+a_{10}(u,v)\quad\vcenter{\hbox{\includegraphics[scale=0.10]{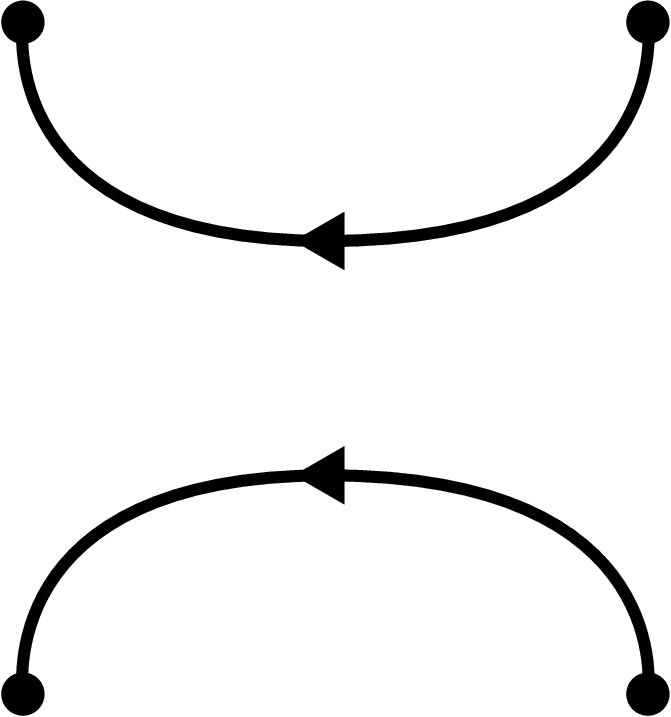}}}\quad+a_{11}(u,v)\quad\vcenter{\hbox{\includegraphics[scale=0.10]{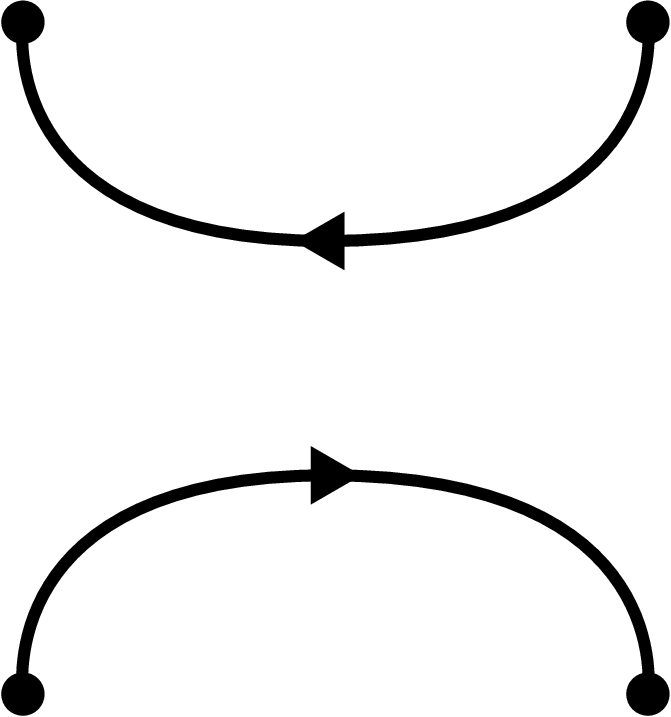}}}\quad+a_{12}(u,v)\quad\vcenter{\hbox{\includegraphics[scale=0.10]{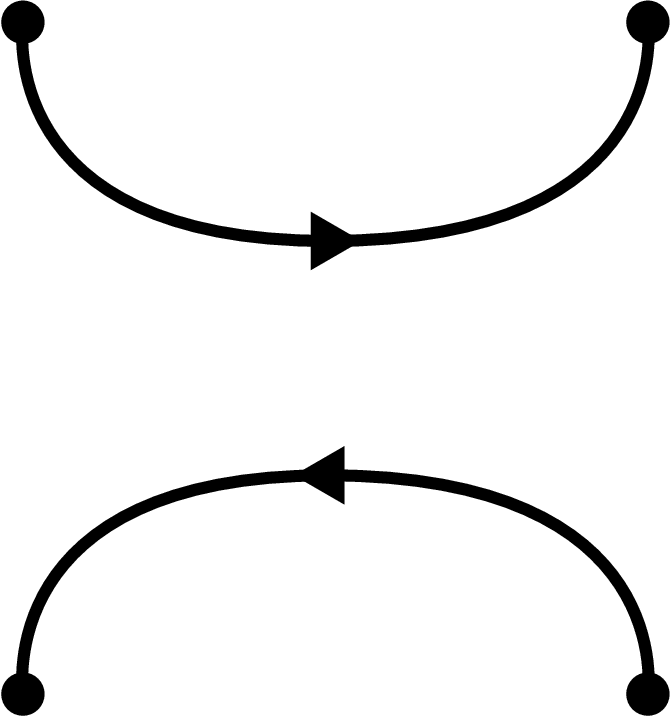}}}\nonumber\\[10pt]
    &+a_{13}(u,v)\quad\vcenter{\hbox{\includegraphics[scale=0.10]{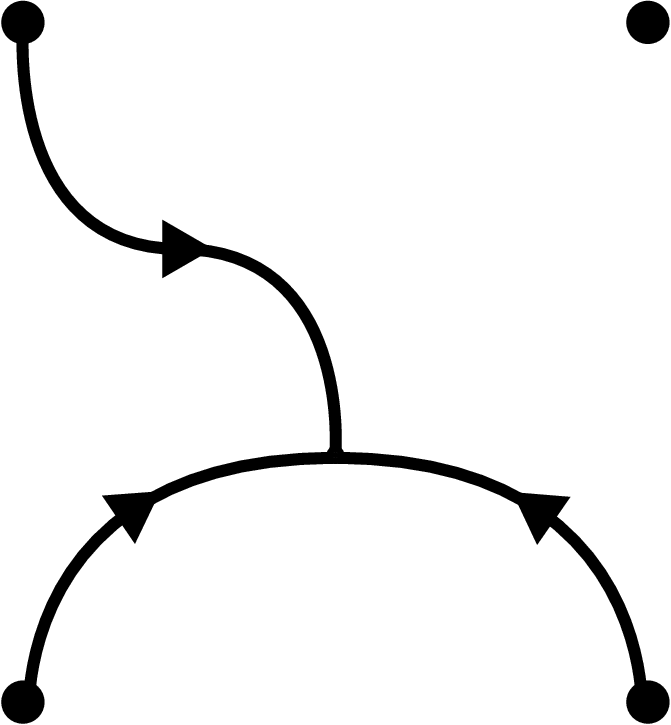}}}\quad+a_{14}(u,v)\quad\vcenter{\hbox{\includegraphics[scale=0.10]{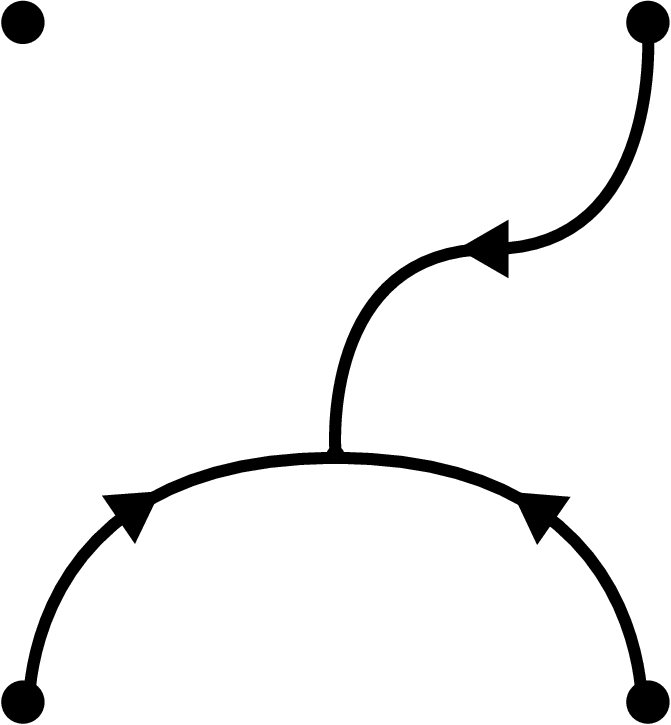}}}\quad+a_{15}(u,v)\quad\vcenter{\hbox{\includegraphics[scale=0.10]{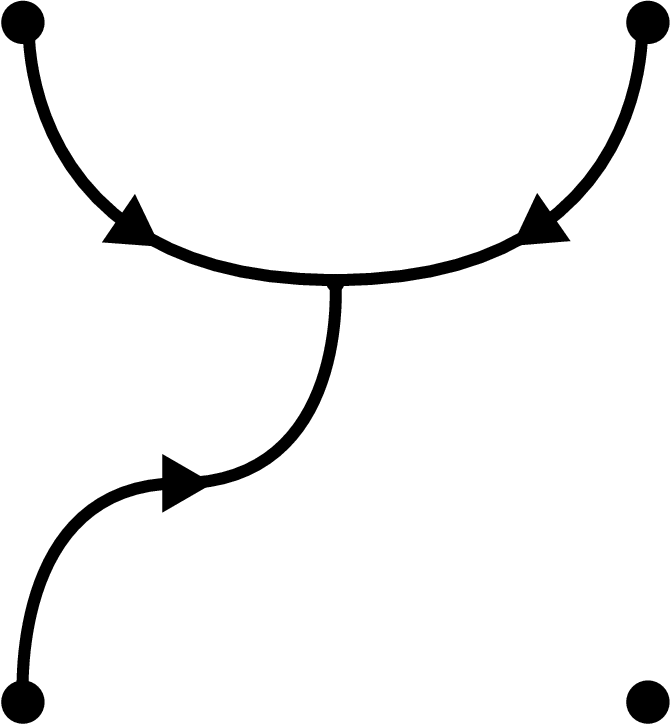}}}\quad+a_{16}(u,v)\quad\vcenter{\hbox{\includegraphics[scale=0.10]{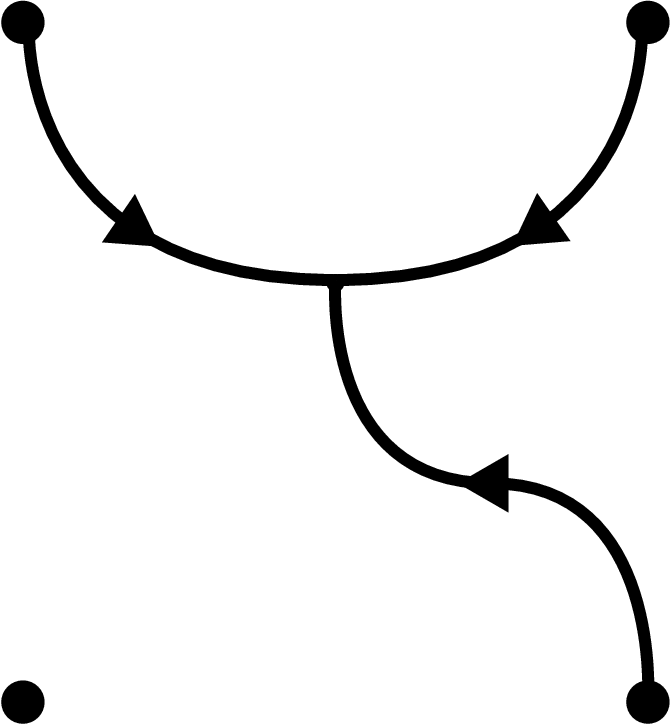}}}\nonumber\\[10pt]
    &+a_{17}(u,v)\quad\vcenter{\hbox{\includegraphics[scale=0.10]{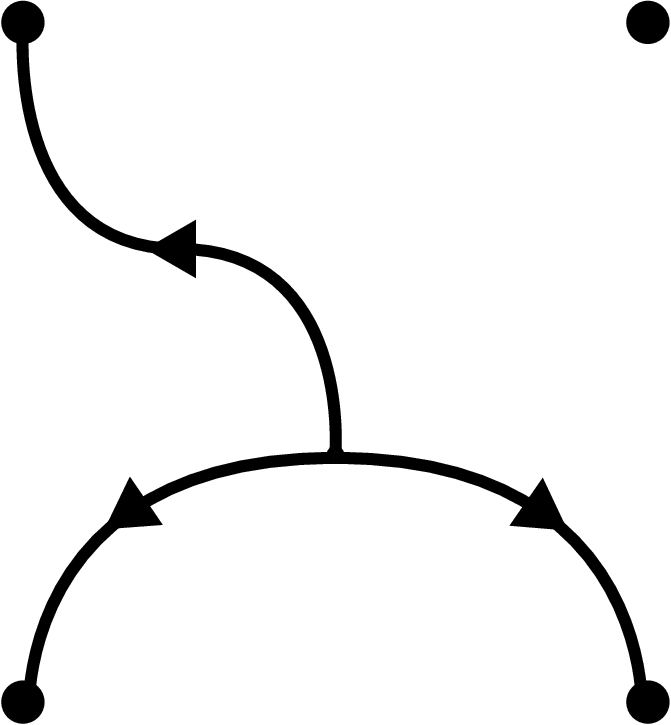}}}\quad+a_{18}(u,v)\quad\vcenter{\hbox{\includegraphics[scale=0.10]{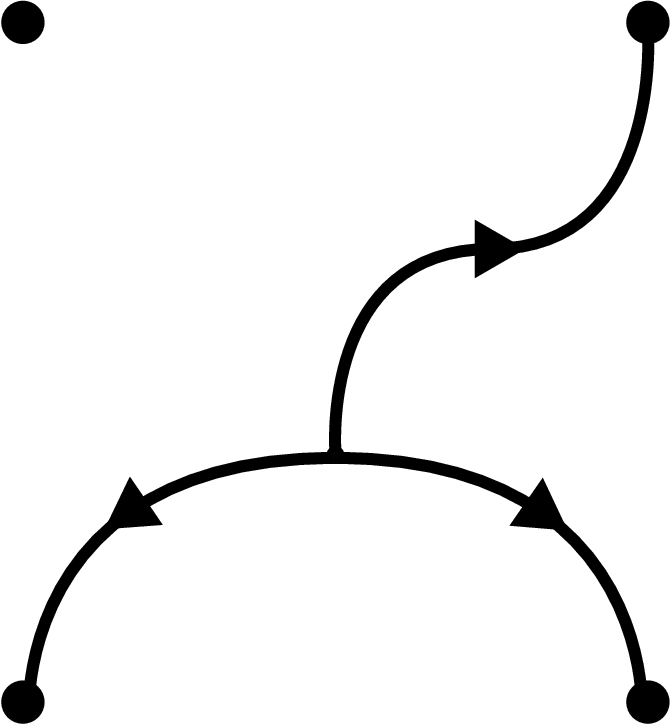}}}\quad+a_{19}(u,v)\quad\vcenter{\hbox{\includegraphics[scale=0.10]{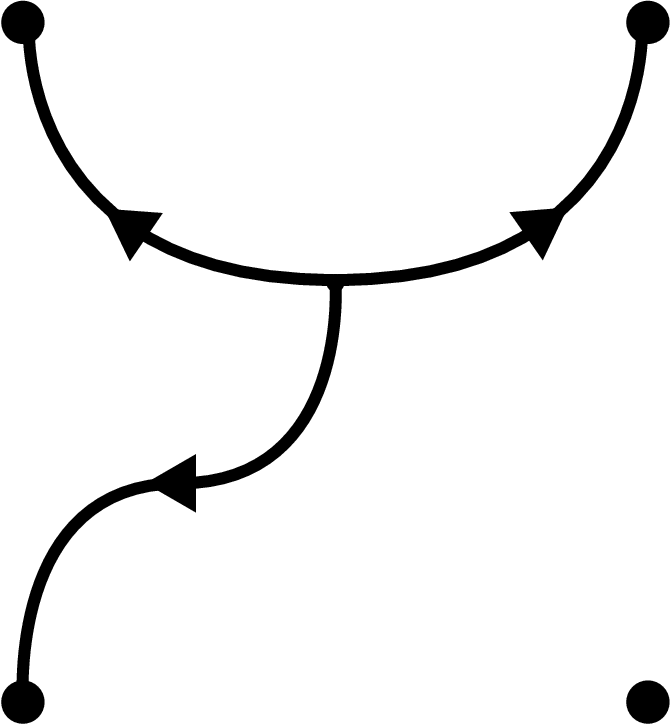}}}\quad+a_{20}(u,v)\quad\vcenter{\hbox{\includegraphics[scale=0.10]{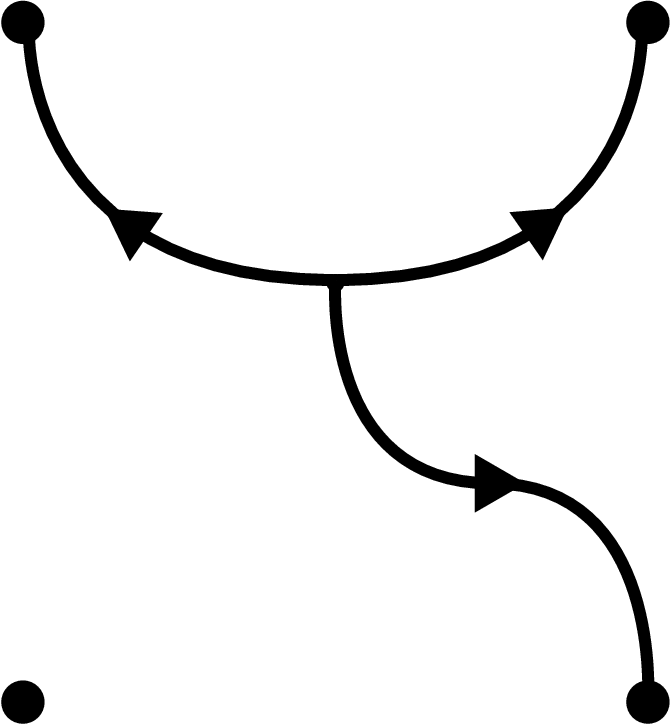}}}\nonumber\\[10pt]
    &+a_{21}(u,v)\quad\vcenter{\hbox{\includegraphics[scale=0.10]{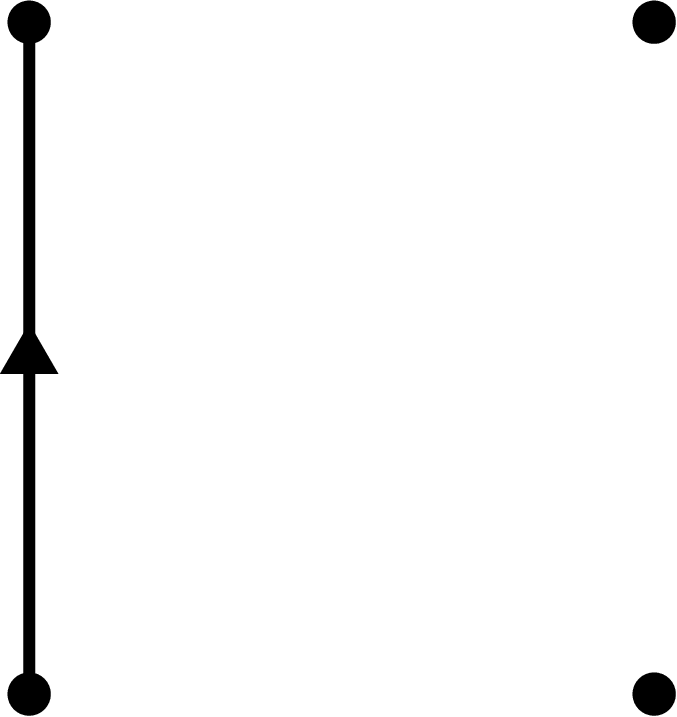}}}\quad+a_{22}(u,v)\quad\vcenter{\hbox{\includegraphics[scale=0.10]{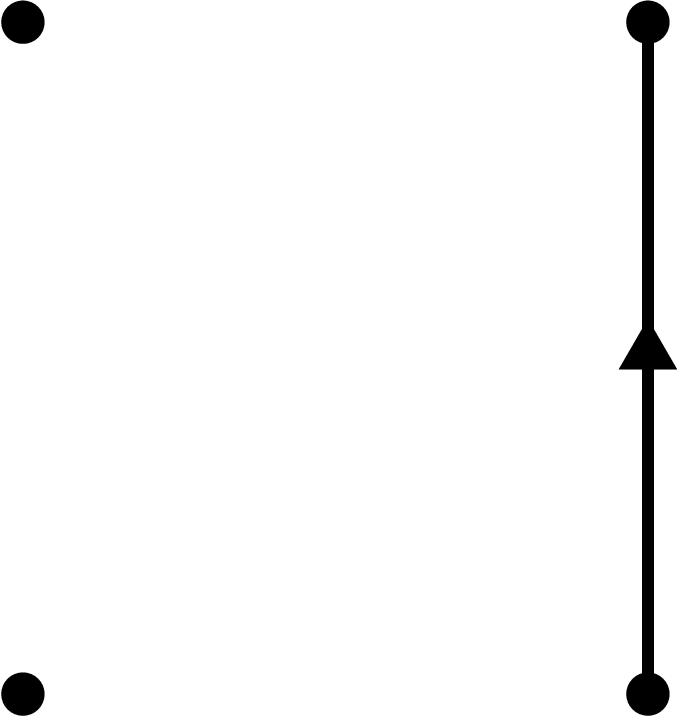}}}\quad+a_{23}(u,v)\quad\vcenter{\hbox{\includegraphics[scale=0.10]{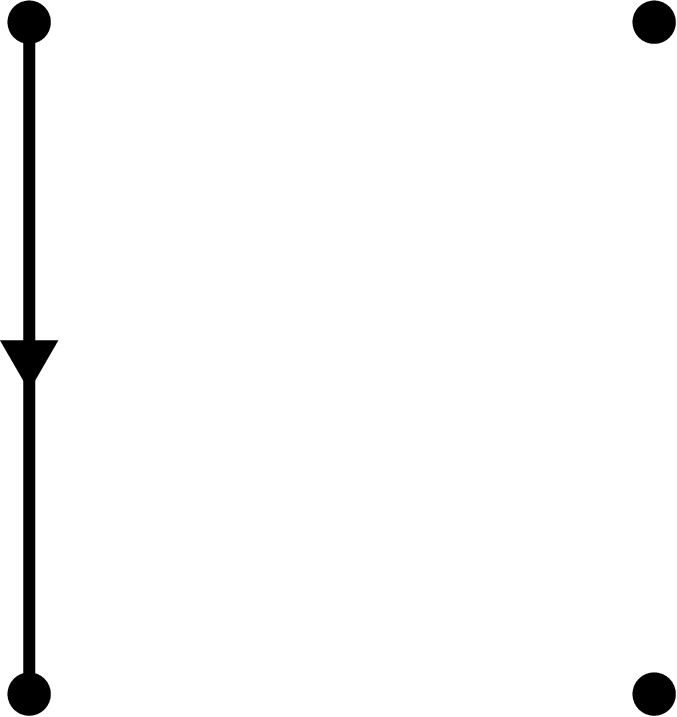}}}\quad+a_{24}(u,v)\quad\vcenter{\hbox{\includegraphics[scale=0.10]{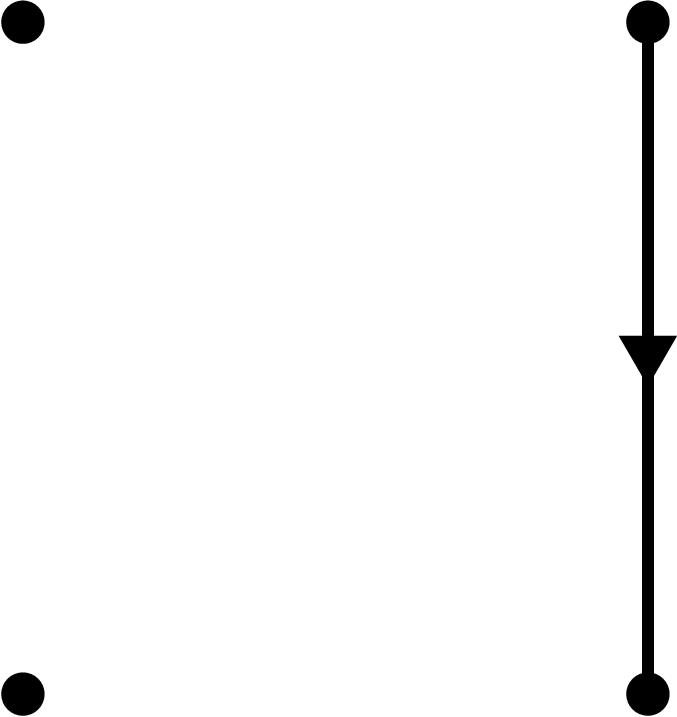}}}\nonumber\\[10pt]
    &+a_{25}(u,v)\quad\vcenter{\hbox{\includegraphics[scale=0.10]{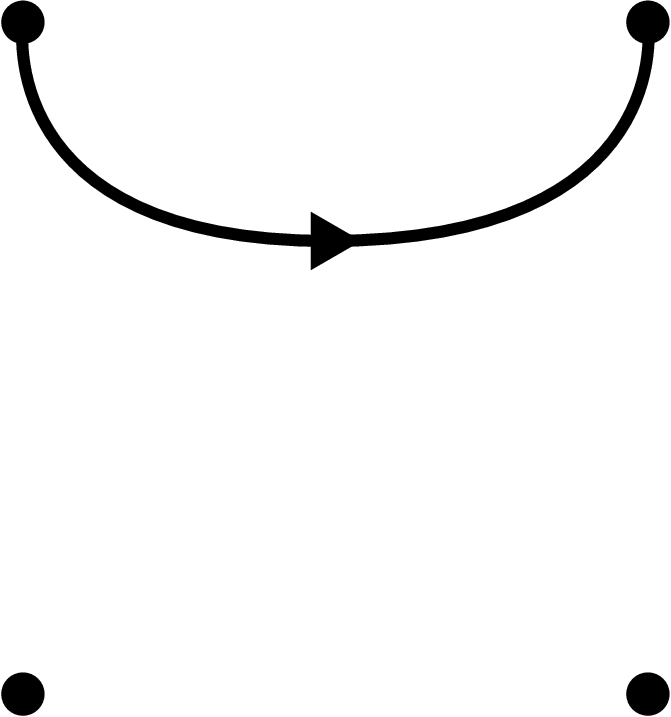}}}\quad+a_{26}(u,v)\quad\vcenter{\hbox{\includegraphics[scale=0.10]{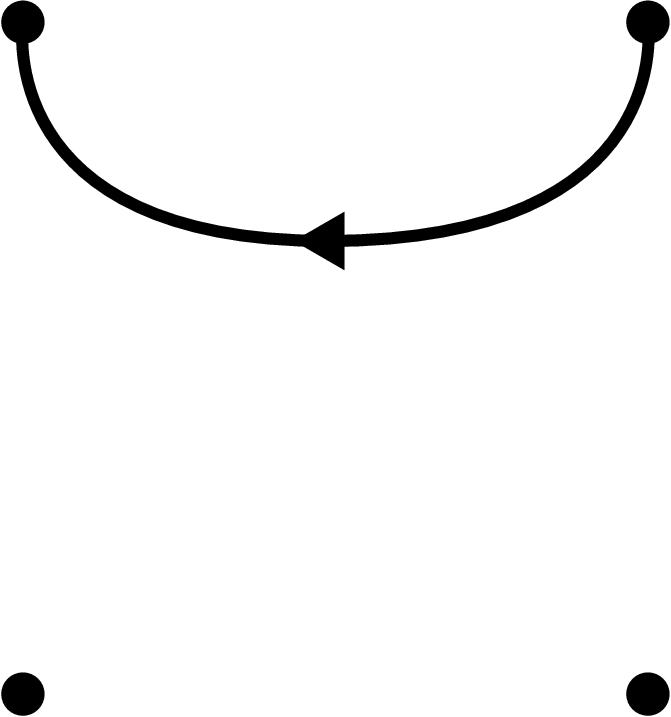}}}\quad+a_{27}(u,v)\quad\vcenter{\hbox{\includegraphics[scale=0.10]{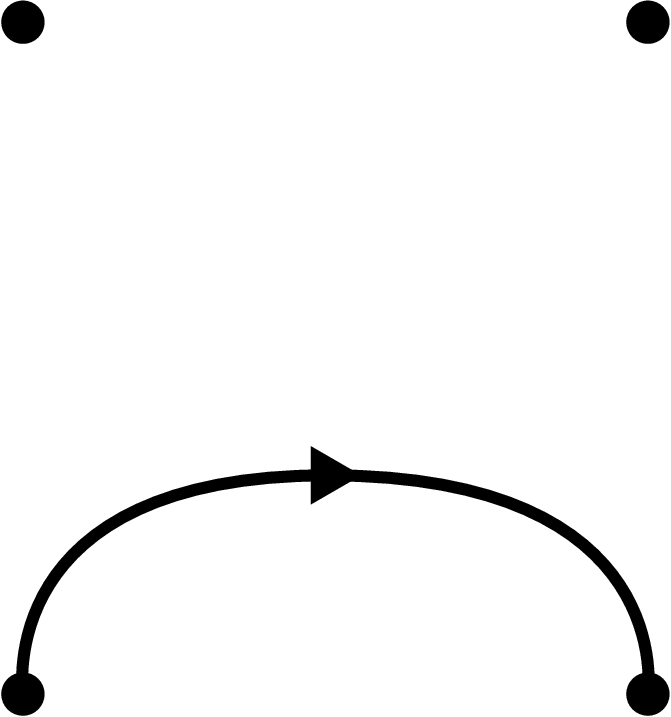}}}\quad+a_{28}(u,v)\quad\vcenter{\hbox{\includegraphics[scale=0.10]{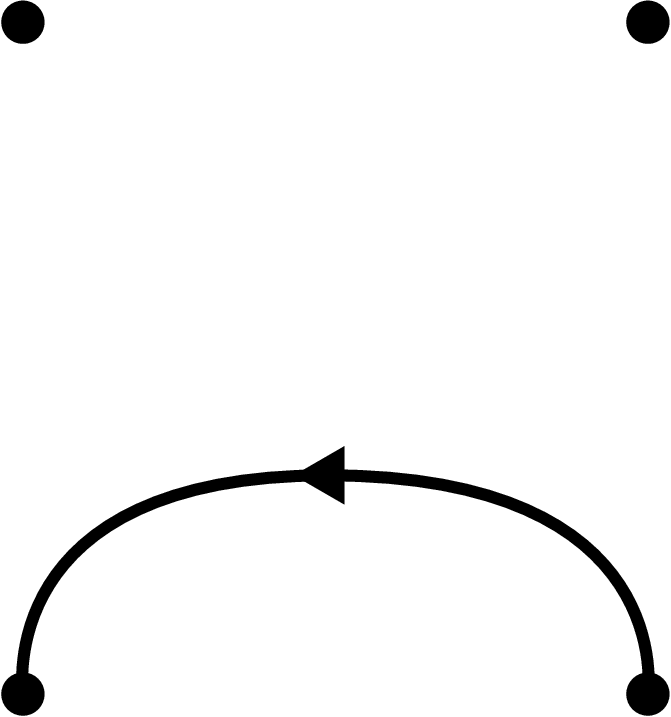}}}\nonumber\\[10pt]
    &+a_{29}(u,v)\quad\vcenter{\hbox{\includegraphics[scale=0.10]{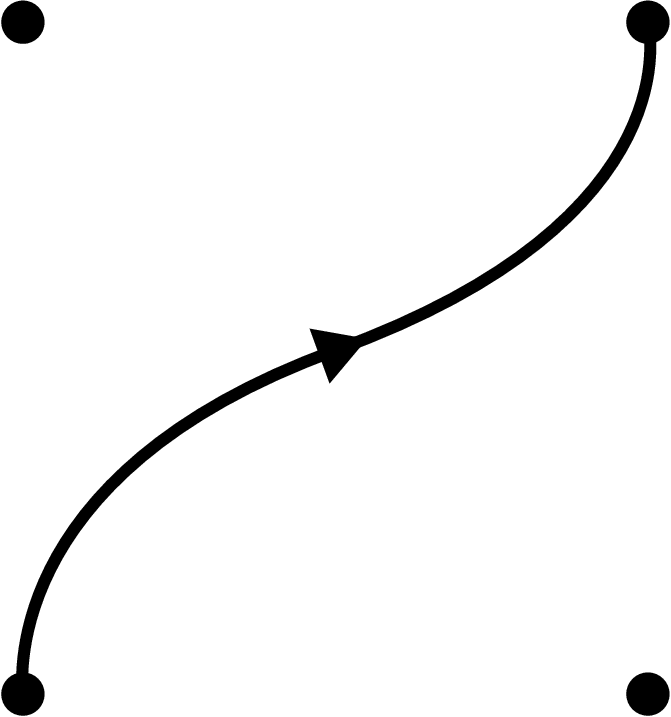}}}\quad+a_{30}(u,v)\quad\vcenter{\hbox{\includegraphics[scale=0.10]{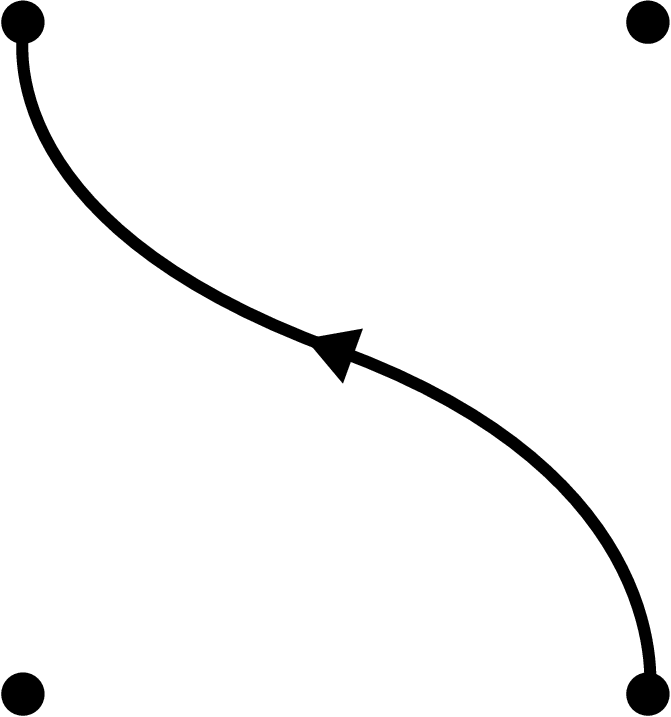}}}\quad+a_{31}(u,v)\quad\vcenter{\hbox{\includegraphics[scale=0.10]{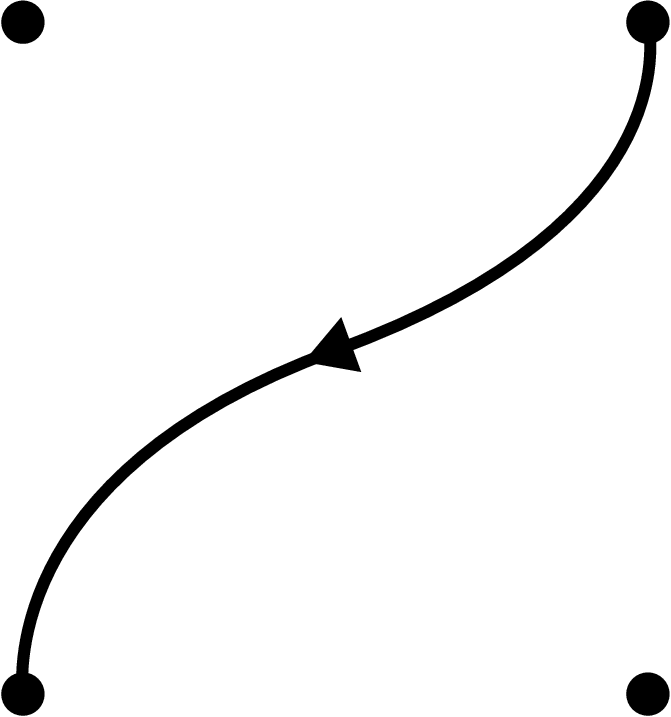}}}\quad+a_{32}(u,v)\quad\vcenter{\hbox{\includegraphics[scale=0.10]{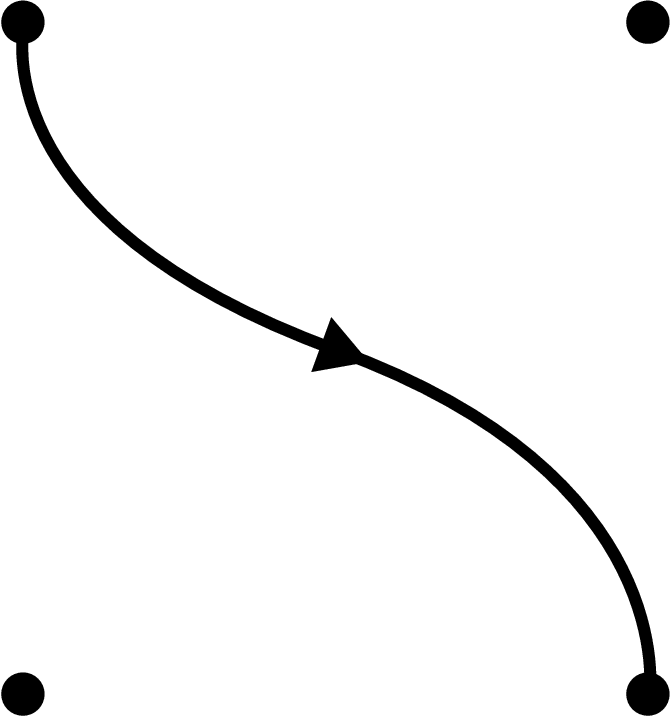}}}\nonumber\\[10pt]
    &+a_{33}(u,v)\quad\vcenter{\hbox{\includegraphics[scale=0.10]{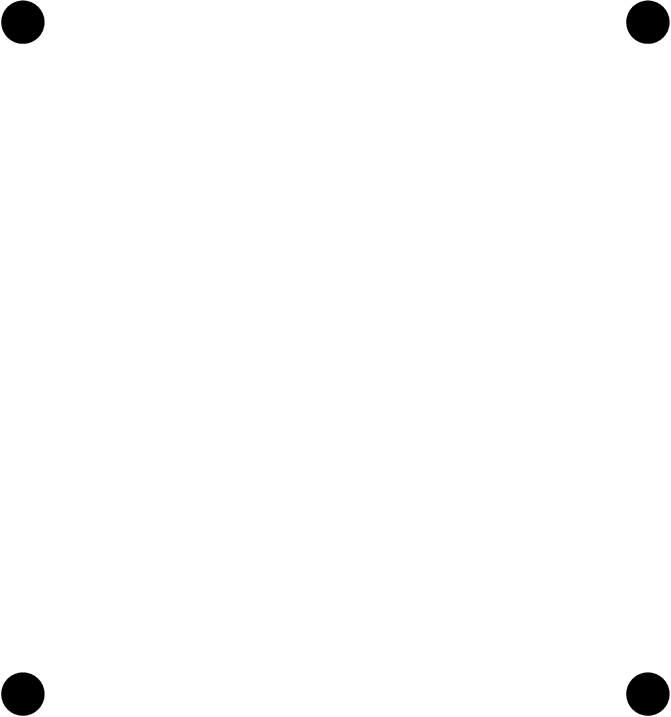}}}
\end{align}
where $a_i(u,v)$ are some coefficients and the corressponding webs span the space of intertwiners $\text{End}_{U_{t^3}(A_2)}\left((V_1\oplus V_2 \oplus \mathbb{C})^2\right)$. Indeed, elements of the latter are in bijection with invariants of $(V_1\oplus V_2 \oplus \mathbb{C})^4$, so upon expanding we get products of $V_1$ and $V_2$ of length $0$ up to $4$. Invariants on these spaces were classified by Kuperbergs \cite{Kuperberg_1996}, our webs are basis elements  of them. 

Asking for $\check{R}(u,v)$ to commute with the remaining generators of $U_t(G_2^{(1)})$, we see that it may depend only on the ratio $s=\frac{u}{v}$. 
We find
\begin{alignat*}{2}
    a_1(s)=&-\frac{\left(t^2-s^3\right)\left(t^8-s^3\right)}{t^2 \sqrt{\frac{1}{t}+t}\left(-1+t^2\right) s^2\left(-1+s^3\right)}\qquad
    &&a_2(s)=-\frac{\left(t^2-s^3\right)\left(t^8-s^3\right)}{t^2 \sqrt{\frac{1}{t}+t}\left(-1+t^2\right) s^2\left(-1+s^3\right)}\\
    a_3(s)=&\frac{-t^8+s^3}{t^2 \sqrt{\frac{1}{t}+t} \left(-1+s^3\right)}\qquad
    &&a_4(s)=\frac{-t^8+s^3}{t^2 \sqrt{\frac{1}{t}+t} \left(-1+s^3\right)}\\
    a_5(s)=&\frac{\sqrt{\frac{1}{t}+t}\left(t^6-t^2 s^3\right)}{\left(-1+t^4\right) s^2}\qquad
    &&a_6(s)=\frac{\sqrt{\frac{1}{t}+t}\left(t^6-t^2 s^3\right)}{\left(-1+t^4\right) s^2}\\
    a_7(s)=&\frac{\sqrt{\frac{1}{t}+t}\left(t^8-s^3\right)}{\left(-1+t^4\right) s^2}\qquad
    &&a_8(s)=\frac{\sqrt{\frac{1}{t}+t}\left(t^8-s^3\right)}{\left(-1+t^4\right) s^2}\\
    a_9(s)=&\frac{t^2\left(t^4-s^3\right)\left(t^{10}-s^3\right)}{\sqrt{\frac{1}{t}+t}\left(-1+t^2\right) s^2\left(t^{12}-s^3\right)}\qquad
    &&a_{10}(s)=\frac{t^2\left(t^4-s^3\right)\left(t^{10}-s^3\right)}{\sqrt{\frac{1}{t}+t}\left(-1+t^2\right) s^2\left(t^{12}-s^3\right)}\\
    a_{11}(s)=&\frac{t^8\left(-t^4+s^3\right)}{\sqrt{\frac{1}{t}+t} \left(t^{12}-s^3\right)}\qquad
    &&a_{12}(s)=\frac{t^4\left(-t^4+s^3\right)}{\sqrt{\frac{1}{t}+t} \left(t^{12}-s^3\right)}\\    
    a_{13}(s)=&1\qquad &&a_{14}(s)=\frac{-t^2}{s}\\
    a_{15}(s)=&\frac{-t}{s}\qquad &&a_{16}(s)=\frac{1}{t}\\
    a_{17}(s)=&\frac{t^5}{s}\qquad &&a_{18}(s)=-t^3\\
    a_{19}(s)=&-t^2\qquad &&a_{20}(s)=\frac{t^4}{s}\\
    a_{21}(s)=&\frac{\sqrt{\frac{1}{t}+t}\left(-t^2+t^4-t^6+s^3\right)}{t\left(-1+s^3\right)}\qquad
    &&a_{22}(s)=\frac{-t^8+s^3+t^6\left(-1+s^3\right)}{t^2 \sqrt{\frac{1}{t}+t} s\left(-1+s^3\right)}\\
    a_{23}(s)=&\frac{-t^8+s^3+t^6\left(-1+s^3\right)}{t^2 \sqrt{\frac{1}{t}+t} s\left(-1+s^3\right)}\qquad
    &&a_{24}(s)=\frac{\sqrt{\frac{1}{t}+t}\left(-t^2+t^4-t^6+s^3\right)}{t\left(-1+s^3\right)}\\
    a_{25}(s)=&\frac{-t^{11}(1+t^2)+t(1+t^6)s^3}{\sqrt{\frac{1}{t}+t}s(t^{12}-s^3)}\qquad
    &&a_{26}(s)=\frac{t^2\sqrt{\frac{1}{t}+t}(-t^6+t^8-t^{10}+s^3)}{t^{12}-s^3}\\
    a_{27}(s)=&\frac{t^4\sqrt{\frac{1}{t}+t}(-t^6+t^8-t^{10}+s^3)}{t^{12}-s^3}\qquad
    &&a_{28}(s)=\frac{-t^{13}(1+t^2)+t^3(1+t^6)s^3}{\sqrt{\frac{1}{t}+t}s(t^{12}-s^3)}\\
    a_{29}(s)=&\frac{t^6-s^3}{\sqrt{\frac{1}{t}+t}(-1+t^2)s^2}\qquad
    &&a_{30}(s)=\frac{t^6-s^3}{\sqrt{\frac{1}{t}+t}(-1+t^2)s^2}\\
    a_{31}(s)=&\frac{t^6-s^3}{\sqrt{\frac{1}{t}+t}(-1+t^2)s^2}\qquad
    &&a_{32}(s)=\frac{t^6-s^3}{\sqrt{\frac{1}{t}+t}(-1+t^2)s^2}\
\end{alignat*}
\begin{align*}
    a_{33}(s)=&-\frac{\left(t^2-s\right) \left(t^4+t^2 s+s^2\right) \left(t^2 \left(t^{12}+\left(t^{14}-2 t^{12}+t^8-2 t^6+t^4-2\right) s^3+s^6\right)+s^3\right)}{t^2 \sqrt{t+\frac{1}{t}} \left(t^2-1\right) s^2 \left(s^3-1\right) \left(t^{12}-s^3\right)}
\end{align*}
by plugging the linear system for the functions $a_i(s)$ into {\sc Mathematica} (or some other formal calculus software).
One can then show, again using {\sc Mathematica}, that the correponding operator $\check{R}(s)$ satisfies the multiplicative spectral-parameter dependant Yang-Baxter equation. 

We can obtain a manifestly PT-invariant $R$-matrix%
\footnote{We here understand PT-symmetry as the invariance under the rotation of the diagrams through an angle $\pi$.}
by using the following gauge transformation 
\begin{align}
\label{A2Gauge}
    D&=\text{Diag}(\alpha,\alpha,\alpha,\beta,\beta,\beta,1)\nonumber\\
    \check{R}(s)&\rightarrow \left(D^{-1}\otimes D^{-1}\right) \check{R}(s) \left(D\otimes D \right)
\end{align}
The form of $D$ corresponds to rescaling subrepresentations of $V$ in \eqref{Vu_A2_decomp} independently.
Renormalizing the $R$-matrix, we obtain
\begin{align*}
    &a_1(s)=a_2(s)=-\frac{\left(t^2-s^3\right) \left(t^8-s^3\right)}{t \left(t^2-1\right) s^2 \left(s^3-1\right)}\\
    &a_3(s)=\frac{t^8-s^3}{t s-t s^4}\\
    &a_4(s)=s\frac{t^8-s^3}{ts-t s^4}\\
    &a_5(s)=a_6(s)=\frac{t^6-t^2 s^3}{\left(t^2-1\right) s^2}\\
    &a_7(s)=a_8(s)=\frac{t^8-s^3}{\left(t^2-1\right) s^2}\\
    &a_9(s)=a_{10}(s)=\frac{t^3 \left(t^4-s^3\right) \left(t^{10}-s^3\right)}{\left(t^2-1\right) s^2 \left(t^{12}-s^3\right)}\\
    &a_{11}(s)=t^9\frac{ t^4-s^3}{s^4-t^{12} s}\\
    &a_{12}(s)=t^5s\frac{ t^4-s^3}{s^4-t^{12} s}\\
    &a_{13}(s)=a_{16}(s)=a_{18}(s)=a_{19}(s)=-i t^2 \sqrt{t+\frac{1}{t}}\\
    &a_{14}(s)=a_{15}(s)=a_{17}(s)=a_{20}(s)=\frac{i t^4 \sqrt{t+\frac{1}{t}}}{s}\\
    &a_{21}(s)=a_{24}(s)=\frac{-t^8+t^2 \left(s^3-1\right)+s^3}{t \left(s^3-1\right)}\\
    &a_{22}(s)=a_{23}(s)=\frac{t^8-\left(t^6+1\right) s^3+t^6}{t \left(s-s^4\right)}\\
    &a_{25}(s)=a_{28}(s)=\frac{t^3 \left(t^{12}+t^{10}-\left(t^6+1\right) s^3\right)}{s^4-t^{12} s}\\
    &a_{26}(s)=a_{27}(s)=\frac{t^3 \left(\left(t^2+1\right) s^3-t^6 \left(t^6+1\right)\right)}{t^{12}-s^3}\\
    &a_{29}(s)=a_{30}(s)=a_{31}(s)=a_{32}(s)=\frac{t \left(t^6-s^3\right)}{\left(t^2-1\right) s^2}\\
    &a_{33}(s)=-\frac{\left(t^2-s\right) \left(t^4+t^2 s+s^2\right) \left(t^2 \left(t^{12}+\left(t^{14}-2 t^{12}+t^8-2 t^6+t^4-2\right) s^3+s^6\right)+s^3\right)}{t \left(t^2-1\right) s^2 \left(s^3-1\right) \left(t^{12}-s^3\right)}
\end{align*}

It is apparent that by setting the spectral parameter $s=t^{\frac{8}{3}}$, $\check{R}(s)$ will be decomposed only in terms of webs appearing in the local transfer matrix of the $A_2$ web model. In order to put it in the form of \eqref{A2transfermatrix}, we renormalise the PT-invariant $R$-matrix so as to recover the local transfer matrix \eqref{A2transfermatrix} with the following parametrisation, setting $t=e^{i\psi}$:
\begin{subequations}
\begin{align}
    q=&-e^{3i\psi}\\
    x=&\frac{1}{2\sin(\psi)}\\
    y=&z=\sqrt{2\sin(2\psi)}\\
    e^{i\phi}=&ie^{i\frac{\psi}{3}} 
\end{align}
\end{subequations}
Remark that points related by $\psi \rightarrow \pi - \psi$ satisfy
\begin{subequations}
\label{paramsym}
\begin{align}
    q &\rightarrow -q^{-1}\\
    x &\rightarrow x\\
    y &\rightarrow iy\\
    z &\rightarrow iz\\
    e^{i\phi} &\rightarrow e^{-i\phi}\tau
\end{align}
\end{subequations}
with $\tau^3=1$. Moreover the points related by 
$\psi \rightarrow - \psi$ satisfy
\begin{subequations}
\begin{align}
    &q\rightarrow q^{-1}\\
    &x\rightarrow -x\\
    &y\rightarrow iy\\
    &z\rightarrow iz\\
    &e^{i\phi}\rightarrow -e^{-i\phi}\tau
\end{align}
\end{subequations}
These transformations are combinations of the symmetries mentioned in section \ref{sec:defmodels}.
They imply that it suffices to focus on the interval $\psi\in [0,\frac{\pi}{2}]$.

\subsubsection{Central charge and phase diagram}

We have numerically diagonalised the transfer matrix for cylinders of circumference of sizes $L=3,4,\ldots,9$. Estimates $c_L$ for the central charge can then be extracted
from the finite-size scaling of the largest eigenvalue for two different sizes. The computations were made for 100 equally-spaced values of $\psi\in [0,\frac{\pi}{2}]$,
and we show here the curves $c_L(\psi)$ obtained by applying {\sc Mathematica}'s interpolation function to these values.

\begin{figure}
\begin{center}
 \includegraphics[width=0.45\textwidth]{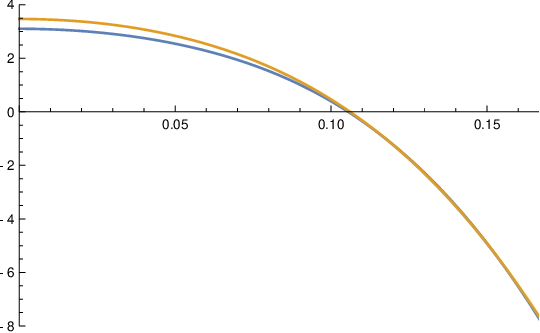} \qquad
 \includegraphics[width=0.45\textwidth]{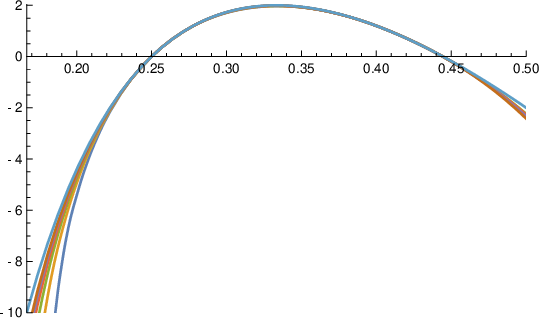}
\end{center}
\caption{Two-point fits $c_L(\psi)$ for the central charge in the $A_2$ web model, plotted against $\tfrac{\psi}{\pi}$. Regime 2 (left panel) is for $\psi \in [0,\tfrac{\pi}{6}[$,
using sizes $\{ L,L-3 \}$ with $L=6$ (dark blue) and $L=9$ (orange). Regime 1 (right panel) is for $\psi \in ]\tfrac{\pi}{6},\tfrac{\pi}{2}]$, using sizes $\{ L,L-1 \}$ with $L=4,5,\ldots,9$.
The smallest size corresponds to the lowest curve (dark blue). The topmost curve (light blue) is the exact analytical result.}
\label{fig:ccA2}
\end{figure}

As shown in Figure~\ref{fig:ccA2} we find two different regimes, defined as follows:
\begin{align}
 \text{Regime 1:} & \quad \psi \in ]\tfrac{\pi}{6},\tfrac{\pi}{2}] \,, \nonumber \\
 \text{Regime 2:} & \quad \psi \in [0,\tfrac{\pi}{6}] \,.
\end{align}
The central charge throughout Regime 1 is shown in the right panel of the figure. A Coulomb gas computation, which will be published elsewhere \cite{LGJ}, gives the exact result
\begin{align}
 c(\psi) = 2-24\frac{(1-g)^2}{g} \,, \qquad g = \frac{3 \psi}{\pi} \,,
 \label{cA2CG}
\end{align}
where $g$ is the Coulomb gas coupling constant.
The agreement between the numerical values $c_L(\psi)$ and the analytical result $c(\psi)$ is seen to be excellent.

We define two distinct phases inside Regime 1:
\begin{align}
 \text{Dense phase:} & \quad \psi \in ]\tfrac{\pi}{6},\tfrac{\pi}{3}] \,, \quad \tfrac12 < g \le 1 \,, \nonumber \\
 \text{Dilute phase:} & \quad \psi \in [\tfrac{\pi}{3},\tfrac{\pi}{2}] \,, \quad 1 \le g \le \tfrac32 \,.
\end{align}
The physical motivation for the names {\em dense} and {\em dilute} comes from properties of the full phase diagram in the two-dimensional space
of bond and vertex fugacities ($x$ and $y=z$ respectively), for a fixed value of $\psi$ (or $q$). This is discussed in more detail in \cite{Lafay:2021wyf}, but we recall here the
salient features. Starting from the trivial empty phase, upon increasing the density of bonds and vertices one first hits a critical line---a
one-dimensional critical sub-manifold of the parameter space---on which the model is in the dilute universality class. The fraction of links
which are covered by a bond is zero. The whole dilute critical line is governed by an attractive renormalisation-group fixed point.
We believe that the integrable point in the dilute phase---which, we recall, is for the $A_2$ model modified by the inclusion of bending weights, but we think
that bending is unlikely to change the critical behaviour---is
in the same universality class as this dilute fixed point. Increasing further the density of bonds and vertices one enters a critical region---a two-dimensional critical sub-manifold
of the parameter space---throughout which the model is in the dense universality class. The fraction of links which are covered by a
bond is now finite, and the whole critical region is governed by a certain fixed point. We believe that the
integrable point in the dense phase is in the same universality class as this dense fixed point.

The central charge is given by the same analytic function of $g$ throughout Regime 1, and the same holds true for each critical exponent \cite{LGJ}.
This is why the two phases are grouped within the same regime. The dense and dilute phases intersect in the point $\psi = \tfrac{\pi}{3}$, for which
the central charge is $c=2$, the rank of the $A_2$ algebra. At this point the field theory is a CFT of two free bosons.

We do not yet have an analytic understanding of Regime 2. Our numerical results for $c_L(\psi)$ are shown in the left panel of Figure~\ref{fig:ccA2}.
It is clear that $c(\psi)$ cannot be given by the same analytic expression as (\ref{cA2CG}), so we are indeed in a different regime.
The numerical results clearly show that the ground state sector---then one determining $c$---is only present for $L$ a multiple of $3$. This is a
hint of a higher symmetry, as is the fact, that $c$ takes larger values than in Regime 1. We have indeed $c \gtrapprox 3$ for small $\psi$,
and possibly even $c(\psi) \to 4$ for $\psi \to 0$. It seems possible that the limit $c(\psi \to \tfrac{\pi}{6}^-)$ is equal
to $c(\psi \to \tfrac{\pi}{6}^+) = -10$, obtained from (\ref{cA2CG}). In any case, it is obvious that the finite-size effects are much larger in Regime 2 than in Regime 1.
Such slow convergence is usually the hallmark of a non-compact continuum limit. Notice that in the $A_1$ loop model there is indeed a
regime III for which the continuum limit has one compact and on non-compact boson. This leads us to conjecture that Regime 2 of the $A_2$ web model
has one or two non-compact bosons.

Further investigations of Regime 2 would require the access to larger sizes. This could be achieved, e.g., by setting up the Bethe Ansatz equations and
studying them numerically or even analytically. We leave such developments for future work.

\subsubsection{Special points}
\label{sec:specptsA2}

We now discuss a number of points of particular interest.

\paragraph{Case of $q=e^{\pm i \frac{\pi}{4}}$.}

These are the values of $q$ for which one has a mapping to a 3-state Potts models (see Section \ref{sec:A2mapping}).  
The corresponding integrable points are given by 
\bgroup
\def\arraystretch{1.5}
\begin{center}
\begin{tabular}{|c|ccc|}
\hline
$\psi$    & $\frac{\pi}{4}$ & $\frac{5\pi}{12}$ & $\frac{11\pi}{12}$ \\[5pt] \hline
$x$        & $\frac{1}{\sqrt{2}}$ & $\sqrt{2-\sqrt{3}}$ & $\sqrt{\sqrt{3}+2}$ \\[5pt]
$y=z$     &  $\sqrt{2}$ & $1$ & $i$ \\[5pt]
$e^{i\phi}$&  $e^{i\frac{7\pi}{12}}$ & $e^{i\frac{23\pi}{36}}$ & $e^{i\frac{29\pi}{36}}$ \\[5pt]
$c$& $0$ & $\tfrac{4}{5}$ & $\approx 1.5$ \\
\hline
\end{tabular}
\end{center}
\egroup
The point $\psi=\tfrac{5\pi}{12}$ in the dilute phase of Regime 1 is likely to be in the same universality class as the analogous point in the dilute phase of the $A_2$ web models considered in \cite{Lafay:2021wyf}. It was argued there that this point is in the ferromagnetic 3-state Potts model class. Recall that the work in \cite{Lafay:2021wyf} did not include the bending weight $\phi$,
but we do not think this changes the universality class. Indeed $c(\psi=\tfrac{5\pi}{12}) = \tfrac{4}{5}$ from (\ref{cA2CG}) as expected.

It seems worth pointing out that the integrable 3-state Potts model described in this paper is not the same as the one
considered in \cite{wu_triangular_1980}, although both include plaquette interactions. 

Similarly, we believe that the point $\psi=\tfrac{\pi}{4}$ in the dense phase of Regime 1 is in the universality class of the analogous point of \cite{Lafay:2021wyf},
which can in turn be identified with the infinite-temperature 3-state Potts model. The latter has obviously $c_\text{eff} = 0$, in agreement
with $c(\psi=\tfrac{\pi}{4}) = 0$ from (\ref{cA2CG}).

Finally we identify the point $\psi=\tfrac{11\pi}{12}$ of the above table with the point $\psi=\tfrac{\pi}{12}$, due to the symmetry (\ref{paramsym}).
For the latter, our numerical results are $c_6 \simeq 1.367$ from sizes $L=3,6$, and $c_9 \simeq 1.516$ from sizes $L=6,9$.
To our best knowledge, no previous study has found such a high value of $c$ for a 3-state spin model.

For completeness we mention that yet other universality classes of a 3-state Potts model on the triangular lattice have been reported in \cite{Jacobsen_2017}.

\paragraph{Case of $q=\pm i$.}

When $q=\pm i$, $[2]_q=0$, so that any web containing a digon has weight $0$. Actually any web that is not a collection of loops has vanishing weight. This can be shown by induction on the number of vertices. It is clearly true for webs with $2$ vertices. Without loss of generality, consider a web that does not contain loops with $V$ vertices, $E$ edges and $F$ faces. Suppose the statement is true for webs with strictly less than $V$ vertices. By the Euler relation and the hand shake lemma we have
\begin{align*}
    &F-E+V=2\\
    &2E=3V
\end{align*}
If the web contains a digon, its weight is $0$. If not, any face is surrounded by at least $4$ edges and 
\begin{align*}
    2E\geq 4F
\end{align*}
This implies a lower bound on the number of vertices
\begin{align*}
    V\geq 8
\end{align*}
If we now reduce the web by the square rule, i.e the only rule applicable, we get a linear combination of webs with a number of vertices $V'=V-4>0$. By the induction hypothesis their weights are $0$ hence also is the weight of the original web.

We thus obtained a model of oriented loops of topological weight $[3]_q=-1$. If we sum over orientation taking into account the fugacities corresponding to the bendings of web edges we obtain the familiar $O(N)$ model of unoriented loops with contractible loop weight 
\begin{align*}
    N=-e^{6i\phi}-e^{-6i\phi}=e^{2i\psi}+e^{-2i\psi},
\end{align*}
non contractible loop weight $-2$ and bond fugacity 
\begin{align*}
    x=&\frac{1}{2\sin(\psi)}.
\end{align*}
$q=\pm i$ is attained for the following values of $\psi$
\begin{align*}
    \psi=\ \frac{\pi}{6},\ \frac{\pi}{2}
\end{align*}
The first point gives $N=1$ and $x= 1$. These values correspond to site percolation on the triangular lattice, a point in the dense phase of the loop model. Since the non contractible loop weight is equal to $-2$, the effective central charge is given by $-\infty$, which is certainly different from the analytical result $c(\psi \to \tfrac{\pi}{6}^+) = -10$, and possibly also incompatible with the numerical result $c(\psi \to \tfrac{\pi}{6}^-)$ for Regime 2. In any case, there is a discontinuity at the junction between Regimes 1 and 2.

The second point, $\psi = \tfrac{\pi}{2}$, gives $N=-2$ and $x=\frac{1}{2}$. This value corresponds to the loop-erased random walk, in agreement with the value $c=-2$ from (\ref{cA2CG}).

\paragraph{Case of $\psi=0$.}

From Figure~\ref{fig:ccA2} this is the most remarkable point in Regime 2, so we investigate here its lattice realisation in some more detail.
For $\psi=0$, one obtains $y=z=0$ and $x\rightarrow \infty$. To make sense of the model, one needs to first renormalise the local transfer matrices \eqref{A2transfermatrix} by $\frac{1}{x}$ and then send $\psi$ to zero. One then obtains
\begin{subequations}
\begin{align}
    t^{A_2}_{(1)}=&\sqrt{2}\vcenter{\hbox{\includegraphics[scale=0.2]{diagrams/kupvertex1.eps}}}+\sqrt{2}\vcenter{\hbox{\includegraphics[scale=0.2]{diagrams/kupvertex2.eps}}}+ e^{i\phi}\vcenter{\hbox{\includegraphics[scale=0.2]{diagrams/kupvertex3.eps}}}+ e^{-i\phi}\vcenter{\hbox{\includegraphics[scale=0.2]{diagrams/kupvertex4.eps}}}\nonumber\\ &+ e^{-i\phi}\vcenter{\hbox{\includegraphics[scale=0.2]{diagrams/kupvertex5.eps}}}+ e^{i\phi}\vcenter{\hbox{\includegraphics[scale=0.2]{diagrams/kupvertex6.eps}}}+ e^{-i\phi}\vcenter{\hbox{\includegraphics[scale=0.2]{diagrams/kupvertex7.eps}}}+ e^{i\phi}\vcenter{\hbox{\includegraphics[scale=0.2]{diagrams/kupvertex8.eps}}}\\[20pt]
    t^{A_2}_{(2)}=&\sqrt{2}\vcenter{\hbox{\includegraphics[scale=0.2]{diagrams/kupvertex10.eps}}}+ \sqrt{2}\vcenter{\hbox{\includegraphics[scale=0.2]{diagrams/kupvertex11.eps}}}+ e^{i\phi}\vcenter{\hbox{\includegraphics[scale=0.2]{diagrams/kupvertex14.eps}}}+ e^{-i\phi}\vcenter{\hbox{\includegraphics[scale=0.2]{diagrams/kupvertex15.eps}}}\nonumber\\ &+ e^{-i\phi}\vcenter{\hbox{\includegraphics[scale=0.2]{diagrams/kupvertex12.eps}}}+ e^{i\phi}\vcenter{\hbox{\includegraphics[scale=0.2]{diagrams/kupvertex13.eps}}}+ e^{-i\phi}\vcenter{\hbox{\includegraphics[scale=0.2]{diagrams/kupvertex16.eps}}}+ e^{i\phi}\vcenter{\hbox{\includegraphics[scale=0.2]{diagrams/kupvertex17.eps}}}
\end{align}
\end{subequations}
with $q=-1$ and $e^{i\phi}=i$.
It is not clear to us at present why this lattice model has the special properties (slow convergence and the largest central charge) that we observe numerically.

\subsection{The $G_2$ web model}
We now consider the third line of table \ref{table:qg}. Let $V_u$, $u\in \mathbb{C}^*$, be the representation of $U_q(D_4^{(3)})$ given in Appendix \ref{sec:explicit} which is actually isomorphic to the one considered in \cite{Tak_cs_1997}\footnote{Beware of a typo in the representation matrices of \cite{Tak_cs_1997}.}. We are looking for an operator $\check{R}(u,v)$ intertwining $V_u\otimes V_v$ and $V_v\otimes V_u$. Remark that, in $U_q(D_4^{(3)})$, $E_i$, $F_i$ and $H_i$ for $i=0,1$ generate a Hopf subalgebra isomorphic to $U_{q}(G_2)$. Under the action of this subalgebra, $V_u$ decomposes as:
\begin{align}
    V_u= \mathbb{C} \oplus V
   \label{Vu_G2_decomp}
\end{align}

Hence $\check{R}(u,v)$ will decompose as a sum of $U_{q}(G_2)$ intertwiners:
\begin{align}
    \check{R}(u,v)=&\ a_1(u,v)\quad\vcenter{\hbox{\includegraphics[scale=0.12]{diagrams/dTLRmat1.eps}}}\quad+a_2(u,v)\quad\vcenter{\hbox{\includegraphics[scale=0.12]{diagrams/dTLRmat6.eps}}}\quad+a_3(u,v)\quad\vcenter{\hbox{\includegraphics[scale=0.12]{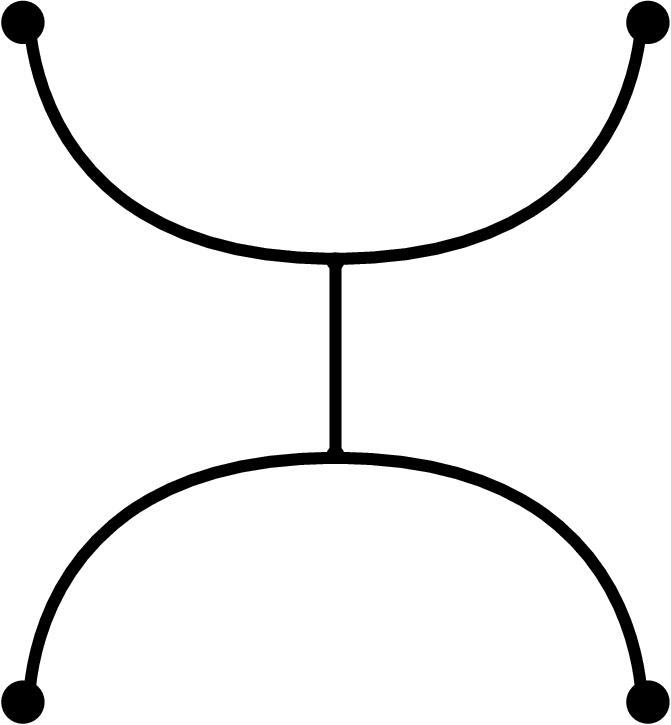}}}\quad
    +a_4(u,v)\quad\vcenter{\hbox{\includegraphics[scale=0.12]{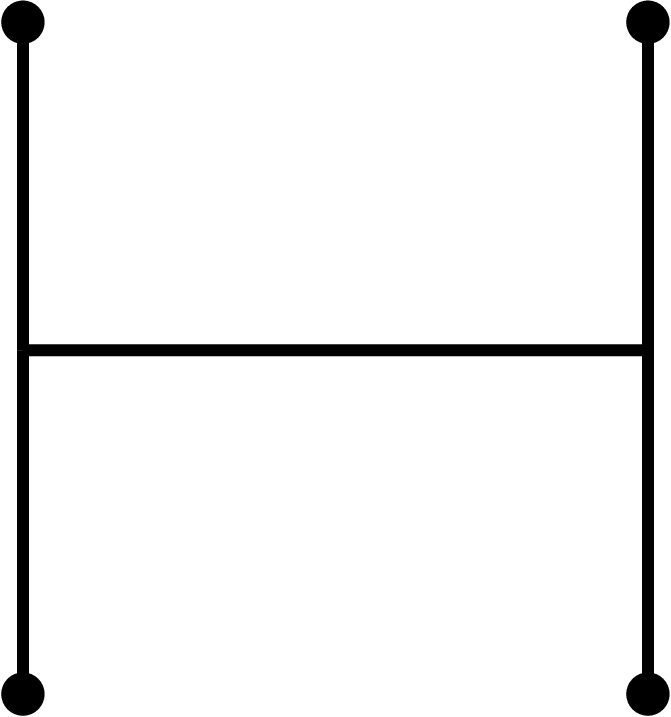}}}\nonumber\\[10pt]
    &+a_5(u,v)\quad\vcenter{\hbox{\includegraphics[scale=0.12]{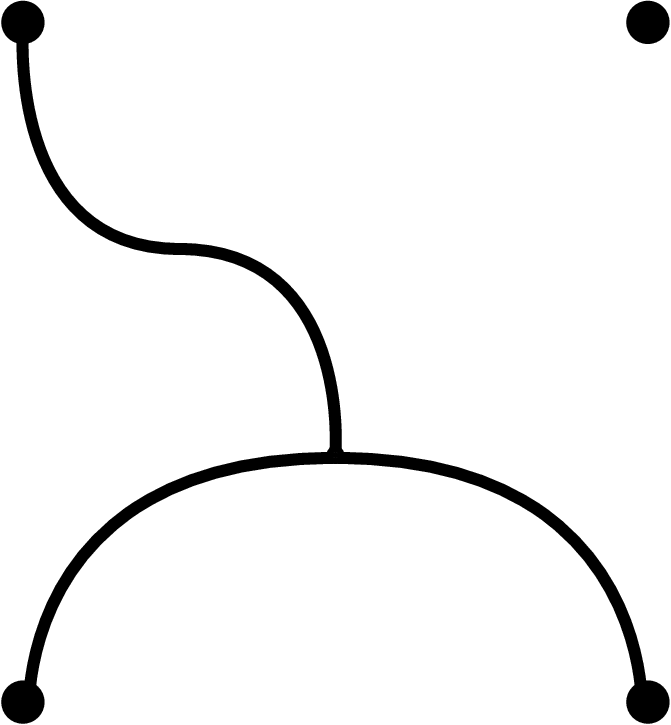}}}\quad+a_6(u,v)\quad\vcenter{\hbox{\includegraphics[scale=0.12]{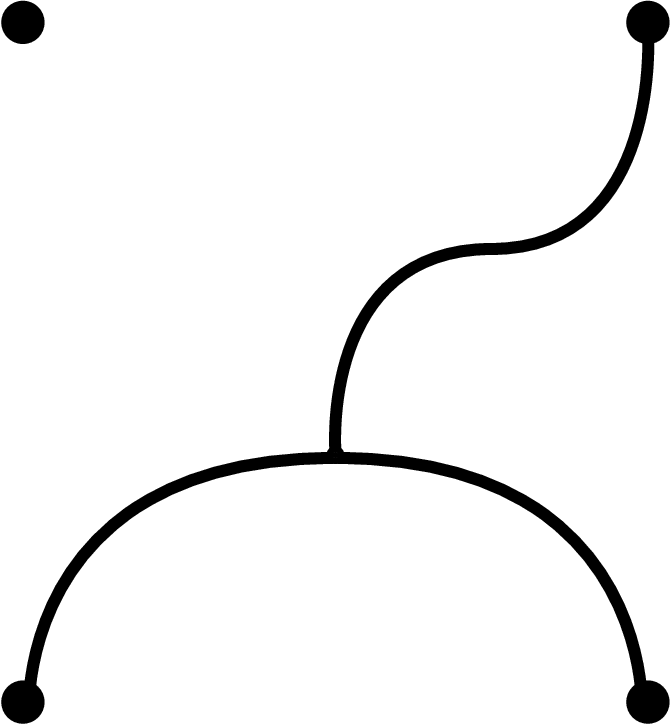}}}\quad +a_7(u,v)\quad\vcenter{\hbox{\includegraphics[scale=0.12]{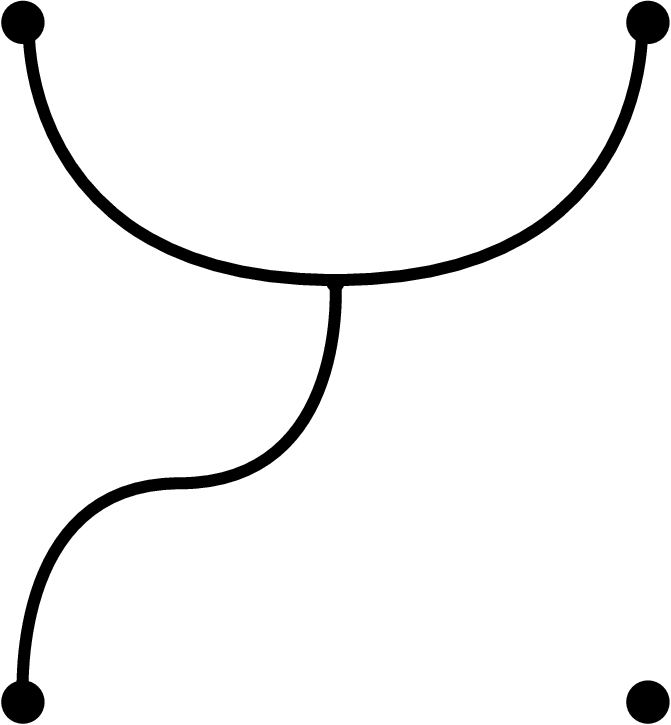}}} \quad +a_8(u,v)\quad\vcenter{\hbox{\includegraphics[scale=0.12]{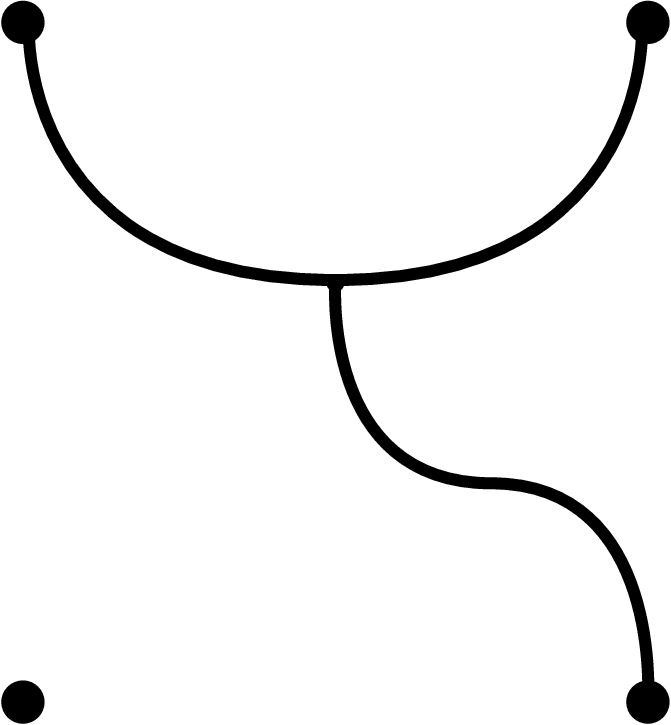}}}\nonumber\\[10pt]
    &+a_9(u,v)\quad\vcenter{\hbox{\includegraphics[scale=0.12]{diagrams/dTLRmat7.eps}}}\quad+a_{10}(u,v)\quad\vcenter{\hbox{\includegraphics[scale=0.12]{diagrams/dTLRmat8.eps}}}\quad+a_{11}(u,v)\quad\vcenter{\hbox{\includegraphics[scale=0.12]{diagrams/dTLRmat2.eps}}}\quad+a_{12}(u,v)\quad\vcenter{\hbox{\includegraphics[scale=0.12]{diagrams/dTLRmat4.eps}}}\nonumber\\[10pt]
    &+a_{13}(u,v)\quad\vcenter{\hbox{\includegraphics[scale=0.12]{diagrams/dTLRmat3.eps}}}\quad+a_{14}(u,v)\quad\vcenter{\hbox{\includegraphics[scale=0.12]{diagrams/dTLRmat5.eps}}}\quad+a_{15}(u,v)\quad\vcenter{\hbox{\includegraphics[scale=0.12]{diagrams/dTLRmat9.eps}}}
\end{align}
where $a_i(u,v)$ are some coefficients and the webs span the space of intertwiners $\text{End}_{U_{q}(G_2)}\left(( \mathbb{C}\oplus V)^2\right)$.

Asking for $\check{R}(u,v)$ to commute with the remaining generators, we see that it depends only on the ratio $s=\frac{u}{v}$. Plugging this linear system for the functions $a_i(s)$ into Mathematica, we find
\begin{align*}
    a_1(s)=&\frac{s^3-q^{12}}{q^4 (s-1)}\\
    a_2(s)=&\frac{q^8-q^2 s^3}{q^6-s}\\
    a_3(s)=&q^4+q^2 s+s^2\\
    a_4(s)=&\frac{q^8+q^4 s+s^2}{q^2}\\
    a_5(s)=&q^4 \sqrt{[3]_q} \left(q^2-1\right) s\\
    a_6(s)=&q^4 \sqrt{[3]_q} \left(q^2-1\right) s\\
    a_7(s)=&-\frac{\sqrt{[3]_q} \left(q^2-1\right) s}{q^2}\\
    a_8(s)=&-\frac{\sqrt{[3]_q} \left(q^2-1\right) s}{q^2}\\
    a_9(s)=&\frac{q^8 \left(q^6-1\right) s (s+1)}{q^6-s}\\
    a_{10}(s)=&\frac{\left(q^6-1\right) s (s+1)}{q^4 \left(q^6-s\right)}\\
    a_{11}(s)=&-\frac{\left(q^6-1\right) s \left(q^6+s\right)}{q^4 (s-1)}\\
    a_{12}(s)=&-\frac{\left(q^6-1\right) s \left(q^6+s\right)}{q^4 (s-1)}\\  
    a_{13}(s)=&q^6+q^4 (s+1)+\frac{s^2}{q^2}+s^2+s\\
    a_{14}(s)=&q^6+q^4 (s+1)+\frac{s^2}{q^2}+s^2+s\\
    a_{15}(s)=&q^{14}-q^8+q^6+q^4 s+q^4+\frac{1}{q^4}+\left(\frac{1}{q^2}+1\right) s^2-q^2+\frac{1-q^{12}}{q^4 (s-1)}+\frac{q^8-q^{20}}{q^6-s}+s
\end{align*}

One can then show, using for instance Mathematica, that the corresponding operator $\check{R}(s)$ satisfy the multiplicative spectral parameter dependant Yang-Baxter equation. Using the following gauge transformation (which is just an elementary rescaling of the irreducible components in the decomposition \eqref{Vu_G2_decomp})
\begin{align}
    D&=\text{Diag}(1,\alpha,\alpha,\alpha,\alpha,\alpha,\alpha,\alpha)\nonumber\\
    \check{R}(s)&\rightarrow \left(D^{-1}\otimes D^{-1}\right) \check{R}(s) \left(D\otimes D \right)
\end{align}
for well chosen $\alpha$, we obtain
\begin{align*}
    a_1(s)=&\frac{s^3-q^{12}}{q^4 (s-1)}\\
    a_2(s)=&\frac{q^8-q^2 s^3}{q^6-s}\\
    a_3(s)=&q^4+q^2 s+s^2\\
    a_4(s)=&\frac{q^8+q^4 s+s^2}{q^2}\\
    a_5(s)=& a_6(s)= a_7(s)= a_8(s)=-iq \sqrt{[3]_q} \left(q^2-1\right) s\\
    a_9(s)=&a_{10}(s)=-\frac{q^2 \left(q^6-1\right) s (s+1)}{q^6-s}\\
    a_{11}(s)=&a_{12}(s)=-\frac{\left(q^6-1\right) s \left(q^6+s\right)}{q^4 (s-1)}\\
    a_{13}(s)=&a_{14}(s)=q^6+q^4 (s+1)+\frac{s^2}{q^2}+s^2+s\\
    a_{15}(s)=&q^{14}-q^8+q^6+q^4 s+q^4+\frac{1}{q^4}+\left(\frac{1}{q^2}+1\right) s^2-q^2+\frac{1-q^{12}}{q^4 (s-1)}+\frac{q^8-q^{20}}{q^6-s}+s
\end{align*}

It is apparent that by setting the spectral parameter $s=e^{-\frac{2i\pi}{3}}q^4$, $\check{R}(s)$ will be decomposed only in terms of webs appearing in the local transfer matrix of the $G_2$ web model. By renormalizing the $R$-matrix,
we recover the local transfer matrix \eqref{transfermatrixG2} with the following parametrisation, setting $q=e^{i\gamma}$:
\begin{subequations}
\begin{align}
    x=&\frac{1}{2\cos(2\gamma+\frac{2\pi}{3})}\\
    y=&2\cos(\gamma-\frac{\pi}{6})\sqrt{\frac{2\cos(2\gamma+\frac{2\pi}{3})}{[3]_q}}
\end{align}
\end{subequations}

Note that points related by $\gamma\rightarrow \gamma + \pi$ have the same bond fugacity but a vertex fugacity related by $y\rightarrow -y$. They are equivalent as the partition function only depend on $y^2$ and $q^2$. We could focus for instance on the interval $\gamma \in [0,\pi]$. The Kuperberg weight only depend on $\pm 2\gamma$ and for each value of the former, there are exactly two integrable points. Hence, the model describes two phases.

\subsubsection{Central charge and phase diagram}

Also for the $G_2$ web model have we numerically diagonalised the transfer matrix on cylinders of circumference $L$. Recall that for the $A_2$ case we found
$L$ mod $3$ parity effects in one of the regimes. For the $G_2$ model such effects are found to depend on $L$ mod $2$, so in this case we perform the diagonalisation
for sizes $L=2, 4, 6, 8$, from which estimates for the central charge $c_L$ can be extracted. The computations were made for 50 equally-spaced values of
$\gamma \in [0,\pi]$ and we show again curves $c_L(\gamma)$ obtained from extrapolation of these values.

The number of regimes is now larger. The numerical results, shown in Figure~\ref{fig:ccG2}, combined with analytical considerations on the weights (see below) lead
us to define four regimes:
\begin{align}
 \text{Regime 1:} & \quad \gamma \in [\tfrac{\pi}{6},\tfrac{\pi}{3}] \,, \nonumber \\
 \text{Regime 2:} & \quad \gamma \in [\tfrac{\pi}{3},\tfrac{\pi}{2}] \,, \nonumber \\
 \text{Regime 3:} & \quad \gamma \in [\tfrac{\pi}{2},\tfrac{2\pi}{3}] \,, \nonumber \\
 \text{Regime 4:} & \quad \gamma \in [\tfrac{2\pi}{3},\tfrac{7\pi}{8}] \,.
\end{align}
The numerical results alone suggest that Regime 1 might have a larger extent, $\gamma = ]\tfrac{\pi}{8},\tfrac{\pi}{3}]$, but in any case the point $\gamma=\tfrac{\pi}{8}$ is special
(see below) and should be excluded. The remaining two pieces of the interval $[0,\pi]$ do not allow for convincing numerical results for $c_L(\gamma)$, and since
$\gamma$ is defined modulo $\pi$ it is not clear whether these pieces should be considered one or two extra regimes. In the following we shall focus only on Regimes
1--4.

\begin{figure}
\begin{center}
 \includegraphics[width=0.45\textwidth]{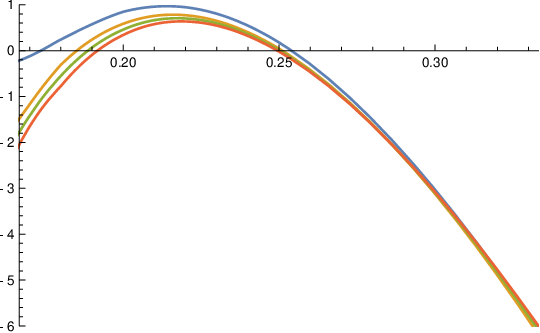} \qquad
 \includegraphics[width=0.45\textwidth]{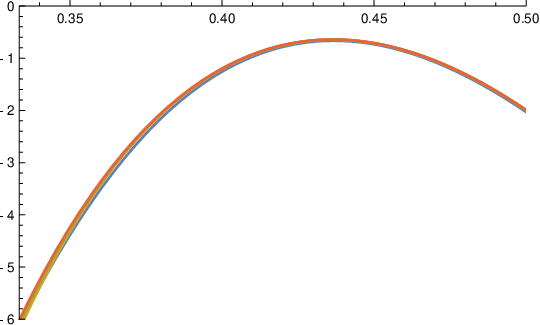} \\[0.5cm]
 \includegraphics[width=0.45\textwidth]{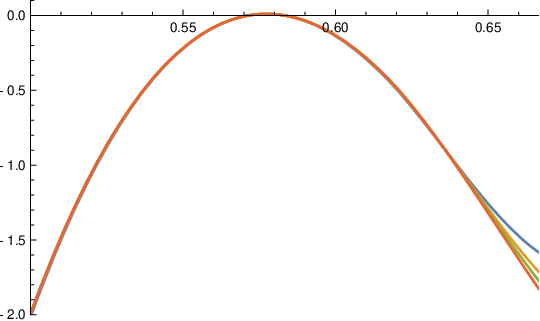} \qquad
 \includegraphics[width=0.45\textwidth]{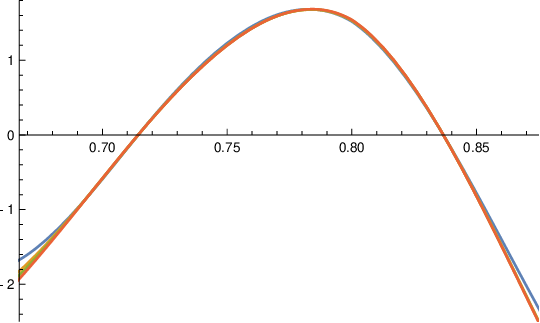} 
\end{center}
\caption{Fits $c_L(\gamma)$ for the central charge in the $G_2$ web model, plotted against $\tfrac{\gamma}{\pi}$. We show two-point fits using sizes $\{L,L-2\}$ with
$L=4$ (blue), $L=6$ (orange) and $L=8$ (green), as well as a three-point fit using sizes $\{L,L-2,L-4\}$ with $L=8$ (red).
Regime 1 (upper left panel) is for $\gamma \in [\tfrac{\pi}{6},\tfrac{\pi}{3}]$, Regime 2 (upper right panel) is for $\gamma \in [\tfrac{\pi}{3},\tfrac{\pi}{2}]$,
Regime 3 (lower left panel) is for $\gamma \in [\tfrac{\pi}{2},\tfrac{2\pi}{3}]$, and finally Regime 4 (lower right panel) is for $\gamma \in [\tfrac{2\pi}{3},\tfrac{7\pi}{8}]$.}
\label{fig:ccG2}
\end{figure}

\subsubsection{Special points}
\label{sec:specptG2}

We discuss again a number of points of special interest.

\paragraph{Case of $\gamma=\frac{\pi}{6},\ \frac{5\pi}{6}$.}

For these values of $\gamma$, we have described in Section \ref{sec:G2mapping} a mapping to the 3-state Potts model on the dual triangular lattice.
We obtain the following integrable points:
\bgroup
\def\arraystretch{1.5}
\begin{center}
\begin{tabular}{|c|cc|}
\hline
$\gamma$   & $\frac{\pi}{6}$ & $\frac{5\pi}{6}$ \\ \hline
$x$         & $-\frac{1}{2}$ & $1$ \\[5pt]
$y$       & $2i$ & $-\frac{1}{\sqrt{2}}$ \\[5pt]
\hline
\end{tabular}
\end{center}
\egroup

The point $\gamma=\frac{5\pi}{6}$ corresponds to the infinite-temperature limit of the 3-state Potts model, with bond fugacity $x=1$.
This identification is consistent with the observed value $c(\tfrac{5\pi}{6}) = 0$ in Regime 4.
On the other hand, the numerical results for $\gamma = \tfrac{\pi}{6}$ in Regime 1 lead us to conjecture that $c(\tfrac{\pi}{6}) = -2$, which
is an unusual and presently unexplained result for a $3$-state model. Notice that the weight of each web configuration depends on $y$
via the combination $y^2$. The fact that both $x$ and $y^2$ are negative in this case is presumably at the root of the observed unusual behaviour.

\paragraph{Case of $\gamma=\frac{\pi}{2}$.}

For this value of $\gamma$, one has $x=1$ and $y=i$. As shown in Section \ref{sec:G2mapping2}, this web model describes the uniform probability measure on spanning trees of the dual lattice. This is known to have $c=-2$, in perfect agreement with the numerical results at the boundary between Regimes 2 and 3.

\paragraph{Case of $\gamma=\frac{2\pi}{3}$.}

For this values of $\gamma$, we have $y=0$ and the model becomes simply the $O(N)$ loop model. For this value, we have $N=-2$ and $x=\frac{1}{2}$,
which corresponds to the dilute phase of the loop model, hence to loop-erased random walks. Also this point is known to have $c=-2$, a value that agrees
perfectly with the numerical results at the boundary between Regimes 3 and 4.

\paragraph{Case of $\gamma=\frac{5\pi}{12}, \frac{11\pi}{12}$.}

For these values of $\gamma$, the bond fugacity is infinite. By renormalising the weights, we obtain the following local transfer matrices
\begin{align*}
    t^{G_2}_{(1)}&=(-1\pm \sqrt{3})\vcenter{\hbox{\includegraphics[scale=0.15]{diagrams/G2vertex1.eps}}}+\vcenter{\hbox{\includegraphics[scale=0.15]{diagrams/G2vertex2.eps}}}+\vcenter{\hbox{\includegraphics[scale=0.15]{diagrams/G2vertex3.eps}}}+\vcenter{\hbox{\includegraphics[scale=0.15]{diagrams/G2vertex4.eps}}}
\end{align*}
and similarly for $t^{G_2}_{(2)}$. The $+$ sign in $(-1\pm \sqrt{3})$ corresponds to $\gamma= \frac{5\pi}{12}$, and the $-$ sign to $\gamma= \frac{11\pi}{12}$.
For the first case, $\gamma= \frac{5\pi}{12}$, the numerical results in Regime 2 lend strong credence to the conjecture $c(\frac{5\pi}{12}) = -\tfrac{4}{5}$, but
we have no theoretical understanding of this value. The second case, $\gamma= \frac{11\pi}{12}$, is outside the four regimes defined above, and we refrain
from giving any numerical estimate for $c$.

\paragraph{Case of vanishing digon weight.}

We now look at values of $\gamma$ such that the digon factor $-(q^6+q^4+q^2+q^{-2}+q^{-4}+q^{-6})$ vanishes. This happens for $\gamma=\frac{\pi}{3},\frac{2\pi}{3},\frac{\pi}{8},\frac{3\pi}{8},\frac{5\pi}{8},\frac{7\pi}{8}$.

For $\gamma=\frac{\pi}{3}$, the vertex fugacity is infinite, whereas the bond fugacity stays finite---and is in fact trivial up to a sign
($x=-1$). Hence the model contains only one configuration, the hexagonal lattice being completely covered by the web.
The numerical results in Regime 2 give strong support for the conjecture $c(\frac{\pi}{3})=-6$. This is again an unusual value for which we have
presently no analytical explanation.

The case $\gamma=\frac{2\pi}{3}$ was treated above. 

For the other four values of $\gamma$, we have a loop weight of $N=-1$ and a bond fugacity
\begin{align*}
    x= \left( \epsilon_1\sqrt{2+\epsilon_2\sqrt{3}} \right)^{-1}
\end{align*}
for some signs $\epsilon_1$, $\epsilon_2$. Moreover, when the digon factor is $0$, all webs that are not a collection of loops have vanishing Kuperberg weight. This can be shown along the lines of the analogous situation for $A_2$ webs, by induction on the number of vertices. It is clearly true for a web having $2$ vertices. Let $V$ be the number of vertices of a given web $G$ and suppose that the statement is true for webs having less than $V$ vertices. Without loss of generality, we may assume that $G$ does not contain any loop. Suppose that $G$ does not contain digons either. We then have, thanks to the Euler relation and the hand-shake lemma,
\begin{align*}
    V\geq 4 \,.
\end{align*}
Suppose there is a trigon, then $G$ is proportional to a web with $V-2$ vertices.
If there is no trigon, we have the better bound
\begin{align*}
    V\geq 8 \,.
\end{align*}
Then we can use the square or pentagon rule and obtain a linear combination of webs. In all cases, we have that $G$ can be written as a linear combination of webs with a strictly smaller, yet non-vanishing number of vertices and the induction hypothesis applies.

The resulting models are in fact $O(-1)$ loops in their dense ($\epsilon_2=-1)$ or dilute ($\epsilon_2=1$) phase. The case $\gamma=\tfrac{\pi}{8}$ is outside of the four
regimes for which we have good numerical results. In fact, the numerical evidence alone is in favour of Regime 1 having extent $]\tfrac{\pi}{8},\tfrac{\pi}{3}]$, but assuming
this, the numerical results indicate that $c(\gamma) \to -\infty$ as $\gamma \to \left.\tfrac{\pi}{8}\right.^+$. This is not compatible with the analytical result $c=-\tfrac{3}{5}$ for the dilute
$O(-1)$ model, so at least the point $\tfrac{\pi}{8}$ cannot be contained in the definition of Regime 1.
For the case $\gamma = \tfrac{5\pi}{8}$, on the other hand, the numerical results in Regime 3 are in perfect agreement with $c=-\tfrac{3}{5}$.

It remains to discuss the two cases $\gamma = \tfrac{3\pi}{8}$ and $\tfrac{7\pi}{8}$ for which we should find a dense $O(-1)$ model with $c=-7$ by the above analytical 
reasoning. The first of these points, $\gamma = \tfrac{3\pi}{8}$, is inside Regime 2, while the other, $\gamma = \tfrac{7\pi}{8}$ is at the boundary of Regime 4.
The numerical results are well-behaved in both cases, finding $c \simeq -2.33$ for the former and $c \simeq -2.50$ for the latter. 
We do not presently know how to reconcile this diagreement with the analytical result and suspect that some non-commutativity of limits might be at play.
In any case, we stress that because of the very high dimension of the transfer matrix, the numerical results are obtained by interpolation from a set of 50 values of
$\gamma$, and we did not examine the diagonalisation problem directly at the points $\gamma = \tfrac{3\pi}{8}$ and $\tfrac{7\pi}{8}$.

\paragraph{Case of $\gamma=\frac{\pi}{4}$.} This point is situated inside Regime 1, and based on the numerical results we conjecture that
$c(\frac{\pi}{4}) = 0$. We have, however, no analytical argument in support of this value.

\subsection{The $B_2$ web model}

We now consider the fourth line of table \ref{table:qg}. Let $V_u$, $u\in \mathbb{C}^*$, be the first representation of $U_t(A_4^{(2)})$ given in Appendix \ref{sec:evalrepB2}. We are looking for an operator $\check{R}(u,v)$ intertwining $V_u\otimes V_v$ and $V_v\otimes V_u$. Remark that $E_i$, $F_i$ and $H_i$ for $i=1,2$ generate a Hopf subalgebra isomorphic to $U_{t^2}(B_2)$. Under the action of this subalgebra, $V_u$ decomposes as:
\begin{align}
    V_u=\mathbb{C}\oplus V_1\oplus V_2 
    \label{Vu_B2_decomp}
\end{align}

Hence $\check{R}(u,v)$ will decompose as a sum of $U_{t^2}(B_2)$ intertwiners:
\begin{align*}
    \check{R}(u,v)=&\ a_1(u,v)\quad\vcenter{\hbox{\includegraphics[scale=0.10]{diagrams/dTLRmat1.eps}}}\quad+a_2(u,v)\quad\vcenter{\hbox{\includegraphics[scale=0.10]{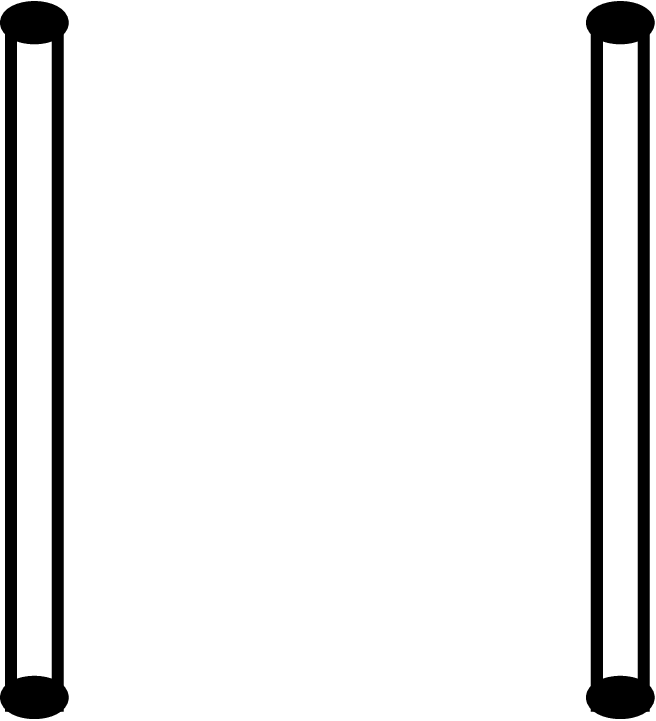}}}\quad+a_3(u,v)\quad\vcenter{\hbox{\includegraphics[scale=0.10]{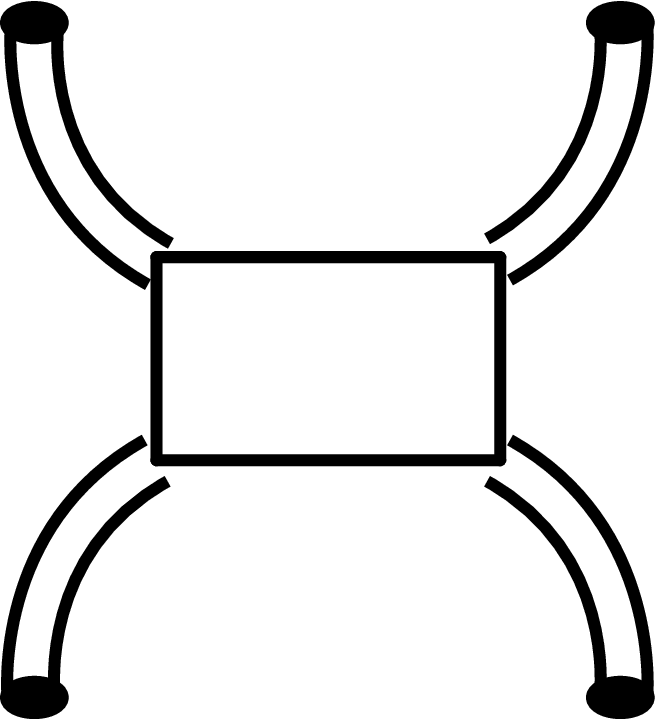}}}\quad
    +a_4(u,v)\quad\vcenter{\hbox{\includegraphics[scale=0.10]{diagrams/dTLRmat9.eps}}}\nonumber\\[10pt]
    &+a_5(u,v)\quad\vcenter{\hbox{\includegraphics[scale=0.10]{diagrams/dTLRmat2.eps}}}\quad+a_6(u,v)\quad\vcenter{\hbox{\includegraphics[scale=0.10]{diagrams/dTLRmat4.eps}}}\quad +a_7(u,v)\quad\vcenter{\hbox{\includegraphics[scale=0.10]{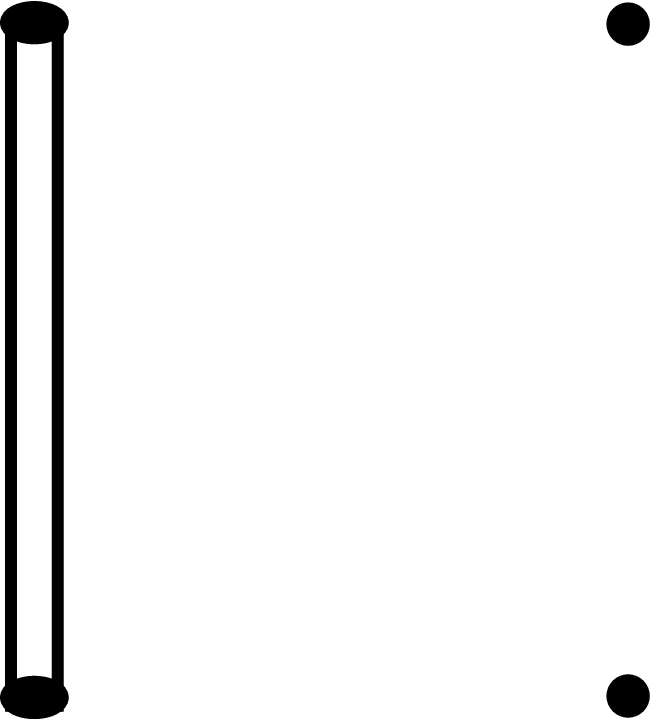}}} \quad +a_8(u,v)\quad\vcenter{\hbox{\includegraphics[scale=0.10]{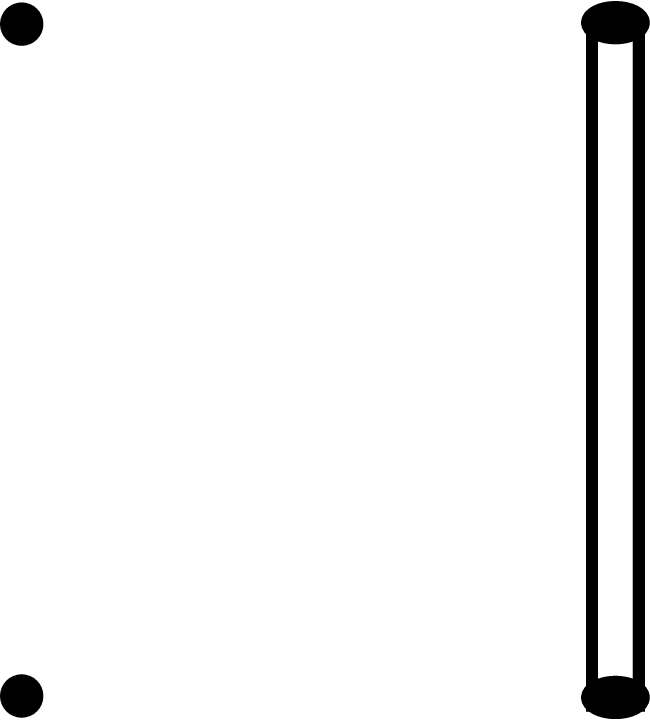}}}\nonumber\\[10pt]
    &+a_9(u,v)\quad\vcenter{\hbox{\includegraphics[scale=0.10]{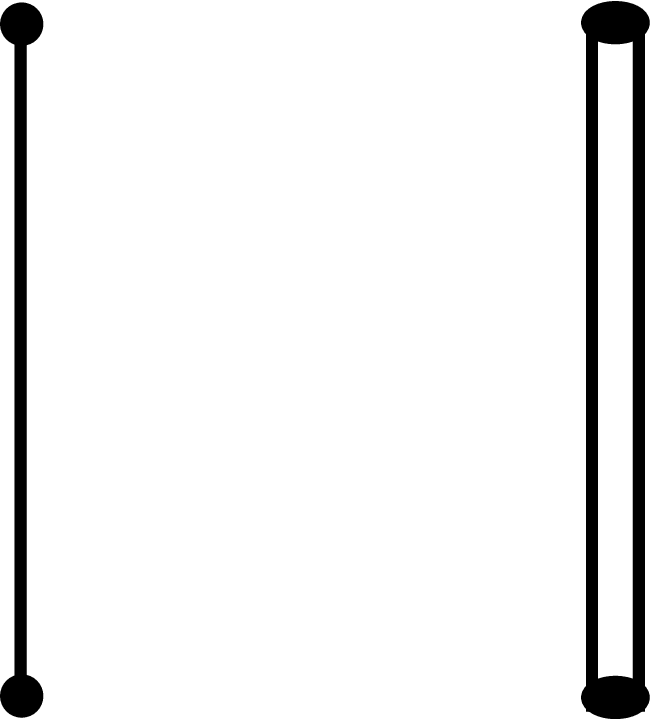}}}\quad+a_{10}(u,v)\quad\vcenter{\hbox{\includegraphics[scale=0.10]{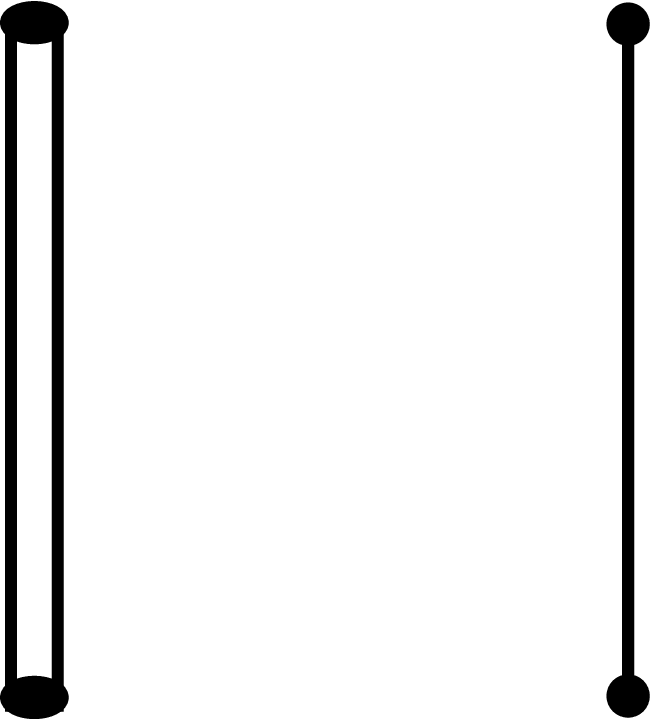}}}\quad+a_{11}(u,v)\quad\vcenter{\hbox{\includegraphics[scale=0.10]{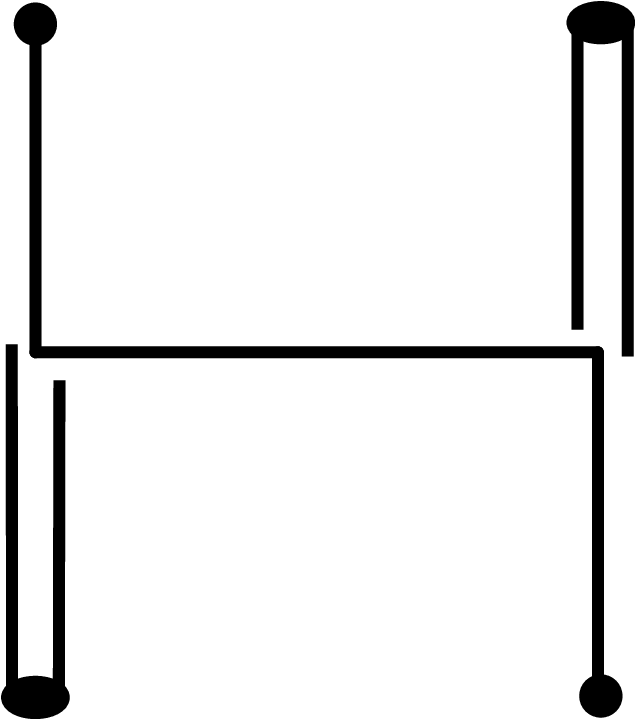}}}\quad+a_{12}(u,v)\quad\vcenter{\hbox{\includegraphics[scale=0.10]{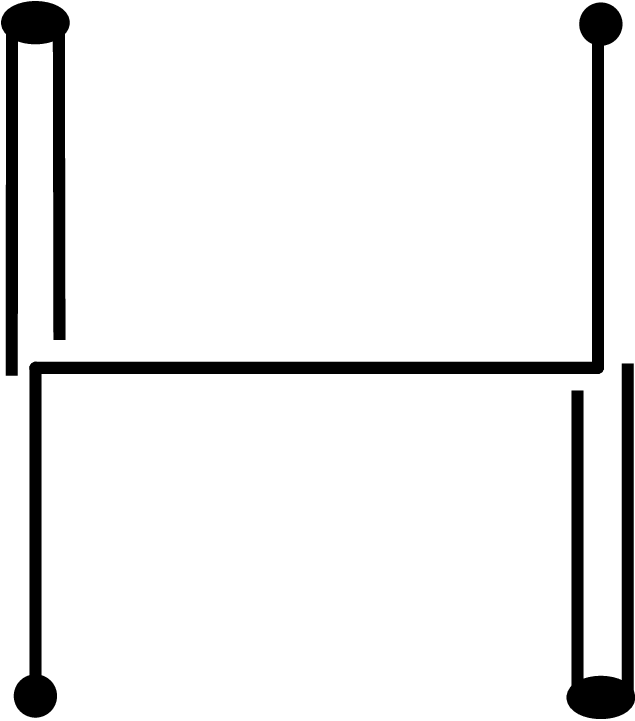}}}\nonumber\\[10pt]
    &+a_{13}(u,v)\quad\vcenter{\hbox{\includegraphics[scale=0.10]{diagrams/dTLRmat6.eps}}}\quad+a_{14}(u,v)\quad\vcenter{\hbox{\includegraphics[scale=0.10]{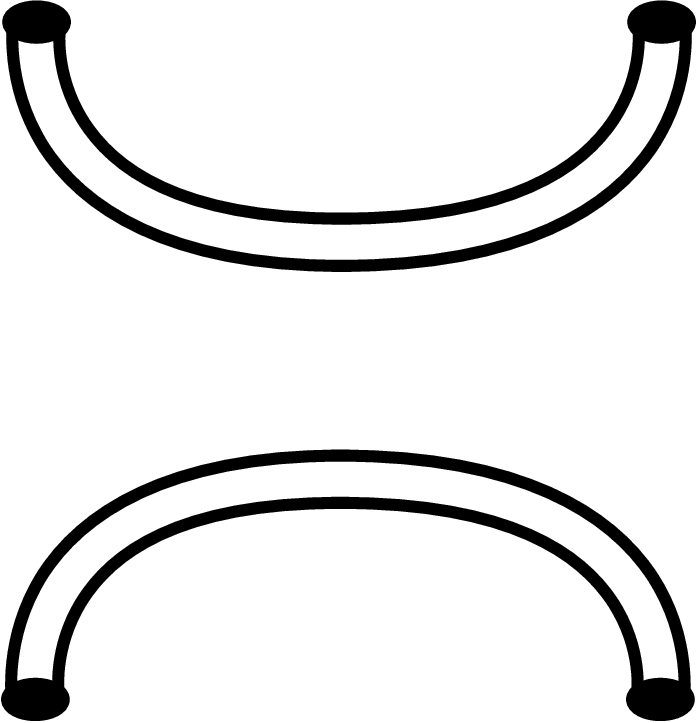}}}\quad+a_{15}(u,v)\quad\vcenter{\hbox{\includegraphics[scale=0.10]{diagrams/dTLRmat7.eps}}}\quad+a_{16}(u,v)\quad\vcenter{\hbox{\includegraphics[scale=0.10]{diagrams/dTLRmat8.eps}}}\\[10pt]
    &+a_{17}(u,v)\quad\vcenter{\hbox{\includegraphics[scale=0.10]{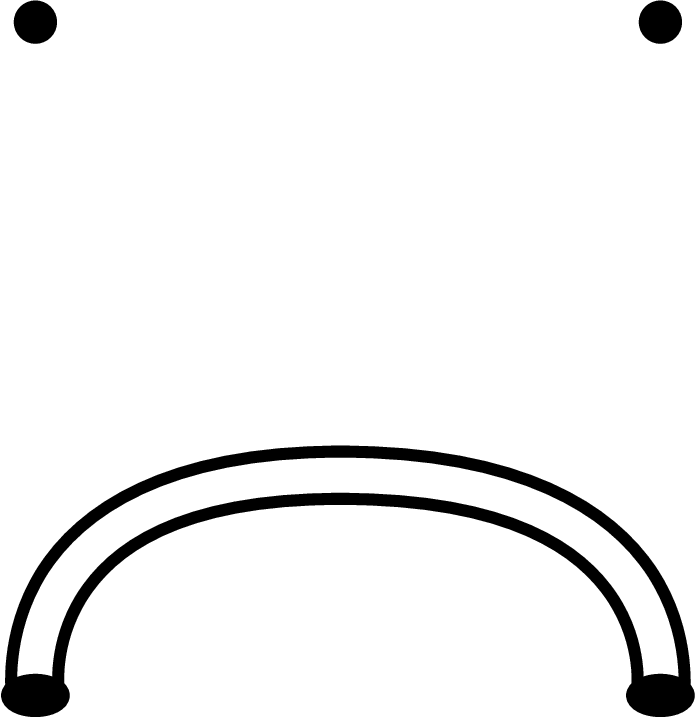}}}\quad+a_{18}(u,v)\quad\vcenter{\hbox{\includegraphics[scale=0.10]{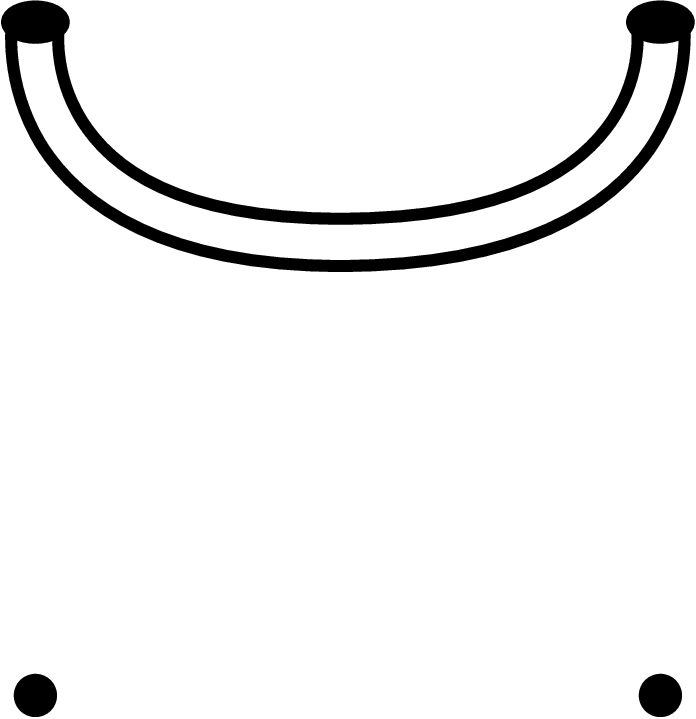}}}\quad+a_{19}(u,v)\quad\vcenter{\hbox{\includegraphics[scale=0.10]{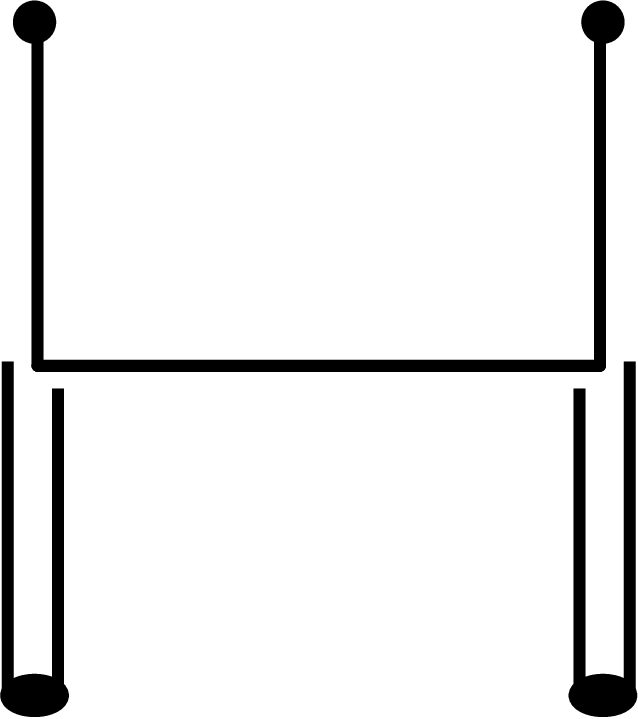}}}\quad+a_{20}(u,v)\quad\vcenter{\hbox{\includegraphics[scale=0.10]{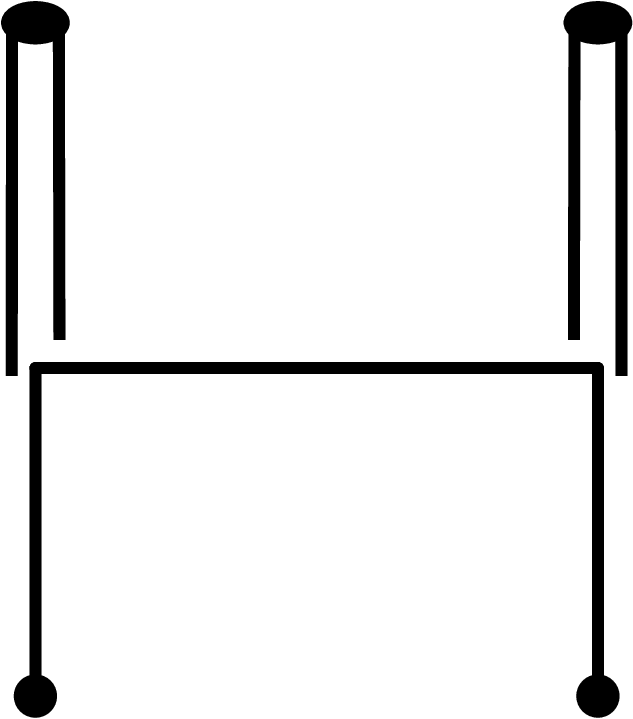}}} \\[10pt]
    &+a_{21}(u,v)\quad\vcenter{\hbox{\includegraphics[scale=0.10]{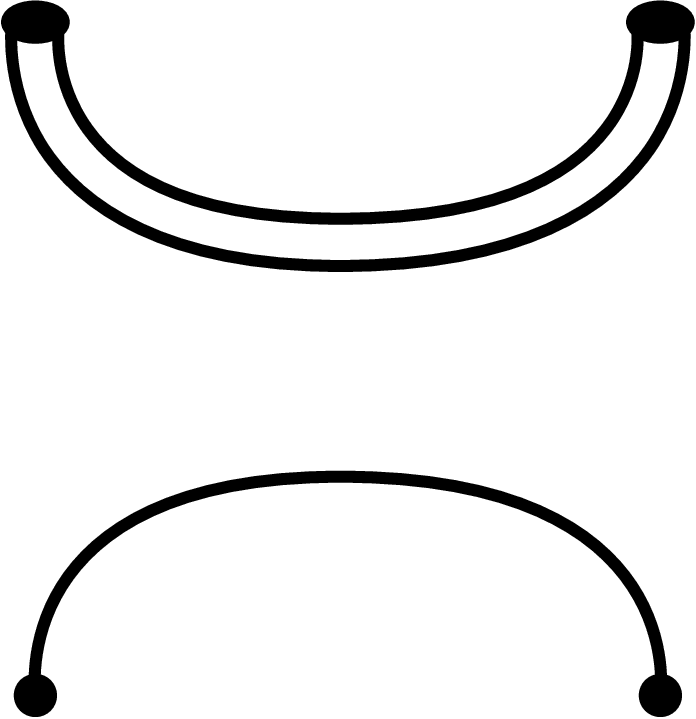}}}\quad+a_{22}(u,v)\quad\vcenter{\hbox{\includegraphics[scale=0.10]{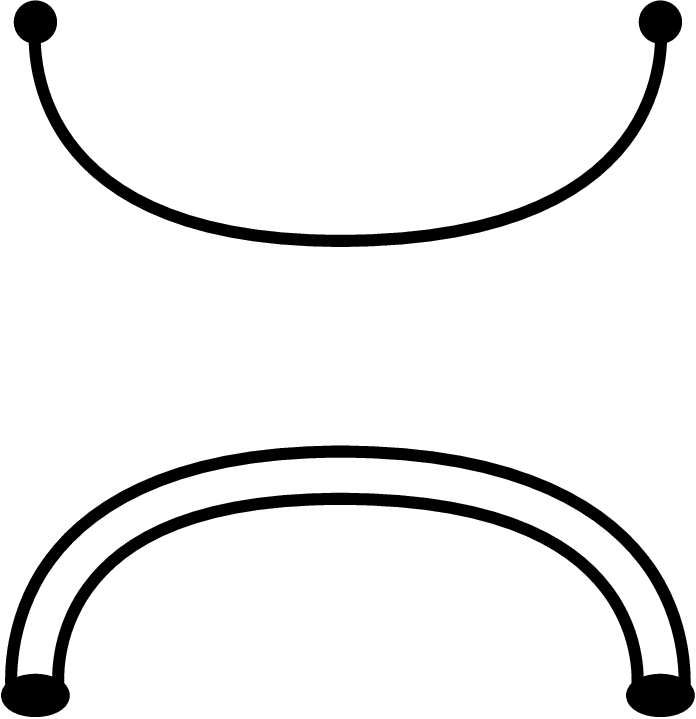}}}\quad+a_{23}(u,v)\quad\vcenter{\hbox{\includegraphics[scale=0.10]{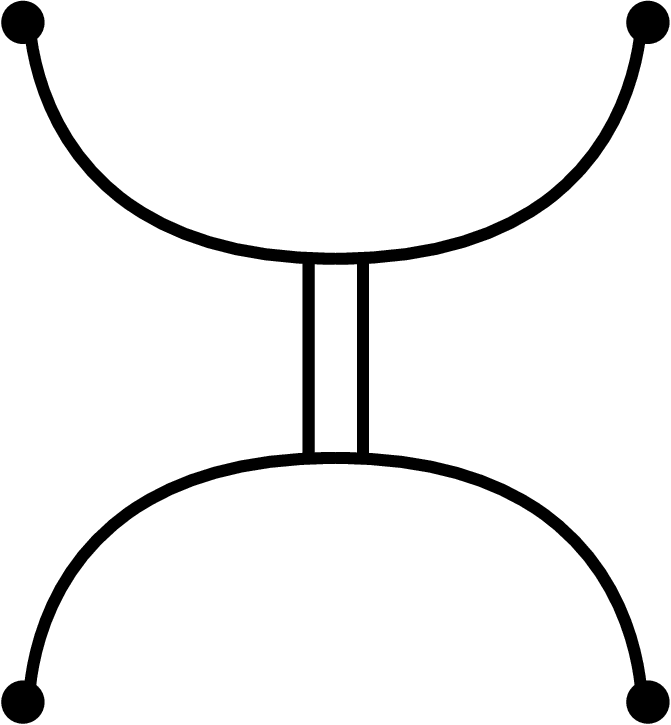}}}\quad+a_{24}(u,v)\quad\vcenter{\hbox{\includegraphics[scale=0.10]{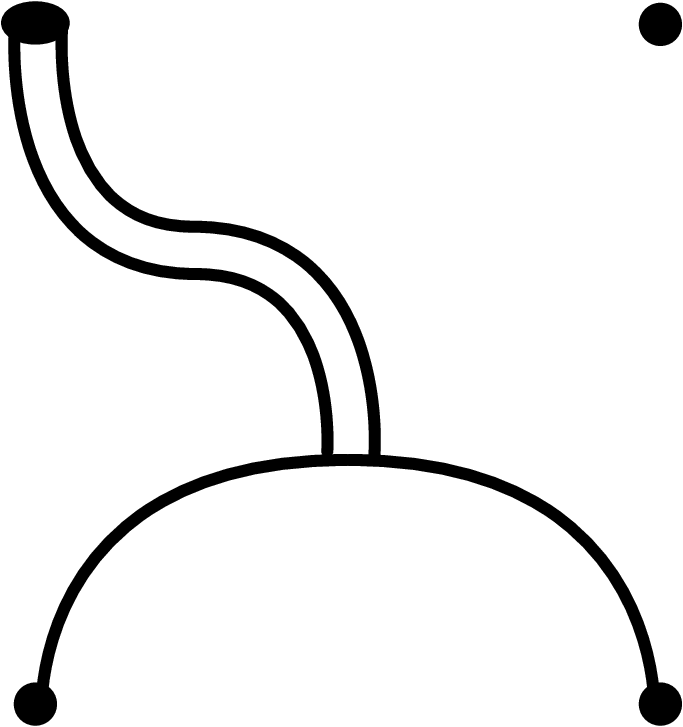}}}\\[10pt]
    &+a_{25}(u,v)\quad\vcenter{\hbox{\includegraphics[scale=0.10]{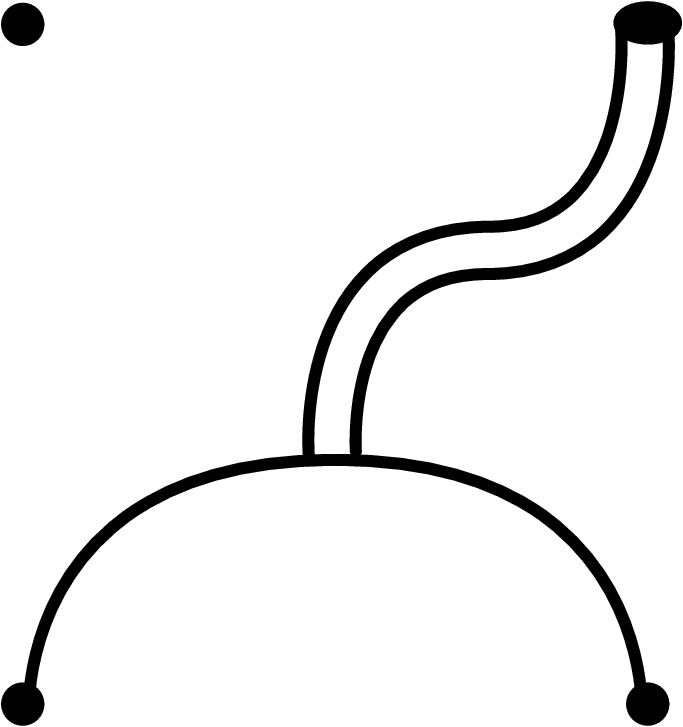}}}\quad+a_{26}(u,v)\quad\vcenter{\hbox{\includegraphics[scale=0.10]{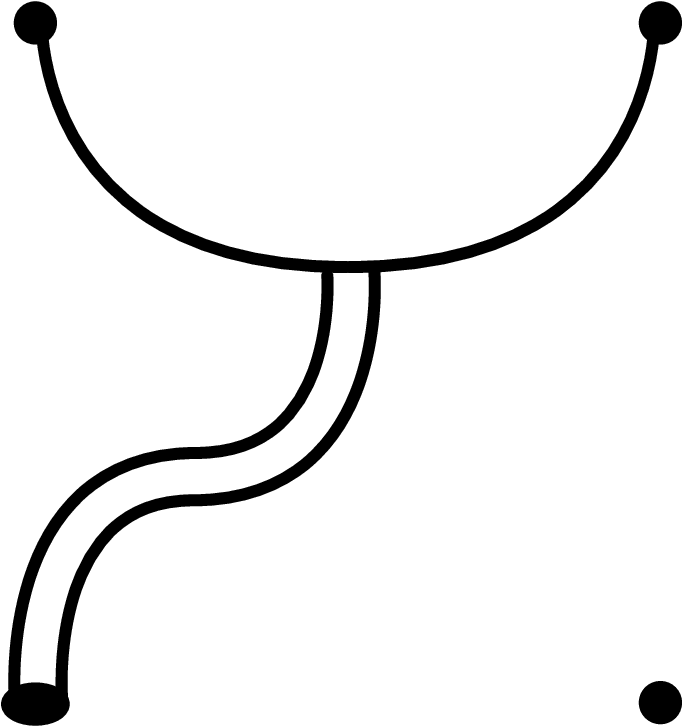}}}\quad+a_{27}(u,v)\quad\vcenter{\hbox{\includegraphics[scale=0.10]{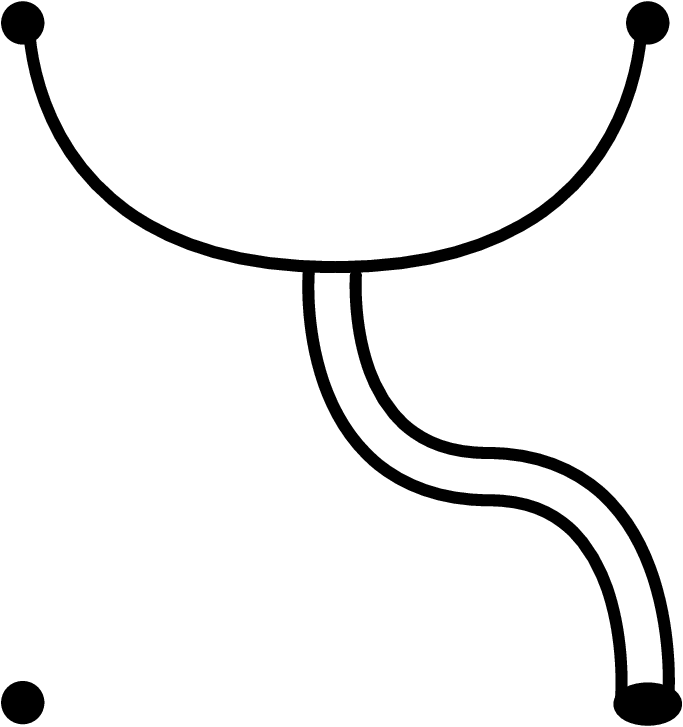}}}\quad+a_{28}(u,v)\quad\vcenter{\hbox{\includegraphics[scale=0.10]{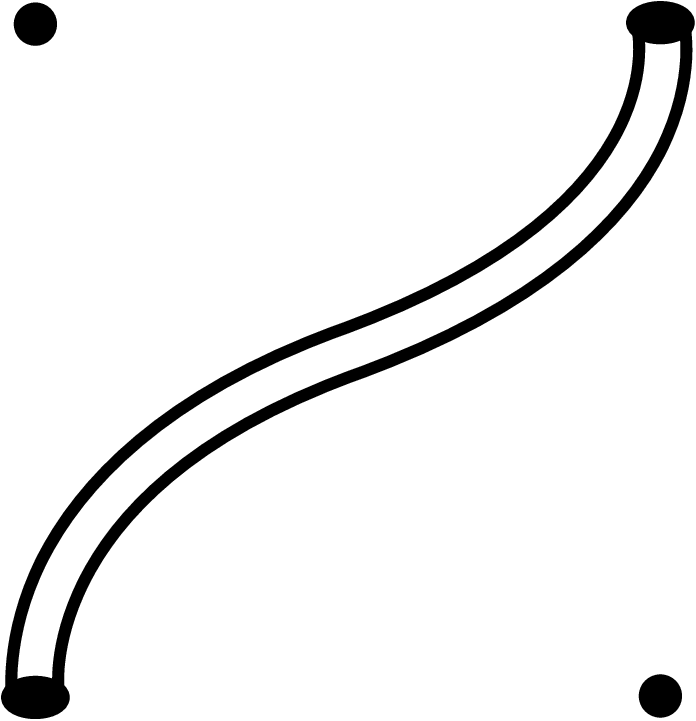}}}\\[10pt]
    &+a_{29}(u,v)\quad\vcenter{\hbox{\includegraphics[scale=0.10]{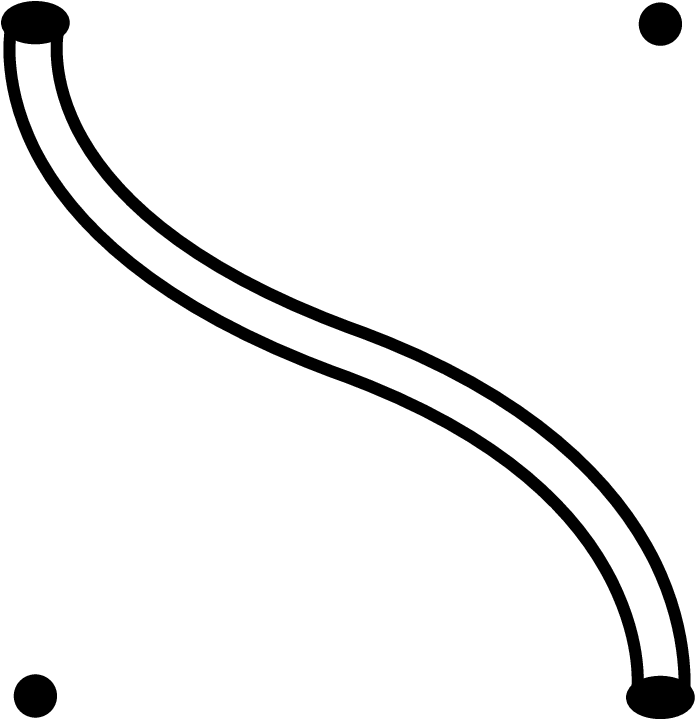}}}\quad+a_{30}(u,v)\quad\vcenter{\hbox{\includegraphics[scale=0.10]{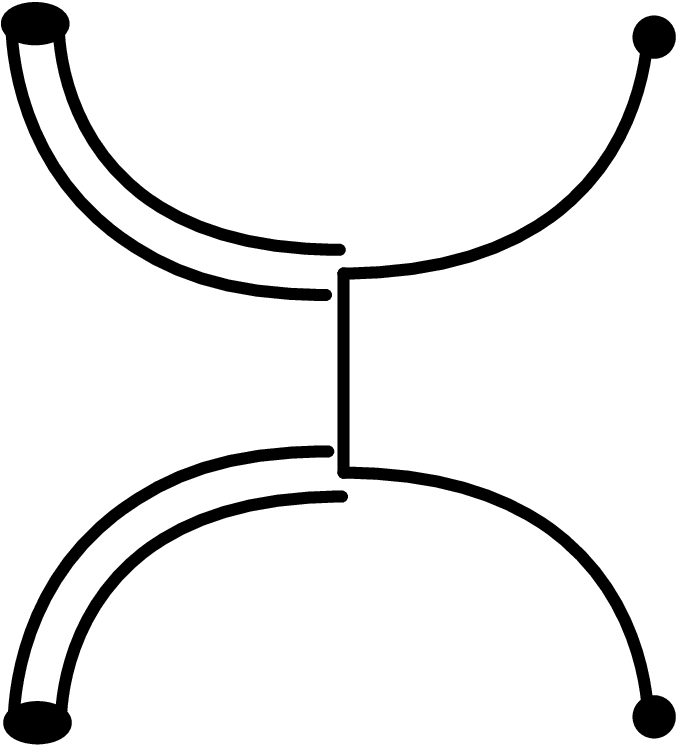}}}\quad+a_{31}(u,v)\quad\vcenter{\hbox{\includegraphics[scale=0.10]{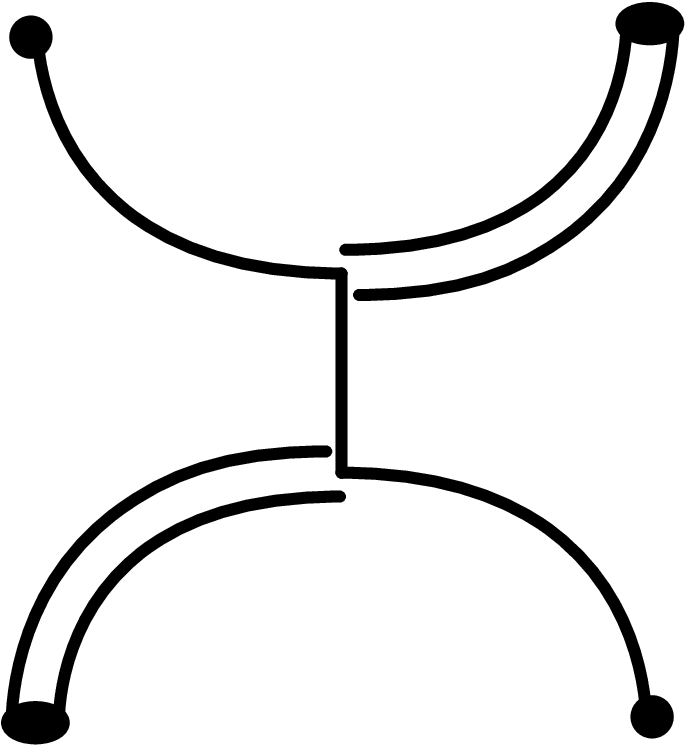}}}\quad+a_{32}(u,v)\quad\vcenter{\hbox{\includegraphics[scale=0.10]{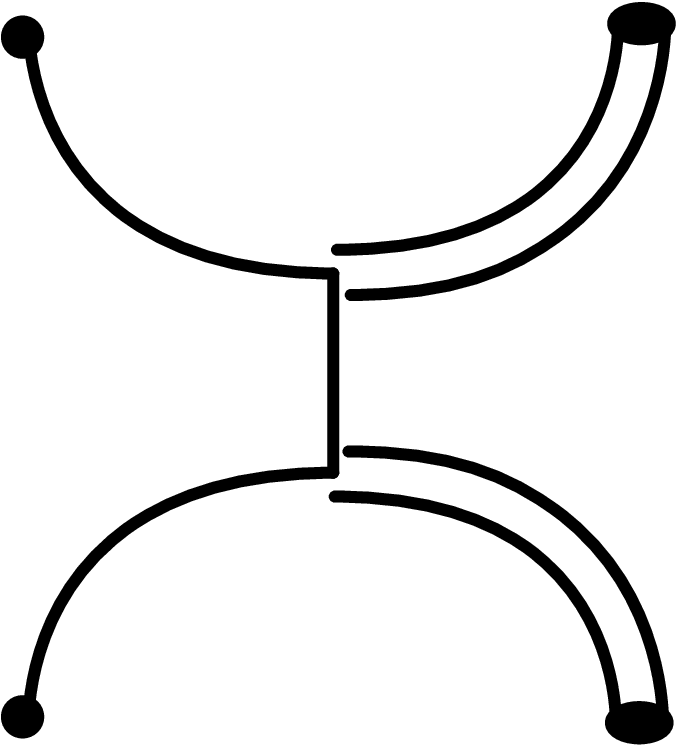}}}\\[10pt]
    &+a_{33}(u,v)\quad\vcenter{\hbox{\includegraphics[scale=0.10]{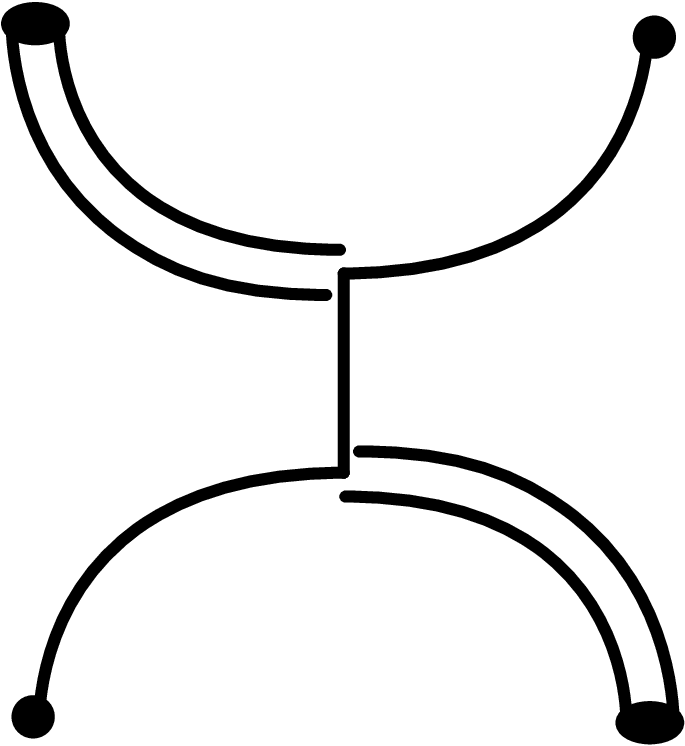}}}\quad+a_{34}(u,v)\quad\vcenter{\hbox{\includegraphics[scale=0.10]{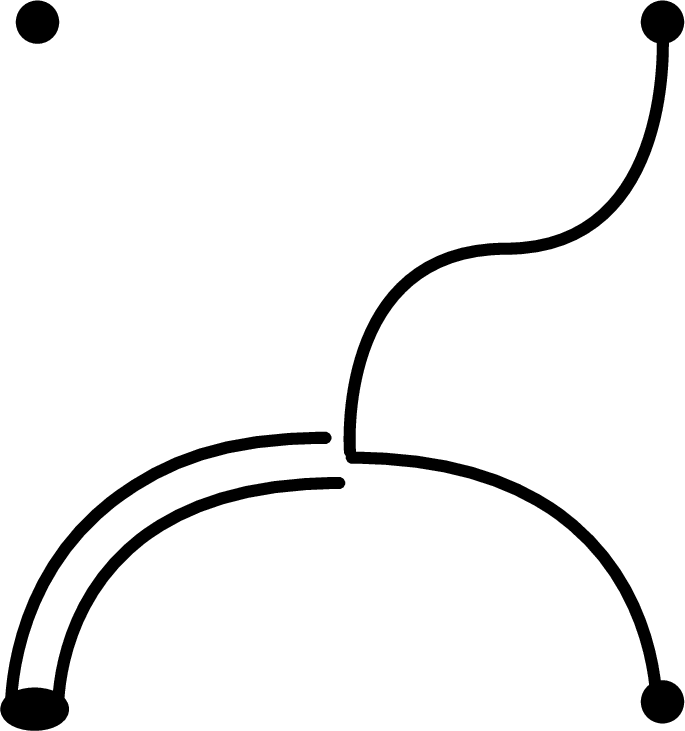}}}\quad+a_{35}(u,v)\quad\vcenter{\hbox{\includegraphics[scale=0.10]{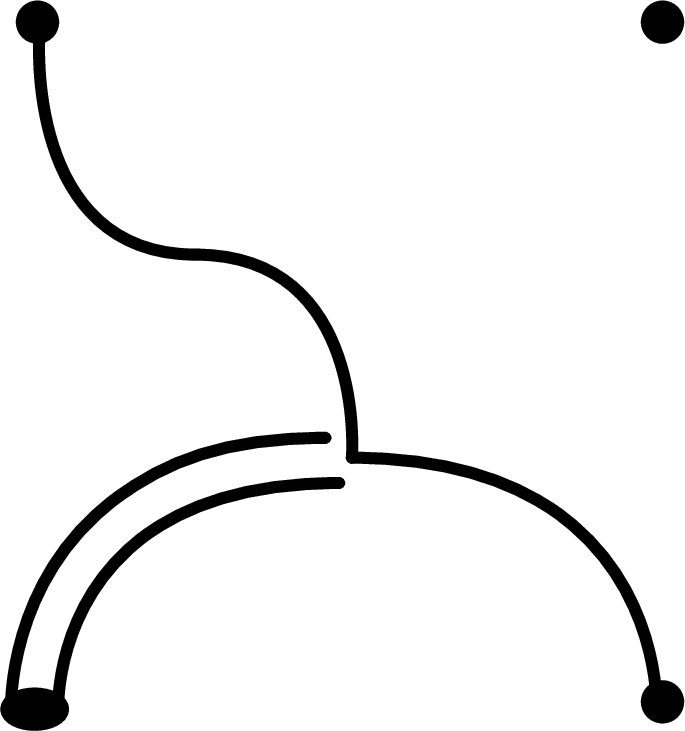}}}\quad+a_{36}(u,v)\quad\vcenter{\hbox{\includegraphics[scale=0.10]{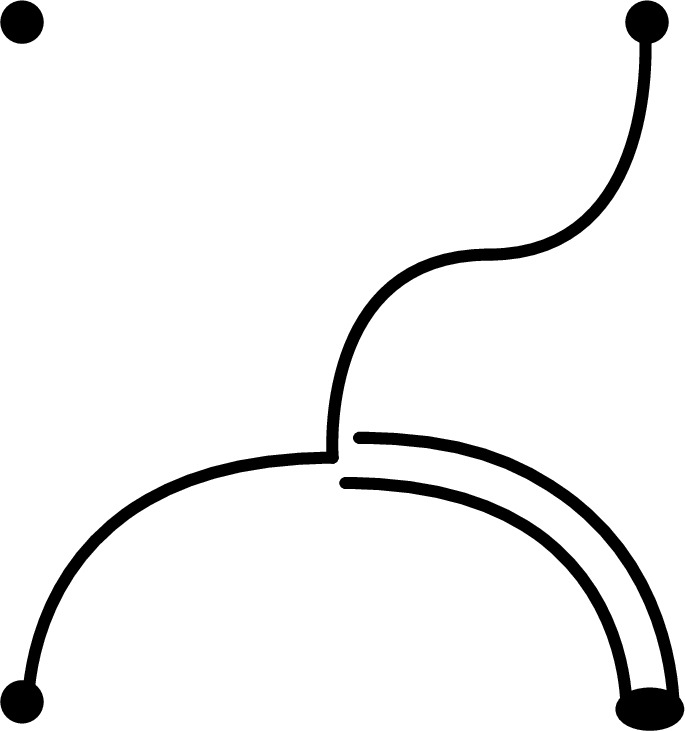}}}\\[10pt]
    &+a_{37}(u,v)\quad\vcenter{\hbox{\includegraphics[scale=0.10]{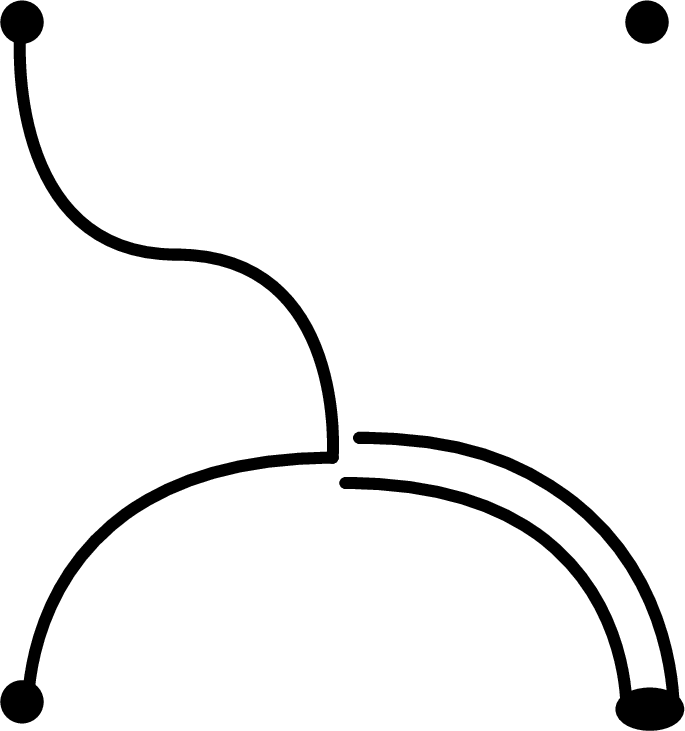}}}\quad+a_{38}(u,v)\quad\vcenter{\hbox{\includegraphics[scale=0.10]{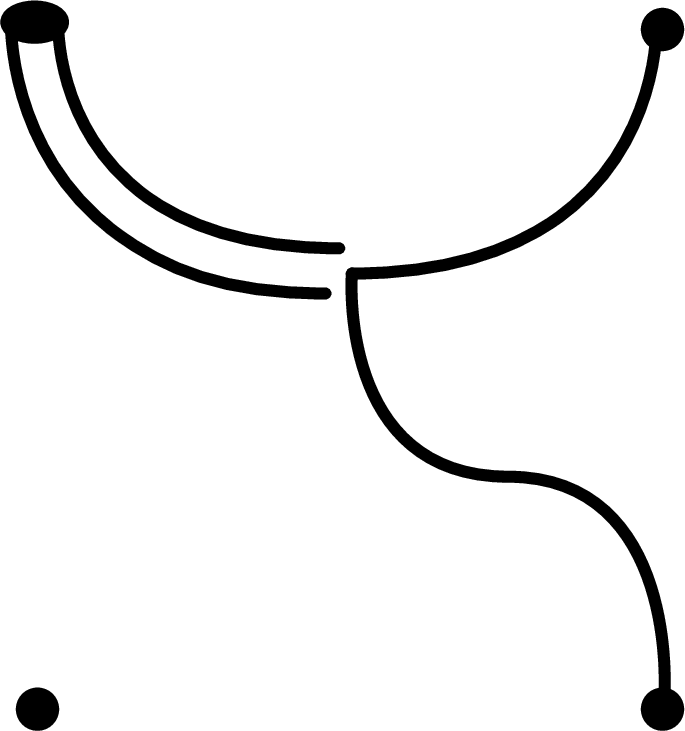}}}\quad+a_{39}(u,v)\quad\vcenter{\hbox{\includegraphics[scale=0.10]{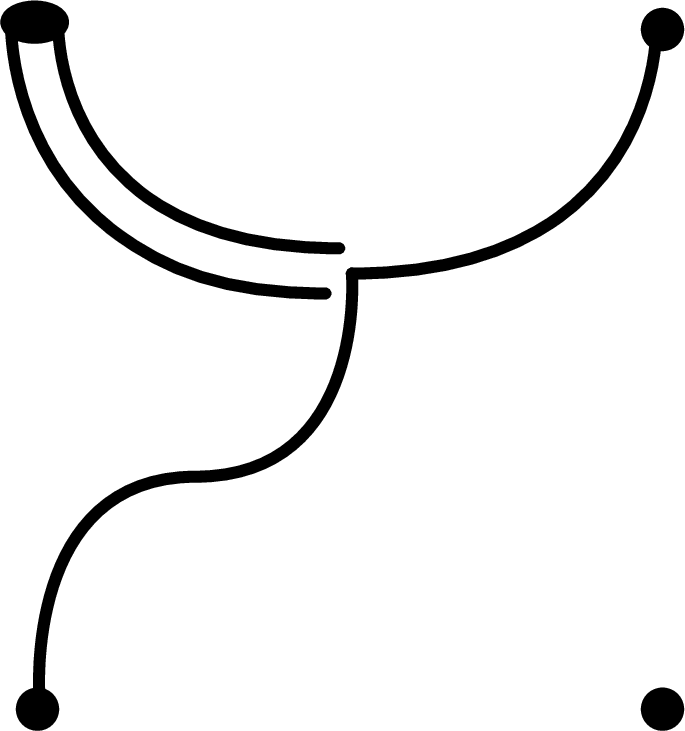}}}\quad+a_{40}(u,v)\quad\vcenter{\hbox{\includegraphics[scale=0.10]{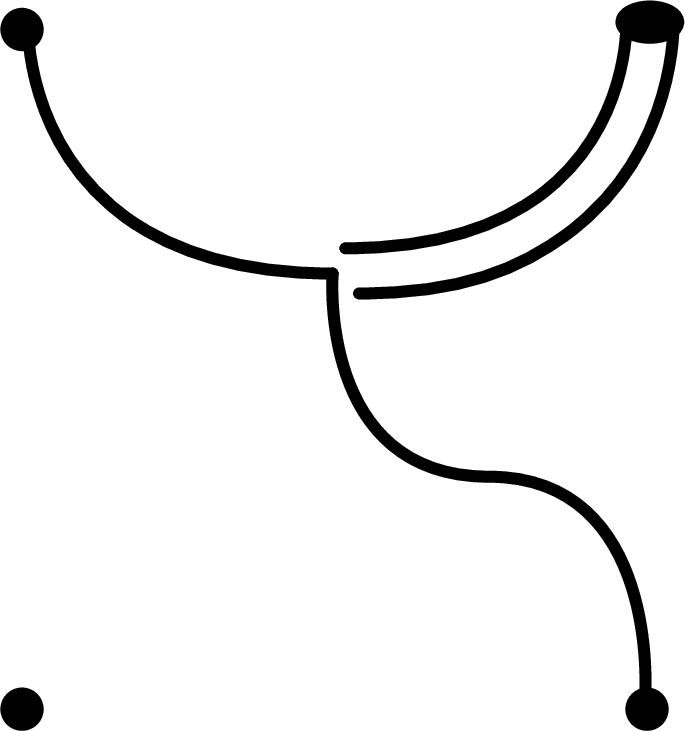}}}\\[10pt]
    &+a_{41}(u,v)\quad\vcenter{\hbox{\includegraphics[scale=0.10]{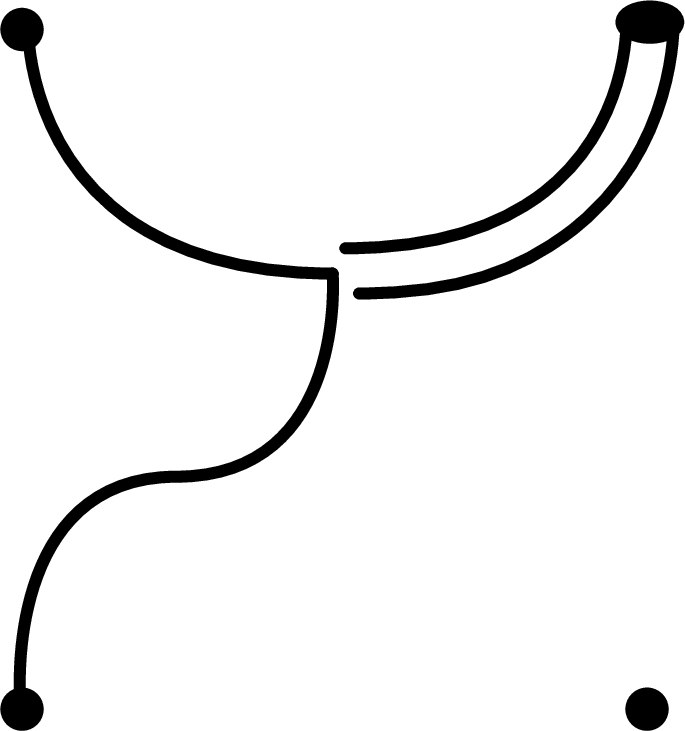}}}\quad+a_{42}(u,v)\quad\vcenter{\hbox{\includegraphics[scale=0.10]{diagrams/dTLRmat3.eps}}}\quad+a_{43}(u,v)\quad\vcenter{\hbox{\includegraphics[scale=0.10]{diagrams/dTLRmat5.eps}}}
\end{align*}
where $a_i(u,v)$ are some coefficients and the webs span the space of intertwiners $\text{End}_{U_{t^2}(B_2)}\left((\mathbb{C} \oplus V_1\oplus V_2)^2\right)$.

Asking for $\check{R}(u,v)$ to commute with the remaining generators, we see that it depends only on the ratio $s=\frac{u}{v}$. Plugging this linear system for the functions $a_i(s)$ into Mathematica, we find
\begin{align*}
    a_1(s)=&\left(t^4+1\right)^2 \left(t^8-s^2\right) \left(t^{10}+s^2\right) \left(-t^8+t^2 s^4-(t-1) (t+1) \left(t^8-t^4+1\right) s^2\right)\\
    a_2(s)=&-\left(t^4+1\right)^2 \left(t^2-s\right) \left(t^2+s\right) \left(t^8-s^2\right) \left(t^6+s^2\right) \left(t^{10}+s^2\right)\\
    a_3(s)=&t^8 \left(s^2-1\right) \left(t^8-s^2\right) \left(t^2+s^2\right) \left(t^{10}+s^2\right)\\
    a_4(s)=&\left(t^4+1\right)^2 [-t^{24}-t^4 s^8-\left(t^4-t^2+1\right) \left(\left(t^{16}-4 t^8-t^6+6 t^4-t^2-4\right) t^8+1\right) s^4\\
    &+\left(t^{14}-3 t^{12}+t^{10}+3 t^8-3 t^6-t^4+3 t^2-1\right) t^2 s^6\\
    &-\left(t^{14}-3 t^{12}+t^{10}+3 t^8-3 t^6-t^4+3 t^2-1\right) t^{12} s^2]\\
    a_5(s)=&-\left(\left(t^{12}-t^{10}+t^8-t^4+t^2-1\right) s \left(t^{10}+s^2\right) \left(t^{12}-t^2 s^4+\left(t^{14}+t^{10}+t^8-t^6-t^4-1\right) s^2\right)\right)\\
    a_6(s)=&-\left(\left(t^{12}-t^{10}+t^8-t^4+t^2-1\right) s \left(t^{10}+s^2\right) \left(t^{12}-t^2 s^4+\left(t^{14}+t^{10}+t^8-t^6-t^4-1\right) s^2\right)\right)\\
    a_7(s)=&-\left(\left(t^6-2 t^4+2 t^2-1\right) \left(t^6+t^4+t^2+1\right)^2 s^2 \left(t^8-s^2\right) \left(t^{10}+s^2\right)\right)\\
    a_8(s)=&-\left(\left(t^6-2 t^4+2 t^2-1\right) \left(t^6+t^4+t^2+1\right)^2 s^2 \left(t^8-s^2\right) \left(t^{10}+s^2\right)\right)\\
    a_9(s)=&-\left(\left(t^4-1\right) \left(t^4+1\right)^2 s \left(t^8-s^2\right) \left(t^6+s^2\right) \left(t^{10}+s^2\right)\right)\\
    a_{10}(s)=&-\left(\left(t^4-1\right) \left(t^4+1\right)^2 s \left(t^8-s^2\right) \left(t^6+s^2\right) \left(t^{10}+s^2\right)\right)\\
    a_{11}(s)=&-t^4 \left(t^4+1\right) \left(s^2-1\right) \left(t^8-s^2\right) \left(t^6+s^2\right) \left(t^{10}+s^2\right)\\
    a_{12}(s)=&-t^4 \left(t^4+1\right) \left(s^2-1\right) \left(t^8-s^2\right) \left(t^6+s^2\right) \left(t^{10}+s^2\right)\\    
    a_{13}(s)=&t^8 \left(t^4+1\right) \left(s^2-1\right) \left(t^2+s^2\right) \left(t^{12}-t^2 s^4+\left(t^{14}-t^{12}+t^8-t^6+t^2-1\right) s^2\right)\\
    a_{14}(s)=&t^8 \left(t^4+1\right)^2 \left(s^2-1\right) \left(t^2-s\right) \left(t^2+s\right) \left(t^2+s^2\right) \left(t^6+s^2\right)\\
    a_{15}(s)=&t^8 \left(t^4-1\right) \left(t^5-t^3+t\right)^2 s \left(s^2-1\right) \left(t^{12}-t^2 s^4+\left(t^{14}+t^{10}+t^8-t^6-t^4-1\right) s^2\right)\\
    a_{16}(s)=&-\left(\left(t^4-1\right) \left(t^4+1\right)^2 s \left(s^2-1\right) \left(t^{12}-t^2 s^4+\left(t^{14}+t^{10}+t^8-t^6-t^4-1\right) s^2\right)\right)\\
    a_{17}(s)=&t^{14} \left(t^2-1\right) \left(t^{10}+t^6+t^4+1\right)^2 s^2 \left(s^2-1\right) \left(t^2+s^2\right)\\
    a_{18}(s)=&t^2 \left(t^2-1\right) \left(t^6+t^4+t^2+1\right)^2 s^2 \left(s^2-1\right) \left(t^2+s^2\right)\\
    a_{19}(s)=&-t^6 \left(t^4-1\right) \left(t^4+1\right)^2 s \left(s^2-1\right) \left(t^8-s^2\right) \left(t^{10}+s^2\right)\\
    a_{20}(s)=&t^4 \left(t^4-1\right) s \left(s^2-1\right) \left(t^8-s^2\right) \left(t^{10}+s^2\right)\\
    a_{21}(s)=&t^8 \left(t^8-1\right) s \left(s^2-1\right) \left(t^4-s^2\right) \left(t^2+s^2\right)\\
    a_{22}(s)=&-t^{10} \left(t^4-1\right) \left(t^4+1\right)^3 s \left(s^2-1\right) \left(t^4-s^2\right) \left(t^2+s^2\right)\\
    a_{23}(s)=&-t^6 \left(t^4+1\right) \left(s^2-1\right) \left(t^8-s^2\right) \left(t^2+s^2\right) \left(t^{10}+s^2\right)\\
    a_{24}(s)=&-i t^6 \left(t^8-t^6+t^2-1\right) s \left(s^2-1\right) \left(t^8-s^2\right) \left(t^{10}+s^2\right)\\
    a_{25}(s)=&-i t^6 \left(t^8-t^6+t^2-1\right) s \left(s^2-1\right) \left(t^8-s^2\right) \left(t^{10}+s^2\right)\\
    a_{26}(s)=&i \left(t^4-1\right) \left(t^5+t\right)^2 s \left(s^2-1\right) \left(t^8-s^2\right) \left(t^{10}+s^2\right)\\
    a_{27}(s)=&i \left(t^4-1\right) \left(t^5+t\right)^2 s \left(s^2-1\right) \left(t^8-s^2\right) \left(t^{10}+s^2\right)\\
    a_{28}(s)=&t^4 \left(t^4+1\right)^2 \left(s^2-1\right) \left(t^8-s^2\right) \left(t^2+s^2\right) \left(t^{10}+s^2\right)\\
    a_{29}(s)=&t^4 \left(t^4+1\right)^2 \left(s^2-1\right) \left(t^8-s^2\right) \left(t^2+s^2\right) \left(t^{10}+s^2\right)\\
    a_{30}(s)=&t^8 \left(t^8-1\right) s \left(s^2-1\right) \left(t^2+s^2\right) \left(t^{10}+s^2\right)\\
    a_{31}(s)=&-t^8 \left(t^4+1\right) \left(s^2-1\right) \left(t^4-s^2\right) \left(t^2+s^2\right) \left(t^{10}+s^2\right)\\
    a_{32}(s)=&t^8 \left(t^8-1\right) s \left(s^2-1\right) \left(t^2+s^2\right) \left(t^{10}+s^2\right)\\
    a_{33}(s)=&-t^8 \left(t^4+1\right) \left(s^2-1\right) \left(t^4-s^2\right) \left(t^2+s^2\right) \left(t^{10}+s^2\right)\\
    a_{34}(s)=&-i t^{10} \left(t^{12}-t^{10}+t^8-t^4+t^2-1\right) s \left(s^2-1\right) \left(t^2+s^2\right) \left(t^{10}+s^2\right)\\
    a_{35}(s)=&-i t^8 \left(t^2-1\right) \left(t^4+1\right) \left(t^6+1\right)^2 s^2 \left(s^2-1\right) \left(t^{10}+s^2\right)\\
    a_{36}(s)=&-i t^8 \left(t^2-1\right) \left(t^4+1\right) \left(t^6+1\right)^2 s^2 \left(s^2-1\right) \left(t^{10}+s^2\right)\\
    a_{37}(s)=&-i t^{10} \left(t^{12}-t^{10}+t^8-t^4+t^2-1\right) s \left(s^2-1\right) \left(t^2+s^2\right) \left(t^{10}+s^2\right)\\
    a_{38}(s)=&-i t^4 \left(t^8-1\right) s \left(s^2-1\right) \left(t^2+s^2\right) \left(t^{10}+s^2\right)\\
    a_{39}(s)=&-i t^2 \left(t^{14}+t^8-t^6-1\right) s^2 \left(s^2-1\right) \left(t^{10}+s^2\right)\\
    a_{40}(s)=&-i t^2 \left(t^{14}+t^8-t^6-1\right) s^2 \left(s^2-1\right) \left(t^{10}+s^2\right)\\
    a_{41}(s)=&-i t^4 \left(t^8-1\right) s \left(s^2-1\right) \left(t^2+s^2\right) \left(t^{10}+s^2\right)\\
    a_{42}(s)=&-t^4 \left(t^4+1\right) \left(s^2-1\right) \left(t^{10}+s^2\right) \left(\left(t^4+1\right) s^4-t^{10} \left(t^4+1\right)+\left(t^{14}-2 t^{12}+t^8-t^6+2 t^2-1\right) s^2\right)\\
    a_{43}(s)=&-t^4 \left(t^4+1\right) \left(s^2-1\right) \left(t^{10}+s^2\right) \left(\left(t^4+1\right) s^4-t^{10} \left(t^4+1\right)+\left(t^{14}-2 t^{12}+t^8-t^6+2 t^2-1\right) s^2\right)
\end{align*}
One can then show, using for instance Mathematica, that the corresponding operator $\check{R}(s)$ satisfy the multiplicative spectral parameter dependant Yang-Baxter equation.

\subsubsection{Integrable $B_2$ web model on the hexagonal lattice}

It is apparent that by setting the spectral parameter $s=t^4$, $\check{R}(s)$ will be decomposed only in terms of webs appearing in the local transfer matrix of the $B_2$ web model on the hexagonal lattice. In order to put it in the form of \eqref{transfermatrixB2}, we renormalise the R matrix and take the following transformation 
\begin{align}
\label{gaugeB2}
    D&=\text{Diag}(1,\alpha,\alpha,\alpha,\alpha,\beta,\beta,\beta,\beta,\beta)\nonumber\\
    \check{R}(s)&\rightarrow \left(D^{-1}\otimes D^{-1}\right) \check{R}(s) \left(D\otimes D \right)
\end{align} 
which is just an elementary rescaling of the irreducible components in the decomposition \eqref{Vu_B2_decomp}.
For a well-chosen normalisation constant and parameters $\alpha$, $\beta$, we recover the local transfer matrix \eqref{transfermatrixB2} with the following parametrisation, setting $t=e^{i\psi}$: 
\begin{subequations}
\begin{align}
    q=&e^{2i\psi}\\
    x_{t;1}=&\frac{2\sin(\psi)}{4\sin^2(\psi)-1}\\
    x_{t;2}=&\frac{1}{4\sin^2(\psi)-1}\\
    x_{v;1}=&-\frac{1}{4\sin(\psi)\cos(2\psi)}\\
    x_{v;2}=&0\\
    y=&1
\end{align}
\end{subequations}

\subsubsection{Integrable $B_2$ web model on the square lattice}

Let us now turn to the model on the square lattice. When one tunes the spectral parameter as $s=e^{i\frac{\pi}{4}}t^{\frac{5}{2}}$ and uses the gauge freedom \eqref{gaugeB2}, one obtains the $B_2$ web model on the square lattice defined in Section \ref{sec:B2websquare} with the following Boltzmann weights:
\begin{align*}
    b_1=&-\frac{i (-1+t (t-i)) (1+(t-1) t (t+1) (t-i))}{t^5+t}\\
    b_2=&-\frac{1+t (t-i) \left(-2+t \left(t^3-t+i\right) (-2+t (t-i))\right)}{t^4}\\
    b_3=&\frac{e^{i\frac{3\pi}{4}} (-1+t (t-i)) \left(1+i \left(t^3-t+i\right) t^2\right) \left(-1+t (t-i) \left(1+(t-1) (t+1) \left(t^4+i t+2\right) t^2\right)\right)}{t^{13/2} \left(t^4+1\right)}\\
    b_4=&-\frac{i (-1+t (t-i)) (1+(t-1) t (t+1) (t-i)) \left(1+\left(t^6-2 t^4+2 i t^3-2 i t-2\right) t^2\right)}{t^9+t^5}\\
    b_5=&\frac{t \left(i t^2+t-i\right) (1+(t-1) t (t+1) (t-i))}{\left(t^4+1\right)^2}\\
    b_6=&1\\
    b_7=&-\frac{(t-i)^2 \left(t^6-2 t^4+2 t^2-1\right)}{t^4}\\
    b_8=&\frac{\left(i t^2+t-i\right) (1+(t-1) t (t+1) (t-i))}{t^3}\\
    b_9=&\frac{e^{i\frac{3\pi}{4}} (t-1) (t-i) (t+1)}{t^{3/2}}\\
    b_{10}=&-\frac{1+(t-1) t (t+1) (t-i)}{t^4+1}\\
    b_{11}=&\frac{e^{i\frac{\pi}{4}} (1+i t) (t-1) (t+1) (1+(t-1) t (t+1) (t-i))}{t^{3/2} \left(t^4+1\right)}\\
    b_{12}=&\frac{\left(t^2-1\right) \left(t^3+i\right)^2 (-1+t (t-i)) (1+(t-1) t (t+1) (t-i))}{\left(t^5+t\right) \sqrt{t^{10}-t^8+t^6}}\\
    b_{13}=&\frac{e^{i\frac{\pi}{4}} (t-1) (t-i) (t+1) \sqrt{t^{10}-t^8+t^6} (1+(t-1) t (t+1) (t-i))}{t^{11/2} \left(t^4+1\right)}\\
    b_{14}=&-i t^{-9}(-1+t (t-i)) [1\\
    &+t (t-i) (-1+t ((t-1) t (t+1) (-3+t (t (3+t (t-i) (-3+t ((-2+t (t+i)) t^3+2 t-i)))-2 i))-i))]
\end{align*}

\subsubsection{An integrable dilute BMW model}

Another integrable solution for the $B_2$ web model on the square lattice can be obtained. 
This model is much simpler than the preceeding one, since it uses only single lines.
Let $V'_u$, $u\in \mathbb{C}^*$, be the second representation of $U_t(A_4^{(2)})$ given in Appendix \ref{sec:evalrepB22}. We are looking for an operator $\check{R}(u,v)$ intertwining $V'_u\otimes V'_v$ and $V'_v\otimes V'_u$. Remark that $E_i$, $F_i$ and $H_i$ for $i=1,2$ generate a Hopf subalgebra isomorphic to $U_{t^2}(B_2)$. Under the action of this subalgebra, $V'_u$ decomposes as:
\begin{align}
  \label{Vu_decomp}
    V'_u=\mathbb{C}\oplus V_1
\end{align}

Hence $\check{R}(u,v)$ will decompose as a sum of $U_{t^2}(B_2)$ intertwiners: 
\begin{align}
    \check{R}(u,v)=&\ a_1(u,v)\quad\vcenter{\hbox{\includegraphics[scale=0.12]{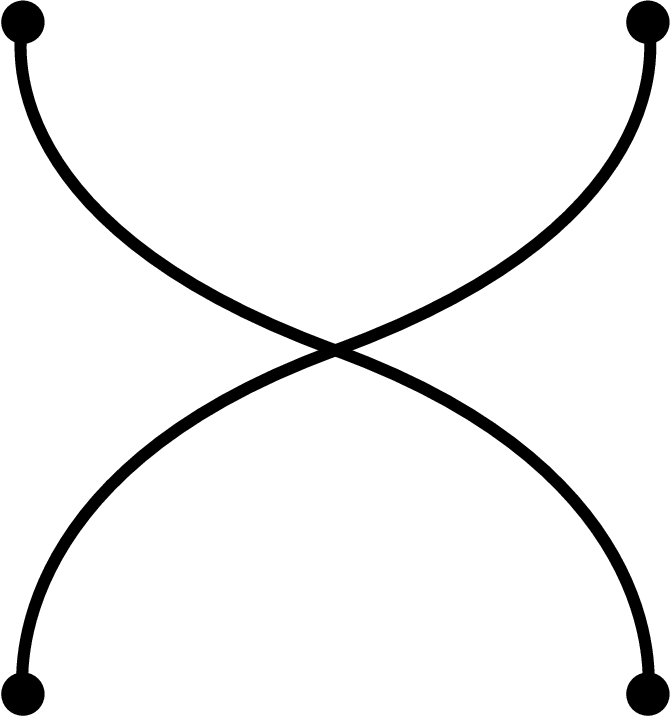}}}\quad+a_2(u,v)\quad\vcenter{\hbox{\includegraphics[scale=0.12]{diagrams/dTLRmat1.eps}}}\quad+a_3(u,v)\quad\vcenter{\hbox{\includegraphics[scale=0.12]{diagrams/dTLRmat6.eps}}}\quad
    +a_4(u,v)\quad\vcenter{\hbox{\includegraphics[scale=0.12]{diagrams/dTLRmat2.eps}}}\nonumber\\[10pt]
    &+a_5(u,v)\quad\vcenter{\hbox{\includegraphics[scale=0.12]{diagrams/dTLRmat4.eps}}}\quad+a_6(u,v)\quad\vcenter{\hbox{\includegraphics[scale=0.12]{diagrams/dTLRmat7.eps}}}\quad +a_7(u,v)\quad\vcenter{\hbox{\includegraphics[scale=0.12]{diagrams/dTLRmat8.eps}}} \quad +a_8(u,v)\quad\vcenter{\hbox{\includegraphics[scale=0.12]{diagrams/dTLRmat3.eps}}}\nonumber\\[10pt]
    &+a_9(u,v)\quad\vcenter{\hbox{\includegraphics[scale=0.12]{diagrams/dTLRmat5.eps}}}\quad+a_{10}(u,v)\quad\vcenter{\hbox{\includegraphics[scale=0.12]{diagrams/dTLRmat9.eps}}}
\label{RchBMW}
\end{align}
where $a_i(u,v)$ are some coefficients and the webs span the space of intertwiners $\text{End}_{U_{t^2}(B_2)}\left((\mathbb{C} \oplus V_1)^2\right)$, with the first web defined in \eqref{4valentvertex}.\footnote{Remark that this map is not the flip map.}

Asking for $\check{R}(u,v)$ to commute with the remaining generators, we see that it depends only on the ratio $s=\frac{u}{v}$. Plugging this linear system for the functions $a_i(s)$ into Mathematica, we find

\begin{align*}
    a_1(s)=&\frac{(t^5s^{-1}+t^{-5}s)(s-s^{-1})}{t^2+t^{-2}}\\
    a_2(s)=&(t^2s^{-1}-t^{-2}s)(t^5s^{-1}+t^{-5}s)\\
    a_{3}(s)=&-(t^3s^{-1}+t^{-3}s)(s-s^{-1})\\
    a_{4}(s)=&(t^2-t^{-2})(t^5s^{-1}+t^{-5}s)\\
    a_{5}(s)=&(t^2-t^{-2})(t^5s^{-1}+t^{-5}s)\\
    a_{6}(s)=&t^5(t^2-t^{-2})(s-s^{-1})\\
    a_{7}(s)=&-\frac{1}{t^5}(t^2-t^{-2})(s-s^{-1})\\
    a_{8}(s)=&-(t^5s^{-1}+t^{-5}s)(s-s^{-1})\\
    a_{9}(s)=&-(t^5s^{-1}+t^{-5}s)(s-s^{-1})\\
    a_{10}(s)=&(t^5+t^{-5})(t^2-t^{-2})-(t^5s^{-1}+t^{-5}s)(s-s^{-1})
\end{align*}
One can then show, using for instance Mathematica, that the corresponding operator $\check{R}(s)$ satisfies the multiplicative spectral parameter dependant Yang-Baxter equation.

Remark that the diagrams appearing in the decomposition of $\check{R}(u,v)$ are generators of the dilute BMW algebra once one considers the braiding operator\cite{Kuperberg_1996}:
\begin{align*}
    \vcenter{\hbox{\includegraphics[scale=0.12]{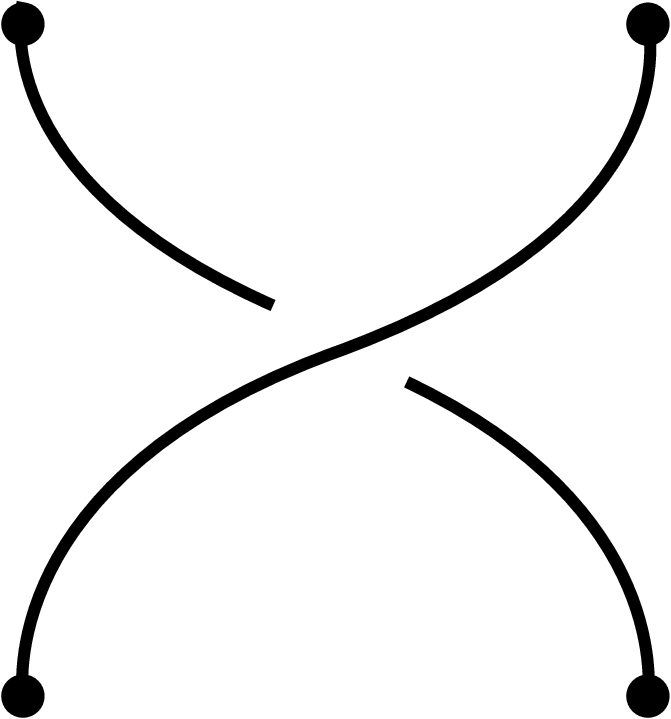}}}=-t^2\quad \vcenter{\hbox{\includegraphics[scale=0.12]{diagrams/dTLRmat1.eps}}} -t^{-2}\quad \vcenter{\hbox{\includegraphics[scale=0.12]{diagrams/dTLRmat6.eps}}} + \frac{1}{[2]_{t^2}} \quad \vcenter{\hbox{\includegraphics[scale=0.12]{diagrams/B2Rmat42.eps}}}
\end{align*}
The dilute BMW algebra has been baxterised\cite{Grimm_1995} and the above integrable $\check{R}$-matrix \eqref{RchBMW} can then be recovered from \cite[eq.~(3.9)]{Grimm_1995}
(with the sign $\sigma=-1$ in the notations of that reference).

We now tune the spectral parameter as $s=-e^{i\frac{\pi}{4}}t^{\frac{5}{2}}$ and use the gauge freedom 
 \begin{align*}
    D&=\text{Diag}(1,\alpha,\alpha,\alpha,\alpha)\nonumber\\
    \check{R}(s)&\rightarrow \left(D^{-1}\otimes D^{-1}\right) \check{R}(s) \left(D\otimes D \right)
\end{align*}
which is just an elementary rescaling, thanks to the decomposition \eqref{Vu_decomp}.
One then obtains the $B_2$ web model on the square lattice defined in Section \ref{sec:B2websquare} with the following Boltzmann weights:
\begin{align*}
    b_1=&\frac{\cos^2(\frac{5}{2}\psi-\frac{\pi}{4})}{2\cos(2\psi)}\\
    b_2=&\sin(\frac{1}{2}\psi+\frac{\pi}{4})\cos(\frac{5}{2}\psi-\frac{\pi}{4})\\
    b_3=&-\sin(2\psi)\cos(\frac{5}{2}\psi-\frac{\pi}{4})\\
    b_4=&-\cos^2(\frac{5}{2}\psi-\frac{\pi}{4})\\
    b_{14}=&\cos(5\psi)\sin(2\psi)-\cos^2(\frac{5}{2}\psi-\frac{\pi}{4})\\
    b_i=&0 \qquad \text{for  }i\in \llbracket5,13\rrbracket
\end{align*}
where we have parametrised as $t=e^{i\psi}$. Notice that the last condition on vanishing weights is the same as \eqref{nodoubleedge} that forbids double edges in the corresponding spin model.

\section{Discussion}
\label{sec:disc}

In this paper, we have defined web models as lattice models based on all the rank-two spiders introduced by Kuperberg in \cite{kuperberg1991quantum,Kuperberg_1996}. We have exhibited specific root-of-unity values of the deformation parameter $q$ for which the weight of closed webs can be used to count the number of colourings of their dual graphs, possibly with some constraints.
These combinatorial properties allowed us to relate the web models at these special values of $q$ to three- and four-state spins model with certain plaquette interactions and
specific global symmetries.

We have then used the central results of \cite{Kuperberg_1996} which connect open webs with invariants of quantum group representations in order to write explicit local transfer matrices of the web models.
Finally, we derived integrable $R$-matrices, which for a well-chosen value of their spectral parameter reproduce these transfer matrices. In particular, we have exhibited new integrable points for the spin models.

Although this paper begins with the definition of certain web models and then demonstrates how they are obtained from affine $R$-matrices of well-chosen integrable models, another point of view would be that the definitions of those particular models are motivated by their integrability. Once the integrable $R$-matrix has been expressed in terms of webs, we may then define a related web model on the lattice.

As in the well known case of loop models, the most convenient models are obtained on the square lattice or the hexagonal lattice.
We have chosen to prioritise the simplest choice, which is the hexagonal lattice $\mathbb{H}$. The configurations are easily described as embeddings of webs in $\mathbb{H}$.
Moreover, the local patterns around a vertex of $\mathbb{H}$ are so few that it is always possible to write rotationally invariant local Boltzmann weights in terms of local fugacities. The fugacities account for links or nodes to be covered by specific types of edges of a given web. 

By contrast, the models on the square lattice $\mathbb{S}$ are a bit more involved, and the local patterns at each vertex are more complicated, decorated graph embeddings.
In particular, we have chosen to define the $B_2$ web models on $\mathbb{S}$, because the integrable solutions on $\mathbb{H}$ are not rotational invariant. 
We might of course also have defined integrable $A_2$ and $G_2$ web models on $\mathbb{S}$. In the $A_2$ case,
for instance, this would have led to interactions of the form
\begin{center}
    \includegraphics[scale=0.12]{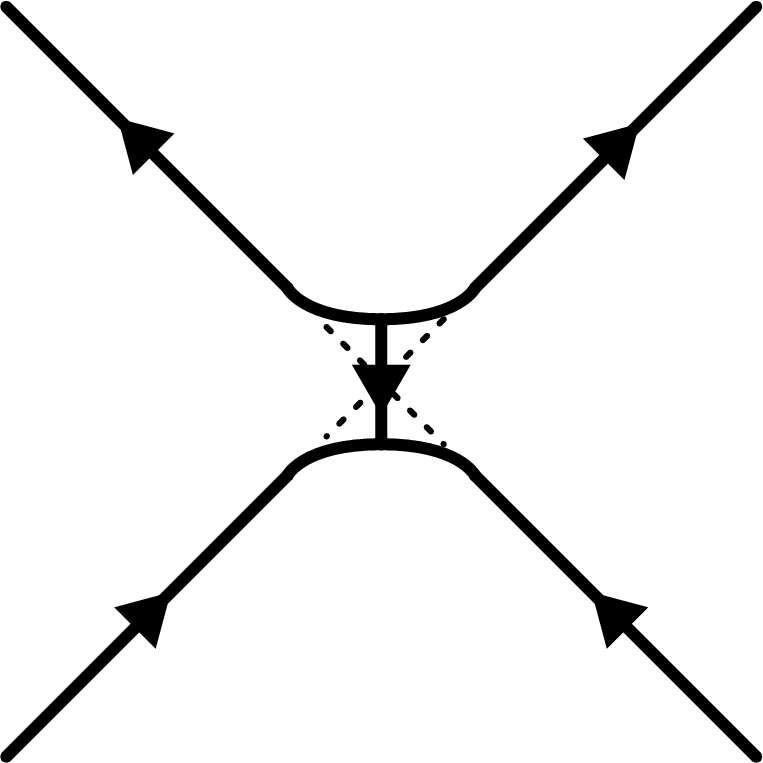}
\end{center}

It is worth mentionning that the web models on $\mathbb{S}$ could potentially have a richer phase diagram than their counterparts on $\mathbb{H}$.
For instance, the loop model on $\mathbb{S}$ presents three different integrable regimes \cite{nienhuis_critical_1990}, one of which has a
non-compact continuum limit \cite{vernier_new_2015}, but only one of those regimes survives the restriction of the model to the simpler lattice $\mathbb{H}$.
Remarkably, our numerical work shows that the $A_2$ and $G_2$ web models defined on $\mathbb{H}$ have two and (at least) four regimes, respectively.
The corresponding models defined on $\mathbb{S}$ might conceivably possess an even greater number of regimes.

\medskip

We now discuss some possible directions for future work.
First and foremost, one might solve the integrable web models by the Bethe ansatz.
The fact that $R$-matrices are obtained from quantum group symmetries suggests the use of the algebraic Bethe ansatz.
But even the more modest goal of setting up the Bethe ansatz equations might be worthwhile, since this would give access
to larger system sizes than can be attained by exact diagonalisation of the transfer matrix.

Another direction is to study the geometrical nature of the excitations present in the spectra of the transfer matrices. This has been discussed in \cite{Lafay:2021wyf} for the $A_2$ case, where it was suggested that such excitations correspond to lattice precursors of electromagnetic operators in a Coulomb 
Gas description of the continuum limit. This $A_2$ Coulomb Gas will be defined in a future paper \cite{LGJ}. One might also hope to find analogous Coulomb Gas descriptions
for the other $G_2$ and $B_2$ web models, although it seems more challenging when the relevant Dynkin diagram is not simply laced (see the discussion in \cite{lafay:tel-03966966}).

Some of the special points that we have identified in the phase diagrams of the web models are of particular interest.
For instance, the dense $G_2$ web model at $q=e^{i\frac{5\pi}{6}}$ describes a three-colour analogue of percolation
which has not been studied before, to our best knowledge (see Section~\ref{sec:specptG2}).
It is also interesting that some instances of the {\it loop} models, such as percolation or LERW, appear at specific points of the rank-$2$ {\it web} models (see Sections~\ref{sec:specptsA2} and \ref{sec:specptG2}).
These give access to operators which are not present in the rank-$1$ realisations of these models, and may hence provide new details about
their critical behaviour. The challenge would then be to describe the geometrical interpretation of such operators.

Other special points indicate that the three-state Potts model with plaquette interactions can give rise to universality classes which are unusual for this model.
For instance we have found cases of central charge $c=-2$ and $c \approx 1.5$. It would be interesting to study further the possible RG flows between these models,
and in particular whether some of the flows are integrable.

We did not present here the numerical analysis of the $B_2$ models, due to the large dimension of its transfer matrix. We also did not study the special points of this model.
However, by the structure of its symmetries at $q=i$,
the $B_2$ web model is a good candidate for realising symplectic fermions at $c=-4$.

Finally, one can define discrete holomorphic observables from the intertwining relations \eqref{LinearSys} and give them a combinatorial formulation thanks to webs,
as was already done in the loop model case \cite{Ikhlef_2009,Ikhlef_2013,Riva_2006,smirnov2007conformal}. One might hope that some of these observables
would enable us to rigorously derive conformal invariance of the corresponding lattice models, analogously to what was already done for
the Ising model in its dilute loop representation \cite{chelkak2011universality}, or for percolation \cite{smirnov2009critical,khristoforov2021percolation}.

\subsubsection*{Acknowledgments}

Part of this work was done when Augustin Lafay was affiliated with the LPENS. His research is now supported by the Academy of Finland grant number 340461,
entitled ``Conformal invariance in planar random geometry''.
The work of Azat M.\ Gainutdinov was supported by the CNRS, and partially by the ANR grant JCJC ANR-18-CE40-0001.
AMG is also grateful to LPENS, Paris for its kind hospitality in 2022 and 2023.
This work of Jesper L.\ Jacobsen was supported by the French Agence Nationale de la Recherche (ANR) under grant ANR-21-CE40-0003 (project CONFICA).

\appendix
\newcommand{\balance}{\boldsymbol{g}}
\newcommand{\kk}{\mathbb{C}}
\newcommand{\coev}{\mathrm{coev}}
\label{sec:appendix}
\section{Some symmetries of the $A_2$ web models}
\label{sec:A2sym}
The Boltzmann weight of the $A_2$ web models is of the form $x^{N}(yz)^{N_V}e^{iM\phi}w_{\rm K}(G)$ for a given configuration $G$. Denote by $w_{\text{loc}}(G)=x^{N}(yz)^{N_V}e^{iM\phi}$ the product of its local fugacities. We will show by induction on the number of edges that $N+N_V+M\equiv 0 \text{ mod }2$, $M\equiv 0 \text{ mod }3$.

Consider a web $G$ embedded in the hexagonal lattice $\mathbb{H}$. If $G$ is a collection of loops, the result is clear. Supose it is not the case. Let $F$ be a face of $G$ other than the exterior one that is not surrounded by a loop. Remove the edges surrounding $F$ going clockwise around it and reverse the orientation of the others. It is clear that the resulting graph $G'$ embedded in $\mathbb{H}$ is again a web. Moreover $G'$ has strictly less edges and the induction hypothesis applies.

Around each node of $\mathbb{H}$ surrounding $F$, there are, up to rotations, $6$ possibilities. We draw them below as well as their transformed counterparts, dotted lines meaning empty links.
\begin{subequations}
    \begin{align}
        &\vcenter{\hbox{\includegraphics[scale=0.2]{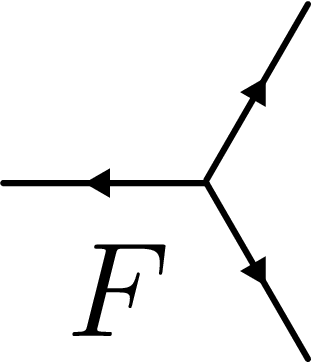}}}\quad \rightarrow \quad \vcenter{\hbox{\includegraphics[scale=0.2]{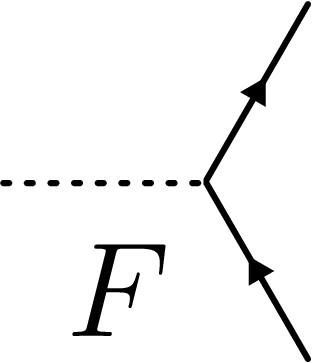}}}\quad : \quad x^{3/2}y\rightarrow xe^{i\phi}\\
        &\vcenter{\hbox{\includegraphics[scale=0.2]{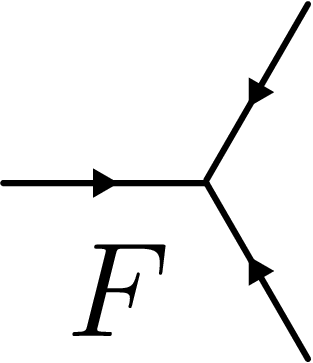}}}\quad \rightarrow \quad \vcenter{\hbox{\includegraphics[scale=0.2]{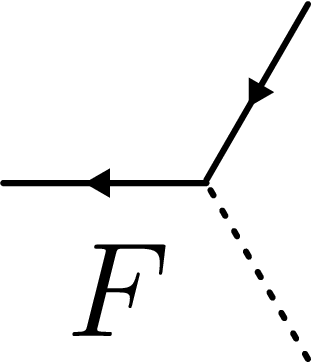}}}\quad : \quad x^{3/2}z\rightarrow xe^{i\phi}\\
        &\vcenter{\hbox{\includegraphics[scale=0.2]{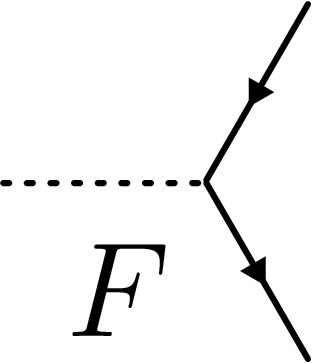}}}\quad \rightarrow \quad \vcenter{\hbox{\includegraphics[scale=0.2]{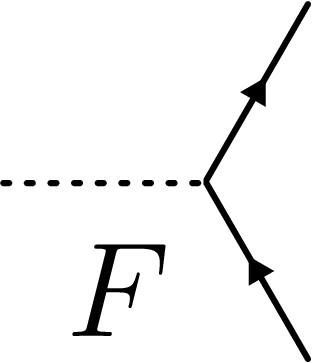}}}\quad : \quad xe^{-i\phi}\rightarrow xe^{i\phi}\label{A2symm1}\\
        &\vcenter{\hbox{\includegraphics[scale=0.2]{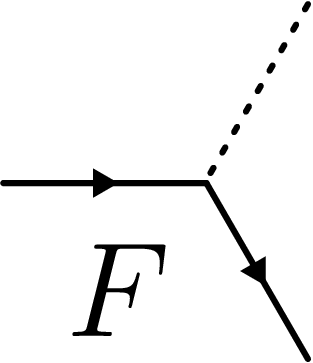}}}\quad \rightarrow \quad \vcenter{\hbox{\includegraphics[scale=0.2]{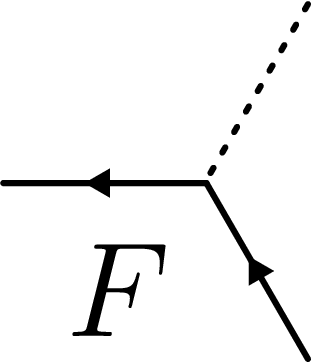}}}\quad : \quad xe^{i\phi}\rightarrow xe^{-i\phi}\label{A2symm2}\\
        &\vcenter{\hbox{\includegraphics[scale=0.2]{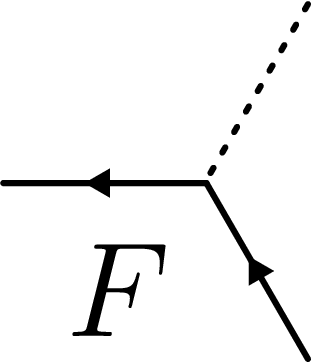}}}\quad \rightarrow \quad \vcenter{\hbox{\includegraphics[scale=0.2]{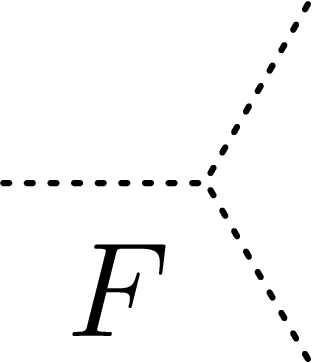}}}\quad : \quad xe^{-i\phi}\rightarrow 1\label{A2symn1}\\
        &\vcenter{\hbox{\includegraphics[scale=0.2]{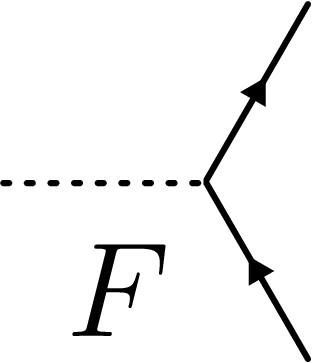}}}\quad \rightarrow \quad \vcenter{\hbox{\includegraphics[scale=0.2]{diagrams/A2sym10.eps}}}\quad : \quad xe^{i\phi}\rightarrow 1\label{A2symn2}
    \end{align}
\end{subequations}
We also wrote the corresponding change in the contribution to the Boltzmann weight from local fugacities, understanding that the fugacity of a half bond is $x^{1/2}$. Denote by $l$ the number of source/sink pairs surrounding $F$, by $m_1$ the number of nodes as in \eqref{A2symm1}, by $m_2$ the number of nodes as in \eqref{A2symm2}, by $n_1$ the number of nodes as in \eqref{A2symn1} and by $n_2$ the number of nodes as in \eqref{A2symn2}.

We have that 
\begin{align*}
    w_{\text{loc}}(G)=&(xyze^{i2\phi})^l(e^{i\phi})^{m_1}(e^{-i\phi})^{m_2}(xe^{i\phi})^{n_1}(xe^{-i\phi})^{n_2}w_{\text{loc}}(G')\\
    &=x^{l+n_1+n_2}(yz)^{l}(e^{i\phi})^{2l+2m_1-2m_2+n_1-n_2}w_{\text{loc}}(G')
\end{align*}
As $F$ is surrounded by an even number of edges, 
\begin{align*}
    2l+m_1+m_2+n_1+n_2\equiv 0 \text{ mod }2
\end{align*}
from which it follows, using the induction hypothesis, that
\begin{align*}
    N+N_V+M\equiv 0 \text{ mod }2
\end{align*}

Moreover, one has that the number of nodes where $F$ is locally convex minus the number of nodes where $F$ is not is equal to $6$. Thus
\begin{align*}
    2l+m_2+n_1-m_1-n_2=6\equiv 0 \text{ mod }3
\end{align*}
which implies that
\begin{align*}
    2l+2m_1-2m_2+n_1-n_2\equiv 0 \text{ mod }3
\end{align*}
and then, using the induction hypothesis, that
\begin{align*}
    M\equiv 0 \text{ mod }3
\end{align*}

\section{Data and conventions for root systems}
\label{sec:appendix1}
Here, we give Cartan matrix conventions for the quantum groups used in the main text, in the order that the big quantum group follows the small one, for instance, $G_2^{(1)}$ follows  $A_2$. We describe also the symmetrisation of the Cartan matrix $A_{ij}$ in terms of the relatively prime positive integers $d_i$ such that $d_iA_{ij}$ is symmetric. We also give the fundamental weights in the non-affine cases.

\subsection{$A_2$ web models}
\subsubsection{$A_2$}
Let $\rootone$ and $\roottwo$ be the two simple roots of $A_2$. We normalise them such that $(\bm{\alpha}_1\bm{\alpha}_1)=2$. The Cartan matrix is given by:

\begin{equation}
    A_{ij}=2\frac{(\bm{\alpha}_i,\bm{\alpha}_j)}{(\bm{\alpha}_i,\bm{\alpha}_i)}=
    \begin{pmatrix}
        2 & -1 \\
        -1 & 2 
    \end{pmatrix}
\end{equation}

The simple coroots are given by 
\begin{subequations}
\begin{align}
    \rootone^\vee=\ &\rootone\\
    \roottwo^\vee=\ &\roottwo
\end{align}
\end{subequations}

The fundamental weights are given by 
\begin{subequations}
\begin{align}
    \weightone=\ &\frac{2}{3}\rootone+\frac{1}{3}\roottwo\\
    \weighttwo=\ &\frac{1}{3}\rootone+\frac{2}{3}\roottwo
\end{align}
\end{subequations}

The Weyl vector and dual Weyl vector are
\begin{subequations}
\begin{align}
    \bm{\rho}=\ &\weightone+\weighttwo=\rootone+\roottwo\\
    \bm{\rho}^\vee=\ &\bm{\rho}
\end{align}
\end{subequations}

\subsubsection{$G_2^{(1)}$}
The Cartan matrix is given by:

\begin{equation}
    A_{ij}=
    \begin{pmatrix}
        2 & -1 & 0\\
        -1 & 2 & -1\\
        0  & -3 & 2
    \end{pmatrix}
\end{equation}

\begin{equation*}
    d_1=3,\qquad d_2=3,\qquad d_3=1
\end{equation*}
\subsection{$G_2$ web models}
\subsubsection{$G_2$}
Let $\rootone$ and $\roottwo$ be the two simple roots of $G_2$ with $\rootone$ the smallest one. We normalise them such that $(\bm{\alpha}_1\bm{\alpha}_1)=2$. The Cartan matrix is given by:

\begin{equation}
    A_{ij}=2\frac{(\bm{\alpha}_i,\bm{\alpha}_j)}{(\bm{\alpha}_i,\bm{\alpha}_i)}=
    \begin{pmatrix}
        2 & -3 \\
        -1 & 2 
    \end{pmatrix}
\end{equation}

\begin{equation*}
    d_1=1,\qquad d_2=3
\end{equation*}

The simple coroots are given by 
\begin{subequations}
\begin{align}
    \rootone^\vee=\ &\rootone\\
    \roottwo^\vee=\ &\frac{1}{3}\roottwo
\end{align}
\end{subequations}

The fundamental weights are given by 
\begin{subequations}
\begin{align}
    \weightone=\ &2\rootone+\roottwo\\
    \weighttwo=\ &3\rootone+2\roottwo
\end{align}
\end{subequations}

The Weyl vector and dual Weyl vector are
\begin{subequations}
\begin{align}
    \bm{\rho}=\ &\weightone+\weighttwo=5\rootone+3\roottwo\\
    \bm{\rho}^\vee=\ &3\rootone+\frac{5}{3}\roottwo
\end{align}
\end{subequations}

\subsubsection{$D_4^{(3)}$}
The Cartan matrix is given by:

\begin{equation}
    A_{ij}=
    \begin{pmatrix}
        2 & -1 & 0\\
        -3 & 2 & -1\\
        0  & -1 & 2
    \end{pmatrix}
\end{equation}

\begin{equation*}
    d_1=3,\qquad d_2=1,\qquad d_3=1
\end{equation*}

\subsection{$B_2$ web models}
\subsubsection{$B_2$}
Let $\rootone$ and $\roottwo$ be the two simple roots of $B_2$ with $\rootone$ the smallest one. We normalise them such that $(\bm{\alpha}_1\bm{\alpha}_1)=2$. The Cartan matrix is given by:

\begin{equation}
    A_{ij}=2\frac{(\bm{\alpha}_i,\bm{\alpha}_j)}{(\bm{\alpha}_i,\bm{\alpha}_i)}=
    \begin{pmatrix}
        2 & -2 \\
        -1 & 2 
    \end{pmatrix}
\end{equation}

\begin{equation*}
    d_1=1,\qquad d_2=2
\end{equation*}

The simple coroots are given by 
\begin{subequations}
\begin{align}
    \rootone^\vee=\ &\rootone\\
    \roottwo^\vee=\ &\frac{1}{2}\roottwo
\end{align}
\end{subequations}

The fundamental weights are given by 
\begin{subequations}
\begin{align}
    \weightone=\ &\rootone+\frac{1}{2}\roottwo\\
    \weighttwo=\ &\rootone+\roottwo
\end{align}
\end{subequations}

The Weyl vector and dual Weyl vector are
\begin{subequations}
\begin{align}
    \bm{\rho}=\ &\weightone+\weighttwo=2\rootone+\frac{3}{2}\roottwo\\
    \bm{\rho}^\vee=\ &\frac{3}{2}\rootone+\roottwo
\end{align}
\end{subequations}

\subsubsection{$A_4^{(2)}$}
The Cartan matrix is given by:

\begin{equation}
    A_{ij}=
    \begin{pmatrix}
        2 & -2 & 0\\
        -1 & 2 & -2\\
        0  & -1 & 2
    \end{pmatrix}
\end{equation}

\begin{equation*}
    d_1=1,\qquad d_2=2,\qquad d_3=4
\end{equation*}

\section{Conventions for quantum groups}
\label{sec:quantumgroupconventions}
Let $q \in \mathbb{C}$ be an arbitrary non-zero
 complex number (but $q \neq \pm 1$). The statistical weights of the web models will be defined in terms of so-called
$q$-numbers $[k]_q$, with $k \in \mathbb{N}$, defined by
\begin{equation}
    [k]_q =\frac{q^k-q^{-k}}{q-q^{-1}} \,.
\end{equation}
Note that the $q$-numbers reduce to the ordinary integers, $[k]_q \to k$, in the limit $q \to 1$.
We shall also need the corresponding $q$-factorial and $q$-binomial coefficients:
\begin{equation*}
    [k]_q!=\prod_{1\leq i\leq k} [i]_q \;, \qquad
    \qbin{n}{k}=\frac{[n]_q!}{[k]_q![n-k]_q!} \,,
\end{equation*}
with the convention $[0]_q! = 1$ and $\footnotesize\qbin{n}{0}=1$.

We recall here a definition of the Hopf algebra $U_q(X)$ and its pivotal structure. Let $d_i$ be the relatively prime positive integers such that $d_iA_{ij}$ is symmetric.
Then, the $\mathbb{C}(q)$-algebra $U_q(X)$ is  generated by $E_i$, $F_i$, $q^{H_i}$ for $i\in\llbracket1,2\rrbracket$ satisfying the following relations:
\begin{subequations}
\begin{align}
    q^{H_i}q^{H_j} &= q^{H_j}q^{H_i} \,, \\
    q^{H_i}E_jq^{-H_i} = q^{A_{ij}}E_j& \,, \qquad
    q^{H_i}F_jq^{-H_i} = q^{-A_{ij}}F_j \,, \\
    [E_i,F_j] &= \delta_{ij}\frac{q^{d_iH_i}-q^{-d_iH_i}}{q^{d_i}-q^{-d_i}} \,, \\
    \sum_{m=0}^{1-A_{ij}} (-1)^m \qbini{1-A_{ij}}{m} &E_i^{1-A_{ij}-m}E_jE_i^m =0\,, \quad \text{if } i\neq j \,, \\
    \sum_{m=0}^{1-A_{ij}} (-1)^m \qbini{1-A_{ij}}{m} &F_i^{1-A_{ij}-m}F_jF_i^m =0\,, \quad \text{if } i\neq j \,,  \,.
\end{align}
\end{subequations}
It is a Hopf algebra with the coproduct
\begin{align}\label{eq:coprod}
    \Delta(E_i)=E_i\otimes q^{d_iH_i} + 1\otimes E_i \,, \qquad \Delta(F_i)=F_i\otimes 1 + q^{-d_iH_i}\otimes F_i \,, \qquad \Delta(q^{H_i})=q^{H_i}\otimes q^{H_i} \,,
\end{align}
the antipode
\begin{align}
    S(E_i)=-E_iq^{-d_iH_i}\,, \qquad S(F_i)=-q^{d_iH_i}F_i \,, \qquad S(q^{H_i})=q^{-H_i} \,,
\end{align}
and the counit
\begin{align}
    \epsilon(E_i)=0 \,, \qquad \epsilon(F_i)=0 \,, \qquad \epsilon(q^{H_i})=1 \,.
\end{align}

We use the notation $H_{\sum_i c_i\bm{\alpha}_i} := \sum_i c_id_iH_i$.

\section{Explicit matrix elements of representations}
\label{sec:explicit}
We give in this section the matrices of our representations of interest of the generators of quantum groups in our chosen bases.

\subsection{The $A_2$ web models}
\subsubsection{$U_{-q}(A_2)$ representations}
\label{sec:UqA2reps}
In the basis $\{u_i, i\in \llbracket 1, 3 \rrbracket\}$ of $V_1$ introduced in Section \ref{sec:TMA2}, we have the following representation of generators of $U_{-q}(A_2)$:
\begin{alignat*}{2}
    &E_1= \begin{bmatrix}
0 & 1 & 0  \\
0 & 0 & 0  \\
0 & 0 & 0  
\end{bmatrix}  \qquad &&E_2 = \begin{bmatrix}
0 & 0 & 0 \\
0 & 0 & 1 \\
0 & 0 & 0 
\end{bmatrix} \\
    &F_1= \begin{bmatrix}
0 & 0 & 0  \\
1 & 0 & 0  \\
0 & 0 & 0 
\end{bmatrix}  \qquad &&F_2 = \begin{bmatrix}
0 & 0 & 0 \\
0 & 0 & 0 \\
0 & 1 & 0 
\end{bmatrix} \\
    &H_1= \begin{bmatrix}
1 & 0 & 0  \\
0 & -1 & 0  \\
0 & 0 & 0
\end{bmatrix}  \qquad &&H_2 = \begin{bmatrix}
0 & 0 & 0  \\
0 & 1 & 0  \\
0 & 0 & -1 
\end{bmatrix} 
\end{alignat*}

In the basis $\{v_i, i\in \llbracket 1, 3 \rrbracket\}$ of $V_2$ introduced in Section \ref{sec:TMA2}, we have the following representation of generators of $U_{-q}(A_2)$:
\begin{alignat*}{2}
    &E_1= \begin{bmatrix}
0 & 0 & 0  \\
0 & 0 & 1  \\
0 & 0 & 0  
\end{bmatrix}  \qquad &&E_2 = \begin{bmatrix}
0 & 1 & 0 \\
0 & 0 & 0 \\
0 & 0 & 0 
\end{bmatrix} \\
    &F_1= \begin{bmatrix}
0 & 0 & 0  \\
0 & 0 & 0  \\
0 & 1 & 0 
\end{bmatrix}  \qquad &&F_2 = \begin{bmatrix}
0 & 0 & 0 \\
1 & 0 & 0 \\
0 & 0 & 0 
\end{bmatrix} \\
    &H_1= \begin{bmatrix}
0 & 0 & 0  \\
0 & 1 & 0  \\
0 & 0 & -1
\end{bmatrix}  \qquad &&H_2 = \begin{bmatrix}
1 & 0 & 0  \\
0 & -1 & 0  \\
0 & 0 & 0
\end{bmatrix} 
\end{alignat*}

\subsubsection{$U_{t}(G_2^{(1)})$ evaluation representation}\label{G21rep}
Let $u\in \mathbb{C}^*$. Consider the following representation $V_u$ of $U_t(G_2^{(1)})$ given in the basis $\{ u_1,u_2,u_3,v_1,v_2,v_3,1 \}$ where $1$ denotes the basis vector of the trivial representation of $U_{-q}(A_2)$, 
\begin{alignat*}{2}
    &E_0= \begin{bmatrix}
0 & 1 & 0 & 0 & 0 & 0 & 0 \\
0 & 0 & 0 & 0 & 0 & 0 & 0 \\
0 & 0 & 0 & 0 & 0 & 0 & 0 \\
0 & 0 & 0 & 0 & 0 & 0 & 0 \\
0 & 0 & 0 & 0 & 0 & 1 & 0 \\
0 & 0 & 0 & 0 & 0 & 0 & 0 \\
0 & 0 & 0 & 0 & 0 & 0 & 0 
\end{bmatrix} \qquad && F_0= \begin{bmatrix}
0 & 0 & 0 & 0 & 0 & 0 & 0 \\
1 & 0 & 0 & 0 & 0 & 0 & 0 \\
0 & 0 & 0 & 0 & 0 & 0 & 0 \\
0 & 0 & 0 & 0 & 0 & 0 & 0 \\
0 & 0 & 0 & 0 & 0 & 0 & 0 \\
0 & 0 & 0 & 0 & 1 & 0 & 0 \\
0 & 0 & 0 & 0 & 0 & 0 & 0 
\end{bmatrix}\\
&E_1=\begin{bmatrix}
0 & 0 & 0 & 0 & 0 & 0 & 0 \\
0 & 0 & 1 & 0 & 0 & 0 & 0 \\
0 & 0 & 0 & 0 & 0 & 0 & 0 \\
0 & 0 & 0 & 0 & 1 & 0 & 0 \\
0 & 0 & 0 & 0 & 0 & 0 & 0 \\
0 & 0 & 0 & 0 & 0 & 0 & 0 \\
0 & 0 & 0 & 0 & 0 & 0 & 0 
\end{bmatrix} \qquad && F_1= \begin{bmatrix}
0 & 0 & 0 & 0 & 0 & 0 & 0 \\
0 & 0 & 0 & 0 & 0 & 0 & 0 \\
0 & 1 & 0 & 0 & 0 & 0 & 0 \\
0 & 0 & 0 & 0 & 0 & 0 & 0 \\
0 & 0 & 0 & 1 & 0 & 0 & 0 \\
0 & 0 & 0 & 0 & 0 & 0 & 0 \\
0 & 0 & 0 & 0 & 0 & 0 & 0 
\end{bmatrix}\\
&E_2=u\begin{bmatrix}
0 & 0 & 0 & 0 & 0 & 0 & 0 \\
0 & 0 & 0 & 0 & 0 & 0 & 0 \\
0 & 0 & 0 & 0 & 0 & 0 & \sqrt{[2]_t} \\
0 & 0 & 0 & 0 & 0 & 0 & 0 \\
1 & 0 & 0 & 0 & 0 & 0 & 0 \\
0 & 1 & 0 & 0 & 0 & 0 & 0 \\
0 & 0 & 0 & \sqrt{[2]_t} & 0 & 0 & 0 
\end{bmatrix} \qquad && F_2= \frac{1}{u}\begin{bmatrix}
0 & 0 & 0 & 0 & 1 & 0 & 0 \\
0 & 0 & 0 & 0 & 0 & 1 & 0 \\
0 & 0 & 0 & 0 & 0 & 0 & 0 \\
0 & 0 & 0 & 0 & 0 & 0 & \sqrt{[2]_t} \\
0 & 0 & 0 & 0 & 0 & 0 & 0 \\
0 & 0 & 0 & 0 & 0 & 0 & 0 \\
0 & 0 & \sqrt{[2]_t} & 0 & 0 & 0 & 0 
\end{bmatrix}\\
 &H_0= \begin{bmatrix}
1 & 0 & 0 & 0 & 0 & 0 & 0 \\
0 & -1 & 0 & 0 & 0 & 0 & 0 \\
0 & 0 & 0 & 0 & 0 & 0 & 0 \\
0 & 0 & 0 & 0 & 0 & 0 & 0 \\
0 & 0 & 0 & 0 & 1 & 0 & 0 \\
0 & 0 & 0 & 0 & 0 & -1 & 0 \\
0 & 0 & 0 & 0 & 0 & 0 & 0 
\end{bmatrix} \qquad && H_1= \begin{bmatrix}
0 & 0 & 0 & 0 & 0 & 0 & 0 \\
0 & 1 & 0 & 0 & 0 & 0 & 0 \\
0 & 0 & -1 & 0 & 0 & 0 & 0 \\
0 & 0 & 0 & 1 & 0 & 0 & 0 \\
0 & 0 & 0 & 0 & -1 & 0 & 0 \\
0 & 0 & 0 & 0 & 0 & 0 & 0 \\
0 & 0 & 0 & 0 & 0 & 0 & 0 
\end{bmatrix}\\
&H_2= \begin{bmatrix}
-1 & 0 & 0 & 0 & 0 & 0 & 0 \\
0 & -1 & 0 & 0 & 0 & 0 & 0 \\
0 & 0 & 2 & 0 & 0 & 0 & 0 \\
0 & 0 & 0 & -2 & 0 & 0 & 0 \\
0 & 0 & 0 & 0 & 1 & 0 & 0 \\
0 & 0 & 0 & 0 & 0 & 1 & 0 \\
0 & 0 & 0 & 0 & 0 & 0 & 0 
\end{bmatrix}
\end{alignat*}

\subsection{The $G_2$ web models}
\subsubsection{$U_q(G_2)$ representations}

In the basis $\{e_i, i\in \llbracket 1, 7 \rrbracket\}$ of $V$ introduced in Section \ref{sec:TMG2}, we have the following representation of generators of $U_q(G_2)$:
\begin{alignat*}{2}
    &E_1= \begin{bmatrix}
0 & 1 & 0 & 0 & 0 & 0 & 0 \\
0 & 0 & 0 & 0 & 0 & 0 & 0 \\
0 & 0 & 0 & [2]_q & 0 & 0 & 0 \\
0 & 0 & 0 & 0 & [2]_q & 0 & 0 \\
0 & 0 & 0 & 0 & 0 & 0 & 0 \\
0 & 0 & 0 & 0 & 0 & 0 & 1 \\
0 & 0 & 0 & 0 & 0 & 0 & 0 
\end{bmatrix}  \qquad &&E_2 = \begin{bmatrix}
0 & 0 & 0 & 0 & 0 & 0 & 0 \\
0 & 0 & 1 & 0 & 0 & 0 & 0 \\
0 & 0 & 0 & 0 & 0 & 0 & 0 \\
0 & 0 & 0 & 0 & 0 & 0 & 0 \\
0 & 0 & 0 & 0 & 0 & 1 & 0 \\
0 & 0 & 0 & 0 & 0 & 0 & 0 \\
0 & 0 & 0 & 0 & 0 & 0 & 0 
\end{bmatrix} \\
    &F_1= \begin{bmatrix}
0 & 0 & 0 & 0 & 0 & 0 & 0 \\
1 & 0 & 0 & 0 & 0 & 0 & 0 \\
0 & 0 & 0 & 0 & 0 & 0 & 0 \\
0 & 0 & 1 & 0 & 0 & 0 & 0 \\
0 & 0 & 0 & 1 & 0 & 0 & 0 \\
0 & 0 & 0 & 0 & 0 & 0 & 0 \\
0 & 0 & 0 & 0 & 0 & 1 & 0 
\end{bmatrix}  \qquad &&F_2 =\begin{bmatrix}
0 & 0 & 0 & 0 & 0 & 0 & 0 \\
0 & 0 & 0 & 0 & 0 & 0 & 0 \\
0 & 1 & 0 & 0 & 0 & 0 & 0 \\
0 & 0 & 0 & 0 & 0 & 0 & 0 \\
0 & 0 & 0 & 0 & 0 & 0 & 0 \\
0 & 0 & 0 & 0 & 1 & 0 & 0 \\
0 & 0 & 0 & 0 & 0 & 0 & 0 
\end{bmatrix}\\
    &H_1= \begin{bmatrix}
1 & 0 & 0 & 0 & 0 & 0 & 0 \\
0 & -1 & 0 & 0 & 0 & 0 & 0 \\
0 & 0 & 2 & 0 & 0 & 0 & 0 \\
0 & 0 & 0 & 0 & 0 & 0 & 0 \\
0 & 0 & 0 & 0 & -2 & 0 & 0 \\
0 & 0 & 0 & 0 & 0 & 1 & 0 \\
0 & 0 & 0 & 0 & 0 & 0 & -1 
\end{bmatrix} \qquad &&H_2 = \begin{bmatrix}
0 & 0 & 0 & 0 & 0 & 0 & 0 \\
0 & 1 & 0 & 0 & 0 & 0 & 0 \\
0 & 0 & -1 & 0 & 0 & 0 & 0 \\
0 & 0 & 0 & 0 & 0 & 0 & 0 \\
0 & 0 & 0 & 0 & 1 & 0 & 0 \\
0 & 0 & 0 & 0 & 0 & -1 & 0 \\
0 & 0 & 0 & 0 & 0 & 0 & 0 
\end{bmatrix}
\end{alignat*}

\subsubsection{$U_{q}(D_4^{(3)})$ evaluation representation}
\label{sec:evalrepG2}
Let $u\in \mathbb{C}^*$. Consider the following representation $V_u$ of $U_q(D_4^{(3)})$ given in the basis $\{ 1,e_1,e_2,e_3,e_4,e_5,e_6,e_7 \}$ where $1$ denotes the basis vector of the trivial representation of $U_{q}(G_2)$, 
\begin{alignat*}{2}
    &E_0= \begin{bmatrix}
0 & 0 & 0 & 0 & 0 & 0 & 0 & 0 \\
0 & 0 & 0 & 0 & 0 & 0 & 0 & 0 \\
0 & 0 & 0 & 1 & 0 & 0 & 0 & 0 \\
0 & 0 & 0 & 0 & 0 & 0 & 0 & 0 \\
0 & 0 & 0 & 0 & 0 & 0 & 0 & 0 \\
0 & 0 & 0 & 0 & 0 & 0 & 1 & 0 \\
0 & 0 & 0 & 0 & 0 & 0 & 0 & 0 \\
0 & 0 & 0 & 0 & 0 & 0 & 0 & 0 
\end{bmatrix} \qquad && F_0= \begin{bmatrix}
0 & 0 & 0 & 0 & 0 & 0 & 0 & 0 \\
0 & 0 & 0 & 0 & 0 & 0 & 0 & 0 \\
0 & 0 & 0 & 0 & 0 & 0 & 0 & 0 \\
0 & 0 & 1 & 0 & 0 & 0 & 0 & 0 \\
0 & 0 & 0 & 0 & 0 & 0 & 0 & 0 \\
0 & 0 & 0 & 0 & 0 & 0 & 0 & 0 \\
0 & 0 & 0 & 0 & 0 & 1 & 0 & 0 \\
0 & 0 & 0 & 0 & 0 & 0 & 0 & 0 
\end{bmatrix}\\
&E_1=\begin{bmatrix}
0 & 0 & 0 & 0 & 0 & 0 & 0 & 0 \\
0 & 0 & 1 & 0 & 0 & 0 & 0 & 0 \\
0 & 0 & 0 & 0 & 0 & 0 & 0 & 0 \\
0 & 0 & 0 & 0 & [2]_q & 0 & 0 & 0 \\
0 & 0 & 0 & 0 & 0 & [2]_q & 0 & 0 \\
0 & 0 & 0 & 0 & 0 & 0 & 0 & 0 \\
0 & 0 & 0 & 0 & 0 & 0 & 0 & 1 \\
0 & 0 & 0 & 0 & 0 & 0 & 0 & 0 
\end{bmatrix} \qquad && F_1= \begin{bmatrix}
0 & 0 & 0 & 0 & 0 & 0 & 0 & 0 \\
0 & 0 & 0 & 0 & 0 & 0 & 0 & 0 \\
0 & 1 & 0 & 0 & 0 & 0 & 0 & 0 \\
0 & 0 & 0 & 0 & 0 & 0 & 0 & 0 \\
0 & 0 & 0 & 1 & 0 & 0 & 0 & 0 \\
0 & 0 & 0 & 0 & 1 & 0 & 0 & 0 \\
0 & 0 & 0 & 0 & 0 & 0 & 0 & 0 \\
0 & 0 & 0 & 0 & 0 & 0 & 1 & 0 
\end{bmatrix}\\
&E_2=u\begin{bmatrix}
0 & -\frac{\sqrt{[3]_q}}{[2]_q} & 0 & 0 & 0 & 0 & 0 & 0 \\
0 & 0 & 0 & 0 & 0 & 0 & 0 & 0 \\
0 & 0 & 0 & 0 & 0 & 0 & 0 & 0 \\
0 & 0 & 0 & 0 & 0 & 0 & 0 & 0 \\
0 & \frac{1}{[2]_q} & 0 & 0 & 0 & 0 & 0 & 0 \\
0 & 0 & \frac{1}{[2]_q}  & 0 & 0 & 0 & 0 & 0 \\
0 & 0 & 0 & \frac{1}{[2]_q}  & 0 & 0 & 0 & 0 \\
-\frac{\sqrt{[3]_q}}{[2]_q} & 0 & 0 & 0 & \frac{1}{[2]_q}  & 0 & 0 & 0
\end{bmatrix} \qquad && F_2= \frac{1}{u}\begin{bmatrix}
0 & 0 & 0 & 0 & 0 & 0 & 0 & -\sqrt{[3]_q} \\
-\sqrt{[3]_q} & 0 & 0 & 0 & 1 & 0 & 0 & 0 \\
0 & 0 & 0 & 0 & 0 & [2]_q & 0 & 0 \\
0 & 0 & 0 & 0 & 0 & 0 & [2]_q & 0 \\
0 & 0 & 0 & 0 & 0 & 0 & 0 & 1 \\
0 & 0 & 0 & 0 & 0 & 0 & 0 & 0 \\
0 & 0 & 0 & 0 & 0 & 0 & 0 & 0 \\
0 & 0 & 0 & 0 & 0 & 0 & 0 & 0 
\end{bmatrix}\\
 &H_0= \begin{bmatrix}
0 & 0 & 0 & 0 & 0 & 0 & 0 & 0 \\
0 & 0 & 0 & 0 & 0 & 0 & 0 & 0 \\
0 & 0 & 1 & 0 & 0 & 0 & 0 & 0 \\
0 & 0 & 0 & -1 & 0 & 0 & 0 & 0 \\
0 & 0 & 0 & 0 & 0 & 0 & 0 & 0 \\
0 & 0 & 0 & 0 & 0 & 1 & 0 & 0 \\
0 & 0 & 0 & 0 & 0 & 0 & -1 & 0 \\
0 & 0 & 0 & 0 & 0 & 0 & 0 & 0 
\end{bmatrix} \qquad && H_1= \begin{bmatrix}
0 & 0 & 0 & 0 & 0 & 0 & 0 & 0 \\
0 & 1 & 0 & 0 & 0 & 0 & 0 & 0 \\
0 & 0 & -1 & 0 & 0 & 0 & 0 & 0 \\
0 & 0 & 0 & 2 & 0 & 0 & 0 & 0 \\
0 & 0 & 0 & 0 & 0 & 0 & 0 & 0 \\
0 & 0 & 0 & 0 & 0 & -2 & 0 & 0 \\
0 & 0 & 0 & 0 & 0 & 0 & 1 & 0 \\
0 & 0 & 0 & 0 & 0 & 0 & 0 & -1 
\end{bmatrix}\\
&H_2= \begin{bmatrix}
0 & 0 & 0 & 0 & 0 & 0 & 0 & 0 \\
0 & -2 & 0 & 0 & 0 & 0 & 0 & 0 \\
0 & 0 & -1 & 0 & 0 & 0 & 0 & 0 \\
0 & 0 & 0 & -1 & 0 & 0 & 0 & 0 \\
0 & 0 & 0 & 0 & 0 & 0 & 0 & 0 \\
0 & 0 & 0 & 0 & 0 & 1 & 0 & 0 \\
0 & 0 & 0 & 0 & 0 & 0 & 1 & 0 \\
0 & 0 & 0 & 0 & 0 & 0 & 0 & 2 
\end{bmatrix}
\end{alignat*}

\subsection{The $B_2$ web  models}
\subsubsection{$U_{q}(B_2)$ representations}

In the basis $\{e_i, i\in \llbracket 1, 4 \rrbracket\}$ of $V_1$ introduced in Section \ref{sec:B2TM}, we have the following representation of generators of $U_q(B_2)$:
\begin{alignat*}{2}
    &E_1= \begin{bmatrix}
0 & 1 & 0 & 0 \\
0 & 0 & 0 & 0 \\
0 & 0 & 0 & 1 \\
0 & 0 & 0 & 0 
\end{bmatrix}  \qquad &&E_2 = \begin{bmatrix}
0 & 0 & 0 & 0 \\
0 & 0 & 1 & 0 \\
0 & 0 & 0 & 0 \\
0 & 0 & 0 & 0 
\end{bmatrix} \\
    &F_1= \begin{bmatrix}
0 & 0 & 0 & 0 \\
1 & 0 & 0 & 0 \\
0 & 0 & 0 & 0 \\
0 & 0 & 1 & 0 
\end{bmatrix}  \qquad &&F_2 = \begin{bmatrix}
0 & 0 & 0 & 0 \\
0 & 0 & 0 & 0 \\
0 & 1 & 0 & 0 \\
0 & 0 & 0 & 0 
\end{bmatrix} \\
    &H_1= \begin{bmatrix}
1 & 0 & 0 & 0 \\
0 & -1 & 0 & 0 \\
0 & 0 & 1 & 0 \\
0 & 0 & 0 & -1 
\end{bmatrix}  \qquad &&H_2 = \begin{bmatrix}
0 & 0 & 0 & 0 \\
0 & 1 & 0 & 0 \\
0 & 0 & -1 & 0 \\
0 & 0 & 0 & 0 
\end{bmatrix} 
\end{alignat*}

In the basis $\{v_i, i\in \llbracket 1, 5 \rrbracket\}$ of $V_2$ introduced in Section \ref{sec:B2TM}, we have the following representation of generators of $U_q(B_2)$:
\begin{alignat*}{2}
    &E_1= \begin{bmatrix}
0 & 0 & 0 & 0 & 0 \\
0 & 0 & [2]_q & 0 & 0 \\
0 & 0 & 0 & [2]_q & 0 \\
0 & 0 & 0 & 0 & 0 \\
0 & 0 & 0 & 0 & 0 
\end{bmatrix}  \qquad &&E_2 = \begin{bmatrix}
0 & 1 & 0 & 0 & 0 \\
0 & 0 & 0 & 0 & 0 \\
0 & 0 & 0 & 0 & 0 \\
0 & 0 & 0 & 0 & 1 \\
0 & 0 & 0 & 0 & 0 
\end{bmatrix} \\
    &F_1= \begin{bmatrix}
0 & 0 & 0 & 0 & 0 \\
0 & 0 & 0 & 0 & 0 \\
0 & 1 & 0 & 0 & 0 \\
0 & 0 & 1 & 0 & 0 \\
0 & 0 & 0 & 0 & 0 
\end{bmatrix}  \qquad &&F_2 = \begin{bmatrix}
0 & 0 & 0 & 0 & 0 \\
1 & 0 & 0 & 0 & 0 \\
0 & 0 & 0 & 0 & 0 \\
0 & 0 & 0 & 0 & 0 \\
0 & 0 & 0 & 1 & 0 
\end{bmatrix}\\
    &H_1= \begin{bmatrix}
0 & 0 & 0 & 0 & 0 \\
0 & 2 & 0 & 0 & 0 \\
0 & 0 & 0 & 0 & 0 \\
0 & 0 & 0 & -2 & 0 \\
0 & 0 & 0 & 0 & 0 
\end{bmatrix} \qquad &&H_2 = \begin{bmatrix}
1 & 0 & 0 & 0 & 0 \\
0 & -1 & 0 & 0 & 0 \\
0 & 0 & 0 & 0 & 0 \\
0 & 0 & 0 & 1 & 0 \\
0 & 0 & 0 & 0 & -1 
\end{bmatrix} 
\end{alignat*}

\subsubsection{First $U_{t}(A_4^{(2)})$ evaluation representation}
\label{sec:evalrepB2}
Let $u\in \mathbb{C}^*$. Consider the following representation $V_u$ of $U_t(A_4^{(2)})$ given in the basis $\{ 1,e_1,e_2,e_3,e_4,v_1,v_2,v_3,v_4,v_5 \}$ where $1$ denotes the basis vector of the trivial representation of $U_{q}(B_2)$, 
\begin{alignat*}{2}
    &E_0= u\begin{bmatrix}
0 & \frac{\sqrt{[2]_t}(-t^2+1-t^{-2})}{t^2+t^{-2}} & 0 & 0 & 0 & 0 & 0 & 0 & 0 & 0\\
0 & 0 & 0 & 0 & 0 & 0 & 0 & 0 & 0 & 0 \\
0 & 0 & 0 & 0 & 0 & \sqrt{[2]_t} & 0 & 0 & 0 & 0 \\
0 & 0 & 0 & 0 & 0 & 0 & \sqrt{[2]_t} & 0 & 0 & 0 \\
-\sqrt{[2]_t} & 0 & 0 & 0 & 0 & 0 & 0 & \sqrt{[2]_t} & 0 & 0 \\
0 & 0 & 0 & 0 & 0 & 0 & 0 & 0 & 0 & 0 \\
0 & 0 & 0 & 0 & 0 & 0 & 0 & 0 & 0 & 0 \\
0 & \frac{\sqrt{[2]_t}}{t^2+t^{-2}} & 0 & 0 & 0 & 0 & 0 & 0 & 0 & 0 \\
0 & 0 & \frac{\sqrt{[2]_t}}{t^2+t^{-2}} & 0 & 0 & 0 & 0 & 0 & 0 & 0 \\
0 & 0 & 0 & \frac{\sqrt{[2]_t}}{t^2+t^{-2}} & 0 & 0 & 0 & 0 & 0 & 0  
\end{bmatrix}\\
& F_0= \frac{1}{u}\begin{bmatrix}
0 & 0 & 0 & 0 & \frac{\sqrt{[2]_t}(-t^2+1-t^{-2})}{t^2+t^{-2}} & 0 & 0 & 0 & 0 & 0 \\
-\sqrt{[2]_t} & 0 & 0 & 0 & 0 & 0 & 0 & \sqrt{[2]_t} & 0 & 0 \\
0 & 0 & 0 & 0 & 0 & 0 & 0 & 0 & \sqrt{[2]_t}(t^2+t^{-2}) & 0 \\
0 & 0 & 0 & 0 & 0 & 0 & 0 & 0 & 0 & \sqrt{[2]_t}(t^2+t^{-2}) \\
0 & 0 & 0 & 0 & 0 & 0 & 0 & 0 & 0 & 0 \\
0 & 0 & \sqrt{[2]_t} & 0 & 0 & 0 & 0 & 0 & 0 & 0 \\
0 & 0 & 0 & \sqrt{[2]_t} & 0 & 0 & 0 & 0 & 0 & 0 \\
0 & 0 & 0 & 0 & \frac{\sqrt{[2]_t}}{t^2+t^{-2}} & 0 & 0 & 0 & 0 & 0 \\
0 & 0 & 0 & 0 & 0 & 0 & 0 & 0 & 0 & 0 \\
0 & 0 & 0 & 0 & 0 & 0 & 0 & 0 & 0 & 0 
\end{bmatrix}\\
&H_0= \begin{bmatrix}
0 & 0 & 0 & 0 & 0 & 0 & 0 & 0 & 0 & 0 \\
0 & -2 & 0 & 0 & 0 & 0 & 0 & 0 & 0 & 0 \\
0 & 0 & 0 & 0 & 0 & 0 & 0 & 0 & 0 & 0 \\
0 & 0 & 0 & 0 & 0 & 0 & 0 & 0 & 0 & 0 \\
0 & 0 & 0 & 0 & 2 & 0 & 0 & 0 & 0 & 0 \\
0 & 0 & 0 & 0 & 0 & -2 & 0 & 0 & 0 & 0 \\
0 & 0 & 0 & 0 & 0 & 0 & -2 & 0 & 0 & 0 \\
0 & 0 & 0 & 0 & 0 & 0 & 0 & 0 & 0 & 0 \\
0 & 0 & 0 & 0 & 0 & 0 & 0 & 0 & 2 & 0 \\
0 & 0 & 0 & 0 & 0 & 0 & 0 & 0 & 0 & 2 
\end{bmatrix}\\
&E_1=\begin{bmatrix}
0 & 0 & 0 & 0 & 0 & 0 & 0 & 0 & 0 & 0 \\
0 & 0 & 1 & 0 & 0 & 0 & 0 & 0 & 0 & 0 \\
0 & 0 & 0 & 0 & 0 & 0 & 0 & 0 & 0 & 0 \\
0 & 0 & 0 & 0 & 1 & 0 & 0 & 0 & 0 & 0 \\
0 & 0 & 0 & 0 & 0 & 0 & 0 & 0 & 0 & 0 \\
0 & 0 & 0 & 0 & 0 & 0 & 0 & 0 & 0 & 0 \\
0 & 0 & 0 & 0 & 0 & 0 & 0 & t^2+t^{-2} & 0 & 0 \\
0 & 0 & 0 & 0 & 0 & 0 & 0 & 0 & t^2+t^{-2} & 0 \\
0 & 0 & 0 & 0 & 0 & 0 & 0 & 0 & 0 & 0 \\
0 & 0 & 0 & 0 & 0 & 0 & 0 & 0 & 0 & 0 
\end{bmatrix}\\
& F_1= \begin{bmatrix}
0 & 0 & 0 & 0 & 0 & 0 & 0 & 0 & 0 & 0 \\
0 & 0 & 0 & 0 & 0 & 0 & 0 & 0 & 0 & 0 \\
0 & 1 & 0 & 0 & 0 & 0 & 0 & 0 & 0 & 0 \\
0 & 0 & 0 & 0 & 0 & 0 & 0 & 0 & 0 & 0 \\
0 & 0 & 0 & 1 & 0 & 0 & 0 & 0 & 0 & 0 \\
0 & 0 & 0 & 0 & 0 & 0 & 0 & 0 & 0 & 0 \\
0 & 0 & 0 & 0 & 0 & 0 & 0 & 0 & 0 & 0 \\
0 & 0 & 0 & 0 & 0 & 0 & 1 & 0 & 0 & 0 \\
0 & 0 & 0 & 0 & 0 & 0 & 0 & 1 & 0 & 0 \\
0 & 0 & 0 & 0 & 0 & 0 & 0 & 0 & 0 & 0 
\end{bmatrix}\\
& H_1= \begin{bmatrix}
0 & 0 & 0 & 0 & 0 & 0 & 0 & 0 & 0 & 0 \\
0 & 1 & 0 & 0 & 0 & 0 & 0 & 0 & 0 & 0 \\
0 & 0 & -1 & 0 & 0 & 0 & 0 & 0 & 0 & 0 \\
0 & 0 & 0 & 1 & 0 & 0 & 0 & 0 & 0 & 0 \\
0 & 0 & 0 & 0 & -1 & 0 & 0 & 0 & 0 & 0 \\
0 & 0 & 0 & 0 & 0 & 0 & 0 & 0 & 0 & 0 \\
0 & 0 & 0 & 0 & 0 & 0 & 2 & 0 & 0 & 0 \\
0 & 0 & 0 & 0 & 0 & 0 & 0 & 0 & 0 & 0 \\
0 & 0 & 0 & 0 & 0 & 0 & 0 & 0 & -2 & 0 \\
0 & 0 & 0 & 0 & 0 & 0 & 0 & 0 & 0 & 0 
\end{bmatrix}\\
&E_2=\begin{bmatrix}
0 & 0 & 0 & 0 & 0 & 0 & 0 & 0 & 0 & 0 \\
0 & 0 & 0 & 0 & 0 & 0 & 0 & 0 & 0 & 0 \\
0 & 0 & 0 & 1 & 0 & 0 & 0 & 0 & 0 & 0 \\
0 & 0 & 0 & 0 & 0 & 0 & 0 & 0 & 0 & 0 \\
0 & 0 & 0 & 0 & 0 & 0 & 0 & 0 & 0 & 0 \\
0 & 0 & 0 & 0 & 0 & 0 & 1 & 0 & 0 & 0 \\
0 & 0 & 0 & 0 & 0 & 0 & 0 & 0 & 0 & 0 \\
0 & 0 & 0 & 0 & 0 & 0 & 0 & 0 & 0 & 0 \\
0 & 0 & 0 & 0 & 0 & 0 & 0 & 0 & 0 & 1 \\
0 & 0 & 0 & 0 & 0 & 0 & 0 & 0 & 0 & 0 
\end{bmatrix}\\
& F_2= \begin{bmatrix}
0 & 0 & 0 & 0 & 0 & 0 & 0 & 0 & 0 & 0 \\
0 & 0 & 0 & 0 & 0 & 0 & 0 & 0 & 0 & 0 \\
0 & 0 & 0 & 0 & 0 & 0 & 0 & 0 & 0 & 0 \\
0 & 0 & 1 & 0 & 0 & 0 & 0 & 0 & 0 & 0 \\
0 & 0 & 0 & 0 & 0 & 0 & 0 & 0 & 0 & 0 \\
0 & 0 & 0 & 0 & 0 & 0 & 0 & 0 & 0 & 0 \\
0 & 0 & 0 & 0 & 0 & 1 & 0 & 0 & 0 & 0 \\
0 & 0 & 0 & 0 & 0 & 0 & 0 & 0 & 0 & 0 \\
0 & 0 & 0 & 0 & 0 & 0 & 0 & 0 & 0 & 0 \\
0 & 0 & 0 & 0 & 0 & 0 & 0 & 0 & 1 & 0 
\end{bmatrix}\\
&H_2= \begin{bmatrix}
0 & 0 & 0 & 0 & 0 & 0 & 0 & 0 & 0 & 0 \\
0 & 0 & 0 & 0 & 0 & 0 & 0 & 0 & 0 & 0 \\
0 & 0 & 1 & 0 & 0 & 0 & 0 & 0 & 0 & 0 \\
0 & 0 & 0 & -1 & 0 & 0 & 0 & 0 & 0 & 0 \\
0 & 0 & 0 & 0 & 0 & 0 & 0 & 0 & 0 & 0 \\
0 & 0 & 0 & 0 & 0 & 1 & 0 & 0 & 0 & 0 \\
0 & 0 & 0 & 0 & 0 & 0 & -1 & 0 & 0 & 0 \\
0 & 0 & 0 & 0 & 0 & 0 & 0 & 0 & 0 & 0 \\
0 & 0 & 0 & 0 & 0 & 0 & 0 & 0 & 1 & 0 \\
0 & 0 & 0 & 0 & 0 & 0 & 0 & 0 & 0 & -1 
\end{bmatrix}
\end{alignat*}
\subsubsection{Second $U_{t}(A_4^{(2)})$ evaluation representation}
\label{sec:evalrepB22}
Let $u\in \mathbb{C}^*$. Consider the following representation $V'_u$ of $U_t(A_4^{(2)})$ given in the basis $\{ 1,e_1,e_2,e_3,e_4 \}$ where $1$ denotes the basis vector of the trivial representation of $U_{q}(B_2)$, 

\begin{alignat*}{3}
    &E_0= u\begin{bmatrix}
0 & -\sqrt{[2]_t} & 0 & 0 & 0 \\
0 & 0 & 0 & 0 & 0 \\
0 & 0 & 0 & 0 & 0 \\
0 & 0 & 0 & 0 & 0 \\
\sqrt{[2]_t} & 0 & 0 & 0 & 0 
\end{bmatrix}  \qquad &&F_0 = \frac{1}{u}\begin{bmatrix}
0 & 0 & 0 & 0 & \sqrt{[2]_t} \\
-\sqrt{[2]_t} & 0 & 0 & 0 & 0 \\
0 & 0 & 0 & 0 & 0 \\
0 & 0 & 0 & 0 & 0 \\
0 & 0 & 0 & 0 & 0 
\end{bmatrix} \qquad &&
    H_0= \begin{bmatrix}
0 & 0 & 0 & 0 & 0 \\
0 & -2 & 0 & 0 & 0 \\
0 & 0 & 0 & 0 & 0 \\
0 & 0 & 0 & 0 & 0 \\
0 & 0 & 0 & 0 & 2 
\end{bmatrix}  \\ &E_1 = \begin{bmatrix}
0 & 0 & 0 & 0 & 0 \\
0 & 0 & 1 & 0 & 0 \\
0 & 0 & 0 & 0 & 0 \\
0 & 0 & 0 & 0 & 1 \\
0 & 0 & 0 & 0 & 0 
\end{bmatrix}\qquad &&
    F_1= \begin{bmatrix}
0 & 0 & 0 & 0 & 0 \\
0 & 0 & 0 & 0 & 0 \\
0 & 1 & 0 & 0 & 0 \\
0 & 0 & 0 & 0 & 0 \\
0 & 0 & 0 & 1 & 0 
\end{bmatrix} \qquad &&H_1 = \begin{bmatrix}
0 & 0 & 0 & 0 & 0 \\
0 & 1 & 0 & 0 & 0 \\
0 & 0 & -1 & 0 & 0 \\
0 & 0 & 0 & 1 & 0 \\
0 & 0 & 0 & 0 & -1 
\end{bmatrix} \\ &E_2 = \begin{bmatrix}
0 & 0 & 0 & 0 & 0 \\
0 & 0 & 0 & 0 & 0 \\
0 & 0 & 0 & 1 & 0 \\
0 & 0 & 0 & 0 & 0 \\
0 & 0 & 0 & 0 & 0 
\end{bmatrix}\qquad &&
    F_2= \begin{bmatrix}
0 & 0 & 0 & 0 & 0 \\
0 & 0 & 0 & 0 & 0 \\
0 & 0 & 0 & 0 & 0 \\
0 & 0 & 1 & 0 & 0 \\
0 & 0 & 0 & 0 & 0 
\end{bmatrix} \qquad &&H_2 = \begin{bmatrix}
0 & 0 & 0 & 0 & 0 \\
0 & 0 & 0 & 0 & 0 \\
0 & 0 & 1 & 0 & 0 \\
0 & 0 & 0 & -1 & 0 \\
0 & 0 & 0 & 0 & 0 
\end{bmatrix} 
\end{alignat*}

\bibliographystyle{utphys}
\bibliography{refs} 

\end{document}